\documentclass[11pt]{article}
\pdfoutput=1
\usepackage{jheppub}
\usepackage{tabu}
\usepackage[vcentermath]{youngtab}
\usepackage[usenames,dvipsnames,table]{xcolor}
\usepackage{graphicx,subfig}
\usepackage{amsmath,amssymb,amsthm,amsfonts,multirow,array,bm,bbm,esint}
\usepackage[mathscr]{eucal}
\usepackage[bbgreekl]{mathbbol}
\usepackage{epsf,grffile}
\usepackage{slashed}
\usepackage[numbers,sort&compress]{natbib}
\usepackage{array,tikz-cd}
\usepackage{braket}
\usepackage{multirow}
\usepackage{enumerate}
\usetikzlibrary{fadings,patterns}
\usetikzlibrary{decorations.pathmorphing}
\usepackage{pgfplots}
\pgfplotsset{compat=1.18}

\usepackage{subfig}
\usepackage{float}
\usetikzlibrary{fadings}

\definecolor{labelkey}{rgb}{0.4,0.4,0.4}

\DeclareMathAlphabet{\mathpzc}{OT1}{pzc}{m}{it}

\newcommand{\Tr}{{\rm {Tr}}}
\newcommand{\normord}[1]{:\mathrel{#1}:}
\newcommand{\sech}{\,\mathrm{sech}\,}
\newcommand{\csch}{\,\mathrm{csch}\,}
 
 \newcommand{\bea}{\begin{eqnarray}}
\newcommand{\eea}{\end{eqnarray}}
\newcommand{\be}{\begin{equation}}
\newcommand{\ee}{\end{equation}}
\newcommand{\ba}{\begin{align}}
\newcommand{\ea}{\end{align}}

\newcommand{\K}{\mathcal{K}}

\title{Generalized entropy of gravitational fluctuations}

\author[a]{Sean Colin-Ellerin,}
\author[a]{Guanda Lin,}
\author[a]{Geoff Penington}

\affiliation[a]{Center for Theoretical Physics and Department of Physics,  \\
University of California, Berkeley, California 94720, USA.}

\emailAdd{scolinellerin@berkeley.edu, geoff\textunderscore guanda\textunderscore lin@berkeley.edu, geoffp@berkeley.edu}

\abstract{The corrections to holographic entanglement entropy from bulk quantum fields in a classical gravitational background are now well understood. They lead, in particular, to unitary Page curves for evaporating black holes. However, the correct treatment of quantum fluctuations of the metric, including graviton excitations, is a longstanding problem. We provide a gauge-invariant prescription for the generalized entropy of gravitons in anti-de Sitter space in terms of areas and bulk entanglement entropy, generalizing the quantum extremal surface prescription to accommodate fluctuations in the semiclassical spacetime geometry. This task requires a careful treatment of the area operator on the graviton Hilbert space and the definition of a ``quantum extremal gauge'' in which the extremal surface is unperturbed. It also requires us to determine the correct vacuum modular Hamiltonian for the graviton field, which we fix by requiring that it doesn't contain a boundary term in extremal gauge. We check our prescription with an explicit computation of the vacuum-subtracted generalized entropy of states containing a graviton in an AdS-Rindler background. Our results exactly match vacuum-subtracted von Neumann entropies for  stress-tensor excited states in holographic conformal field theory with $d>2$ dimensions. We also use covariant phase space techniques to give a partial proof of our prescription when the entanglement wedge for the background spacetime has a bifurcate Killing horizon. Along the way, we identify a class of perturbative graviton states that have parametrically larger generalized entropy, in the small $G_{N}$ expansion, than any low-energy excitations of an ordinary quantum field.
}

\begin{document}

\maketitle


\section{Introduction}
\label{sec:intro}

A fundamental problem in quantum gravity is how to define the quantum entropy associated to a subregion of spacetime. Without knowledge of the explicit Hilbert space of quantum gravity, one is restricted to semiclassical arguments to try to find a formula for entropy. For black hole spacetimes, the first and second laws of thermodynamics naturally lead to the generalized entropy, defined as the area of a cut of the event horizon plus the matter entropy outside, as the appropriate candidate for the semiclassical entropy of gravity coupled to quantum fields \cite{PhysRevD.7.2333}. Indeed, a generalized second law of thermodynamics can be proven for the generalized entropy of causal horizons for very general matter configurations \cite{Wall:2011hj}. 

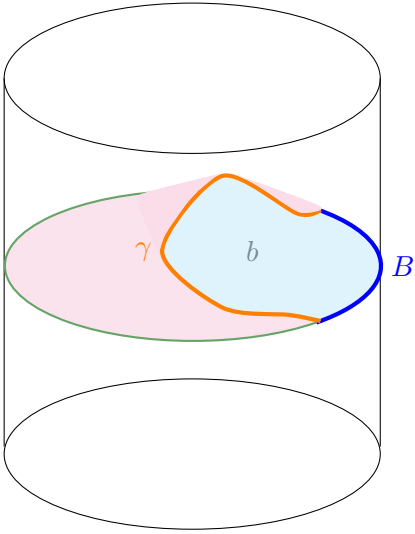
\begin{figure}[h]
\begin{center}
\begin{tikzpicture}

\draw (0,2.5) ellipse (2.5cm and 1cm);
\draw (-2.5,-2.4) -- (-2.5,2.5);
\draw (2.5,-2.4) -- (2.5,2.5);
\draw (0,-2.5) ellipse (2.5cm and 1cm);

\draw[draw=blue,ultra thick,fill=cyan!15!white!75] (1.65,-0.75) arc(311:409:2.5cm and 1cm);
\node at (2.8,0) {$\color{blue}B$};
\draw[draw=green!40!black!60,thick,fill=magenta!15!white!85] (1.65,0.76) arc(49:311:2.5cm and 1cm);

\draw[draw=orange,ultra thick,fill=cyan!15!white!75] plot [smooth] coordinates {(1.705,0.75) (1.4,0.7) (0.4,1.2) (-0.4,0.2) (0.4,-0.55) (1.3,-0.65) (1.705,-0.73)};
\fill[magenta!20!white!80] plot [smooth cycle] coordinates {(1.695,0.8) (1.45,0.71) (1.18,0.86) (1.05,0.951) (1,0.985) (0.8,1.11) (0.6,1.21)}; 
\fill[magenta!20!white!80] plot [smooth cycle] coordinates {(0.35,1.225) (0.318,1.2) (0.119,1.035) (-0.11,0.844) (-0.15,0.88) (-0.5,0.8) (-0.7,0.95)};
\fill[magenta!20!white!80] plot [smooth cycle] coordinates {(-0.03,0.91) (-0.085,0.82) (-0.355, 0.425) (-0.44,0.2) (-0.5,0.28) (-0.73,0.91) (-0.36,0.85)};
\node at (-0.65,0.2) {$\color{orange}\gamma$};
\node at (0.8,0.2) {$\color{cyan!30!black!70}b$};
\end{tikzpicture}
\end{center}
\caption{A spatial subregion $B$ (blue) of the asymptotic boundary of AdS with codimension-2 surface $\gamma$ (orange) anchored on $\partial B$ and homology surface $b$ (cyan) bounded by $B$ and $\gamma$. The CFT vN entropy for $B$ in a state $\psi$ is equal to the extremized generalized entropy for $b$ in the dual state $\Psi$.}
\label{fig:QES}
\end{figure}

The situation is better in asymptotically AdS spacetimes because the generalized entropy can be related to a quantum entropy via the AdS/CFT correspondence, where the dual CFT provides the requisite microscopic theory. This is made precise by the quantum extremal surface (QES) formula. Consider any spatial subregion $B$ in a holographic CFT dual to weakly-coupled Einstein gravity and consider an AdS subregion $b$ bounded by $B$ and a codimension-2 surface $\gamma$ anchored on $\partial B$, as illustrated in figure \ref{fig:QES}. The formula states that the CFT von Neumann (vN) entropy for $B$ in a state $\psi$ is equal to the generalized entropy $S_{\mathrm{gen}}$ for the subregion $b$ in the dual state $\Psi$ that gives the minimal extremum of such generalized entropy:
\begin{equation}\label{eqn:QES}
S^{\mathrm{CFT}}(\rho_{B}^{\psi}) = \underset{\gamma}{\mathrm{min\;ext}}\underbrace{\left[\frac{A[\gamma]_{\Psi}}{4G_{N}}+S\left(\rho_{b}^{\Psi}\right)\right]}_{S_{\mathrm{gen}}},
\end{equation}
where $A[\gamma]_{\Psi}$ is the area of the surface $\gamma$ in the geometry $\Psi$ dual to $\psi$, including the backreaction of the bulk quantum fields, and $S(\rho_{b}^{\Psi})$ is the von Neumann entropy of the bulk quantum fields in $b$ in the state $\Psi$. The minimal extremal surface $\gamma_{\mathrm{ext}}$ is known as the quantum extremal surface. 

This formula has been proven at leading order in small $G_{N}$ using the gravitational path integral \cite{Lewkowycz:2013nqa, Dong:2016hjy}, where only the area term contributes. The first quantum correction in $G_{N}$ can be similarly proven for fluctuating quantum fields on a classical gravitational background using path integral techniques \cite{Faulkner:2013ana, Dong:2017xht, Penington:2019kki, Almheiri:2019qdq} and has undergone many explicit checks \cite{Belin:2018juv,Agon:2015ftl,Belin:2019mlt,Agon:2020fqs,Chowdhury:2024fpd,Colin-Ellerin:2024npf}. It has been conjectured to hold order-by-order in $G_{N}$ \cite{Engelhardt:2014gca,Dong:2017xht}, but only a partial check exists beyond the first subleading order \cite{Belin:2021htw}. Remarkably, when applied to the radiation of an evaporating black hole in AdS coupled to a bath, this formula produces a unitary Page curve \cite{Penington:2019npb,Almheiri:2019psf}, which has led to major progress on the black hole information paradox \cite{PhysRevD.14.2460}.\footnote{Notably, the quantum extremal surface for an evaporating black hole is not close to a classical extremal surface, where the area term in \eqref{eqn:QES} is extremised on its own. As a result, the inclusion of quantum effects is crucial the calculation even at leading order in $G_N$. This is possible because, in an evaporating black hole, the Page time scales as $O(1/G_N)$ and so (in contrast to the situations considered in this paper) both the spacetime geometry and the state of the quantum fields cannot be held completely fixed as $G_N \to 0$.}

While the QES formula for ordinary quantum fields\footnote{In this work, ``ordinary quantum fields'' will always refer to any quantum field that is not the graviton.} is now well understood, the correct treatment of perturbative graviton fluctuations is much murkier. In particular, the area of a surface in the classical background geometry should receive corrections from quantum fluctuations in that geometry associated to the graviton modes. It therefore cannot be treated classically and must be promoted to an operator acting on the graviton Hilbert space. However, a naive definition of such an area operator will transform nontrivially at $O(\sqrt{G_N})$ and beyond under perturbative diffeomorphisms of the graviton fluctuations and is therefore not gauge-invariant. If the bulk entropy term in \eqref{eqn:QES} did not exist, one could try to define a gauge-invariant extremal-area operator acting on the graviton Hilbert space.\footnote{Because the surface needs to be extremal in both time and space, there are operator-ordering ambiguities that need to be worried about here.} But quantum extremal surfaces depend nonlinearly on the quantum state and so cannot be described by a linear operator.

The goal of this work is to resolve these issues and generalize the QES formula beyond fixed classical background by including perturbative metric fluctuations. Working order-by-order in $\kappa = \sqrt{16\pi G_{N}}$, we incorporate the generalized entropy of gravitons into the QES prescription in a way that:
\begin{enumerate}[(1)]
    \item Is manifestly gauge invariant with respect to perturbative diffeomorphisms.
    \item Takes the form of an area plus a bulk von Neumann entropy.
    \item Agrees with the von Neumann entropy of the dual CFT state.
\end{enumerate}
To check this last claim, we compute vacuum-subtracted generalised entropies for states containing a single graviton excitation (and superpositions of a single graviton state with the vacuum) and show that they match a CFT computation at $O(G_N^0)$.

To understand how our prescription works, let $G_{\mu\nu}$ be a fixed classical background metric and, for the moment let
\be\label{eq:classicalperturb}
g_{\mu\nu} = G_{\mu\nu} + \kappa h_{\mu\nu} + O(\kappa^{2})
\ee
be a small classical perturbation of that metric. If $\gamma^{(0)}$ is a stable classical extremal surface with respect to the metric $G_{\mu\nu}$, there will exist a classical extremal surface in the perturbed metric $g_{\mu\nu}$ that can be written as
\begin{align}\label{eq:perturbedclassicalextremal}
    \gamma = \gamma^{(0)} + \kappa \gamma^{(1)} + O(\kappa^{2}).
\end{align}
The perturbed extremal surface is a gauge-invariant object; however, the expansion \eqref{eq:perturbedclassicalextremal} describing its location is not. In particular, we can always find a perturbative diffeomorphism that maps $\gamma^{(1)} \to 0$ so that the unperturbed surface $\gamma^{(0)}$ remains extremal; we shall refer to this condition as (classical) extremal gauge.

What about quantum fluctuations
\be
\hat g_{\mu\nu} = G_{\mu\nu} + \kappa \hat h_{\mu\nu} + O(\kappa^{2})
\ee
in the metric $G_{\mu\nu}$? For a QES prescription to be gauge invariant, it needs to respect the equivalence relation\footnote{If the background metric $G_{\mu\nu}$ has isometries that act trivially at any asymptotic boundaries then those (nonperturbative) isometries also need to be imposed as gauge constraints. Such isometries arise in e.g. global de Sitter space, but do not generally appear in the asymptotically-AdS spacetimes we will consider.}
\be\label{eq:coinvariantequiv}
\hat P_i \Phi \cong 0
\ee
where $\Phi$ is any quantum state (that does not by itself need to be gauge invariant) and $\hat P_i$ is any generator of a perturbative diffeomorphism.\footnote{Perhaps the most principled approach to quantizing gauge theory, at least when the gauge group is noncompact, involves explicitly imposing the relation \eqref{eq:coinvariantequiv} on the space of quantum states and then defining a gauge-invariant inner product on the quotiented space. This approach goes under various names including refined algebraic quantisation, the group-averaging method, or the method of coinvariants. See \cite{Marolf:2000iq} for a detailed review or \cite{Chandrasekaran:2022cip, Held:2024rmg} for more recent discussions. However, even when \eqref{eq:coinvariantequiv} is not imposed explicitly on the space of states, it still underlies the quantisation of gauge theories using any fixed choice of gauge.} This equivalence relation is much stronger than the corresponding relation in the classical theory: not only do we have 
\be
\Psi \cong  (\mathbb{1} + \varepsilon \hat P_i) \Psi,
\ee
for any gauge transformation $(\mathbb{1} + \varepsilon P_i)$ acting on the quantum state $\Psi$ but also
\be\label{eq:statedependdiffs}
\Psi \cong \Psi + \varepsilon \hat P_i \Phi
\ee
for any other quantum state $\Phi$. Transformations of the form \eqref{eq:statedependdiffs} are sometimes known as ``state-dependent gauge transformations''; they allow us to act with perturbative diffeomorphisms on only part of the quantum state that depends on, e.g., the state of matter fields or the graviton fluctuations $\hat h_{\mu\nu}$.

This larger space of gauge transformations present in a quantum theory means that, extremizing over classical perturbations of the surface $\gamma^{(0)}$ will not be sufficient to produce a gauge-invariant prescription. Instead, we need to allow different perturbations for different parts of the quantum state. The cleanest and most general way to do this seems to involve extremizing not over surfaces but over the space of gauge-equivalent states $\Psi$.\footnote{We expect that it should be possible to instead extremize over the space of ``operator-valued surfaces'' in some appriopriate sense, but it is less clear how to define such objects beyond linear order.} We say that a state $\Psi$ is in \emph{quantum extremal gauge} if
\begin{align}\label{eq:quantumextremalgauge}
    \left[\hat P_i, \frac{\hat A[\gamma^{(0)}]}{4G_N} - \log \rho_b^\Psi\right] \Psi = 0
\end{align}
where the operator $\hat A[\gamma^{(0)}]$ describes the area of the unperturbed surface $\gamma^{(0)}$, $\rho_b^\Psi$ is the density matrix of the bulk quantum fields and gravitons on the region $b$ bounded by $\gamma^{(0)}$, and $\hat P_i$ is the generator of an arbitrary diffeomorphism. We then define the QES prescription, as a perturbative expansion in $G_N$, to be 
\be\label{eq:genentfluct}
S^{\mathrm{CFT}}(\rho_B^{\psi}) = \underset{\gamma^{(0)}, \tilde\Psi \cong \Psi}{\mathrm{min\;ext}}\left[\frac{\braket{\hat A[\gamma^{(0)}]}_{\tilde\Psi}}{4G_{N}}+S\left(\rho_{b}^{\tilde\Psi}\right)\right],
\ee
where $\gamma^{(0)}$ is the minimal area classical extremal surface in the background metric $G_{\mu\nu}$ and $\tilde\Psi$ is a state in quantum-extremal gauge that is gauge-equivalent to $\Psi$.

There are two primary distinctions between \eqref{eq:genentfluct} and the naive QES prescription \eqref{eqn:QES}. The first is that the quantum extremal gauge condition \eqref{eq:quantumextremalgauge} has to be satisfied as an equality between quantum states and not just as a statement about expectation values. To be able to satisfy this stricter condition, there is a second distinction from \eqref{eqn:QES}: we are allowed to optimize not just over classical perturbations of $\gamma^{(0)}$, but over all gauge-equivalent quantum states $\tilde\Psi$. As a result, \eqref{eq:genentfluct} is manifestly gauge invariant with respect to perturbative diffeomorphisms. 

In principle, we are hopeful that \eqref{eq:quantumextremalgauge} and \eqref{eq:genentfluct} can be defined to all orders in $G_N$. However, there are a number of subtle issues that would need to be resolved to demonstrate this and that we do not attempt to tackle. Instead, we content ourselves with showing that \eqref{eq:quantumextremalgauge} and \eqref{eq:genentfluct} at least make sense when considering $O(1)$ and larger corrections to \eqref{eq:genentfluct}, and that they are consistent with CFT calculations at that level of precision. 

If we are only interested in $O(1)$ contributions to generalized entropy, it is sufficient to determine the location of the quantum extremal surface at $O(G_N^{-1/2})$. At this order, the $\log \rho_b^\Psi$ term in \eqref{eq:quantumextremalgauge} can be ignored and quantum extremal gauge reduces to the requirement that $A[\gamma^{(0)}]$ should remain extremal at linear order in $h_{\mu\nu}$. In practice, it turns out to be easier to do an equivalent computation where we work in a gauge where $\gamma^{(0)}$ is not extremal but replace $A[\gamma^{(0)}]$ by the area $A[\gamma^{(0)} + \kappa \hat \gamma^{(1)}]$ of the perturbed extremal surface. Here the operator-valued perturbation $\hat \gamma^{(1)}$ is a linear function of $\hat h_{\mu\nu}$. We can then expand the area as
\begin{equation}\label{eq:Aopschematic}
        A[\hat{g},\hat\gamma] = A[G,\gamma^{(0)}] + \kappa \frac{\delta A}{\delta g}\hat{h} + \kappa^{2}\Big(\frac{\delta^2 A}{\delta g^2}(\hat{h})^2 + \frac{\delta^2 A}{\delta g \delta \gamma} \hat{h} \hat{\gamma}^{(1)}  + \frac{\delta^2 A}{{\delta \gamma}^2} (\hat \gamma^{(1)})^2 + \frac{\delta A}{\delta g}\hat{g}^{(2)} \Big) + \cdots \,,
    \end{equation}
    where the perturbed metric is $\hat g_{\mu\nu} = G_{\mu\nu} + \kappa  \hat h_{\mu\nu} + \kappa^2 \hat{g}^{(2)}_{\mu\nu} +\dots $ and we have dropped indices for notational convenience. Up to terms that vanish as $G_N \to 0$, \eqref{eq:genentfluct} then becomes
    \begin{align}\label{eq:QEStoO1}
       S^{\mathrm{CFT}}(\rho_B^{\psi}) = \frac{\braket{A[\hat{g},\hat\gamma]}_{\Psi}}{4G_{N}}+S\left(\rho_{b}^{\Psi}\right) + \dots.
    \end{align}

    To match \eqref{eq:QEStoO1} to CFT calculations, we consider perturbative graviton excitations around vacuum-AdS and take the boundary subregion $B$ to be a polar cap subregion of the spatial sphere with angular size $\theta_{0}$. On the CFT side, a bulk state containing a single graviton is dual to the state created by an insertion of the stress-energy tensor $T_{\mu\nu}$. Both sides of \eqref{eq:QEStoO1} can then be computed in a controlled manner in the limit of small $\theta_0$ and we show that they match precisely. The only condition that we need to impose on the CFT is the existence of a twist gap $1 \ll \Delta_{\mathrm{gap}} \ll C_{T}$ \cite{Heemskerk:2009pn}, where $C_{T}$ is the central charge and $\Delta_{\mathrm{gap}}$ is the twist gap.\footnote{More precisely, $C_{T}$ is the coefficient of the stress-tensor two point function and $\Delta_{\mathrm{gap}}$ is the conformal dimension of the lightest single-trace primary with spin $J>2$. In all known examples of holographic CFTs, $C_{T} \sim N^{a}$ for some $a>0$ so large $C_{T}$ corresponds to large $N$.} Such a gap should exist in any holographic CFT dual to Einstein gravity.

    The vacuum-subtracted entropy $\Delta S^{\mathrm{CFT}}(\rho_B^\psi)$ can be written as a linear combination
    \begin{align}
\Delta S^{\rm CFT} = \braket{\Delta K_{\rm CFT}}_\psi - S^{\rm rel}_{\rm CFT}(\rho_B^{\psi}|\omega_B)
    \end{align}
    of the expectation value of the vacuum modular Hamiltonian $\Delta K_{\rm CFT}$ (defined so that $\braket{\Delta K_{\rm CFT}}_\omega = 0$) and the relative entropy $S^{\rm rel}_{\rm CFT}(\rho_B^{\psi}|\omega_B)$ of $\rho_B^{\psi}$ relative to the vacuum reduced density matrix $\omega_B$. It follows from \eqref{eq:QEStoO1} that
    \begin{align}\label{eqn:JLMSintro}
    \Delta K_{\rm CFT}^{\psi} & = \frac{\Delta \langle \hat{A} \rangle_{\Psi}}{4G_N} - \Delta \langle \log \omega_b \rangle_{\Psi} \quad \& \quad   S^{\rm rel}_{\rm CFT}(\rho_B^{\psi}|\omega_{B})  = S^{\rm rel}_{\rm bulk}(\rho_b^{\Psi}|\omega_b)\,,
\end{align} 
where $\omega_b$ is the reduced density matrix of the bulk vacuum and $\Delta \langle \hat{A} \rangle_{\Psi} = \langle \hat{A} \rangle_{\Psi} - \langle \hat{A} \rangle_{\omega}$.

At $O(\kappa^{-2})$, \eqref{eq:QEStoO1} is determined entirely by the background geometry and so $\Delta S^{CFT}$ vanishes for all small perturbations of the vacuum. At $O(\kappa^{-1})$, $S^{\rm rel}_{\rm bulk}$ is zero, while
\be
\Delta K_{\rm CFT}^{\psi} = \frac{\Delta \langle \hat{A} \rangle_{\Psi}}{4G_N} = \frac{\kappa}{4 G_N} \frac{\delta A}{\delta g_{\mu\nu}} \braket{\hat h_{\mu\nu}}_\Psi.
\ee
This vanishes if $\psi$ is a single graviton state but is nonzero for superpositions of the vacuum and a single graviton state. This is in sharp contrast to ordinary matter excitations, which can only lead to $O(1)$ changes in generalized entropy, but we show that it exactly matches the CFT entropy for superpositions of the vacuum and states with a single stress-energy insertion.

Finally, at $O(1)$, there are nonzero contributions from the last four terms of \eqref{eq:Aopschematic}, along with $\Delta \langle \log \omega_b \rangle$ and $S^{\rm rel}_{\rm bulk}$. The last two contributions can be found by choosing a basis of Rindler graviton modes and calculating their Bogoliubov coefficients. 
On the CFT side, we can compute both $\Delta K_{\rm CFT}^{\psi}$ and $S^{\rm rel}_{\rm CFT}(\rho_B^{\psi}|\omega_B)$ at leading order in the state perturbation $\delta\rho \equiv (\rho^{\Psi}_b-\omega_b)\sim \theta_0^{2d}$.
 In both cases, we find a perfect match of the formula \eqref{eqn:JLMSintro} at orders up to and including $O(\theta_0^{4d})$.\footnote{In principle, we see no reason why it wouldn't be possible to also check higher orders, but the calculations become somewhat involved beyond this point and we did not attempt to carry them out.}

The layout of the paper is as follows. We begin in \S\ref{sec:prelims} with a brief review of generalized entropy for ordinary matter fields and some preliminary details about gravitons. We then motivate and explain our prescription  in \S\ref{sec:prescriptgrav}. A partial justification for our prescription using covariant phase space techniques is given in \S\ref{sec:HW}. Finally, \S\ref{sec:ex} presents the explicit example where we compute the vacuum-subtracted vN entropy in any holographic CFT in $d>2$ dimensions for stress-tensor excited states and match the result to the vacuum-subtracted generalized entropy of graviton excited states in AdS using our prescription. We end in \S\ref{sec:discussion} with some discussion and open questions. The Appendices contain various technical details.

\section{Preliminaries}
\label{sec:prelims}

Before proceeding to our prescription for generalized entropy of gravitons, we introduce the necessary background material. We begin with a review of how the procedure works for ordinary matter fields. The details of how to treat the quantum theory of gravitons will be then presented, along with a precise definition of what is meant by perturbative diffeomorphisms.

\subsection{Generalized entropy for ordinary matter fields: review}
\label{sec:genS}

Let us review how to obtain the generalized entropy for ordinary quantum fields coupled to classical gravity in asymptotically AdS spacetimes, including $O(G_{N}^{0})$ effects, which was first derived via the path integral in \cite{Faulkner:2013ana}. Explicit examples can be found in AdS$_{3}$ \cite{Belin:2018juv,Belin:2021htw} and in higher-dimensional AdS \cite{Colin-Ellerin:2024npf} for various matter theories.

Consider a minimally-coupled\footnote{For non-minimally coupled field theories, the area term in the generalized entropy must be replaced with the Wald entropy \cite{Wald:1993nt}.} quantum field theory defined in an asymptotically AdS spacetime $(\mathcal{M},G_{\mu\nu})$ and a state $\ket{\psi}$. Take the spacetime to be determined by classical Einstein gravity and let the quantum fields backreact on the spacetime. The metric can be expanded in $G_{N}$ as
\begin{equation}\label{eqn:fullmetric}
    g_{\mu\nu} = G_{\mu\nu}+\kappa^{2} g_{\mu\nu}^{(2)} + O(\kappa^{3})
\end{equation}
where $G_{\mu\nu}$ solves the vacuum Einstein equation (possibly sourced by some classical fields) and $g_{\mu\nu}^{(2)}$ is the leading backreaction to the metric,\footnote{Since both the matter fields and the spacetime geometry on which they backreact are quantum mechanical, $g_{\mu\nu}^{(2)}$ is really the expectation value of the backreaction on the metric. However, for the moment it can be consistently treated as a classical object.} which is obtained by solving the semiclassical Einstein equations
\begin{equation}\label{eqn:backreact_matter}
E_{\mu\nu}^{(1)}[\kappa^2 g_{\mu\nu}^{(2)}] = 8\pi G_{N}\langle T_{\mu\nu} \rangle_{\Psi}
\end{equation}
where $E_{\mu\nu}^{(1)}$ is the linearized Einstein tensor with negative cosmological constant and $T_{\mu\nu}$ is the stress-energy tensor of the quantum field theory. The area of a codimension-$2$ surface $\gamma$ in the geometry corresponding to the state $\Psi$ means the area of the same surface in the full metric is 
\begin{align}\label{eq:areaexpansion}
    A[\gamma]_{\Psi} = A[G,\gamma] + \kappa^2  \frac{\partial A[G,\gamma]}{\partial G_{\mu\nu}} g_{\mu\nu}^{(2)} + O(G_N^2).
\end{align}

The position of the surface $\gamma$ can also be expanded order-by-order in $G_N$ as
\begin{align}
    \gamma = \gamma^{(0)} + \sum_n \kappa^{2n} \gamma^{(n)}.
\end{align}
(Even if $\gamma^{(n)} = 0$ for all $n \geq 1$ in some particular gauge, a nontrivial expansion will generically be introduced by perturbative diffeomorphisms.) Generically, the term $\gamma^{(1)}$ would affect the area of the surface $\gamma$ at $O(G_N)$ and so would also need to be included as a correction in \eqref{eq:areaexpansion}. However, if $\gamma^{(0)}$ is classically extremal then this leading correction vanishes and $\gamma^{(1)}$ only affects $A[\gamma]$ at $O(G_N^2)$.

Next, compute the vN entropy $S(\rho_{b}^{\Psi})$ of quantum fields in the state $\Psi$ for the subregion given by the homology surface $b$ bounded by the CFT subregion $B$ and a codimension-2 surface $\gamma$ anchored on $\partial B$, which is homologous to $B$. Then the CFT vN entropy in the dual state $\psi$ for the subregion $B$ is equal to the extremization, order-by-order in $G_{N}$, of the resulting generalized entropy
\begin{equation}\label{eqn:QES2}
S(\rho_{B}^{\psi}) = \underset{\gamma}{\mathrm{min\;ext}}\Bigg[\frac{A[\gamma]_{\Psi}}{4G_{N}}+S\left(\rho_{b}^{\Psi}\right)\Bigg].
\end{equation}
Up to the first subleading order $O(G_N^0)$, we have 
\begin{equation}\label{eqn:QESatO1}
S(\rho_{B}^{\psi})|_{O(N^{0})} = \Bigg[\frac{A[G,\gamma^{(0)}_{\rm min, ext}]}{4G_{N}} +4 \pi \frac{\partial A[G,\gamma^{(0)}_{\rm min, ext}]}{\partial G_{\mu\nu}} g_{\mu\nu}^{(2)}+S\left(\rho_{b}^{\Psi}\right)\Bigg]\Bigg|_{O(G_{N}^{0})},
\end{equation}
where $\gamma^{(0)}_{\rm min, ext}$ is the smallest area classical extremal surface in the unbackreacted geometry. In particular, the extremization over the surface $\gamma$ in \eqref{eqn:QES2} depends on the state $\Psi$ of the quantum fields at $O(G_N)$, but, because the original surface $\gamma^{(0)}_{\rm min, ext}$ was classically extremal, this dependence again only affects the entropy $S(\rho_{B}^{\psi})$ at $O(G_N)$.

One can generalize \eqref{eqn:QES2} to the expectation value of the modular Hamiltonian in other states. Consider two bulk states $\Psi$ and $\Phi$ living in the same code subspace, i.e., perturbative excitations in the same background geometry $G_{\mu\nu}$, with dual CFT states $\psi$ and $\phi$, respectively. 
It can be argued that
\begin{equation}\label{eqn:JLMS}
\Tr\left(\rho_{B}^{\psi}K_{B}^{\phi}\right)|_{O(N^{0})} = \underset{\gamma}{\mathrm{min\;ext}}\Bigg[\frac{A[G+G_{N}\langle g_{\mu\nu}^{(2)} \rangle_{\Psi},\gamma]}{4G_{N}}+\Tr\left(\rho_{B}^{\Psi}K_{B}^{\Phi}\right)\Bigg]\Bigg|_{O(G_{N}^{0})}
\end{equation}
known as the JLMS formula \cite{Jafferis:2015del,Dong:2017xht}. When $\Phi=\Psi$ we recover \eqref{eqn:QES2}.

\subsection{Gravitons in AdS spacetimes}
\label{sec:gravitons}

The quantum field theory of gravitons is most easily described in terms of the background field expansion, commonly used in Yang-Mills theory. Consider a manifold $\mathcal{M}$ and some metric $g_{\mu\nu}$ that we write as a background classical field $G_{\mu\nu}$ plus a quantum fluctuation $h_{\mu\nu}$. The coupling of $h_{\mu\nu}$ to the background $G_{\mu\nu}$ can then be understood as the propagation of a massless spin-$2$ field on the spacetime described by $(\mathcal{M},G_{\mu\nu})$. For simplicity, we will ignore all ordinary matter fields for the remainder of this work, although it is straightforward to include them.

We restrict to asymptotically AdS spacetimes in $d+1$ dimensions described by Einstein gravity with the action
\begin{equation}\label{eqn:gravaction}
    S = \frac{1}{\kappa^{2}}\int_{\mathcal{M}} d^{d+1}x\,\sqrt{g} \left(R^{(g)}-2\Lambda\right)+\frac{2}{\kappa^{2}}\int_{\mathcal{B}} d^{d}x\,\sqrt{\sigma^{(g)}} K^{(g)} + S_{\mathrm{ct}},
\end{equation}
where the first term is the Einstein-Hilbert action with cosmological constant $\Lambda = -d(d-1)/2$, the second term is the Gibbons-Hawking-York term for $\mathcal{B}=\partial \mathcal{M}$ with induced metric $\sigma_{\mu\nu}^{(g)}$ and trace of extrinsic curvature $K^{(g)}$, and the final term consists of local boundary counterterms that make the action finite. 

To obtain the graviton action, we expand the full action \eqref{eqn:gravaction} perturbatively around the background metric. The coupling constant for the graviton for which its action is canonically normalised is given by
\begin{equation}\label{eqn:backgdexp_metric}
    g_{\mu\nu} = G_{\mu\nu}+\kappa h_{\mu\nu}, \qquad \kappa \equiv \sqrt{16\pi G_{N}}.
\end{equation}
At $O(\kappa^{0})$, we obtain the action for the background metric whose Euler-Lagrange equations are the background Einstein equations
\begin{equation}\label{eqn:Einsteintensor}
    E_{\mu\nu}^{(G)} \equiv R_{\mu\nu}^{(G)} - \frac{1}{2}G_{\mu\nu}R^{(G)}+\Lambda G_{\mu\nu} = 0,
\end{equation}
and we will always assume $G_{\mu\nu}$ is a solution to this equation. By the variational principle, the action then vanishes at $O(\kappa)$. The action of the free graviton theory comes from expanding to $O(\kappa^{2})$:
\begin{align}\label{eqn:EHactionexp_2ndorder}
\begin{split}
    S[h]^{(2)} &= \frac{1}{2}\int d^{d+1}x \, \sqrt{-G}\bigg(\frac{1}{2}h^{\nu\alpha}\nabla^{2}h_{\nu\alpha}-\frac{1}{2}h\nabla^{2}h-\nabla_{\nu}h^{\nu\alpha}\nabla_{\alpha}h+\nabla_{\nu}h^{\nu\alpha}\nabla^{\rho}h_{\rho\alpha}
\\  &\qquad -\frac{\Lambda}{(d-1)} h^{2}+h^{\nu\rho}{h^{\beta}}_{\omega}{R_{\nu\beta\rho}}^{\omega(G)}\bigg) + \mathrm{boundary\;terms},
\end{split}
\end{align}
where we use the background metric $G_{\mu\nu}$ to raise and lower indices and the covariant derivative is the background one $\nabla = \nabla^{(G)}$. The expansion of curvatures at quadratic order used in deriving this can be found in App.~\ref{sec:pertcurv}. This is the action for a free massless spin-2 field in the background $G_{\mu\nu}$. 

The higher-order terms in $\kappa$ in the expansion of the full action \eqref{eqn:gravaction} will give rise to an infinite set of interactions for the graviton, viz.,
\begin{equation}
S[h]|_{E_{\mu\nu}^{(G)}=0} = \frac{1}{\kappa^{2}}S[h]^{(0)}+S[h]^{(2)}+\sum_{n=3}^{\infty}\kappa^{n-2}S[h]^{(n)}.
\end{equation}
All of the interactions are unimportant for studying the generalized entropy at $O(\kappa^{0})$, except for the cubic interaction $S^{(3)}[h]$. However, this contribution can be dealt with by backreacting the graviton on the spacetime $g_{\mu\nu} \to g_{\mu\nu} + \kappa^{2}g_{\mu\nu}^{(2)}$. This backreaction is obtained by solving the linearized Einstein equations for $g^{(2)}$ sourced by the quadratic Einstein tensor for $h$. To obtain this, one needs to vary the cubic interaction for $h$ and we will show in \S\ref{sec:obs} that $g^{(2)}$ will capture the contribution from such interactions so all computations involving $h$ can be done in the free theory. 

To quantize this theory, we will always use canonical quantization as we take a strictly Hilbert space approach, ignoring the path integral. This can be achieved by imposing the diffeomorphism constraints and then gauge-fixing to construct the reduced phase space for the graviton. Hamilton's equations on this reduced phase space are the linearized Einstein equations whose solutions give the wavefunctions $h_{q}$, where $q$ labels the different solutions. The quantized graviton field then takes the form 
\begin{equation}\label{eqn:quantizedgraviton}
\hat{h} = \sum_{q}\left(h_{q}a_{q}+h_{q}^{\ast}a_{q}^{\dagger}\right)
\end{equation}
with canonical commutation relations $[a_{q},a_{q'}^{\dagger}]=\delta_{q,q'}$. The graviton Hilbert space is the Fock space $\mathbb{H}_{\mathrm{graviton}}$ constructed from the creation operators $a_{q}^{\dagger}$ acting on the `vacuum' state.\footnote{Any perturbative interactions are dealt with via backreaction to obtain corrections to the background metric, such as $g^{(2)}$, as discussed above. However, for any state comprising of a gas of gravitons with non-perturbatively large mass $M \sim 1/\kappa^{2}$ inside an AdS radius, the gas will reach its Chandresaker limit and collapse into a black hole. The Fock space picture then breaks down as a new background is required. See \cite{El-Showk:2011yvt} for a nice discussion and \cite{Arsiwalla:2010bt} for an explicit example using fermions.} We will explain this whole procedure in great detail for the graviton in global AdS in \S\ref{sec:gravitonglobal}.

Typically, canonical quantization of quantum fields in curved spacetimes is highly non-unique due to infinitely many unitarily inequivalent representations of the commutation relations \cite{Hollands:2014eia}.\footnote{As opposed to quantum mechanics where the Stone-von Neumann theorem guarantees unitary equivalence.}$^{,}$\footnote{This is an advantage of the algebraic formulation of quantum field theory where the Hilbert space is not the fundamental object.} 
However, in asymptotically AdS spacetimes, one can use the fact that both short wavelength modes and the modes localized near the boundary can be unambiguously split into positive and negative frequency so one expects that there will always be a unique Hilbert space although this has not been rigorously established, see \cite{Witten:2021jzq} for a nice discussion.

\subsection{Diffeomorphisms}
\label{sec:diffs}

When using the background field formalism for gauge theories, one needs to distinguish between the background gauge symmetries that affect the classical background field and the perturbative (or quantum) gauge symmetries that leave the background fixed, but change the quantum part of the field. The background diffeomorphisms are generated by a smooth vector field $\epsilon_{\mu}^{(0)} \sim O(\kappa^{0})$:
\begin{equation}\label{eqn:backgddiffs}
    g_{\mu\nu} \to g_{\mu\nu} + \pounds_{\epsilon^{(0)}}g_{\mu\nu} \implies \
    \begin{array}{l} G_{\mu\nu} \to G_{\mu\nu} + \nabla_{(\mu}\epsilon_{\nu)}^{(0)}
    \\[5pt]  h_{\mu\nu} \to h_{\mu\nu} - \Gamma_{\mu\nu}^{\rho(1)}\epsilon_{\rho}^{(0)}.
    \end{array}
\end{equation}
We will always fix these diffeomorphisms by picking some particular coordinates for our background spacetime.

The perturbative diffeomorphisms are instead generated by smooth vector fields $\kappa\epsilon_{\mu}^{(1)}$ so they do not affect the background metric:
\begin{equation}\label{eqn:pertdiffs}
    g_{\mu\nu} \to g_{\mu\nu} + \pounds_{\kappa\epsilon^{(1)}}g_{\mu\nu} \implies 
    \begin{array}{l}
    G_{\mu\nu} \to G_{\mu\nu}
    \\[5pt]  h_{\mu\nu} \to h_{\mu\nu} + \nabla_{(\mu}\epsilon_{\nu)}^{(1)}.
    \end{array}
\end{equation}
One can check that the graviton action \eqref{eqn:EHactionexp_2ndorder} is indeed invariant under both of these types of diffeomorphisms \eqref{eqn:backgddiffs} and \eqref{eqn:pertdiffs}. Invariance under the perturbative diffeomorphisms \eqref{eqn:pertdiffs} is what we mean by gauge-invariance in this work.\footnote{There are also $O(\kappa^{2})$ pieces in the transformation of $g$ under these gauge symmetries which will lead to a transformation of $g^{(2)}$ that we have not written explicitly.}

It is important in a gauge theory to distinguish between gauge symmetries that change the boundary conditions for the fields and those that do not. Any gauge symmetry that preserves the boundary conditions are actual redundancies of the theory while those that do not are physical symmetries of the theory. These are commonly referred to as `small' and `large' gauge transformations, respectively. For subregions $b$ of a Cauchy slice $\Sigma$, the distinction becomes more subtle because for any part $\gamma \subset \partial b$ that does not coincide with $\partial \Sigma$, there are gauge transformations that do not fall off at $\gamma$ so they are large for the gauge theory defined in $b$, but are certainly small for the theory on the whole slice $\Sigma$.

This will play a key role as we seek a gauge-invariant formulation for the generalized entropy. Since we are interested in understanding the generalized entropy for the reduced state on $b$ obtained from the global state on $\Sigma$ by tracing out the complement $b^{c}$, we require that the generalized entropy be invariant under small gauge transformations for $\Sigma$, including those that are large for $b$, where a precise definition of `small' for the global theory will be given in \S\ref{sec:HWgaugemore}.

\section{A prescription for the generalized entropy of gravitons}
\label{sec:prescriptgrav}
Having dealt with all of these preliminary details regarding gravitons and diffeomorphisms, we now turn to our proposal for their generalized entropy. The goal is to find a prescription that satisfies the requirements stated in the Introduction: (1) gauge-invariance; (2) agrees with the CFT entanglement entropy; (3) can be written as a sum of an area plus vN entropy of gravitons.

We begin in \S\ref{sec:obs} with some basic observations about the generalized entropy of gravitons that follow from gauge invariance and from studying our explicit example in \S\ref{sec:ex}. This motivates our prescription, which we provide in \S\ref{sec:prescript}. We first describe this briefly, and somewhat loosely, in a manner that we hope works to all orders in perturbation theory and then describe much more precisely to $O(\kappa^2)$.

\subsection{Some observations}
\label{sec:obs}

Consider first the classical area of a codimension-$2$ extremal surface $\gamma^{\mu}$ anchored on $\partial B$, homologous to $B$, which we can expand order-by-order in the coupling
\begin{equation}\label{eqn:gammapert}
\gamma^{\mu} = \gamma^{(0)}+\kappa\gamma^{(1)\mu}+O(\kappa^{2}),
\end{equation}
and we will always assume that $\gamma^{(0)}$ extremizes the area $A[G,\gamma^{(0)}]$. This leads to an expansion for the area in the metric $g_{\mu\nu} = G_{\mu\nu}+\kappa h_{\mu\nu}+\kappa^{2}g_{\mu\nu}^{(2)}$ given by
\begin{align}\label{eqn:areaexp}
\begin{split}
A[g,\gamma] &= A[G,\gamma^{(0)}]+\kappa A^{\mathrm{lin}}[h,\gamma^{(0)}]
\\	&+\kappa^{2}\left(A^{\mathrm{quad}}[G,\gamma^{(1)}]+A^{\mathrm{quad}}[h,\gamma^{(0)}]+A^{\mathrm{lin}}[h,\gamma^{(1)}]+A^{\mathrm{lin}}[g^{(2)},\gamma^{(0)}]\right),
\end{split}
\end{align}
where `lin' and `quad' denote linear and quadratic expansion in the perturbation, respectively. The expansion \eqref{eqn:gammapert} of the location of the surface, and hence also its area, are gauge dependent. However, the space of surfaces that can be written in the form \eqref{eqn:gammapert} is gauge invariant and, as a result, so is the extremal surface area.

\paragraph{The area must be an operator.} In principle, in a theory of quantum gravity, the area of a surface is always a quantum mechanical object. However, for ordinary matter fields, we can treat changes in the area as classical perturbations sourced by the semiclassical Einstein equations, at least to 1st order in $\kappa^2$, even though this classical perturbation is really only the expectation value of of a quantum perturbation. The same  is not true for graviton excitations. Let us focus on the $O(1/\kappa)$ part of $A[g,\gamma]/\kappa^{2}$, which comes from $(\kappa A^{\mathrm{lin}}[h,\gamma^{(0)}])/\kappa^2$. For perturbative states, the bulk vN entropy can never give a contribution of this order so the area term is the only possible source of $O(1/\kappa)$ contributions to the generalized entropy. A simple CFT calculation, done in \S\ref{sec:ex}, shows that, for the stress-tensor state $\ket{\epsilon_{q} \cdot T}$ dual to a graviton single-particle state $\ket{\mathfrak{g}_{q}} = a_{q}^{\dagger}\ket{0}$, the CFT entanglement entropy vanishes at this order, but not for $\ket{S_{q}}=(\ket{0}+\ket{\epsilon_{q} \cdot T})/\sqrt{2}$. This can never be true if the area is treated classically, but only if the area is promoted to an operator by
\begin{equation}\label{eqn:quantumarea}
A^{\mathrm{lin}}[h,\gamma^{(0)}] \xrightarrow{\mathrm{promotion}} A^{\mathrm{lin}}[\hat{h},\gamma^{(0)}].
\end{equation}

The fact that a superposition of the vacuum and a single-graviton state can have parametrically larger entanglement entropy than either state on its own is perhaps somewhat surprising. However, it matches the results one obtains from a direct CFT calculation. We will explain that calculation in detail in \S\ref{sec:ex}, but the basic story is as follows.
 Consider the CFT thermofield double state $\ket{\mathrm{TFD}} = \frac{1}{\sqrt{Z}}\sum_{n}e^{-\beta E_{n}/2}\ket{n}_{L}\ket{n}_{R}$ at inverse temperature $\beta$.\footnote{The polar cap subregion on the cylinder considered throughout this work can be conformally mapped to a Rindler wedge in Minkowski space with the vacuum state on the cylinder mapped to the TFD state so all the arguments here carry over.} As $N \to \infty$, the right Hamiltonian $H_{R}$, has expectation value $\braket{H_R} \sim N^2$ and $\Delta H_R = H_R - \braket{H_R}$ has approximately Gaussian fluctuations with $\langle \Delta H_R^2\rangle_{\beta} \sim N^{2}$. The normalized state $\ket{\Delta H_R} = \Delta H_R \ket{\mathrm{TFD}}/ N$ is dual to a bulk state containing a single graviton with respect to the Hartle-Hawking vacuum. Since the energy fluctuations in $\ket{\mathrm{TFD}}$ were almost Gaussian, with  $\braket{\Delta H_R^3} \sim N^2$, the average energy (and hence also the entropy) of $\ket{\Delta H_R}$ differs from $\ket{\mathrm{TFD}}$ only at $O(1)$.
However, as shown in figure \ref{fig:TFD}, the superposition state $(\ket{\mathrm{TFD}} +\ket{\Delta H_R})/\sqrt{2}$ has an $O(N)$ shift in its entropy and average energy. This large increase in entanglement entropy was possible precisely because the original state $\ket{\mathrm{TFD}}$ (like typical semiclassical bulk states but unlike e.g. a purification of the microcanonical ensemble) already had large $O(N)$ fluctuations in its modular Hamiltonian. The effect of the superposition is essentially just to project onto positive fluctuations of this modular Hamiltonian, thereby increasing its expectation value.

\paragraph{Classical diffeomorphisms and classical surface perturbations are equivalent.} Consider the classical perturbed surface $\gamma = \gamma^{(0)} + \kappa \gamma^{(1)}$, where we can think of $\gamma^{(1)}$ as a section of the normal bundle on $\gamma^{(0)}$. Now let $\epsilon_\mu^{(1)}$ be a vector field whose restriction to $\gamma^{(0)}$ is $-\gamma^{(1)}$. The perturbative diffeomorphism generated by $\kappa \epsilon_\mu^{(1)}$ maps $\gamma$ to $\gamma^{(0)}$ and hence puts the perturbation in classical extremal gauge. A further diffeomorphism $\kappa \tilde\epsilon_\mu^{(1)}$ will then preserve extremal gauge if and only if its restriction to $\gamma^{(0)}$ is contained in the tangent bundle to $\gamma^{(0)}$, i.e. it preserves $\gamma^{(0)}$. Finally, we will need the fact that a change in the surface perturbation $\gamma^{(1)}$ is linear in the perturbative diffeomorphism that causes it. It follows immediately from these observations that extremizing over diffeomorphisms is equivalent to extremizing over surfaces (and hence leads to extremal gauge), except that the extremal surface corresponds to an entire ``extremal submanifold'' of diffeomorphisms, namely those that map $\gamma$ to $\gamma^{(0)}$. Furthermore, this extremal submanifold is always isomorphic to the space of diffeomorphisms that preserve $\gamma^{(0)}$.

\begin{figure}
\centering
\begin{tikzpicture}
  \begin{axis}[
    domain=0:10,
    samples=100,
    axis lines=middle,
    xlabel={$E$},
    ylabel={$\alpha(E)$},
    xmin=0, xmax=10,
    ymin=-1, ymax=1,
    xtick={5},
    xticklabels={},
    ytick=\empty,
    width=10cm, height=6cm,
    enlargelimits=false,
    clip=false,
    axis line style={->},
    xlabel style={at={(axis cs:10,0)},anchor=west},
    ylabel style={at={(axis description cs:0,1)},anchor=south east},
  ]
    \addplot[thick,blue,domain=0:10] {exp(-((x-5)^2)/2)};
    \addplot[thick,red,domain=0:10] {(x-5)*exp(-((x-5)^2)/2)};
    \addplot[thick,green,domain=0:10] {0.707(1+(x-5))*exp(-((x-5)^2)/2)};
    \node[below] at (axis cs:5.2,0) {$E_0$};
  \end{axis}
\end{tikzpicture}
    \caption{Wavefunctions $\ket{\Psi} = \sum_{n} \alpha(E_n) e^{-S(E_n)/2} \ket{n}_L \ket{n}_R$ in the energy basis for $\ket{\mathrm{TFD}}$ (blue), $\ket{\Delta H_R}$ (red) and the superposition (green). The exponential of the microcanonical entropy $S(E)$ is factored out so that at large $N$ the distribution over energies becomes approximately $p(E) = |\alpha(E)|^2$ with no prefactor. Acting with $\Delta H_R$ changes the Gaussian TFD distribution and creates a relative phase between large and small energy eigenstate, but only shifts the average energy by a small $O(1)$ amount. However, in the superposition state, large energies constructively interfere, while small energies are projected out, increasing the average energy by $O(N)$. Since $\ket{\Psi}$ has entanglement entropy $S = \braket{S(E)-\log p(E)} $ and $dS/dE = \beta = O(1)$, this also increases the entropy by $O(N)$.}
    \label{fig:TFD}
\end{figure}

\paragraph{Classical perturbations are not enough.}
Consider again the single graviton state $\ket{\mathfrak{g}_{q}}$. At $O(1/\kappa)$, the expectation value of the area gradient at the extremal surface vanishes (in any gauge) because free-graviton vacuum three-point functions vanish. So, if we are only supposed to extremize over classical perturbations to the surface $\gamma$, we can safely use the unperturbed surface $\gamma^{(0)}$ when computing generalized entropy to $O(1)$. However, it is easy to check that the $O(\kappa^{2})$ corrections to the area $A[\gamma^{(0)}]$ of this unperturbed surface are not gauge-invariant. The problem here, as explained in the introduction, is that quantum gauge-equivalence condition
\begin{align}\label{eq:equivrelation}
    \Psi \cong \Psi + \hat P_i \Phi
\end{align}
leads to too large a space of gauge-equivalent quantum states: it involves not only a choice of perturbative diffeomorphism $\hat P_i$ but also the arbitrary quantum state $\Phi$. To produce a gauge-invariant QES prescription, we therefore need to, in some manner, allow \emph{quantum} perturbations of the surface $\gamma$, whether by directly extremizing over the space of gauge-equivalent quantum states or by somehow extremizing over a space of operator-valued surface perturbations.

\paragraph{Backreaction accounts for graviton interactions.}
In \S\ref{sec:gravitons},  we explained that the perturbative graviton action includes a cubic term $\kappa S[h]^{(3)}$. As a result, one might expect that the expectation value of \eqref{eqn:quantumarea} in the single-graviton state $\ket{\mathfrak{g}_{q}}$ is nonzero at $O(\kappa^2)$. This would indeed be the case, if we defined the graviton field $\hat h_{\mu\nu}$ to be the full and exact all-orders metric perturbation, as we did in \eqref{eqn:backgdexp_metric}. When working perturbatively to $O(\kappa^2)$, however, it is convenient to instead include an additional classical $O(\kappa^2)$ perturbation $g_{\mu\nu}^{(2)}$, as we did when working with ordinary matter fields in \eqref{eqn:fullmetric}, so that
\begin{align}
    \hat g_{\mu\nu} = G_{\mu\nu} + \kappa \hat h_{\mu\nu} + \kappa^2 g^{(2)}_{\mu\nu}.
\end{align}
Here, $g^{(2)}_{\mu\nu}$ is defined so that the semiclassical Einstein equations hold at $O(\kappa^{2})$, i.e.
\begin{equation}\label{eqn:g2eqn}
E_{\mu\nu}^{(1)}[g^{(2)}] = -\braket{E_{\mu\nu}^{(2)}[h]}
\end{equation}
where $E_{\mu\nu}^{(1)}$ and $E_{\mu\nu}^{(2)}$ are the linearized and quadratic Einstein tensors, respectively. This uniquely determines $g^{(2)}$ so long as we also require it to be orthogonal to all solutions to the homogeneous linearized Einstein equations $E_{\mu\nu}^{(1)} = 0$. We can solve for $g^{(2)}$ using the Green's function $\mathfrak{G}_{\mu\nu}^{\alpha\beta}(x-y)$ for the graviton $h_{\mu\nu}$ obtained by inverting the differential operator $E_{\mu\nu}^{(1)}$. To wit,
\begin{equation}\label{eqn:g2_formalsoln}
g_{\mu\nu}^{(2)}(x) = -\int d^{d+1}y\,\mathfrak{G}_{\mu\nu}^{\alpha\beta}(x-y)E_{\alpha\beta}^{(2)}[h](y).
\end{equation}

For the generalized entropy at $O(\kappa^{0})$, we only need to worry about the interaction contribution to $A^{\mathrm{lin}}[\hat{h},\gamma^{(0)}]$. Its expectation value in $n$-particle states is a $(2n+1)$-point function, but it is sufficient to focus on the three-point function, i.e., single particle states, as the argument trivially generalizes to higher-points.\footnote{At $O(\kappa)$, $(2n+ 1)$-point functions contain only products of a single three-point and $(n-1)$ free two-points functions.} The linearized area is an integral of $h_{\mu\nu}(x)$ over the codimension-$2$ surface which takes the form
\begin{equation}
A^{\mathrm{lin}}[\hat{h},\gamma^{(0)}] = \int_{\gamma^{(0)}}d^{d-1}x\,\sqrt{q^{(0)}}k^{\mu\nu}\hat{h}_{\mu\nu}
\end{equation}
for some kernel $k^{\mu\nu}$ and $q_{\mu\nu}^{(0)}$ is the induced metric on $\gamma^{(0)}$ in the background metric $G_{\mu\nu}$. The expectation value of the area integrand in the interacting theory thus contains
\begin{align}
\begin{split}
\bra{0}a_{q}\hat{h}_{\mu\nu}(x)a_{q}^{\dagger}\ket{0}_{\mathrm{int.}} &= \underbrace{\bra{0}a_{q}\hat{h}_{\mu\nu}(x)a_{q}^{\dagger}\ket{0}_{\mathrm{free}}}_{\color{red}=0}+i\kappa \bra{0}a_{q}\hat{h}_{\mu\nu}(x)S^{(3)}[\hat{h}]a_{q}^{\dagger}\ket{0}_{\mathrm{free}}+O(\kappa^{2})
\\	&= -\kappa\bra{0}a_{q}\int d^{d+1}y\,\mathfrak{G}_{\mu\nu}^{\alpha\beta}(x-y)\normord{E_{\alpha\beta}^{(2)}[h](y)}a_{q}^{\dagger}\ket{0}_{\mathrm{free}}+\mathrm{tadpoles}+O(\kappa^{2})
\\	&= \kappa\bra{0}a_{q}\normord{g_{\mu\nu}^{(2)}(x)}a_{q}^{\dagger}\ket{0}_{\mathrm{free}}+\mathrm{tadpoles}+O(\kappa^{2})
\end{split}
\end{align}
where we used Wick's theorem in the second line to write the product $\hat{h}_{\mu\nu}(x)S^{(3)}[\hat{h}](y)$ as a sum of Wick contractions times normal-ordered products. The tadpoles come from internal contractions of $S^{(3)}[\hat{h]}$, but are renormalised by counterterms in the Lagrangian to ensure that the vacuum one-point function vanishes. Thus, the area operator for the backreacted metric $\normord{{A}^{\mathrm{lin}}[\hat{g}^{(2)},\gamma^{(0)}]}$ accounts for cubic interactions of $\hat{h}_{\mu\nu}$, and hence we can treat $\hat{h}_{\mu\nu}$ as free.

\paragraph{Free-graviton relative entropies are gauge invariant.} In the continuum free-graviton field theory, we can define an algebra of gauge-invariant operators localised within any fixed domain of dependence $\mathcal{D}(b)$. There is no issue with defining the region $b$ in a gauge-invariant way, because in the free theory perturbative diffeomorphisms do not change the background metric used to define $b$. This algebra will be a Type III von Neumann factor, meaning that entropies are divergent. However, relative entropies are well defined using Tomita-Takesaki theory. 

 \paragraph{Vacuum-subtracted entropies are somewhat more complicated.} To define vacuum-subtracted entropies, we also need to be able to define a one-sided vacuum modular Hamiltonian as a densely defined quadratic form. The two-sided modular Hamiltonian $\log \Delta$, which is a true densely defined Hilbert space operator, can again be defined using Tomita-Takesaki theory. We can then define the one-sided modular Hamiltonian by splitting $\log \Delta$ into two quadratic forms that each commute with operators on the opposite side (i.e. one commutes with operators outside $b$ and one with operators inside $b$). But this splitting is ambiguous because we can always add a local operator at the edge of the region $b$ to one side and subtract it from the other. (Since the one-sided modular Hamiltonians are already only quadratic forms, doing so does not make them ``worse'' objects in any obvious algebraic sense.) In ordinary quantum field theories, we typically define the ``true'' modular Hamiltonian (the one-sided boost generator in the case of Rindler space) as the unique quadratic form that does not include any such boundary term. But, in gravity, the question of whether the canonical energy, which is the classical analogue of the vacuum modular Hamiltonian, includes a boundary term is gauge-dependent, as we show in \S \ref{sec:HW}. To match CFT calculations, the right choice for the vacuum modular Hamiltonian turns out to be the canonical energy that contains no boundary term in extremal gauge. It is perfectly satisfactory that extremal gauge should play an explicit role here, since the prescription \eqref{eq:genentfluct} picks out extremal gauge as the ``right'' gauge to compute generalized entropy in. But it would be nice to understand better whether the role of extremal gauge is truly necessary or not. In other words, whether the correct interpretation here is a) that graviton entanglement entropy is truly gauge-dependent (even at leading order), but that to match CFT entropies we are supposed to compute it in extremal gauge, or b) that vacuum-subtracted entropies are gauge invariant after all, with some other principle picking out the true modular Hamiltonian to be the one with no boundary term in extremal gauge.

\subsection{Our prescription}
\label{sec:prescript}
Since classical perturbations of the surface $\gamma^{(0)}$ are insufficient to produce a gauge-invariant prescription, we need a larger space to search over. Since extremizing over classical surface perturbations is equivalent to extremizing over classical gauge transformations, the obvious approach is to, in some manner, ``extremize'' over the space of all quantum states that are equivalent to the state of interest under the relation \eqref{eq:equivrelation}. If we do so, the resulting prescription is guaranteed to be gauge-invariant.

However, it is not completely clear what it means for a quantum state to be extremized. For classical perturbations, we searched for a gauge where diffeomorphisms left the area of $\gamma^{(0)}$ invariant at linear order. So an obvious definition is that in quantum extremal gauge the commutator of $\hat A[\gamma^{(0)}]/4G - \log \rho_b$ with $P_i$ to vanish. But the expectation value of this commutator vanishing is equivalent to generalized entropy being extremal under the classical perturbation generated by $P_i$, and we already saw that this condition could not be enough.\footnote{Previously we argued that this would lead to a prescription that is not gauge-invariant. Since we are now explicitly searching over the space of all gauge-equivalent states, it will instead lead, at least at $O(\kappa^2)$, to a generalized entropy that is not bounded from below and hence to a sick QES prescription.} Instead, we propose
that, in quantum extremal gauge, the commutator should entirely annihilate the quantum state $\Psi$ so that
\begin{align}\label{eq:quantumextremalmaintext}
    \left[\hat P_i, \frac{\hat A[\gamma^{(0)}]}{4G_N} - \log \rho_b^\Psi\right] \Psi = 0
\end{align}
for all diffeomorphism generators $\hat P_i$.\footnote{As usual, when discussing the QES prescription, the two terms on the left-hand side of \eqref{eq:quantumextremalmaintext} are individually UV-divergent, but it is expected that their sum is UV finite in the same way that generalized entropy is UV finite \cite{Susskind:1994sm, Akers:2023fqr} .} 
The QES prescription can then be stated as
\be\label{eq:genentfluctmaintext}
S^{\mathrm{CFT}}(\rho_B^{\psi}) = \underset{\gamma^{(0)}, \tilde\Psi \cong \Psi}{\mathrm{min\;ext}}\,\,\Big\langle\tilde\Psi\Big|\left[\frac{\hat A[\gamma^{(0)}]}{4G_N} - \log \rho_b^{\tilde\Psi}\right] \tilde\Psi\Big\rangle
\ee
where we first find the minimal area classical extremal surface $\gamma^{(0)}$ with respect to the unperturbed metric $G_{\mu\nu}$ and then extremize and minimize over all quantum extremal states $\tilde\Psi \cong \Psi$.

Let us make a few comments about how we think \eqref{eq:quantumextremalmaintext} and \eqref{eq:genentfluctmaintext} should in theory be defined to arbitrary orders in $G_N$, before focusing on their restriction to $O(1)$ precision. While we are hopeful that the former is possible, it is important to emphasize that this is by far the most speculative part of this paper. It is not at all clear that \emph{any} version of the QES prescription can be defined to all orders in $G_N$, except very formally as the analytic continuation of a sequence of gravitational replica trick saddle points plus their perturbative fluctuations. 

Firstly, we expect that the expectation value in \eqref{eq:genentfluctmaintext} should be defined using a gauge-invariant, group-averaged inner product on the perturbative graviton Hilbert space \cite{Marolf:2000iq, Chandrasekaran:2022cip, Held:2024rmg}. Similarly, the density matrix $\rho_b^\Psi$ should be defined in an algebra of observables that are invariant under diffeomorphisms preserving $b$, so that $[\rho_b^\Psi,\hat P_i] = 0$ whenever $\hat P_i$ preserves $\gamma^{(0)}$.

Now, for \eqref{eq:genentfluctmaintext} to make sense, two crucial conditions must be satisfied: a) at least one state satisfying \eqref{eq:quantumextremalmaintext} needs to exist in any gauge-equivalence class of states and b) there cannot be so many gauge-equivalent states satisfying \eqref{eq:quantumextremalmaintext} that the minimization in \eqref{eq:genentfluctmaintext} is not bounded from below. Just like classical extremal gauge was invariant under diffeomorphisms that preserved $\gamma^{(0)}$, one might expect that \eqref{eq:quantumextremalmaintext} will be preserved under the equivalence relation \eqref{eq:equivrelation} when the diffeomorphism $\hat P_i$ preserves $\gamma^{(0)}$. Indeed, this seems to be true so long as the state $\Phi$ is itself in quantum extremal gauge. We then have
\begin{align}
\left[\hat P_j, \frac{\hat A[\gamma^{(0)}]}{4G_N} - \log \rho_b^\Psi\right] \hat P_i \Phi = \hat P_i \left[\hat P_j, \frac{\hat A[\gamma^{(0)}]}{4G_N} - \log \rho_b^\Psi\right]  \Phi -  \left[[\hat P_i,\hat P_j], \frac{\hat A[\gamma^{(0)}]}{4G_N} - \log \rho_b^\Psi\right]  \Phi  = 0.
\end{align}
However, these gauge transformations also satisfy
\be\label{eq:sgenunchanged}
\left(\frac{\hat A[\gamma^{(0)}]}{4G_N} - \log \rho_b^\Psi\right) \hat P_i \Phi = \hat P_i \left(\frac{\hat A[\gamma^{(0)}]}{4G_N} - \log \rho_b^\Psi\right)  \Phi \cong 0
\ee
and so leave \eqref{eq:genentfluctmaintext} unchanged. Just like classical extremal gauge is perturbatively unique up to diffeomorphisms preserving $\gamma^{(0)}$, we expect (or at least hope) that in fact the gauge \eqref{eq:quantumextremalmaintext} is unique up to gauge equivalences satisfying \eqref{eq:sgenunchanged}, and hence that the minimization over quantum extremal states in \eqref{eq:genentfluctmaintext} is well defined and is in fact essentially trivial.

\paragraph{Generalized entropy of gravitons at $O(1)$ precision.} With those somewhat speculative comments made, let us now restrict ourselves to studying graviton fluctuations at $O(1)$ precision. It follows from extremality that to determine \eqref{eq:genentfluctmaintext} to $O(\varepsilon)$ precision, we only need to solve for \eqref{eq:quantumextremalmaintext} to $O(\varepsilon^{\frac{1}{2}} \kappa^{-1})$ precision, where the factor of $\kappa^{-1}$ comes from the $O(\kappa^{-2})$ Hessian of the area term. At $O(\kappa^{-2})$, \eqref{eq:quantumextremalmaintext} is a $c$-number, which vanishes because $\gamma^{(0)}$ is classically extremal. At $O(\kappa^{-1})$, the gradient of $\log \rho_b^\Psi$ still vanishes, but $[\hat P_i, \hat A[\gamma^{(0)}]/4G_N]$ is a nontrivial linear function of the graviton field $\hat h_{\mu\nu}$. When computing \eqref{eq:genentfluctmaintext} to $O(1)$ precision, quantum extremal gauge therefore becomes a linear constraint on the graviton field $\hat h_{\mu\nu}$, which is equivalent to first imposing classical extremal gauge and then quantizing the theory.

Our prescription then has reduced to the following: we work in classical extremal gauge, quantize the theory and then take the quantum extremal surface to be unperturbed.

However, there is another way to state the same prescription that does not require us to be in extremal gauge. Classically, we can write the linear-order perturbation $\gamma^{(1)}$ to the extremal surface as a gauge-invariant linear function of $h_{\mu\nu}$.\footnote{The location of the surface is not gauge-invariant, because $h_{\mu\nu}$ is not gauge-invariant, but the map is invariant.} Then we can write the area of this perturbed surface to quadratic order as
    \begin{equation}\label{eq:Aopschematicmaintext}
        A[g,\gamma] = A[G,\gamma^{(0)}] + \kappa \frac{\delta A}{\delta g} h + \kappa^{2}\Big(\frac{\delta^2 A}{\delta g^2}(h)^2 + \frac{\delta^2 A}{\delta g \delta \gamma} h {\gamma}^{(1)}  + \frac{\delta^2 A}{{\delta \gamma}^2} (\gamma^{(1)})^2 + \frac{\delta A}{\delta g}{g}^{(2)} \Big) + \cdots \,,
    \end{equation}
where $\gamma^{(1)}$ is the linear function of $h_{\mu\nu}$ described above. Clearly, this should be a gauge-invariant function. But then we can promote this function to a gauge-invariant operator $A[\hat g,\hat \gamma]$ on the quantum Hilbert space by promoting $h_{\mu\nu}$ and $\gamma^{(1)}$ to operators (with the latter still a linear function of the former). There are minor ordering ambiguities in doing so, but these can be removed by normal ordering because at quadratic order they give $c$-numbers that will drop out after vacuum subtraction. We then have
    \begin{align}\label{eq:QEStoO1maintext}
       S^{\mathrm{CFT}}(\rho_B^{\psi}) = \frac{\braket{\normord{A[\hat{g},\hat\gamma]}}_{\Psi}}{4G_{N}}+S\left(\rho_{b}^{\Psi}\right) + O(\kappa)
    \end{align}
where the calculation can now be done in any gauge.

\paragraph{An alternative prescription.} There is one final point of view on the prescription \eqref{eq:QEStoO1maintext} that we include because it is conceptually interesting and may lead to an alternative way to define the QES prescription to all orders in $G_N$ while including graviton fluctuations. Let us assume that we have completely fixed to some gauge, which may or may not be extremal, and we want to determine the correct $\hat \gamma^{(1)}$. The idea is that rather than fixing $\hat \gamma^{(1)}$, a priori, to be a particular function of $\hat h_{\mu\nu}$ that describes the perturbation to the extremal surface, we take 
\be
\hat \gamma^{(1)} = \sum_q \gamma_q^{(1)} a_q + \gamma_q^{(1)*} a_q^\dagger
\ee
to be an \emph{arbitrary} linear function of $\hat h_{\mu\nu}$, or equivalently of the graviton creation and annihilation operators $a_q^\dagger, a_q$, parameterized by the normal-bundle valued coefficients $\gamma_q^{(1)}$. Then we extremize \eqref{eq:QEStoO1maintext}, without the normal-ordering, over the coefficients $\gamma_q^{(1)}$.\footnote{Although the expectation value of the full quadratic area operator will be divergent since it contains terms like $\langle h(x)^{2} \rangle_{\Psi}$ and derivatives thereof, we can extremize mode-by-mode since the contribution of each mode is finite.} In doing so, we are extremizing over the space of operator-valued surface perturbations, where the operator in question needs to be linear in $\hat h_{\mu\nu}$. The expectation value of the area operator in a particular state will pick out some linear combination of the surfaces $\gamma_q^{(1)}$, e.g., for a single-particle state $\psi=\mathfrak{g}_{q_{0}}$ only $\gamma_{q_{0}}^{(1)\mu}$ gets picked out, and then extremization of this expectation value gives the same extremal value for the surface that appears in \eqref{eq:QEStoO1maintext}. Hence, this alternative prescription should give equivalent results to \eqref{eq:QEStoO1maintext}. 

At higher orders, one should presumably allow $\hat \gamma$ to depend not only on $\hat h_{\mu\nu}$ but also on matter operators etc. It is not clear exactly how such an approach would work and whether it would end up being equivalent to the quantum extremal gauge prescription described above (if either make sense). Its primary disadvantage relative to that prescription is that some significant work would be required to show that the result is truly gauge-invariant.

\section{Covariant phase space analysis of the prescription}
\label{sec:HW}

Some justification of our prescription can be obtained by analyzing the classical phase space of gravitational perturbations. This phase space was studied for perturbations of stationary, asymptotically flat black hole spacetimes by Hollands and Wald \cite{Hollands:2012sf}. We will adapt their analysis to asymptotically AdS spacetimes and make their results gauge-invariant since they worked in a specific gauge. 
The upshot is that the resulting phase space identity relates the bulk to boundary charge for a Killing vector $\xi^{\mu}$ by 
\begin{equation}
\begin{aligned}
   O(\kappa^{-1}): \quad \delta K_{(\xi)}^{\partial} & = \frac{2\mathfrak{s}}{\kappa^2} \delta A[g,\gamma]\\
    O(1): \quad \delta^{2}K_{(\xi)}^{\partial} & = \frac{2\mathfrak{s}}{\kappa^{2}}\delta^{2}A[g,\gamma] + \Omega\left[g;\delta g,\pounds_{\xi}\delta g\right].
\end{aligned}
\end{equation}
We show that these relations can be promoted to an expectation value of quantum operators, which is actually the JLMS formula
\begin{equation}\label{eqn:JLMS0}
\Delta \hat{K}_{B}^{\psi} = \frac{\Delta\langle{A}[\hat{g},\hat{\gamma}]\rangle_\Psi}{4G_N}+\Delta\hat{K}_{b}^{\Psi}\,.
\end{equation}
Here $\Delta$ means the difference of expectation values for the excited state density matrix $\rho=\omega+\delta\rho$ and the vacuum one $\omega$.
This gives the leading contribution in $\delta \rho$ to the bulk and boundary entanglement entropies by the first law of entanglement, and hence the leading contribution to the QES formula for gravitons.

We begin in \S\ref{sec:HWgauge} by describing the Hollands-Wald gauge choice and then use their results in \S\ref{sec:HWformula} to derive a relation between the bulk and boundary charges for the Killing vector.
It is then shown how to make this result gauge-invariant in \S\ref{sec:restoregaugeinv}.
We also give an alternative form which makes the graviton formulas particularly similar to those for ordinary matter in \S\ref{sec:altprescript}.
Then we show how to relate the original form to the JLMS formula in \S\ref{sec:JLMSgraviton}.
Some of these ideas were already present in \cite{Jafferis:2015del}, but we make them more precise and clarify them significantly.
We end with a discussion of some important subtleties in using the Hollands-Wald gauge for asymptotically AdS spacetimes in \S\ref{sec:HWgaugemore}.

\subsection{Choice of gauge}
\label{sec:HWgauge}

Consider a family of $(d+1)$-dimensional asymptotically AdS spacetimes $g_{\mu\nu}(\lambda)$ with $G=g_{\mu\nu}(0)$ being the metric for a stationary asymptotically AdS black hole.\footnote{We allow for non-compact horizons, such as the AdS black brane or hyperbolic black hole.} The bifurcation surface $\gamma^{(0)}$ of the stationary black hole extremizes the area of all codimension-$2$ surfaces, which can be characterized by the vanishing of the trace of the extrinsic curvatures for the two null normals $l^{\mu}$ and $k^{\mu}$ to this codimension-$2$ surface:
\begin{equation}\label{eqn:extremality}
K_{(l)}[G]|_{\gamma^{(0)}}=K_{(k)}[G]|_{\gamma^{(0)}}=0.
\end{equation}
With this in mind, Hollands and Wald chose a convenient gauge to make their analysis tractable, consisting of two conditions. 

First, to linear order in $\lambda$, the extremal surface for $g_{\mu\nu}(\lambda)$ is still located at $\gamma^{(0)}$:
\begin{equation}\label{eqn:HWgauge1}
\delta K_{(l)}|_{\gamma^{(0)}}=\delta K_{(k)}|_{\gamma^{(0)}}=0
\end{equation}
where $\delta = \frac{d}{d\lambda}|_{\lambda=0}$. This means choosing the diffeomorphism that takes the perturbed extremal surface and maps it to the background extremal surface $\gamma^{(0)}$. This is the extremal gauge described in the Introduction.

Second, they chose Gaussian null coordinates \cite{Hollands:2006rj} in a neighborhood of the future event horizon $\mathfrak{H}^{+}$ of the black hole such that this surface is null for all $\lambda$. One foliates $\mathfrak{H}^{+}$ by codimension-$2$ surfaces $\gamma(k)$ with affine parameter $k$ and fires null geodesics orthogonally into the past with affine parameter $l$, all in a $\lambda$-independent way, leading to the metric
\begin{equation}\label{eqn:GNC_metric}
    ds^{2} = 2dl dk+a(l,k,x^{i})dk^{2}+b_{i}(l,k,x^{i})dk dx^{i}+q_{ij}(l,k,x^{i})dx^{i}dx^{j}
\end{equation}
where $\mathfrak{H}^{+}$ lies at $l=0$ on which $a(l=0,k,x^{i})=b_{i}(l=0,k,x^{i})=0$, $\gamma^{(0)}$ lies at $l=k=0$, and $x^{i}$ are $d-1$ coordinates on the surfaces $\gamma(l,k)$. A simple characterization of this gauge choice can be found \cite{Lashkari:2015hha}: it was shown by Hollands and Wald that in these coordinates the background Killing vector takes the form
\begin{equation}\label{Killingvec_GNC}
    \xi^{\mu} = k\left(\frac{\partial}{\partial k}\right)^{\mu}-l\left(\frac{\partial}{\partial l}\right)^{\mu},
\end{equation}
and so, since $\xi^{\mu}|_{\gamma^{(0)}}=0$, we have $\nabla_{\mu}\xi_{\nu}|_{\gamma^{(0)}}=\partial_{\mu}\xi_{\nu}|_{\gamma^{(0)}}$ which gives
\begin{equation}\label{eqn:HWcondition2}
    \pounds_{\xi}g_{\mu\nu}|_{\gamma^{(0)}} = \nabla_{(\mu}\xi_{\nu)}|_{\gamma^{(0)}} = 0.
\end{equation}
That is, the background Killing vector still satisfies the Killing equation on the bifurfaction surface $\gamma^{(0)}$. 
The gauge satisfying the above two conditions is often referred to as Hollands-Wald gauge.

We will give a detailed discussion in \S\ref{sec:restoregaugeinv} on the role played by these two gauge conditions. In particular, the second condition is not inherently crucial, while the first one contains the key ingredient of making the graviton entropy simple.


\subsection{Relation between bulk and boundary charges}
\label{sec:HWformula}
Using covariant phase space techniques to analyse perturbations of stationary black holes, Wald derived the first law of black hole mechanics and showed that the entropy is the Noether charge for the stationary Killing field \cite{Wald:1993nt}. The gauge choice described above can be used to go beyond first-order variations of charges and understand their second-order variations. We briefly review this covariant phase space formalism and then use it to derive the desired second-order variation of charges following \cite{Hollands:2012sf}.

\paragraph{Covariant phase space.}

Consider the Einstein-Hilbert Lagrangian as the $(d+1)$-form
\begin{equation}\label{eqn:Lagrangianform}
    \mathbf{L} = \frac{1}{\kappa^{2}}\left(R-2\Lambda\right)\boldsymbol{\epsilon}
\end{equation}
where $\boldsymbol{\epsilon}$ is the (positively-oriented) volume form. Variation of the Lagrangian gives the Einstein equations plus a boundary term
\begin{equation}\label{eqn:varL}
    \frac{d}{d\lambda}\mathbf{L}[g(\lambda)] = \boldsymbol{E}^{\mu\nu}[g(\lambda)]\frac{d}{d\lambda}g_{\mu\nu}(\lambda) + \mathbf{d}\boldsymbol{\theta}\left[g(\lambda);\frac{d}{d\lambda}g(\lambda)\right]
\end{equation}
where
\begin{equation}\label{eqn:Einsteinform}
  \boldsymbol{E}^{\mu\nu}[g(\lambda)] = -\frac{1}{\kappa^{2}}\left(R^{\mu\nu}-\frac{1}{2}g^{\mu\nu}R+\Lambda g^{\mu\nu}\right)\boldsymbol{\epsilon}
\end{equation}
and $\boldsymbol{\theta}$ is the symplectic potential current $d$-form given by
\begin{equation}\label{eqn:symppotcurrent}
    \boldsymbol{\theta} = \frac{1}{\kappa^{2}}\iota_{y}\boldsymbol{\epsilon}
\end{equation}
with
\begin{equation}\label{eqn:vvecdef}
    y^{\mu} = g^{\mu\rho}g^{\nu\sigma}\left(\nabla_{\sigma}\frac{d}{d\lambda}g_{\nu\rho}-\nabla_{\rho}\frac{d}{d\lambda}g_{\nu\sigma}\right).
\end{equation}
We can now define the symplectic current density $d$-form
\begin{equation}
    \boldsymbol{\omega}\left[g;\frac{d}{d\lambda_{1}}g,\frac{d}{d\lambda_{2}}g\right] = \frac{d}{d\lambda_{1}}\boldsymbol{\theta}\left[g(\lambda);\frac{d}{d\lambda_{2}}g(\lambda)\right]-\frac{d}{d\lambda_{2}}\boldsymbol{\theta}\left[g(\lambda);\frac{d}{d\lambda_{1}}g(\lambda)\right]
\end{equation}
where we consider here a two-parameter family of metrics $g(\lambda_{1},\lambda_{2})$. To see that $\boldsymbol{\omega}$ is a closed form when the perturbations satisfy the linear Einstein equations, we use \eqref{eqn:varL} to obtain
\begin{equation}\label{eqn:omegaclosed}
   \mathbf{d}\boldsymbol{\omega}\left[g;\nabla_{\sigma}\frac{d}{d\lambda_{1}}g,\frac{d}{d\lambda_{2}}g\right] = \frac{d}{d\lambda_{1}}\frac{d}{d\lambda_{2}}\mathbf{L}[g]-\frac{d}{d\lambda_{2}}\frac{d}{d\lambda_{1}}\mathbf{L}[g] = 0.
\end{equation}
The explicit expression for the symplectic current density $d$-form is given by
\begin{equation}\label{eqn:omegaexplicit}
    \boldsymbol{\omega} = \frac{1}{\kappa^{2}}\iota_{w}\boldsymbol{\epsilon}
\end{equation}
where
\begin{equation}\label{eqn:wdef}
    w^{\mu} = P^{\mu\nu\rho\sigma\alpha\beta}\left(\frac{d}{d\lambda_{2}}g_{\nu\rho}\nabla_{\sigma}\frac{d}{d\lambda_{1}}g_{\alpha\beta}-\frac{d}{d\lambda_{1}}g_{\nu\rho}\nabla_{\sigma}\frac{d}{d\lambda_{2}}g_{\alpha\beta}\right)
\end{equation}
with
\begin{equation}\label{eqn:Pdef}
    P^{\mu\nu\rho\sigma\alpha\beta} = g^{\mu\alpha}g^{\beta\nu}g^{\rho\sigma}-\frac{1}{2}g^{\mu\sigma}g^{\nu\alpha}g^{\beta\rho}-\frac{1}{2}g^{\mu\nu}g^{\rho\sigma}g^{\alpha\beta}+\frac{1}{2}g^{\nu\rho}g^{\mu\sigma}g^{\alpha\beta}.
\end{equation}

Finally, we define the symplectic form on a partial Cauchy slice $b$ bounded by $\gamma^{(0)}$ and $B=b \cap \mathcal{B}$  by 
\begin{equation}\label{eqn:sympform}
    \Omega = \int_{b}\boldsymbol{\omega},
\end{equation}
which is independent of the choice of $b$ when evaluated on solutions of the linearized Einstein equations since $\mathbf{d}\boldsymbol{\omega}=0$, provided that $\omega$ falls off sufficiently fast at $\partial b$ for such solutions.

\paragraph{Noether charge for diffeomorphism.}
Diffeomorphism are a gauge symmetry of the theory of Einstein gravity, and the corresponding Noether current is of crucial importance in the following discussion. Given a family of metrics $g(\lambda)$ and a smooth vector field $u^{a}$ (assumed to be $\lambda$ independent at the moment), one can define the Noether current $d$-form
\begin{equation}\label{eqn:Noethercurrent}
    \mathbf{J}_{(u)} = \boldsymbol{\theta}\left[g;\pounds_{u}g_{\mu\nu}\right]-\iota_{u}\mathbf{L}[g],
\end{equation}
which is conserved ($\mathbf{d}\mathbf{J}_{(u)}=0$) when $g_{\mu\nu}$ satisfies the Einstein equations. 

The Noether current can actually be expressed in terms of the constraints of the theory. 
(Components of) Einstein equations can be considered as the constraint of the theory,\footnote{The Hamiltonian and diffeomorphism constraints of general relativity are the $t$ components of the Einstein equations, $E_{\mu t}=0$, which can be seen, for instance, in the ADM formalism by observing that the Lagrangian has no dependence on time derivatives of the time components of the induced metric $\sigma_{\mu t}$ on a spatial slice (up to boundary terms).}
and thus we define the constraint $d$-form
\begin{equation}\label{eqn:constraintform}
    \mathbf{C}_{(u)} = \frac{2}{\kappa^{2}}\iota_{u \cdot E}\boldsymbol{\epsilon}, \qquad u \cdot E \equiv u^{\mu}E_{\mu}^{\nu}\,.
\end{equation}
One can show that $\mathbf{d}\mathbf{J}_{(u)} = \mathbf{d} \mathbf{C}_{(u)}$, so that the Noether current must be written as
\begin{equation}\label{eqn:Noethercurrent_constraints}
    \mathbf{J}_{(u)} = \mathbf{C}_{(u)} + \mathbf{d}\mathbf{Q}_{(u)}.
\end{equation}
where the Noether charge $\mathbf{Q}$ for a diffeomorphism $u$ is a  $(d-1)$-form
\begin{equation}\label{eqn:Noethercharge}
    (\mathbf{Q}_{(u)})_{\alpha_{1}\ldots\alpha_{d-1}} = -\frac{1}{\kappa^{2}}\nabla^{(g)\mu}u^{\nu}(\boldsymbol{\epsilon})_{\mu\nu\alpha_{1}\ldots\alpha_{d-1}}.
\end{equation}

Taking the first variation of \eqref{eqn:Noethercurrent} and using Cartan's magic formula, one finds
\begin{equation}\label{eqn:Noethercurrent_1stordervar}
    \frac{d}{d\lambda}\mathbf{J}_{(u)} = \boldsymbol{\omega}\left[g;\frac{d}{d\lambda}g,\pounds_{u}g\right] + \mathbf{d}\iota_{u}\boldsymbol{\theta}\left[g;\frac{d}{d\lambda}g_{\mu\nu}\right]-\iota_{u}\boldsymbol{E}^{\mu\nu}[g]\frac{d}{d\lambda}g_{\mu\nu}.
\end{equation}
Doing a similar variation for \eqref{eqn:Noethercurrent_constraints} and comparing with \eqref{eqn:Noethercurrent_1stordervar} gives
\begin{equation}\label{eqn:symplecticcurrent_Noetherchargereln}
    \boldsymbol{\omega}\left[g;\frac{d}{d\lambda}g,\pounds_{u}g\right] = \frac{d}{d\lambda}\mathbf{C}_{(u)} + \iota_{u}\boldsymbol{E}^{\mu\nu}[g]\frac{d}{d\lambda}g_{\mu\nu} + \mathbf{d}\left(\frac{d}{d\lambda}\mathbf{Q}_{(u)}-\iota_{u}\boldsymbol{\theta}\left[g(\lambda);\frac{d}{d\lambda}g_{\mu\nu}(\lambda)\right]\right).
\end{equation}
This is the fundamental identity that we will use to relate bulk and boundary charges.

\paragraph{First-order variation of charges.}
As a first application of the Noether charge analysis, we derive (a generalization of) the first law of black hole mechanics. 

We now make the assumption that $g(\lambda)$ is a family of asymptotically AdS metrics all satisfying the Einstein equations, and $g_{\mu\nu}(\lambda=0)$ is a stationary black hole spacetime with bifurcation surface $\gamma^{(0)}$.
Let $u^{\mu}$ be the asymptotically timelike Killing vector $\xi^{\mu}$ of the $\lambda=0$ spacetime. We can integrate \eqref{eqn:symplecticcurrent_Noetherchargereln} over $b$ to obtain
\begin{align}\label{eqn:symplecticform_Noetherchargereln}
\begin{split}
    \Omega\left[g;\frac{d}{d\lambda}g,\pounds_{\xi}g\right] &= \int_{B}\left(\frac{d}{d\lambda}\mathbf{Q}_{(\xi)}-\iota_{\xi}\boldsymbol{\theta}\left[g(\lambda);\frac{d}{d\lambda}g_{\mu\nu}(\lambda)\right]\right) 
    \\	&\qquad - \int_{\gamma^{(0)}}\left(\frac{d}{d\lambda}\mathbf{Q}_{(\xi)}-\iota_{\xi}\boldsymbol{\theta}\left[g(\lambda);\frac{d}{d\lambda}g_{\mu\nu}(\lambda)\right]\right).
\end{split}
\end{align}
The first term on the righthand side is actually the $\lambda$ derivative of the boundary conserved quantity (CFT charge) $K_{(u)}^{\partial}$ for the asymptotic Killing vector $\xi^{\mu}|_{\partial}$ generating an isometry of the asymptotic boundary $\mathcal{B}$, viz.,
\begin{equation}\label{eqn:derivativeofADMcharge}
    \frac{d}{d\lambda}K_{(\xi)} = \int_{B}\left(\frac{d}{d\lambda}\mathbf{Q}_{(\xi)}-\iota_{\xi}\boldsymbol{\theta}\left[g(\lambda);\frac{d}{d\lambda}g_{\mu\nu}(\lambda)\right]\right).
\end{equation}
For example, when the isometry is boundary time translation, then $K_{(\xi)}^{\partial}$ is the CFT Hamiltonian.

For perturbations around $\lambda=0$, the lefthand side of \eqref{eqn:symplecticform_Noetherchargereln} vanishes since $\xi^{\mu}$ is a Killing vector for $g_{\mu\nu}(\lambda=0)$. Furthermore, $\xi^{\mu}$ vanishes on the bifurcation surface (or is tangent in the rotating case) so $\iota_{\xi}\boldsymbol{\theta}|_{\gamma^{(0)}}=0$. Finally, one can show using \eqref{eqn:Noethercharge} that the integral of the Noether charge over the bifurcation surface gives the area of this surface. Therefore, we have derived the first law of black hole mechanics
\begin{equation}\label{eqn:blackhole1stlaw}
    \delta K_{(\xi)}^{\partial} = \frac{2\mathfrak{s}}{\kappa^{2}}\delta A[g,\gamma^{(0)}],
\end{equation}
where $\mathfrak{s}$ is the surface gravity of $\gamma^{(0)}$ for $g_{\mu\nu}(\lambda=0)$. This relates the Noether charge in the bulk, given by the area of $\gamma^{(0)}$, to the CFT charge corresponding to the asymptotic Killing vector $\xi^{\mu}|_{\partial}$.\footnote{Strictly speaking, this is only the first law of black hole mechanics if $g(\lambda=0)+\delta g$ is also a stationary black hole spacetime. Nevertheless, the righthand side compares the area of the same surface $\gamma^{(0)}$ in the two different metrics, even if it is not meaningful to call it the bifurcation surface for $g(\lambda=0)+\delta g$.} For the metric expansion of interest
\begin{equation}\label{eqn:pertmetriclambda}
g_{\mu\nu}(\lambda) = G_{\mu\nu}+\lambda\kappa h_{\mu\nu}^{\mathrm{HW}} + (\lambda\kappa)^{2} g_{\mu\nu}^{(2)} + O\left((\lambda\kappa)^{3}\right),
\end{equation}
where $\mathrm{HW}$ denotes Hollands-Wald gauge, this becomes\footnote{Here $\delta$ represents $\frac{d}{d\lambda}|_{\lambda=0}$ so it produces a tangent vector on the target space of the field.} 
\begin{equation}\label{eqn:blackhole1stlaw_h}
\delta K_{(\xi)}^{\partial} = \frac{2\mathfrak{s}}{\kappa}A^{\mathrm{lin}}[h^{\rm HW},\gamma^{(0)}],
\end{equation}
where $A^{\mathrm{lin}}[h^{\rm HW},\gamma^{(0)}]$ is equal to $A[g,\gamma^{(0)}]$ expanded to linear order in $h$. This is the contribution to generalized entropy that makes it parametrically large for perturbative states.

Our goal is to understand second-order variations of thermodynamic quantities, so we actually need a formula like \eqref{eqn:blackhole1stlaw} for variations around any value of $\lambda$. Our choice of Gaussian null coordinates means that the area of $\gamma$ is the integral of the Noether charge on $\gamma$ for all $\lambda$. However, the symplectic form $\Omega$ does not vanish because $\xi^{\mu}$ is not necessarily a Killing vector for $g(\lambda)$. Therefore, \eqref{eqn:symplecticform_Noetherchargereln} becomes
\begin{equation}\label{eqn:thermoreln1storder}
    \frac{d}{d\lambda}K_{(\xi)}^{\partial} = \frac{2\mathfrak{s}}{\kappa^{2}}\frac{d}{d\lambda} A[g,\gamma] + \Omega\left[g;\frac{d}{d\lambda}g,\pounds_{\xi}g\right].
\end{equation}

\paragraph{Second-order variations and canonical energy.}

Consider another $\lambda$ derivative of \eqref{eqn:thermoreln1storder}  around $\lambda=0$
\begin{equation}\label{eqn:thermoreln2ndorder}
    \delta^{2}K_{(\xi)}^{\partial} = \frac{2\mathfrak{s}}{\kappa^{2}}\delta^{2}A[g,\gamma] + \Omega\left[g;\delta g,\pounds_{\xi}\delta g\right].
\end{equation}
The symplectic inner product appearing on the righthand side is known as the canonical energy
\begin{equation}\label{eqn:AdSRindlercanonicalenergy}
    E_{\mathrm{can}}[\delta g] = \Omega\left[g;\delta g,\pounds_{\xi}\delta g\right].
\end{equation}
Therefore, using our expansion \eqref{eqn:pertmetriclambda}, we arrive at the desired relation between boundary and bulk charges to second-order:
\begin{equation}\label{eqn:bdytobulkcharges_2ndorder}
\delta^{2}K_{(\xi)}^{\partial} = {2\mathfrak{s}}\left(A^{\mathrm{lin}}[g^{(2)},\gamma^{(0)}] +A^{\mathrm{quad}}[h^{\mathrm{HW}},\gamma^{(0)}] \right) + E_{\mathrm{can}}[\kappa h^{\mathrm{HW}}]
\end{equation}
where $A^{\mathrm{lin}}[g^{(2)},\gamma^{(0)}]$ and $A^{\mathrm{quad}}[h^{\mathrm{HW}},\gamma^{(0)}]$ are the expansions of $A[g,\gamma^{(0)}]$ to linear order in $g^{(2)}$ and quadratic order in $h$, respectively. Note that the contribution of the canonical energy $E_{\mathrm{can}}[h]$ is $O(\kappa^{0})$ because it contains a factor of $1/\kappa^{2}$ via \eqref{eqn:omegaexplicit}.

\subsection{Restoring gauge-invariance}
\label{sec:restoregaugeinv}

The relation between bulk and boundary charges \eqref{eqn:thermoreln2ndorder} is not gauge-invariant and depends crucially on the gauge choices made. Neither the area term nor the canonical energy are invariant under gauge transformations that do not fall off sufficiently fast at $\partial b = \gamma^{(0)}\cup B$.\footnote{There is one exception to this statement given by diffeomorphisms $u^{\mu}$ which are tangent to the null generators of $\mathfrak{H}^{+}$, see Lemma 2 in \cite{Hollands:2012sf}.} 
Recall from the discussion in \S\ref{sec:diffs} that small gauge transformations falling off fast enough at asymptotic infinity are redundancies for the graviton theory defined on the full Cauchy slice $\Sigma$.
These small gauge transformations can have non-vanishing profile on $\gamma^{(0)}$ so that the area $A[g,\gamma^{(0)}]$ and the canonical energy $E_{\rm can}[h]=\int_b \omega(G,h,\pounds_\xi h)$ are both affected. We now explain how to make both of these quantities gauge-invariant.

The area of a given surface will naturally be a gauge-invariant quantity as long as the surface is defined in a coordinate-independent way. 
As explained in our prescription for the generalized entropy in \S\ref{sec:prescript}, 
one should extremize the area $A[G+\kappa h, \gamma^{(0)}+\kappa \gamma^{(1)}]$ to find the extremal surface perturbation $\gamma^{(1)}$. This is equivalent to saying that $\gamma^{(0)}+\kappa \gamma^{(1)}$ is extremal in the metric $G+\kappa h$. 
Now the second order area change is 
\begin{equation}
    \delta^2 A[g,\gamma] = A^{\mathrm{lin}}[g^{(2)},\gamma^{(0)}] + A^{\mathrm{quad}}[h,\gamma^{(0)}] + A^{\mathrm{quad}}[G,\gamma^{(1)}] +  A^{\mathrm{lin}}[h,\gamma^{(1)}]\,.
\end{equation}

A gauge-invariant version of the canonical energy can be derived as follows \cite{Lashkari:2015hha}. Consider the diffeomorphism $v^{\mu}$ 
that goes from an arbitrary gauge $h_{\mu\nu}$ to Hollands-Wald gauge $h_{\mu\nu}^{\mathrm{HW}}$
\begin{equation}\label{eqn:goingtoHWgauge}
    h_{\mu\nu}^{\mathrm{HW}} = h_{\mu\nu}+\pounds_{v}G_{\mu\nu}.
\end{equation}
Then the canonical energy in the new gauge becomes
\begin{align}\label{eqn:Ecan_gaugetrans}
\begin{split}
    E_{\mathrm{can}}[h^{\rm HW}] &= \Omega\left(G;h+\pounds_{v}G,\pounds_{\xi}\left(h+\pounds_{v}G\right)\right) 
    \\   &= E_{\mathrm{can}}[h]+\int_{b}\boldsymbol{\omega}\left(G;h+\pounds_{v}G,\pounds_{[\xi,v]}G\right)-\int_{b}\boldsymbol{\omega}\left(G;\pounds_{\xi}h,\pounds_{v}G\right)
    \\  &= E_{\mathrm{can}}[h]+\int_{\gamma^{(0)}}\boldsymbol{\Upsilon}
\end{split}
\end{align}
where in the third line we used \eqref{eqn:symplecticcurrent_Noetherchargereln} and we dropped the boundary term at asymptotic infinity due to the fall off of $h_{\mu\nu}$ at infinity, e.g., $h_{\mu\nu} \sim r^{-(d-2)}$ in global coordinates, and we defined the $(d-1)$-form
\begin{equation}
\\ \boldsymbol{\Upsilon} \equiv \mathbf{Q}_{([\xi,\kappa v])}[\kappa h+\pounds_{\kappa v}G]-\iota_{[\xi,\kappa v]}\boldsymbol{\theta}\left[G;\kappa h+\pounds_{\kappa v}G\right]-\left(\mathbf{Q}_{(\kappa v)}[\pounds_{\xi}\kappa h]-\iota_{\kappa v}\boldsymbol{\theta}\left[G;\pounds_{\xi}\kappa h\right]\right).
\end{equation}
The canonical energy written as \eqref{eqn:Ecan_gaugetrans} is trivially gauge-invariant: in any gauge $h$ it computes $E_{\mathrm{can}}[h^{\rm HW}]$.

Thus, we have arrived at the gauge-invariant formulation of \eqref{eqn:bdytobulkcharges_2ndorder}
\begin{equation}\label{eqn:bdytobulkcharges_2ndorder_gaugeinv}
\delta^{2}K_{(\xi)} = 2\mathfrak{s}\left(A^{\mathrm{lin}}[g^{(2)},\gamma^{(0)}] + A^{\mathrm{quad}}[h,\gamma^{(0)}] + A^{\mathrm{quad}}[G,\gamma^{(1)}] +  A^{\mathrm{lin}}[h,\gamma^{(1)}]\right) + E_{\mathrm{can}}[\kappa h] + \int_{\gamma^{(0)}}\boldsymbol{\Upsilon}
\end{equation}
where $\gamma^{(0)}+\kappa \gamma^{(1)}$ is the extremal surface for $G+\kappa h$, that is, $\gamma^{(1)}$ extremizes the area at $O(\kappa^{2})$. We would like to understand how to promote  this result to the quantum level to obtain a formula for gravitons that is analogous to the JLMS formula \eqref{eqn:JLMS} for ordinary matter.

Some comments on the restoration of gauge-invariance are in order. First, the specialty of the Hollands-Wald gauge conditions 
in making the phase space analysis simple comes from two facts: (1) the Noether charge $\int_{\gamma^{(0)}}\mathbf{Q}$ equals to the area and (2) there is no additional boundary term $\Upsilon$ in defining the canonical energy. Neither of them depend on the Gaussian null coordinate, but the extremal gauge condition $\delta K^{(1)} = 0$ is crucially important.
We provide a proof to this in App.~\ref{sec:detailcanonicalenergy} and \ref{sec:areaequalsNoethercharge}.
Second, it is interesting to observe that the master formula \eqref{eqn:bdytobulkcharges_2ndorder_gaugeinv} holds true even if one performs large diffeomorphisms with respect to asymptotic infinity, as long as some subtleties from large gauge transformations are addressed properly, see \S\ref{sec:HWgaugemore}. 

\subsection{Alternate relation}
\label{sec:altprescript}

Let us explain how, as pointed out by JLMS \cite{Jafferis:2015del}, one can use the Hollands-Wald formalism to derive an alternate formula for the boundary charges.
The expression, which we will justify shortly, is 
\begin{equation}\label{eqn:JLMS2}
\delta^{2}K_{(\xi)} = 2\mathfrak{s} A^{\mathrm{lin}}[g^{(2)},\gamma^{(0)}] + \int_{b} d^{d}x\,\sqrt{-G_{b}}\,T_{\mu\nu}^{\mathrm{grav}} \tau^{\mu}\xi^{\nu}\,,
\end{equation}
where $T^{\rm grav}_{\mu\nu}$ can be thought of as a ``stress-energy tensor'' for the graviton\footnote{We emphasize that this is not a stress-tensor in the usual sense for a quantum field theory: it comes from the variation with respect to $h_{\mu\nu}$ of the cubic interaction term like $h^3$ in the graviton action, not variation of the quadratic action with respect to the background metric as a usual matter field.}
\begin{equation}\label{eqn:Tgravdef}
    T_{\mu\nu}^{\mathrm{grav}} \equiv -E_{\mu\nu}^{\mathrm{quad}}[\kappa h] 
\end{equation}
in the sense that it sources the backreaction, and $g_{\mu\nu}^{(2)}$ is the backreacted metric that solves the linearized Einstein equations sourced by $T^{\rm grav}_{\mu\nu}$. 
%

One can derive \eqref{eqn:Tgravdef} by finding a relationship between $A^{\mathrm{quad}}$, $E_{\mathrm{can}}$, and $\int T^{\mathrm{grav}}$ as was done in \cite{Hollands:2012sf}. However, we will derive it in a more direct way that never uses $A^{\mathrm{quad}}$, and hence never needs the perturbed surface $\gamma^{(1)}$. Consider the diffeomorphism charge for the two metrics: (1) $g_A=G+\lambda \kappa h + (\lambda \kappa)^2 \delta g^{(2)}$, which is on-shell up to $O(\kappa^2)$, and (2) $g_B=G+\lambda \kappa h $, ignoring $g^{(2)}$, which only satisfies the Einstein equations to $O(\kappa)$. 
We start by integrating the general identity \eqref{eqn:symplecticcurrent_Noetherchargereln} over $b$, without yet assuming the metric is on-shell, to obtain an identity for the symplectic form:
\begin{align}\label{eqn:symplecticform_Noetherchargereln_h}
\begin{split}
    {\Omega}\left[g;\frac{d}{d\lambda} g ,\pounds_{\xi}g\right] &= \int_{b}\left(\frac{d}{d\lambda}\mathbf{C}_{(\xi)} + \iota_{\xi}\boldsymbol{E}^{\mu\nu}[g]\frac{d}{d\lambda} g_{\mu\nu}\right) + \int_{B}\left(\frac{d}{d\lambda}\mathbf{Q}_{(\xi)}-\iota_{\xi}\boldsymbol{\theta}\left[g;\frac{d}{d\lambda} g \right]\right) 
    \\  &\qquad -\int_{\gamma^{(0)}}\left(\frac{d}{d\lambda}\mathbf{Q}_{(\xi)}-\iota_{\xi}\boldsymbol{\theta}\left[g;\frac{d}{d\lambda} g\right]\right),
\end{split}
\end{align}
and then to take a $\lambda$ derivative and set $\lambda=0$.

For both $g_A$ and $g_B$, the LHS of \eqref{eqn:symplecticform_Noetherchargereln_h} gives ${\Omega}\left[g;\frac{d}{d\lambda} g ,\pounds_{\xi}\frac{d}{d\lambda}g\right]\big|_{\lambda=0}$. 
The RHS for $g_A$ involves only the charge terms on $\gamma^{(0)}$ and $B$, while for $g_B$ there is an additional $\delta^{2}\mathbf{C}_{(\xi)}[g_B]$ term because of our on-shell assumptions for the two metrics. Given the definition of $\mathbf{C}_{(\xi)}[g_B]$ in \eqref{eqn:constraintform}, we see that 
\begin{equation}
\int_b \delta^{2}\mathbf{C}_{(\xi)}[g_B] = \int_b d^{d}x\,\sqrt{-G_{b}}\, T_{\mu\nu}^{\mathrm{grav}}\tau^{\mu} \xi^{\nu}.
\end{equation}
Taking the difference between the corresponding equations for $g_{A}$ and $g_{B}$, we arrive at
\begin{equation}
	0 = -\int_b d^{d}x\,\sqrt{-G_{b}}\, T_{\mu\nu}^{\mathrm{grav}}\tau^{\mu} \xi^{\nu} + \int_B \delta^2 (\mathbf{Q}_{(\xi)}[g_A]-\mathbf{Q}_{(\xi)}[g_B])   -  \int_{\gamma^{(0)}} \delta^{2} (\mathbf{Q}_{(\xi)}[g_A]-\mathbf{Q}_{(\xi)}[g_B]).
\end{equation}
One can check that the asymptotic boundary charge difference is $\delta^2 K_{\partial}$ because $\int_B\mathbf{Q}_{(\xi)}[g_{B}]$ is actually zero, stemming from the fast fall-off of quadratic terms in $h$ at asymptotic infinity.  
It remains to show that the difference of charges at the bifurcation surface gives $A^{\rm lin}[g^{(2)}]$. 

To prove this, we write down the explicit form of the charge $\mathbf{Q}$ given in \eqref{eqn:Noethercharge}. 
We wish to analyse the difference between the Noether charge $\int\mathbf{Q}$ and the area $A$: 
\begin{equation}
    \int_{\gamma^{(0)}} \mathbf{Q} = \int_{\gamma^{(0)}} \nabla_{\mu} \xi^{\nu} g^{\mu\rho}\sigma_{\rho\nu} \sqrt{q} \quad \mathrm{and} \quad A = \int_{\gamma^{(0)}} \sqrt{q} 
\end{equation}
where $\sigma_{\mu\nu}$ is the binormal of the codimension-2 surface $\gamma^{(0)}$.
After some calculation,  one can show that 
\begin{equation}\label{eqn:secondorderQandA}
    \int_{\gamma^{(0)}} \delta^{2} (\mathbf{Q}_{(\xi)}[g_A]-\mathbf{Q}_{(\xi)}[g_B]) = \int_{\gamma^{(0)}} \sigma^{(0),\nu}_{\mu} \delta_{\kappa^2 g^{(2)}}(\sigma^{\mu}_{~\nu})\sqrt{q} + \delta^{2}\left(A[g_A,\gamma^{(0)}] - A[g_B,\gamma^{(0)}]\right)
\end{equation}
where 
the superscript $(0)$ again denotes the unperturbed ones, and $\delta_{\kappa^2 g^{(2)}}$ represents the contribution from $g^{(2)}$. 
The discrepancy term vanishes because of the unit normal condition (see App.~\ref{sec:detailcanonicalenergy} for more details), 
so that the RHS of \eqref{eqn:secondorderQandA} is nothing but $A^{\rm lin}[g^{(2)},\gamma^{(0)}]$.

Thus, we have shown, using the difference of two phase space identities, that \eqref{eqn:JLMS2} is true. Observe that the validity of \eqref{eqn:JLMS2} is independent\footnote{Although the derivation was done without choosing any specfic gauge, $g^{(2)}_{\mu\nu}$ and $T^{\rm grav}_{\mu\nu}$ both depend on the gauge choice of $h_{\mu\nu}$. Nevertheless, the sum of the two must be invariant under small diffeomorphisms because the boundary charge is invariant. It is more non-trivial to show invariance under large gauge transformations using phase space techniques.} of the gauge choice for $h_{\mu\nu}$ because the derivation erases any gauge dependent piece by taking the difference of two phase space identities. 


\subsection{JLMS formula for gravitons}
\label{sec:JLMSgraviton}
As outlined at the beginning of this section, to promote our classical results \eqref{eqn:blackhole1stlaw_h} and \eqref{eqn:bdytobulkcharges_2ndorder_gaugeinv} to a quantum result \eqref{eqn:JLMS0}, we need to identify all of the terms in the phase space discussion with the expectation values of quantum operators. 
We treat the CFT modular Hamiltonian, the area term and the bulk modular Hamiltonian separately.

\paragraph{CFT modular Hamiltonian.}
First, let us discuss the relation between the conserved charge defined on the asymptotic boundary $K_{(\xi)}^{\partial}$ and the CFT modular Hamiltonian.
At the classical level, we see that the Killing vector $\xi^{\mu}$ asymptotes to a timelike asymptotic Killing vector, and the corresponding charge is $K_{(\xi)}^{\partial}$.
In the quantum theory, such a charge naturally corresponds to a time-independent ``Hamiltonian" operator that generates the timelike asymptotic Killing symmetry along $\xi^{\mu}|_{\partial}$, which is an integral of the CFT stress-tensor. 

It is known that in such cases, this exact operator will be the modular Hamiltonian $K_{B}=-\log\omega_{B}$ for the CFT vacuum state $\psi_{0}$ dual to the background asymptotically AdS geometry \cite{Wong:2013gua}.\footnote{Strictly speaking, the one-sided modular Hamiltonian $\hat{K}_{B}^{\psi}$ is not a well-defined operator in the continuum and thus requires some regulator. The actual well-defined operator in the continuum is the two-sided modular Hamiltonian $\hat{K}_{B}^{\psi}-\hat{K}_{B^{c}}^{\psi}$. However, we will consider vacuum-subtracted expectation values of $\hat{K}_{B}^{\psi}$ which will be finite and manifestly regulator-independent.} 
Examples include (1) the thermal state in the CFT with $B$ equal to the entire spatial slice where $K_{B}^{\mathrm{th}}$ is the Hamiltonian that generates time translation and (2) the vacuum state for a polar cap subregion on the cylinder which can be conformally mapped to hyperbolic space where the modular Hamiltonian is again time translation, which is the case that we are interested in.

We now elaborate on how the expectation value maps to the classical charge. The CFT states dual to bulk graviton excitations are stress-tensor excited states, so the expectation value calculation boils down to stress-tensor correlators. 
A careful analysis of correlators shows that the contribution to $O(\sqrt{C_T})$ and $O(1)$ only involves two- and three-point functions,\footnote{To be more explicit, we can consider normalised multi-stress-tensor state $\prod_{i=1}^k \frac{1}{\sqrt{C_T}} (\epsilon_i\cdot \hat{T})|0\rangle$. 
There is one stress-tensor in the modular Hamiltonian, and $2k$ from the states. There are $(k-1)$ two-point functions and one three-point function, contributing $C_T^k$ to the stress-tensor correlator. There is a $C_T^{-k}$ factor from the normalisation of states. This gives the diconnected, yet leading in $C_T$, piece of the $(2k+1)$-point function of stress-tensor. The connected pieces are more suppressed by $C_T$ so that they do not contribute to $O(C_T^0)$.}
so that only single stress-tensor states (and their superposition) are required to test the relation between the classical charge and quantum expectation value. We explicitly test this in \S\ref{sec:ex}.

\paragraph{Area correction.}
Although an in-depth discussion of the area was already given in the prescription \S\ref{sec:prescript}, for completeness we also give a brief summary here.
The area contribution on the RHS of \eqref{eqn:blackhole1stlaw_h} and \eqref{eqn:bdytobulkcharges_2ndorder_gaugeinv} is the extremal area for any specific perturbation $h_{\mu\nu,q}$ satisfying the linearized Einstein equations. 
The classical area has an expansion in terms of $\kappa$, and it is natural to promote the metric perturbation $h$, $g^{(2)}$ and the surface perturbation $\gamma^{(1)}$ to operators. 
Up to the order that we are interested in, the area contribution is at most quadratic in the graviton modes, so that it is also sufficient to consider single graviton states and their superposition (multi-particle states can be easily dealt with in the same way as multi-stress-tensor states discussed above).
For a single graviton mode, Wick contraction in the quantum operator expectation value can pick out the aforementioned classical result. 
Note that, as given in our prescription, we do not need to specify the gauge of $h$ and the area operator can be written as $A[\hat{g},\hat{\gamma}]$ to maintain gauge invariance. 

\paragraph{Canonical energy.}

Comparing the phase space identity \eqref{eqn:bdytobulkcharges_2ndorder_gaugeinv} with the quantum formula \eqref{eqn:JLMS0}, one would like to identify $ E_{\mathrm{can}}[\kappa h] + \int_{\gamma^{(0)}}\boldsymbol{\Upsilon}$ to the expectation value change of vacuum modular Hamiltonian $\hat{K}_{b}$. 
However, it is not at all clear why this should be the case because $ E_{\mathrm{can}}[\kappa h] + \int_{\gamma^{(0)}}\boldsymbol{\Upsilon}$ is not the integral of a stress-tensor (see \S\ref{sec:altprescript} for further discussion on this) so an argument analogous to the CFT one does not apply. 
We will show shortly that in a particular gauge, it is rather straightforward to show that $E_{\rm can}$ equals to $\Delta \hat{K}_b$. 
Since both the classical quantity and the quantum expectation values are gauge-invariant (as will be explained shortly), the matching holds true for any gauge.  


To summarize how $ E_{\mathrm{can}}[\kappa h] + \int_{\gamma^{(0)}}\boldsymbol{\Upsilon}$ and $\Delta \hat{K}_{b}^{\Psi}$ are related, we have the following procedure
\begin{equation}
\begin{aligned}
      \Delta \hat{K}^{\Psi}_b \equiv - \Tr\big( (\rho_{b}^{\Psi}-\omega_b)\log \omega_b \big) \xrightarrow{\rm Killing\  symm.} \Big\langle \sum_{\lambda}\int d\omega \omega b^{\dagger}_{\omega,\lambda}b_{\omega,\lambda} \Big\rangle_{\Psi} \\
      \downarrow \text{\scalebox{0.8}{orthonormal basis}} \\
     \underbrace{\Omega\left( \sum_{\lambda} \int d\omega \beta^*_{\omega,\lambda} \mathpzc{h}^{\rm R}_{\omega,\lambda} ,\pounds_{\xi}\sum_{\lambda^{\prime}} \int d\omega^{\prime} \beta_{\omega^{\prime},\lambda^{\prime}} \mathpzc{h}^{{\rm R},*}_{\omega^{\prime},\lambda^{\prime}}\right)}_{E_{\rm can}[\kappa h^{\rm HW}]}
\end{aligned}
\end{equation}
Although we will use AdS-Rindler for concreteness, the argument works for any stationary black hole spacetime with an everywhere timelike Killing vector\footnote{For rotating asymptotically AdS black hole spacetimes, this requires that the angular velocities be sufficiently small \cite{Hawking:1999dp}. When such a Killing vector does not exist, there is no thermofield double state.} because none of the key ingredients in the above procedure depends on Rindler.

We first need to be careful about which gauge we choose and which boundary conditions we impose on the horizon. We will choose a Hollands-Wald gauge so that $\Upsilon=0$ and we impose Dirichlet boundary conditions on the horizon $\mathpzc{h}_{\rho \alpha}|_{\rho=0}=0$ where $\rho$ is the Rindler radial direction and $\alpha$ are tangent directions (strictly speaking, we put in a brick wall cut-off at $\rho=\epsilon$). An example of such a gauge that respects these boundary conditions is the Rindler radial gauge where $\mathpzc{h}_{\rho \mu}^{R}=0$ in $b$. The upshot is that we then we have an orthonormal basis with the inner product defined via the symplectic form without additional boundary terms, as explained in detail in App \ref{sec:bdyconditions}.

Starting from the definition of the vacuum modular Hamiltonian $\hat{K}_{b} = - \log \omega_b$ with $\omega_b$ vacuum density matrix for the (right) Rindler wedge $b$, $\hat K$ can be written down using the Rindler mode expansion:
\begin{equation}\label{eqn:modH_sum}
\hat{K}_b=\hat{H}_{\mathrm{ADM}}^{(\xi)}=\sum_\lambda \int d\omega \; \omega b^{\dagger}_{\omega,\lambda} b_{\omega,\lambda}+ \text{const.}
\end{equation}
where $\hat{H}_{\mathrm{ADM}}^{(\xi)}$ is the ADM Hamiltonian generating the Killing symmetry $\xi^{\mu}$. Thus, when $\Psi$ a single graviton excitation state  $a_q^{\dagger}|0\rangle$, the difference of the expectation values $\Delta \hat{K}^{\Psi}_b = -\mathrm{Tr}((\rho^{\Psi}_b - \omega_b) \log \omega_b)$ requires us to compute $\langle a_q b_{\omega,\lambda}^{\dagger} b_{\omega,\lambda} a^{\dagger}_q \rangle$. This computation can then be trivially generalized to any (multiparticle) state in the Fock space by appropriate Wick contractions.  
The expectation value thus requires a calculation of Bogoliubov coefficients $\alpha^{q}_{\omega,\lambda},\beta^q_{\omega,\lambda}$ relating the Rindler modes to global modes, see \S\ref{sec:gravitonvN}. We show in App.~\ref{sec:Bogcoeffs} that these Bogoliubov coefficients can be computed in a gauge-invariant way using the Weyl tensor, and hence $\Delta \hat{K}^{\Psi }_b$ is gauge-invariant.

Now we try to connect $\Delta \hat{K}^{\Psi}_b = \sum_{\lambda} \int d\omega \; \omega (|\alpha_{\omega,\lambda}|^{2}+|\beta_{\omega,\lambda}|^{2})$ with the canonical energy obtained in the phase space language. 
The key point in relating these two quantities is to introduce an orthonormal basis of Rindler wavefunctions $\mathpzc{h}^{\rm R}_{\omega,\lambda}$ satisfying $\pounds_\xi \mathpzc{h}^{\rm R}_{\omega,\lambda}=-i\omega \mathpzc{h}^{\rm R}_{\omega,\lambda}$ and $-i \Omega(\kappa \mathpzc{h}^{\rm R}_{\omega,\lambda},\kappa \mathpzc{h}^{{\rm R},*}_{\omega^{\prime},\lambda^{\prime}})=\delta_{\omega,\omega^{\prime}}\delta_{\lambda, \lambda^{\prime}}$, and at the same time satisfying the Hollands-Wald gauge condition.
Then
\begin{equation}\label{eqn:introduceOmega}
    \sum_{\lambda} \int d\omega \; \omega \left(|\alpha_{\omega,\lambda}|^{2}+|\beta_{\omega,\lambda}|^{2}\right) = \kappa^2 \Omega\left( \sum_{\lambda} \int d\omega \beta^*_{\omega,\lambda} \mathpzc{h}^{\rm R}_{\omega,\lambda} ,\pounds_{\xi}\sum_{\lambda^{\prime}} \int d\omega^{\prime} \beta_{\omega^{\prime},\lambda^{\prime}} \mathpzc{h}^{{\rm R},*}_{\omega^{\prime},\lambda^{\prime}}\right)+(\beta\rightarrow\alpha),
\end{equation}
takes the form of $E_{\rm can}[\kappa {h}^R]$ with ${h}^{\rm R} \equiv \sum_{\lambda} \int d\omega \beta_{\omega,\lambda} \mathpzc{h}^{{\rm R},*}_{\omega,\lambda} + (\beta\rightarrow \alpha)^*$. Thus, we conclude that 
\begin{equation}\label{eqn:EcanequalsmodH}
  \Delta \hat{K}_{b}^{\Psi} =  E_{\rm can}[\kappa {h}^{\rm R}] = E_{\mathrm{can}}[\kappa h] + \int_{\gamma^{(0)}}\boldsymbol{\Upsilon},
\end{equation}
where we have also made $E_{\rm can}$ manifestly gauge invariant in the last equation using the argument in \S\ref{sec:restoregaugeinv}. 



We now can promote the classical phase space identity \eqref{eqn:blackhole1stlaw_h} and \eqref{eqn:bdytobulkcharges_2ndorder_gaugeinv} to the quantum level to get our JLMS formula for gravitons
\begin{equation}\label{eqn:JLMSgraviton}
\Delta \hat{K}_{B}^{\psi} = \frac{\Delta \langle A[\hat{g},\hat{\gamma}]\rangle_{\Psi}}{4G_N}+\Delta\hat{K}_{b}^{\Psi}
\end{equation}
for the difference of expectation values with any state $\Psi$ in the code subspace. This gives a partial justification for our proposal for the generalized entropy because 
of the first law of entanglement \cite{Blanco:2013joa} gives $\Delta S = \Delta K + O((\delta \rho)^{2})$. Therefore, the result \eqref{eqn:JLMSgraviton} justifies our prescription for the generalized entropy of gravitons to leading order in $\delta \rho$ in the case that the classical entanglement wedge $\mathcal{D}[\gamma^{(0)}]$ has a Killing horizon, up to all of the subtleties of how to properly define vacuum-subtraction in light of the divergences which we address in the next section. 

\paragraph{Alternative form.}
We can also promote \eqref{eqn:JLMS2} to a quantum formula
\begin{equation}\label{eqn:JLMSgraviton2}
\Delta \hat{K}^{\psi }_{B}= \Delta \Big\langle A[\hat{g},\gamma^{(0)}] +\int_{b} d^{d}x\,\sqrt{-G_{b}} \, \normord{T_{\mu\nu}^{\mathrm{grav}}[h]}\tau^{\mu}\xi^{\nu} \Big\rangle_\Psi.
\end{equation}
This looks more like the JLMS formula for matter \eqref{eqn:JLMS} in the sense that it takes the form of linearized area for the unperturbed surface $\gamma^{(0)}$ and, for ordinary matter fields for the backgrounds considered here, the integral of the matter stress-tensor will give the modular Hamiltonian for the background state. However, as we have seen, for gravitons the integral of the graviton ``stress-tensor'' is not the vacuum modular Hamiltonian. Nevertheless, in practice, it is much easier than \eqref{eqn:JLMSgraviton} to compute because one does not need to either solve for the perturbed surface $\gamma^{(1)}$ or go to the Hollands-Wald gauge.

\subsection{Large gauge transformations  and IR divergences}
\label{sec:HWgaugemore}

In the argument in the last subsection, the fact that both the classical and the quantum equation are gauge-invariant under small gauge transformations played an important role.
Therefore, it may be worrisome that the gauge transformation between Hollands-Wald gauge and an arbitrary gauge is in general a large diffeomorphism. 
We will argue that the validity of the equations is not jeopardized by large gauge transformations at asymptotic infinity. 
This requires that we also address the subtleties of IR divergences.

The boundary conditions on the graviton at asymptotic infinity are dictated by the extrapolate dictionary that relates the graviton to the CFT stress-tensor: as $r \to \infty$
\begin{equation}\label{eqn:extrapolatedictionary}
\lim_{r \to \infty}r^{d-2}h_{\mu\nu}(t,r,\Omega) = T_{\mu\nu}(t,\Omega), \qquad h_{\mu r} \sim \frac{1}{r^{d-1}}
\end{equation}
so any diffeomorphism that affects these boundary conditions is large.

To give a heuristic understanding of why the large diffeomorphisms appear, we observe that for a perturbation $h$ in a given gauge, the diffeomorphism $v^{\mu}$ that transforms to the Hollands-Wald gauge moves the new extremal surface $\gamma^{(0)}+\kappa \gamma^{(1)}$ for $G+\kappa h$ back to $\gamma^{(0)}$. 
Since $\nabla_{(\mu} v_{\nu)}$ counteracts the effect of $h_{\mu\nu}$ on the extremal surface which is non-compact, they may be comparable at large $r$, in which case $v$ is a large diffeomorphism. Indeed, we will show that this actually happens in a simple case, see  \S\ref{sec:areaop}.

Due to their slow fall-off at infinity, the invariance of boundary Noether charge and the quantum conterpart, the CFT modular Hamiltonian, is not entirely obvious.
For the classical boundary charge, this is purely because of the large $r$ behavior of the metric.\footnote{For order $O(\kappa^{-1})$, we are interested in the $O(\kappa)$ part of the metric which behaves as $O(r^{2-d})$. This part does get affected, but including the shift of the cut-off surface discussed below will cancel the contribution. At $O(\kappa^{0})$, these possible large diffeomorphisms get squared so that they drop-off at large $r$.} 
On the other hand, the bulk large gauge transformation corresponds to a boundary conformal transformation (or boundary diffeomorphism). Without any conformal anomaly, the argument for why $\Delta K$ (or $\Delta S$) is invariant is easy: the conformal transformation acts as a unitary $U (\cdot)U^\dagger$ on operators and such a unitary preserves the trace ($\Delta K = \Delta \Tr(\rho \hat{K})$). The possible anomaly term in even dimensions can be computed and checked to not contribute.

We must also address why the variation of the extremal area does not get affected by the large diffeomorphisms.
Naively, this seems to be trivial since the extremized area should be invariant under all diffeomorphisms. 
However, the area of the extremal surface anchored on the asymptotic boundary has an IR divergence due to the infinite volume of AdS near the boundary. 
Then, when taking a difference, it is subtle to subtract off one infinity from another, and a naive treatment would make the result seemingly gauge dependent.\footnote{Note that in \cite{Sorce:2019zce}, it was argued that the entropy is cut-off covariant. 
However, there mostly the mutual information is considered and the area difference of two extremal surfaces with the same anchoring points is involved. This procedure washes out the anchoring point dependence. 
In our set-up, we are comparing the area of the same surface in different metrics. The anchoring points are different in the different metrics and give a finite contribution.}

To cure this, we introduce a cut-off surface $\mathcal{R}_{\rm cut-off}$ for which the induced metric is fixed. This surface should be regarded as the $d$-dimensional spacetime where the CFT lives so the induced metric is the standard metric on the cylinder $ \mathbb{R} \times S^{d-1}$. 
To each order in perturbations of the bulk metric, we find the new location of the cut-off surface by imposing the induced metric condition. 
It is natural to expect that with this regulator, one should compute the area of the portion of the extremal surface that lies inside $\mathcal{R}_{\rm cut-off}$. 

An extremal surface is uniquely defined when specifying its anchoring point (technically, a codimension-3 surface). 
We take this anchoring point to be $\partial \tilde{B}$, where $\tilde{B}$ is the ``boundary" subregion defined on $\mathcal{R}_{\rm cut-off}$.\footnote{Practically, we first pick a coordinate to write down the metric $ds^2_{\rm CFT}$ of the manifold on which the CFT lives.
Then we specify the actual boundary region $\tilde{B}$ by its coordinate location. 
Afterwards, we make the induced metric of $\mathcal{R}_{\rm cut-off}$ to be $ds^2_{\rm cut}=r_c^2 ds^2_{\rm CFT}$, and define $\tilde{B}$ with the same coordinate as $B$. One sometimes needs to translate this coordinate on $\mathcal{R}_{\rm cut-off}$ back to the bulk coordinate.}
The resulting extremal surface, denoted as $\gamma_{\rm cut}$ due to the cut-off procedure, is then the one that is needed for the area calculation, 
i.e., the area difference is $\Delta A = A\big[g_1,\gamma_{\rm cut}[g_1]\big] - A\big[g_2,\gamma_{\rm cut}[g_2]\big]$. 

This ends our discussion on the subtleties from the non-compact extremal surface, and we show explicitly how to deal with them when we turn to a concrete example in \S\ref{sec:areaop}.



\section{Example: simple excited states}
\label{sec:ex}

We now present a non-trivial example where we can justify all of the claims made in sections \S\ref{sec:prescriptgrav} and \S\ref{sec:HW}. We compute the vacuum-subtracted vN entropy for a polar cap region in the stress-tensor excited state in any holographic CFT dual to weakly-coupled Einstein gravity and then compute the vacuum-subtracted generalized entropy for the dual graviton state in AdS-Rindler using our prescription in \S\ref{sec:prescript}, finding exact agreement.

\subsection{Entanglement entropy in CFT$_{d}$}
\label{sec:CFTvN}

Consider a $d$-dimensional conformal field theory (CFT) on the cylinder $\mathbb{R} \times S^{d-1}$ for $d>2$. For now, we make no assumptions about the CFT and will only introduce them when needed. Every CFT has a stress-tensor operator $T_{\mu\nu}$ from which we can construct a primary excited state
\begin{equation}\label{eqn:stresstensorstate}
    \ket{\epsilon \cdot T}_{\mathbb{R} \times S^{d-1}} = \lim_{t_{E} = -\infty}e^{dt_{E}}\epsilon^{\mu\nu}T_{\mu\nu}(t_{E})\ket{0}_{\mathbb{R} \times S^{d-1}}
\end{equation}
with some symmetric traceless polarization tensor $\epsilon^{\mu\nu}$. The canonically normalised stress-tensor has a non-unit coefficient appearing in its two-point function, e.g., on $\mathbb{R}^{d}$ it is given by
\begin{align}\label{eqn:TTtwoptfn}
\begin{split}
\bra{0}T_{\mu\nu}(x)T_{\rho\sigma}(0)\ket{0}_{\mathbb{R}^{d}} &= \frac{C_{T}}{V_{S^{d-1}}^{2}}\frac{I_{\mu\nu,\rho\sigma}(x)}{x^{2d}}
\\	I_{\mu\nu,\rho\sigma}(x) &= \frac{1}{2}\left(I_{\mu\rho}(x)I_{\nu\sigma}(x)+I_{\mu\sigma}(x)I_{\nu\rho}(x)\right)-\frac{1}{d}\delta_{\mu\nu}\delta_{\rho\sigma}
\\	I_{\mu\nu}(x) &= \delta_{\mu\nu} - 2\frac{x_{\mu}x_{\nu}}{x^{2}},
\end{split}
\end{align}
where $V_{S^{d-1}}$ is the volume of $S^{d-1}$, which means the norm of the state $\ket{\epsilon \cdot T}_{\mathbb{R} \times S^{d-1}}$ is not $1$. Roughly speaking, $C_{T}$ measures the number of degrees of freedom in the CFT. To identify our CFT state with the dual single-particle graviton state in AdS, we need to rescale our state such that it has unit norm so we define
\begin{equation}\label{eqn:graviton_CFTdual}
    \ket{\epsilon \cdot \tilde{T}} = \frac{1}{\sqrt{\epsilon_{\mu\nu}^{\ast}\epsilon^{\mu\nu}}}\frac{V_{S^{d-1}}}{\sqrt{C_{T}}}\ket{\epsilon \cdot T},
\end{equation}
which has unit norm.

\begin{figure}[h]
\begin{centering}
\begin{tikzpicture}

\draw (-3,-2.5) -- (-3,2.5);
\fill[red!70!blue!30] (-3,0) circle (0.8mm);
\node at (-3.8,0) {$t_{E}=0$};
\draw[orange,thick] (-3.1,-2.6) -- (-2.9,-2.4);
\draw[orange,thick] (-3.1,-2.4) -- (-2.9,-2.6);
\draw[green,thick] (-3.1,2.6) -- (-2.9,2.4);
\draw[green,thick] (-3.1,2.4) -- (-2.9,2.6);
\node at (-3.6,-2.5) {$-\infty$};
\node at (-2.5,-2.3) {$\epsilon \cdot T$};
\node at (-3.6,2.5) {$+\infty$};
\node at (-2.5,2.8) {$\epsilon^{\ast} \cdot T$};
\node at (-3,-3) {$\mathbb{R}$};

\draw (0,0) -- (0,2);
\shade[ball color = blue!50, opacity = 0.4] (0,0) circle (2cm);
\draw (0,0) circle (2cm);
\draw (-2,0) arc (180:360:2 and 0.6);
\draw[dashed] (2,0) arc (0:180:2 and 0.6);

\fill[red!40, opacity=0.6] (-1.525,1.3) .. controls (-0.5,0.9) and (0.5,0.9) .. (1.525,1.3) .. controls (0.75,2.25) and (-0.75,2.25) .. (-1.525,1.3);
\draw[cyan,very thick] (-1.525,1.3) .. controls (-0.5,0.9) and (0.5,0.9) .. (1.525,1.3);
\draw[blue!50!black!50] (0,0) -- (1.525,1.3);
\node[red!30!black!70] at (0.2,1.5) {$\mathcal{B}$};

\draw[blue!50!black!50] (0.11,0.1) arc (30:80:0.15cm);
\node[purple!50!black!50] at (0.2,0.5) {$\theta_{0}$};
\node at (-1.5,-3) {$\times$};
\node at (0,-3) {$\mathbb{S}^{d-1}$};

\draw [->] (2.5,0) -- (4.5,0);
\node at (3.5,0.4) {CHM};

\draw (6.8,0) ellipse (0.5cm and 2cm); 
\node at (5.5,0) {$\tau_{H}=0$};
\node at (6.8,-3) {$S^{1}$};
\fill[red!80!blue!30] (6.3,0) circle (0.8mm);
\draw[green,thick] (7.35,-1) -- (7.15,-0.8);
\draw[green,thick] (7.35,-0.8) -- (7.15,-1);
\draw[orange,thick] (7.35,0.8) -- (7.15,1);
\draw[orange,thick] (7.35,1) -- (7.15,0.8);
\node at (7.8,1.4) {$\pi+\theta_{0}$};
\node at (7.8,-1.4) {$\pi-\theta_{0}$};

\fill[red!70!blue!30] (10.5,2) .. controls (7.8,1) and (7.8,-1) ..  (10.5,-2) .. controls (10.9,-1.6) and (10.9,1.6) .. (10.5,2);
\draw[cyan,very thick] (10.5,0) ellipse (0.3cm and 2cm);
\draw[green,thick] (8.38,-0.2) -- (8.58,0);
\draw[green,thick] (8.38,0) -- (8.58,-0.2);
\draw[orange,thick] (8.38,0.1) -- (8.58,-0.1);
\draw[orange,thick] (8.38,-0.1) -- (8.58,0.1);
\node at (8.3,-3) {$\times$};
\node at (10,-3) {$H^{d-1}$};

\end{tikzpicture}
\caption{Casini-Huerta-Myers map from cylinder to thermal hyperbolic space. The subregion $B$ gets mapped to all of hyperbolic space with entangling surface $\partial B$ mapped to infinity ($u=\infty$) on the hyperboloid. The operator insertions get mapped to the origin ($u=0$) on the hyperboloid.}
\label{fig:CHM}
\end{centering}
\end{figure}
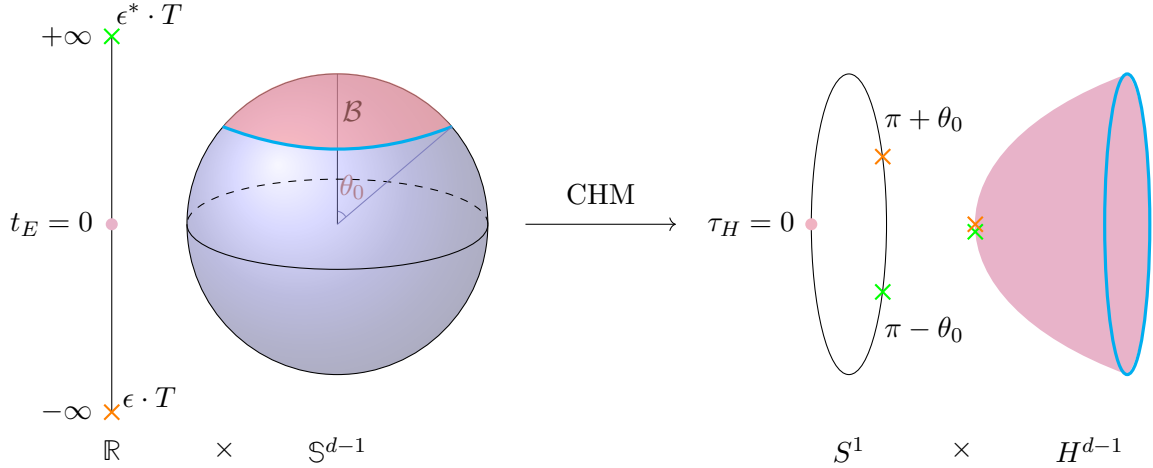

\paragraph{Setup.} Now consider a polar cap subregion $B=\{\theta \leq \theta_{0}\}$ on the spatial sphere at $t=0$, as illustrated in the left side of figure \ref{fig:CHM}. We wish to compute the vacuum-subtracted vN entropy of the reduced density matrix on $B$ for the stress-tensor excited state
\begin{equation}\label{eqn:reduceddensitymatrix}
    \rho_{B}^{\epsilon \cdot \tilde{T}} = \Tr_{B^{c}}\ket{\epsilon \cdot \tilde{T}}\bra{\epsilon^{\ast} \cdot \tilde{T}},
\end{equation}
where the conjugate state is created by $\epsilon^{\ast} \cdot \tilde{T} \equiv \epsilon_{\mu_{1}\mu_{2}}^{\ast}{I^{\mu_{1}}}_{\nu_{1}}{I^{\mu_{2}}}_{\nu_{2}}\tilde{T}^{\nu_{1}\nu_{2}}$. To do this, we use the conformal map constructed by Casini, Huerta, and Myers (CHM) \cite{Casini:2011kv} from the cylinder to hyperbolic space at finite temperature $\mathcal{H} = S^{1} \times H^{d-1}$ whose metric is
\begin{equation}\label{eqn:bdyhyperbolicspacemetric}
ds_{\mathcal{H}}^{2} = d\tau_{H}^{2}+du^{2}+\sinh^{2}u\,d\Omega_{d-2}^{2},
\end{equation}
with $\tau_{H} \sim \tau_{H}+2\pi$, $u \in [0,\infty)$. The explicit map in Euclidean signature is
\begin{equation}
\tanh t_{E} = \frac{\sin\theta_{0}\sin\tau_{H}}{\cosh u + \cos\theta_{0}\cos\tau_{H}}, \qquad \tan\theta = \frac{\sin\theta_{0}\sinh u}{\cos\tau_{H} + \cos\theta_{0}\cosh u},
\end{equation} 
which maps the subregion $B$ to all of hyperbolic space $\{\tau_{H}=0,\; u \in [0,\infty)\}$ and the operator insertions to the origin
\begin{equation}\label{eqn:opinsertionimages}
t_{E} = -\infty \to (u=0,\tau_{H}=\pi+\theta_{0}), \qquad t_{E} = \infty \to (u=0,\tau_{H}=\pi-\theta_{0})
\end{equation}
as illustrated in figure \ref{fig:CHM}.

The key feature of this map is that the (vacuum) reduced density matrix on $B$ gets mapped to the (ground-state) thermal density matrix on $\mathcal{H}$, viz., $\omega_{B} \to \omega_{\mathcal{H}} = e^{-2\pi H_{\mathcal{H}}}$ where $H_{\mathcal{H}}$ is the Hamiltonian on $\mathcal{H}$. This means that the excited state density matrix takes the simple form 
\begin{equation}\label{eqn:rhoHnorm}
\rho_{B}^{\epsilon \cdot \tilde{T}} \to \rho_{\mathcal{H}}^{\epsilon \cdot \tilde{T}} = \frac{e^{-\pi H_{\mathcal{H}}}\epsilon \cdot \tilde{T}(\theta_{0})\epsilon^{\ast} \cdot \tilde{T}(-\theta_{0})e^{-\pi H_{\mathcal{H}}}}{\langle \epsilon\cdot \tilde{T}(\tau_{T})\epsilon^{\ast}\cdot \tilde{T}(\hat{\tau}_{T})\rangle_{\mathcal{H}}}.
\end{equation}
where $\tau_{T} = \pi+\theta_{0}$ and $\hat{\tau}_{T} = \pi-\theta_{0}$. The map here is conjugation by the unitary matrix implementing the conformal transformation. Since the vN entropy is invariant under the action of unitaries, the excited density matrices on the cylinder and thermal hyperboloid give the same vN entropy.

The von Neumann entropy can now be computed in perturbation theory where the small parameter is the angular size $\theta_{0}$. Start by writing the density matrix as a pertubation around the thermal density matrix $\rho_{\mathcal{H}}^{\epsilon \cdot \tilde{T}} = \omega_{\mathcal{H}} + \delta \rho$. The perturbation $\delta \rho$ can be written explicitly in terms of the $T_{\mu\nu} \times T_{\rho\sigma}$ OPE expansion and small $\theta_{0}$ is the OPE limit where the two stress-tensors in \eqref{eqn:rhoHnorm} approach each other so $\delta \rho$ is indeed small when $\theta_{0}$ is small. 

\paragraph{von Neumann entropy.} The modular Hamiltonian for the excited state can be expanded in $\delta \rho$ using the integral representation of the logarithm \cite{Sarosi:2017rsq}
\begin{equation}\label{eqn:logintrep}
K_{\mathcal{H}}^{\epsilon \cdot \tilde{T}} = -\log \rho_{\mathcal{H}}^{\epsilon \cdot \tilde{T}} = \int_{0}^{\infty} d\beta\,\left(\frac{1}{\beta+\rho_{\mathcal{H}}^{\epsilon \cdot \tilde{T}}} - \frac{1}{\beta+1}\right) = K^0_{\mathcal{H}} + \sum_{n=1}^{\infty}(-1)^{n}\delta K_{(n)}^{\epsilon \cdot \tilde{T}}
\end{equation}
where $\K^0_{\mathcal{H}} = -\log \omega_{\mathcal{H}}$ 
\begin{align}\label{eqn:deltaKdef}
\begin{split}
\delta K_{(n)}^{\epsilon \cdot \tilde{T}} &= \int_{-\infty}^{\infty} ds_{1} \ldots ds_{n} \, \mathcal{K}_{n}(s_{1},\ldots,s_{n})\prod_{k=1}^{n} e^{iH_{\mathcal{H}}(s_{k}-i\pi)}\delta\rho e^{-iH_{\mathcal{H}}(s_{k}+i\pi)}
\\ \mathcal{K}_{n}(s_{1},\ldots,s_{n}) &= \frac{(2\pi)^{2}i^{n-1}}{(4\pi)^{n+1}}\frac{1}{\cosh \frac{s_{1}}{2} \cosh \frac{s_{n}}{2} \prod_{k=2}^{n}\sinh\left(\frac{s_{k}-s_{k-1}}{2}\right)}.
\end{split}
\end{align}
Since $\delta \rho$ is small, we will only need the first few terms in the infinite sum. The vacuum-subtracted entanglement entropy can now be expressed as
\begin{equation}\label{eqn:deltaS_scalar}
\Delta S^{\epsilon \cdot \tilde{T}} \equiv S(\rho_{B}^{\epsilon \cdot \tilde{T}})-S(\omega_B) = \Tr\left(\rho_{\mathcal{H}}^{\epsilon \cdot \tilde{T}}K_{\mathcal{H}}^{\epsilon \cdot \tilde{T}}\right)-\Tr\left(\omega_{\mathcal{H}} K^0_{\mathcal{H}}\right) = \Delta K_{0}^{\epsilon \cdot \tilde{T}}  + \sum_{n=1}^{\infty}(-1)^{n}\,\Tr\left(\rho_{\mathcal{H}}^{\epsilon \cdot \tilde{T}}\delta K_{(n)}^{\epsilon \cdot \tilde{T}}\right),
\end{equation}
where $\Delta K_{0}^{\epsilon \cdot \tilde{T}} = \Tr\left((\rho_{\mathcal{H}}^{\epsilon \cdot \tilde{T}}-\rho_{\mathcal{H}}^{0})K_{\mathcal{H}}^{0}\right)$ comes from the first law of entanglement entropy. This modular Hamiltonian is simply the Hamiltonian operator on $\mathcal{H}$ so it is an integral of the stress-tensor, which we will compute on the cylinder where it takes the form (note that the $\frac{(\cos\theta-\cos\theta_0)}{\sin\theta_0}$ simply comes from the Killing vector)
\begin{equation}\label{eqn:modHcyl}
K_{\mathrm{cyl}}^{0} = 2\pi \int d\Omega_{d-2}\,\int_{0}^{\theta_{0}}d\theta\,\sin^{d-2}\theta\frac{\left(\cos\theta-\cos\theta_{0}\right)}{\sin\theta_{0}}T_{tt}+c'
\end{equation}
($c'$ is a constant that is unimportant because it does not contribute to $\Delta K^{0}$). The first law of entanglement can thus be obtained from an integrated stress-tensor three-point function.

Using the following identity for the relative entropy
\begin{equation}\label{eqn:Srel}
S_{\mathrm{rel}}(\rho^{\phi_{1}}|\rho^{\phi_{2}}) \equiv \Tr\left(\rho^{\phi_{1}}(K^{\phi_{2}}-K^{\phi_{1}})\right) = \Delta K_{\phi_{2}}^{\phi_{1}} - \Delta S,
\end{equation}
we see that the infinite set of corrections in \eqref{eqn:deltaS_scalar} beyond the first law of entanglement are equal to (minus) the relative entropy
\begin{equation}\label{eqn:SrelT}
S_{\mathrm{rel}}(\rho_{B}^{\epsilon \cdot \tilde{T}}|\omega_B) = -\sum_{n=1}^{\infty}(-1)^{n}\,\Tr\left(\rho_{\mathcal{H}}^{\epsilon \cdot \tilde{T}}\delta K_{(n)}^{\epsilon \cdot \tilde{T}}\right).
\end{equation}
The terms appearing in the infinite sum can be manipulated to obtain
\begin{align}\label{eqn:trtocorrfn}
\begin{split}
&\Tr\left(\rho_{\mathcal{H}}^{\epsilon \cdot \tilde{T}}\delta K_{(n)}^{\epsilon \cdot \tilde{T}}\right) = \int_{-\infty}^{\infty} ds_{1} \ldots ds_{n} \, \mathcal{K}_{n}(s_{1},\ldots,s_{n})
\\	&\times \bigg\langle\mathcal{T}\left(\frac{\epsilon \cdot \tilde{T}(\tau_{T})\epsilon^{\ast} \cdot \tilde{T}(\hat{\tau}_{T})}{\langle \epsilon\cdot \tilde{T}(\tau_{T})\epsilon^{\ast} \cdot \tilde{T}(\hat{\tau}_{T})\rangle_{\mathcal{H}}}\prod_{k=1}^{n}\left(\frac{\epsilon \cdot \tilde{T}(\theta_{0}+is_{k})\epsilon^{\ast} \cdot \tilde{T}(-\theta_{0}+is_{k})}{\langle \epsilon \cdot \tilde{T}(\theta_{0}+is_{k})\epsilon^{\ast} \cdot \tilde{T}(-\theta_{0}+is_{k})\rangle_{\mathcal{H}}}-1\right)\right)\bigg\rangle_{\mathcal{H}}
\end{split}
\end{align}
where $\mathcal{T}$ indicates that this thermal correlation function must be time-ordered (up to KMS relations) in order to be well-defined. This can be done via the following contour prescription for the $s$ integrals: $s_{k} \to s_{k}-i\epsilon_{k}$ with
\begin{equation}\label{eqn:sintcontour}
-\pi < \epsilon_{k} < \ldots < \epsilon_{1} < \pi, \qquad -\pi+2\theta_{0} < \epsilon_{k} < \pi-2\theta_{0}, \qquad 2\theta_{0} < \epsilon_{k-1} - \epsilon_{k}.
\end{equation}
This contour prescription can only be satisfied for $\theta_{0} < \pi/(n+1)$ so for any fixed $\theta_{0}$ only a finite number of terms in the sum in \eqref{eqn:logintrep} are convergent, nonetheless, it can be argued that the fully resummed answer must be convergent \cite{Sarosi:2017rsq}.

All of the correlation functions that appear in the entropy \eqref{eqn:deltaS_scalar} are theory-dependent so we need to make some assumptions about our CFT. We will assume that the CFT satisfies $1 \ll \Delta_{\mathrm{gap}} \ll C_{T}$, where $\Delta_{\mathrm{gap}}$ is the lightest single-trace primary operator with spin $J>2$. These conditions surpress the string-scale and Planck-scale physics, respectively, and were shown to be necessary conditions for a holographic CFT to be dual to weakly-coupled Einstein gravity \cite{Heemskerk:2009pn}.

\paragraph{First-order contribution in $\delta\rho$.} We first compute the vacuum-subtracted expectation value of the vacuum modular Hamiltonian \eqref{eqn:modHcyl}. The three-point function of the stress-tensor is fixed by conformal symmetry, conservation of the stress-tensor, and Ward identities up to two theory-dependent constants in $d \geq 4$ \cite{Osborn:1993cr}. The $d=3$ case needs to be treated separately which is done in App.~\ref{sec:CFTstresstensor} and so for the rest of this section we will assume that $d \geq 4$. One can write these two constants in terms of $t_{2},t_{4}$ which are the anisotropies of the one-point functions of the average null energy operator in stress-tensor excited states considered by Hofman and Maldacena \cite{Hofman:2008ar}. It has been shown \cite{Afkhami-Jeddi:2016ntf,Afkhami-Jeddi:2018own} that for any CFT satisfying our assumptions of $1 \ll \Delta_{\mathrm{gap}} \ll C_{T}$, crossing symmetry along with causality in the Regge plus bulk point limit requires that $t_{2} \sim t_{4} \sim \Delta_{\mathrm{gap}}^{-2}$. This can be directly related to the AdS action for gravity which has two higher-derivative corrections that can contribute to the three-point vertex
\begin{equation}\label{eqn:AdSgravityaction}
S \sim \int d^{d+1}x\,\sqrt{g}\left(R-2\Lambda + \alpha_{2}W_{\mu\nu\rho\sigma}W^{\mu\nu\rho\sigma} + \alpha_{4}W_{\mu\nu\rho\sigma}W^{\rho\sigma\alpha\beta}{W_{\alpha\beta}}^{\mu\nu}\right) + \ldots
\end{equation}
where $\ldots$ indicates higher-point interactions and $W_{\mu\nu\rho\sigma}$ is the Weyl tensor. The fall-off at $t_{2},t_{4}$ at large $\Delta_{\mathrm{gap}}$ implies $\alpha_{2} \lesssim \Delta_{\mathrm{gap}}^{-2}$ and $\alpha_{4} \lesssim \Delta_{\mathrm{gap}}^{-4}$. Therefore, taking $\Delta_{\mathrm{gap}} \to \infty$ gives $t_{2}=t_{4}=0$ and, at the level of the three-point vertex, we recover Einstein gravity in AdS.

This completely fixes the stress-tensor three-point function and now we can obtain the desired vacuum-subtracted expectation value of $K_{0}$. We consider a special choice of polarization tensor that corresponds to the dual state that we will consider in AdS
\begin{equation}\label{eqn:specialpol}
\tilde{\epsilon}^{\mu\nu} = \varepsilon^{\mu}\varepsilon^{\nu}-\frac{\varepsilon^{2}}{d}\delta^{\mu\nu}, \qquad \theta_{\varepsilon}=0,
\end{equation}
where $\theta_{\varepsilon}$ is the inclination angle of $\varepsilon^{\mu}$, for which we find
\begin{align}\label{eqn:K0dif}
\begin{split}
\left(\Delta K_{B}^{\tilde{\epsilon} \cdot \tilde{T}}\right)_{\Delta_{\mathrm{gap}} \to \infty} &\equiv \frac{\bra{\tilde{\epsilon}^{\ast} \cdot \tilde{T}}K_{\mathrm{cyl}}^{0}\ket{\tilde{\epsilon} \cdot \tilde{T}}|_{\Delta_{\mathrm{gap}} \to \infty}}{\langle\tilde{\epsilon}^{\ast} \cdot \tilde{T}|\tilde{\epsilon} \cdot \tilde{T}\rangle} - \bra{0}K_{\mathrm{cyl}}^{0}\ket{0}
\\	&= \frac{\pi d}{2(d-1)^{3}(d+1)}\frac{V_{S^{d-2}}}{V_{S^{d-1}}}\sin^{d-2}\theta_{0}\bigg[(d+1)(9d^{2}-19d+6)k(d,\theta_{0})
\\	&+d(d-2)\left(-2(3d+1)\sin^{2}\theta_{0}k(d+2,\theta_{0})+(d+2)\sin^{4}\theta_{0}k(d+4,\theta_{0})\right)\bigg],
\end{split}
\end{align}
with 
\begin{equation}\label{eqn:fdef}
k(d,\theta_{0}) \equiv 1-\cos\theta_{0}\,{}_{2}{F}_{1}\left(\frac{1}{2},\frac{d-1}{2},\frac{d+1}{2};\sin^{2}\theta_{0}\right).
\end{equation}
This result is for finite $\theta_{0}$ and can be expanded to any desired order at small $\theta_{0}$. The details of the stress-tensor three point function needed for this result can be found in App.~\ref{sec:CFTstresstensor}.

Observe that $\Delta K_{B}^{\tilde{\epsilon} \cdot \tilde{T}} \sim C_{T}^{0}$. It is interesting to consider the superposition of this state with the vacuum state
\begin{equation}\label{eqn:CFTsuperposstate}
\ket{S} = \frac{1}{\sqrt{2}}\left(\ket{0} + \ket{\tilde{\epsilon} \cdot T}\right)
\end{equation}
because it has a qualitatively different expectation value. The reason for the new behavior is a cross-term that leads to a stress-tensor two-point function instead of a three-point function, viz.,\footnote{Note that the constant $c'$ in \eqref{eqn:modHcyl} still cancels in this expression because it only contributes to the first and last terms in parentheses.}
\begin{align}\label{eqn:DeltaKCFT_superpos}
\begin{split}
(\Delta K^{S}_{B})_{\Delta_{\mathrm{gap}} \to \infty} &= \frac{1}{2}\left(\bra{\tilde{\epsilon}^{\ast} \cdot \tilde{T}}K_{0}\ket{\tilde{\epsilon} \cdot \tilde{T}}|_{\Delta_{\mathrm{gap}} \to \infty}+\bra{\tilde{\epsilon}^{\ast} \cdot \tilde{T}}K_{0}\ket{0}+\bra{0}K_{0}\ket{\tilde{\epsilon} \cdot \tilde{T}}-\bra{0}K_{0}\ket{0}\right)
\\	&= \frac{1}{2}\left(\Delta K_{B}^{\tilde{\epsilon} \cdot \tilde{T}}\right)_{\Delta_{\mathrm{gap}} \to \infty}+\sqrt{C_{T}}\frac{4\pi}{(d+1)\sqrt{d(d-1)}}\frac{V_{S^{d-2}}}{V_{S^{d-1}}}\sin^{d}\theta_{0}
\end{split}
\end{align}
Thus, the state $\ket{S}$ has $\Delta S \sim \sqrt{C_{T}}$ which is parametrically larger than for $\ket{\tilde{\epsilon} \cdot \tilde{T}}$.

This is rather surprising from a quantum information perspective: the superposition of two states, each of which has a parametrically small entanglement entropy, actually has a parametrically large entanglement entropy. It can be understood as coming from the fact that the stress-tensor is a special operator because, unlike other operators, it knows about all the degrees of the freedom in theory, as measured by its two-point function. So in large-$N$ theories where the number of degrees of freedom grows with $N$, there is a class of states built out of $O(1)$ many insertions of the stress-tensor which have larger entanglement entropy than all other states in the theory constructed from an $O(1)$ number of light operators.\footnote{By light operators, we mean those with conformal dimension $\Delta \sim N^{0}$.} We will reproduce this behavior from the generalized entropy in AdS in \S\ref{sec:areaop} where it will come from the area term.

\paragraph{Higher-order corrections in $\delta\rho$.} We now go beyond the first law of entanglement and consider higher-order corrections in $\delta\rho$ to $\Delta S$ coming from the relative entropy \eqref{eqn:SrelT}, focusing again on the state $\ket{\tilde{\epsilon} \cdot \tilde{T}}$. Notice that the correlation functions \eqref{eqn:trtocorrfn} appearing in the relative entropy are $2m$-point functions, which in a large-$N$ theory can be computed by large-$N$ factorization so all correlation functions reduce to those of generalized free fields. 

Therefore, we only need the two-point functions on $\mathcal{H}$:
\begin{equation}\label{eqn:TT2ptfn}
\langle \tilde{\epsilon}^{\ast} \cdot \tilde{T}(\tau_{\mathcal{H}}) \tilde{\epsilon}^{\ast} \cdot \tilde{T}(0) \rangle = \langle \tilde{\epsilon} \cdot \tilde{T}(\tau_{\mathcal{H}}) \tilde{\epsilon} \cdot \tilde{T}(0) \rangle = \langle \tilde{\epsilon}^{\ast} \cdot \tilde{T}(\tau_{\mathcal{H}}) \tilde{\epsilon} \cdot \tilde{T}(0) \rangle = \frac{16}{\left(2\sin\left(\frac{\tau_{\mathcal{H}}}{2}\right)\right)^{2d}}.
\end{equation}
Furthermore, one can argue using the OPE \cite{Sarosi:2017rsq,Colin-Ellerin:2024npf} that larger $n$ terms contribute at higher-order in $\theta_{0}$ to the relative entropy in \eqref{eqn:SrelT} so at leading order in $\theta_{0}$ only the $n=1,2$ terms are needed. For the $n=1$ term, we find by performing the Wick contractions of the resulting four-point function and expanding at small $\theta_{0}$ that (here GFF stands for generalized free field)
\begin{equation}\label{eqn:deltaS_TGFF1}
\Tr\left(\rho_{\mathcal{H}}^{\epsilon \cdot \tilde{T}}\delta K_{(1)}^{\epsilon \cdot \tilde{T}}\right)_{\mathrm{GFF}} = \frac{1}{2}\theta_{0}^{4d}\left(1+O(\theta_{0}^{2})\right)\int_{-\infty}^{\infty} ds\,\frac{1}{\cosh^{4d+2}\frac{s}{2}}.
\end{equation}
The contribution from the $n=2$ term comes from four-point and six-point functions whose Wick contractions give
\begin{align}\label{eqn:deltaS_GFF2}
\begin{split}
\Tr\left(\rho_{\mathcal{H}}^{\epsilon \cdot \tilde{T}}\delta K_{(2)}^{\epsilon \cdot \tilde{T}}\right)_{\mathrm{GFF}}  &= \frac{i}{8\pi}\int_{-\infty}^{\infty} ds_{1}\,ds_{2}\,\frac{1}{\cosh\left(\frac{s_{1}}{2}\right)\cosh\left(\frac{s_{2}}{2}\right)\sinh\left(\frac{s_{2}-s_{1}}{2}\right)}
\\	&\qquad \times \frac{1}{\left(i\sinh\left(\frac{s_{1}-s_{2}}{2}\right)\right)^{4d}}\theta_{0}^{4d}\left(1+O(\theta_{0}^{2})\right)+ O\left(\theta_{0}^{6d}\right)
\end{split}
\end{align}
where we have again expanded at small $\theta_{0}$. The $\theta_{0}^{4d}$ terms come from the $T_{\mu\nu}T^{\mu\nu}$ double-trace operator of stress-tensors in the OPE. We can perform the $s_{1},s_{2}$ integrals by making the change of variables $u=s_{1}+s_{2}$ and $v=s_{2}-s_{1}-i\pi$ and then deforming the $v$ contour to the real line, which is consistent with our contour prescription, leaving us with
\begin{equation}\label{eqn:deltaS_GFF2_final}
\Tr\left(\rho_{\mathcal{H}}^{\epsilon \cdot \tilde{T}}\delta K_{(2)}^{\epsilon \cdot \tilde{T}}\right)_{\mathrm{GFF}} = \frac{1}{4}\theta_{0}^{4d}\left(1+O(\theta_{0}^{2})\right)\int_{-\infty}^{\infty} ds\,\frac{1}{\cosh^{4d+2}\frac{s}{2}}+ O\left(\theta_{0}^{6d}\right).
\end{equation}
Notice that this is proportional to the $n=1$ term in \eqref{eqn:deltaS_TGFF1}.

Combining \eqref{eqn:deltaS_TGFF1} and \eqref{eqn:deltaS_GFF2_final} and performing the $s$ integral gives
\begin{equation}\label{eqn:SrelTfinal}
S_{\mathrm{rel}}(\rho_{B}^{\epsilon \cdot \tilde{T}}|\omega_{B})_{\mathrm{GFF}} = \frac{\sqrt{\pi}\Gamma(2d+1)}{2\Gamma(2d+\frac{3}{2})}\theta_{0}^{4d}\left(1+O(\theta_{0}^{2})\right).
\end{equation}
Thus, we arrive at our final result for the vacuum-subtracted vN entropy
\begin{equation}\label{eqn:deltaS_T_Ein_special}
\left(\Delta S^{\tilde{\epsilon} \cdot \tilde{T}}_B \right)_{1 \ll \Delta_{\mathrm{gap}} \ll C_{T}} = \left(\Delta K_{B}^{\tilde{\epsilon} \cdot T}\right)_{\Delta_{\rm gap}\rightarrow \infty }-S_{\mathrm{rel}}(\rho_{B}^{\tilde{\epsilon} \cdot \tilde{T}}|\omega_B)_{\mathrm{GFF}}
\end{equation}
with these two contributions given in \eqref{eqn:K0dif} and \eqref{eqn:SrelTfinal}. We will match these results to the generalized entropy for the dual graviton state in AdS in \S\ref{sec:areaop} and \S\ref{sec:gravitonvN}.

\subsection{Gravitons in global AdS$_{d+1}$}
\label{sec:gravitonglobal}

We now turn to our analysis of the generalized entropy for a graviton excited state in AdS. This state will be defined as a single-particle state in global AdS so we begin with quantization of the graviton in this spacetime. There are two ways to quantize a gauge theory: one can either quantize all degrees of freedom, including the unphysical ones, and then impose the gauge conditions on the Hilbert space or one can first impose constraints and gauge fix the classical theory and then quantize only the physical degrees of freedom. We will choose to do the latter as it will be simpler for our choice of gauge. This way of quantizing gauge theories can be performed using Dirac's method for the quantization of Hamiltonian systems with constraints \cite{Dirac}. The reduced phase space comes equipped with a Dirac bracket which modifies the usual Poisson bracket such that constraints hold for the bracket, which is then promoted to a commutator to obtain the quantum theory. For nice reviews of this method, see \cite{Henneaux:1992ig,Hanson:1976cn}.

\paragraph{Phase space and constraints.}
Consider the classical, free graviton theory described by the action \eqref{eqn:EHactionexp_2ndorder} with Lagrangian $\mathcal{L}_{\mathrm{graviton}}$ and background metric $G_{\mu\nu}$ given by global AdS$_{d+1}$
\begin{equation}\label{eqn:AdSmetric}
ds^{2} = -(r^{2}+1)dt^{2}+\frac{dr^{2}}{(r^{2}+1)} + r^{2}d\Omega_{d-1}^{2},
\end{equation}
where $d\Omega_{d-1}^{2}$ is the metric on $S^{d-1}$ and we have chosen the AdS radius $L_{\mathrm{AdS}} = 1$.

The full, unconstrainted phase space has canonical coordinates consisting of the space of initial conditions $\mathfrak{P} = \{h_{\mu\nu},\pi^{\mu\nu}\}$ on a fixed-$t$ slice $\Sigma_{t}$, which is $(d+1)(d+2)$-dimensional, with equal-time Poisson brackets
\begin{equation}\label{eqn:PB}
\{h_{\mu\nu}(x),\pi^{\alpha\beta}(y)\}_{\mathrm{P.B.}} = \delta_{(\mu}^{\alpha}\delta_{\nu)}^{\beta}\delta^{d}(x-y),
\end{equation}
where $\delta^{d}(x-x') = \prod_{i=1}^{d}\delta(x_{i}-x_{i}')$. The quadratic action for gravitons does not contain any time derivatives of $h_{\mu t}$ (up to total derivative terms) so they serve as Lagrange multipliers. This implies that their conjugate momenta are zero giving $d+1$ primary constraints on phase space
\begin{equation}\label{eqn:primaryconstraints}
    \mathscr{C}^{\mu} \equiv \pi^{\mu t} \approx 0,
\end{equation}
where we have used the standard notation $\approx 0$ to indicate weakly zero, which means that it vanishes on the primary constraint subspace, but may have non-zero Poisson bracket. The remaining conjugate momenta to the graviton are given by
\begin{equation}\label{eqn:conjmomenta}
\pi^{\mathpzc{i}\mathpzc{j}} = \frac{\partial \mathcal{L}_{\mathrm{graviton}}}{\partial \pounds_{t}h_{\mathpzc{i}\mathpzc{j}}} = \frac{1}{2}\sqrt{-G}\left({-}\nabla^{t}h^{\mathpzc{i}\mathpzc{j}}+G^{\mathpzc{i}\mathpzc{j}}(\nabla^{t}{h_{\mathpzc{k}}}^{\mathpzc{k}}-2\nabla_{\mathpzc{k}}h^{\mathpzc{k} t})+2\nabla^{(\mathpzc{i}}h^{\mathpzc{j})t}\right)
\end{equation}
where $\mathpzc{i},\mathpzc{j}$ are spatial indices.

The canonical Hamiltonian comes from the Legendre transformation of the Lagrangian
\begin{equation}\label{eqn:canonH}
H_{c} = \int_{\Sigma_{t}} d^{d}x\,\left(\pi^{\mu\nu}\pounds_{t}h_{\mu\nu}-\mathcal{L}_{\mathrm{graviton}}\right).
\end{equation}
This is not the unique Hamiltonian on the primary constraint subspace because the primary constraints multiplied by any Lagrange multiplier can be added to this Hamiltonian, giving the primary Hamiltonian $H_{p}$:
\begin{align}\label{primH}
\begin{split}
&H_{p} = \int_{\Sigma_{t}} d^{d}x\,\bigg[{-}\frac{G_{tt}}{\sqrt{-G}}\left(\pi^{\mathpzc{i}\mathpzc{j}}\pi_{\mathpzc{i}\mathpzc{j}}-\frac{1}{(d-1)}(\pi_{\mathpzc{k}}^{\mathpzc{k}})^{2}\right)+\sqrt{-G}\,h^{tt}E_{tt}^{(1)}[h]-2\hat{\nabla}_{\mathpzc{i}}\pi^{\mathpzc{i}\mathpzc{j}}h_{\mathpzc{j}t}+u_{\mu}^{(p)}\pi^{\mu t}
\\  &+\sqrt{G_{\Sigma}}\left(\frac{1}{4}h_{\mathpzc{i}}^{\mathpzc{i}}\hat{\nabla}^{2}h_{\mathpzc{j}}^{\mathpzc{j}}-\frac{1}{4}h^{\mathpzc{i}\mathpzc{j}}\hat{\nabla}^{2}h_{\mathpzc{i}\mathpzc{j}}-\frac{1}{2}h_{\mathpzc{i}}^{\mathpzc{i}}\hat{\nabla}^{\mathpzc{j}}\hat{\nabla}^{\mathpzc{k}}h_{\mathpzc{j}\mathpzc{k}}+\frac{1}{2}h^{\mathpzc{i}\mathpzc{j}}\hat{\nabla}^{\mathpzc{k}}\hat{\nabla}_{\mathpzc{i}}h_{\mathpzc{j}\mathpzc{k}}+(d-1)h_{\mathpzc{i}\mathpzc{j}}h^{\mathpzc{i}\mathpzc{j}}-\frac{(d-1)}{2}h_{\mathpzc{i}}^{\mathpzc{i}}h_{\mathpzc{j}}^{\mathpzc{j}}\right)\bigg]
\end{split}
\end{align}
where $\hat{\nabla}$ is the covariant derivative on $\Sigma_{t}$ and $u_{\mu}^{(p)}$ are the Lagrange multipliers, and we have dropped boundary terms.\footnote{In the ADM formalism of general relativity, the Hamiltonian is purely a function of the constraints up to boundary terms so the Hamiltonian on the constraint subspace is a boundary term. However, the situation is different for linearized gravity because, expanding the ADM Hamiltonian to quadratic order in $h$, one finds linearized constraints and quadratic constraints. Only the linearized constraints vanish for gravitons, while the quadratic constraints give a non-zero bulk Hamiltonian (see Appendix A of \cite{Benedetti:2021lxj} for a discussion).} In deriving this, we have solved \eqref{eqn:conjmomenta} for $\pounds_{t}h_{\mathpzc{i}\mathpzc{j}}$ to obtain
\begin{equation}\label{eqn:hij_tderiv}
    \pounds_{t}h_{\mathpzc{i}\mathpzc{j}} = 2\hat{\nabla}_{(\mathpzc{i}}h_{\mathpzc{j})t}-\frac{2}{\sqrt{-G}}G_{tt}\left(\pi_{\mathpzc{i}\mathpzc{j}}-\frac{1}{(d-1)}\pi_{\mathpzc{k}}^{\mathpzc{k}}G_{\mathpzc{i}\mathpzc{j}}\right).
\end{equation}
Eventually, after gauge-fixing, we will set the primary constraints strongly to zero via the Dirac bracket at which point the distinction between primary Hamiltonian and canonical Hamiltonian will be unimportant, but before gauge-fixing, it is important to use the primary Hamiltonian.

There are also $d+1$ constraints on the phase space that come from requiring that the primary constraints \eqref{eqn:primaryconstraints} hold for all times: $\{\mathscr{C}^{\mu},H_{p}\}=0$. This gives the secondary constraints
\begin{align}\label{eqn:diffconstr}
\begin{split}
    \mathrm{Hamiltonian\;constraint:\;}& \tilde{\mathscr{C}}_{t} \equiv \sqrt{-G}E_{t}^{(1)t} = 0
    \\ \mathrm{Momentum\;constraints:\;}& \tilde{\mathscr{C}}_{r} \equiv \sqrt{-G}E_{r}^{(1)t} = 0, \; \tilde{\mathscr{C}}_{\phi_{i}} \equiv \sqrt{-G}E_{\phi_{i}}^{(1)t} = 0,
\end{split}
\end{align}
where we have used the fact that $\hat{\nabla}^{\mathpzc{j}}\pi_{\mathpzc{i}\mathpzc{j}} = -\sqrt{-G}E_{\mathpzc{i}}^{(1)t}[h]$.
These primary and secondary constraints are first-class constraints, i.e., $\{\mathscr{C}^{\mu},\mathscr{C}^{\nu}\}_{\mathrm{P.B.}}=\{\mathscr{C}^{\mu},\tilde{\mathscr{C}}_{\nu}\}_{\mathrm{P.B.}}=\{\tilde{\mathscr{C}}^{\mu},\tilde{\mathscr{C}}_{\nu}\}_{\mathrm{P.B.}}=0$. They thus generate gauge transformations on $\Sigma_{t}$, that is, for $F_{\epsilon} = \int_{\Sigma_{t}}d^{d}v\,\mathscr{C}^{\mu}\varrho_{\mu}^{(1)}+\tilde{\mathscr{C}}^{\mu}\varrho_{\mu}^{(2)}$ we find
\begin{equation}
\{h_{\mu t},F_{\epsilon}\}_{\mathrm{P.B.}} = \varrho_{\mu}^{(1)}, \qquad \{h_{\mathpzc{i}\mathpzc{j}},F_{\epsilon}\}_{\mathrm{P.B.}} = \nabla_{(\mathpzc{i}}\varrho_{\mathpzc{j})}^{(2)}.
\end{equation}
So for gauge parameter $\epsilon_{\mu}$, we take $\varrho_{\mathpzc{i}}^{(2)} = \epsilon_{\mathpzc{i}}$, $\varrho_{t}^{(1)} = \nabla_{t}\epsilon_{t}$, and $\varrho_{\mathpzc{i}}^{(1)}=\nabla_{\mathpzc{i}}\epsilon_{t}+\nabla_{t}\epsilon_{\mathpzc{i}}$. 

Next, we construct the secondary Hamiltonian $H_{s}$ obtained from the primary Hamiltonian $H_{p}$ by adding the secondary constraints times Lagrange multipliers $u_{\mu}^{(s)}$, but these can always be absorbed into $h_{\mu t}$ since these are also Lagrange mutlipliers so it is sufficient to continue working with the primary Hamiltonian. The requirement that the secondary constraints hold for all times $\{\tilde{\mathscr{C}}^{\mu},H_{p}\}=0$ does not lead to any new constraints as the lefthand side is zero by using the Bianchi identity, so the process of finding new constraints in this way has terminated. The primary Lagrange multipliers get determined by $\pounds_{t}h_{\mu t} = \{h_{\mu t},H_{s}\}_{\mathrm{P.B.}} = u_{\mu}^{(p)}$.

To obtain the reduced phase space, we now gauge-fix this constrained phase space $\mathfrak{C} = \{(h_{\mu\nu},\pi^{\mu\nu})|\mathscr{C}^{\mu}=\tilde{\mathscr{C}}_{\mu}=0\}$. We choose the holographic gauge \cite{Kabat:2013wga}, which gives $d+1$ more constraints\footnote{Strictly speaking, this choice of gauge is singular at $r=0$ because tensor fields are ill-defined there, but this is simply an artifact of working in global coordinates and it does not lead to any issues in our analysis.}
\begin{equation}\label{eqn:gauge-fixing}
    \mathscr{G}_{t} \equiv \pi_{rr}-\frac{1}{(d-1)}G_{rr}\pi_{\mathpzc{j}}^{\mathpzc{j}} = 0, \qquad \mathscr{G}_{\mathpzc{i}} \equiv h_{\mathpzc{i} r} = 0.
\end{equation}
Observe that we can explicitly solve the constraint $\mathscr{G}_{t}=0$ by using \eqref{eqn:conjmomenta} to obtain
\begin{equation}\label{eqn:htr=0}
\mathscr{G}_{t}=0 \implies \partial_{r}h_{rt}+\frac{r}{r^{2}+1}h_{rt}=0 \implies h_{rt} \propto \frac{1}{\sqrt{r^{2}+1}} \xrightarrow{r \to \infty} \frac{1}{r}, 
\end{equation}
but this violates the boundary conditions \eqref{eqn:extrapolatedictionary} for the graviton $h_{\mu r} \sim 1/r^{d-1}$ so we conclude that $h_{rt}=0$.

The requirement that this gauge-fixing \eqref{eqn:gauge-fixing} hold for all times $\{\mathscr{G}_{\mu},H_{p}\}_{\mathrm{P.B.}}=0$ then gives the final $d+1$ constraints
\begin{align}\label{eqn:gauge-fixingalltimes}
\begin{split}
    \tilde{\mathscr{G}}_{t} &\equiv {-}\frac{r}{(r^{2}+1)^{2}}\hat{\nabla}_{r}h_{tt}+\frac{r}{2}\hat{\nabla}_{r}h_{rr}-(d-1)h_{rr}
    \\  &+\frac{1}{(d-1)}\left(\frac{(d-2)}{2}(\hat{\nabla}_{\phi_{i}}\hat{\nabla}^{\phi_{i}}h_{rr}+\hat{\nabla}_{r}\hat{\nabla}_{r}h_{\phi_{i}}^{\phi_{i}}+G_{rr}h_{\phi_{i}}^{\phi_{i}}-2\hat{\nabla}_{\phi_{i}}\hat{\nabla}_{r}\hat{h}^{\phi_{i}}_{r})+\frac{1}{2}G_{rr}\hat{\nabla}_{\phi_{i}}\hat{\nabla}_{\phi_{j}}h^{\phi_{i}\phi_{j}}\right) = 0
    \\  \tilde{\mathscr{G}}_{\mathpzc{i}} &\equiv 2\sqrt{-G}\,\hat{\nabla}_{\mathpzc{(i}}h_{r)t}-G_{tt}\left(\pi_{\mathpzc{i}r}-\frac{1}{(d-1)}\pi_{\mathpzc{k}}^{\mathpzc{k}}G_{\mathpzc{i}r}\right) = 0.
\end{split}
\end{align}
No new constraints come from requiring that these  constraints hold for all times because $\{\tilde{\mathscr{G}}_{\mathpzc{i}},H_{p}\}_{\mathrm{P.B.}} = 0$.

Thus, we have obtained the reduced phase space
\begin{equation}\label{eqn:reducedphasespace}
    \mathfrak{R} = \{(h_{\mu\nu},\pi^{\mu\nu})\,|\,\mathscr{C}^{\mu}=\tilde{\mathscr{C}}_{\mu}=\mathscr{G}_{\mu}=\tilde{\mathscr{G}}^{\mu}=0\}.
\end{equation}
As a sanity check that we have the correct number of physical degrees of freedom, we see that the reduced phase space has dimension $(d+1)(d+2)-4(d+1) = (d+1)(d-2)$, which is equal to $2$ times the number of physical polarizations for the graviton in $d+1$ dimensions.

Now all the constraints are second-class since the gauge-fixing constraints have non-vanishing Poission bracket with the primary and secondary constraints. There is still a problem though with this reduced phase space: the Poisson brackets of functions on this phase space with the constraints can be non-zero. This is what leads to the introduction of the Dirac bracket.

\paragraph{Dirac bracket.}

Collect the full set of constraints defining $\mathfrak{R}$ into a single $4(d+1)$-dimensional vector $C_{a}$. The Dirac bracket is constructed precisely so that these constraints vanish for the bracket, i.e., $\{\mathcal{P},C_{a}\}_{\mathrm{D.B.}}=0$ for any function $\mathcal{P}$ on phase space. Define the constraint matrix $C_{ab}$ from their equal-time Poisson brackets
\begin{equation}\label{eqn:constraintmatrix}
    C_{ab}(x,y) = \{C_{a}(x),C_{b}(y)\}_{\mathrm{P.B.}}.
\end{equation}
This matrix takes the explicit form
\begin{equation}\label{eqn:Poissonbracketmatrix}
C_{ab}(x,y) = \left(\begin{matrix}
                \mathbf{0}_{2(d+1)} & M_{1}(x,y) \\ 
                M_{1}^{T}(x,y) & M_{2}(x,y) \\
                \end{matrix}\right)
\end{equation}
where the upper left quadrant of $C_{ab}$ is the $2(d+1)\times 2(d+1)$ zero matrix due to the fact that the primary and secondary constraints all commute. The explicit results for $M_{1}(x,y)$ and $M_{2}(x,y)$ are given in App.~\ref{sec:Diracbracket}.\footnote{Note that $C_{ab}(x,y)$ is antisymmetric under the double exchange $a \leftrightarrow b$ and $x \leftrightarrow y$.}

The inverse matrix is defined by
\begin{equation}\label{eqn:Poissonbracketmatrixinvrel}
\int d^{d}x_{2}\, C_{ab}^{-1}(x_{1},x_{2})C_{bc}(x_{2},x_{3}) = \delta_{ac}\delta^{d}(x_{1}-x_{3}).
\end{equation}
From \eqref{eqn:Poissonbracketmatrix}, we find the inverse matrix to be
\begin{equation}\label{eqn:Poissonbracketinvmatrix}
C_{ab}^{-1} = \left(\begin{matrix} -(M_{1}^{T})^{-1}M_{2}M_{1}^{-1} & (M_{1}^{T})^{-1} \\ M_{1}^{-1} & \mathbf{0}_{2(d+1)} \end{matrix}\right),
\end{equation}
where $M_{1}^{-1}(x,y)$ can be found in \eqref{eqn:M1inv}. The equal-time Dirac bracket, is now defined for any two functions $A(x)$ and $B(x)$ on phase space by
\begin{equation}\label{eqn:Diracbracket}
\{A(x),B(y)\}_{\mathrm{D.B.}} = \{A(x),B(y)\}_{\mathrm{P.B.}} - \int d^{d}z\,d^{d}w\,\{A(x),C_{a}(z)\}_{\mathrm{P.B.}}C_{ab}^{-1}(z,w)\{C_{b}(w),B(y)\}_{\mathrm{P.B.}},
\end{equation}
which by construction sets the constraints strongly zero. The physical coordinates on the reduced phase space satisfy canonical Dirac brackets
\begin{align}\label{eqn:Diracbracketcanon}
\begin{split}
\{h_{\phi_{i}\phi_{j}}(x),\pi^{\phi_{k}\phi_{l}}(y)\}_{\mathrm{D.B.}} &= {\delta_{(i}}^{k}{\delta_{j)}}^{l}\delta^{d}(x-y)
\\  \{h_{\phi_{i}\phi_{j}}(x),h_{\phi_{k}\phi_{l}}(y)\}_{\mathrm{D.B.}} &= \{\pi^{\phi_{i}\phi_{j}}(x),\pi^{\phi_{k}\phi_{l}}(y)\}_{\mathrm{D.B.}} = 0.
\end{split}
\end{align}
However, $h_{\mu t},\pi^{\mu t}$ do not have canonical brackets since $\{h_{\mu t}(x),\pi^{\nu t}(y)\}_{\mathrm{D.B.}} \neq \delta^{d}(x-y)$, as is necessary for the constraints to hold at the level of the bracket.

Having obtained the reduced phase space with corresponding bracket, we now restrict to this subspace by setting all constraints strongly to zero. Thus, the Hamiltonian becomes the constrained one
\begin{equation}\label{eqn:reducedH}
    H_{\mathrm{red}} = H_{p}|_{C_{a}=0}.
\end{equation}

\paragraph{Symplectic form.}

The presymplectic form can be obtained from variation of the Lagrangian as was done in \S\ref{sec:HWformula} to give its form in \eqref{eqn:sympform}, \eqref{eqn:omegaexplicit}, \eqref{eqn:wdef}. Here we write it more explicitly in terms of gravitons in the unconstrained phase space $\mathfrak{P}$:
\begin{align}\label{eqn:symplecticform_graviton}
\begin{split}
\Omega_{\mathfrak{P}}(h_{1},h_{2}) &= \int_{\Sigma_{t}}d^{d}x\,\sqrt{-G}\bigg(\frac{1}{2} h_{1} \nabla_{\nu} h_{2}^{t\nu}-\frac{1}{2}  h_{1} \nabla^{t} h_{2}+\frac{1}{2} h_{1}^{\mu\nu} \nabla^{t} h_{2,\mu\nu}
\\  &\qquad \qquad \qquad \qquad \qquad \qquad - h_{1,\mu\nu} \nabla^{\mu} h_{2}^{\nu t}+\frac{1}{2} h_{1}^{t\nu} \nabla_{\nu} h_{2} - (1 \leftrightarrow 2)\bigg)
\end{split}
\end{align}
for some fixed-$t$ Cauchy slice $\Sigma_{t}$. As discussed previously, $\Omega_{\mathfrak{P}}$ will be independent of time slice $\Sigma_{t}$ for on-shell gravitons and it will be gauge-invariant for diffeomorphisms that fall off sufficiently quickly at asymptotic infinity. This gauge-invariance comes from the fact that the symplectic form vanishes when applied to any vector tangent to a gauge orbit, and hence it is degenerate. However, the induced symplectic form on the reduced phase space $\Omega_{\mathfrak{R}} = \Omega_{\mathfrak{P}}|_{C_{a}=0}$ is non-degenerate.

\paragraph{Equations of motion.} The equations of motion for the physical polarizations of the graviton are the Hamiltonian equations on the reduced phase space given by the Dirac brackets
\begin{equation}
    \partial_{t}h_{\phi_{i}\phi_{j}} = \{h_{\phi_{i}\phi_{j}},H_{\mathrm{red}}\}_{\mathrm{D.B.}}, \qquad \partial_{t}\pi_{\phi_{i}\phi_{j}} = \{\pi_{\phi_{i}\phi_{j}},H_{\mathrm{red}}\}_{\mathrm{D.B.}}.
\end{equation}
The first set of equations simply give the relation between $\pi_{\phi_{i}\phi_{j}}$ and $h_{\phi_{i}\phi_{j}}$ and the second set of equations give the linearized Einstein equations $E_{\phi_{i}\phi_{j}}^{(1)}[h]=0$. Combined with the constraints \eqref{eqn:diffconstr}, we have all of the linearized Einstein equations except for those for $E_{\mathpzc{i} r}^{(1)}[h]$.\footnote{It may seem strange that we only have a subset of the Einstein equations, but it is simply a consequence of using an ``axial'' gauge. There is no equation of motion for $h_{\mu r}$ because it has been gauge-fixed to be zero for all times. The same is true in Maxwell theory, see Chapter 5.C in \cite{Hanson:1976cn}.} Explicitly, the linearized Einstein tensor in global AdS is given by
\begin{equation}\label{eqn:gravitonEOM}
    E_{\alpha\beta}^{(1)}[h] = -\frac{1}{2}\nabla^{2}h_{\alpha\beta}-\frac{1}{2}\nabla_{\alpha}\nabla_{\beta}h+\nabla_{\rho}\nabla_{(\alpha}{h^{\rho}}_{\beta)}+dh_{\alpha\beta}-\frac{1}{2}G_{\alpha\beta}\left(\nabla^{\mu}\nabla^{\nu}h_{\mu\nu}-\nabla^{2}h+dh\right).
\end{equation}

To solve these linearized Einstein equations, we will use the ``master variable'' formalism \cite{Ishibashi:2004wx,Kodama:2000fa} which reduces the linearized Einstein equations to differential equations for three different scalar fields. The idea is to decompose any symmetric $2$-tensor in terms of representations of the $S^{d-1}$ part of global AdS. To do this, we use the following mathematical result proven by \cite{Ishibashi:2004wx}. Let $\mathcal{C}$ be an $m$-dimensional compact Riemann Einstein space with metric $\eta_{ij}$ (Einstein space means $R_{ij} = c\eta_{ij}$). Any one-form $v_{i}$ on $\mathcal{C}$ has a unique decomposition as
\begin{equation}\label{eqn:vectordecomp}
v_{i} = V_{i} + (\nabla_{\mathcal{C}})_{i}S, \qquad (\nabla_{\mathcal{C}})^{i}V_{i} = 0.
\end{equation}
Furthermore, any symmetric $2$-tensor $t_{ij}$ on $\mathcal{C}$ can be uniquely decomposed as
\begin{equation}\label{eqn:tensordecomp}
    t_{ij} = \mathscr{T}_{ij}+(\nabla_{\mathcal{C}})_{(i}\mathscr{V}_{j)}+\left((\nabla_{\mathcal{C}})_{i}(\nabla_{\mathcal{C}})_{j}-\frac{1}{m}\eta_{ij}\right)\mathscr{S}+\frac{1}{m}\eta_{ij}{t^{k}}_{k}
\end{equation}
where
\begin{equation}
    \nabla_{\mathcal{C}}^{i}\mathscr{T}_{ij} = 0, \qquad {\mathscr{T}^{i}}_{i} = 0, \qquad \nabla_{\mathcal{C}}^{i}\mathscr{V}_{i} = 0. 
\end{equation}
We refer to $T_{ij}$, $\mathcal{V}_{i}$, and $\mathcal{S}$, ${t^{k}}_{k}$ as the tensor, vector, and scalar components, respectively. One can think of this as a version of the Hodge decomposition, but for symmetric tensors rather than antisymmetric tensors.

Let us apply this to the compact manifold of interest $S^{d-1}$ with isometry group $SO(d)$. Denote the covariant derivative by $D_{i}$. A complete basis for normalisable functions (with respect to the $L^{2}$-norm) is given by the scalar spherical harmonics $\mathbb{S}_{\mathbf{k}_{S}}$ satisfying
\begin{equation}\label{eqn:scalarsphharm}
\left(D^{2}+k_{S}^{2}\right)\mathbb{S}_{\mathbf{k}_{S}} = 0, \qquad k_{S}^{2} = \ell(\ell+d-2), \; \ell \geq 0,
\end{equation}
a complete basis for normalisable, divergenceless vector fields is given by the vector spherical harmonics $\mathbb{V}_{\mathbf{k}_{V}}$ satisfying
\begin{equation}\label{eqn:vecsphharm}
\left(D^{2}+k_{V}^{2}\right)\mathbb{V}_{\mathbf{k}_{V},i} = 0, \qquad D^{i}\mathbb{V}_{\mathbf{k}_{V},i} = 0, \qquad k_{V}^{2} = \ell(\ell+d-2)-1, \; \ell \geq 1,
\end{equation}
and a complete basis for normalisable, traceless, divergenceless symmetric $2$-tensor fields is given by the tensor spherical harmonics $\mathbb{T}_{\mathbf{k}_{T}}$ satisfying
\begin{equation}\label{eqn:tenssphharm}
\left(D^{2}+k_{T}^{2}\right)\mathbb{T}_{\mathbf{k}_{T},i} = 0, \qquad D^{i}\mathbb{T}_{\mathbf{k}_{T},ij} = 0, \qquad {\mathbb{T}_{\mathbf{k}_{T},i}}^{i} = 0, \qquad k_{T}^{2} = \ell(\ell+d-2)-2, \; \ell \geq 2.
\end{equation}
It can be argued that the scalar, vector, and tensor spherical harmonics form independent representations of $SO(d)$ \cite{Ishibashi:2004wx}. Therefore, these three types of modes all decouple so we can solve the linearized Einstein equations separately for each.

We now apply this to the gravtion $h_{\mu\nu}$. Let $a,b,c,\ldots$ denote coordinates on the two-dimensional $(t,r)$ part of the global AdS spacetime and let $i,j,k,\ldots$ denote coordinates on the $S^{d-1}$ part. Then we can write the graviton as
\begin{equation}\label{eqn:gravitonexp}
    h_{\mu\nu}dx^{\mu}dx^{\nu} = h_{ab}dx^{a}dx^{b}+2h_{ai}dx^{a}dx^{i}+h_{ij}dx^{i}dx^{j}.
\end{equation}
The $h_{ab}$ part transforms as a scalar under $SO(d)$ transformations so
\begin{equation}\label{eqn:habexp}
    h_{ab}(t,r,\Omega) = \sum_{\mathbf{k}_{S}}\mathfrak{h}_{ab,\mathbf{k}_{S}}(t,r)\mathbb{S}_{\mathbf{k}_{S}}(\Omega),
\end{equation}
the $h_{ai}$ part transforms as a vector under $SO(d)$ transformations so we use \eqref{eqn:vectordecomp} to obtain
\begin{equation}\label{eqn:haiexp}
    h_{ai}(t,r,\Omega) = \sum_{\mathbf{k}_{V}}\mathfrak{h}_{a,\mathbf{k}_{V}}^{V}(t,r)\mathbb{V}_{i,\mathbf{k}_{V}}(\Omega)+\sum_{\mathbf{k}_{S}}\mathfrak{h}_{a,\mathbf{k}_{S}}^{S}(t,r)D_{i}\mathbb{S}_{\mathbf{k}_{S}}(\Omega),
\end{equation}
and $h_{ij}$ transforms as a tensor under $SO(d)$ transformations so we use \eqref{eqn:tensordecomp} to obtain
\begin{align}\label{eqn:hijexp}
\begin{split}
    &h_{ij}(t,r,\Omega) = \sum_{\mathbf{k}_{T}}\mathfrak{h}_{\mathbf{k}_{T}}^{T}(t,r)\mathbb{T}_{ij,\mathbf{k}_{T}}(\Omega) +2\sum_{\mathbf{k}_{V}}\mathfrak{h}_{\mathbf{k}_{V}}^{V}(t,r)D_{(i}\mathbb{V}_{j),\mathbf{k}_{V}}(\Omega)
    \\  &+\sum_{\mathbf{k}_{S}}\left(\mathfrak{h}_{\mathbf{k}_{S}}^{S,\mathrm{tr}}(t,r)(G_{S^{d-1}})_{ij}\mathbb{S}_{\mathbf{k}_{S}}(\Omega)+\mathfrak{h}_{\mathbf{k}_{S}}^{S,\mathrm{trless}}(t,r)\left(D_{i}D_{j}-\frac{1}{d-1}(G_{S^{d-1}})_{ij}D_{m}D^{m}\right)\mathbb{S}_{\mathbf{k}_{S}}(\Omega)\right).
\end{split}
\end{align}
Therefore, the scalar part of the graviton is given by
\begin{align}\label{eqn:gravscalar}
\begin{split}
    h_{\mu\nu}^{S}dx^{\mu}dx^{\nu} &= \sum_{\mathbf{k}_{S}}\bigg[\mathfrak{h}_{ab,\mathbf{k}_{S}}\mathbb{S}_{\mathbf{k}_{S}}dx^{a}dx^{b}+2\mathfrak{h}_{a,\mathbf{k}_{S}}^{S}D_{i}\mathbb{S}_{\mathbf{k}_{S}}dx^{a}dx^{i}
    \\  &+\left(\mathfrak{h}_{\mathbf{k}_{S}}^{S,\mathrm{tr}}(G_{S^{d-1}})_{ij}+\mathfrak{h}_{\mathbf{k}_{S}}^{S,\mathrm{trless}}\left(D_{i}D_{j}-\frac{1}{(d-1)}(G_{S^{d-1}})_{ij}D_{m}D^{m}\right)\right)\mathbb{S}_{\mathbf{k}_{S}}dx^{i}dx^{j}\bigg],
\end{split}
\end{align}
the vector part is given by
\begin{equation}\label{eqn:gravvector}
    h_{\mu\nu}^{V}dx^{\mu}dx^{\nu} = \sum_{\mathbf{k}_{V}}\left(\mathfrak{h}_{a,\mathbf{k}_{V}}^{V}\mathbb{V}_{i,\mathbf{k}_{V}}dx^{a}dx^{i}+2\mathfrak{h}_{\mathbf{k}_{V}}^{V}D_{(i}\mathbb{V}_{j),\mathbf{k}_{V}}dx^{i}dx^{j}\right),
\end{equation}
and the tensor part is given by
\begin{equation}\label{eqn:gravtensor}
    h_{\mu\nu}^{T}dx^{\mu}dx^{\nu} = \sum_{\mathbf{k}_{T}}\mathfrak{h}_{\mathbf{k}_{T}}^{T}\mathbb{T}_{ij,\mathbf{k}_{T}}dx^{i}dx^{j}.
\end{equation}

As we will see momentarily, the CFT stress-tensor state $\ket{\tilde{\epsilon} \cdot \tilde{T}}$ analysed in \S\ref{sec:ex} is dual to the state for the lowest energy mode of the scalar part with no azimuthal angular momenta so we will restrict our attention to this sector. The solutions to the equations of motion are labelled by a non-negative integer $n$ and the angular momenta $\mathbf{k}_{S}$. It can be proven that there are no dynamical modes with total angular momentum $\ell=0,1$ \cite{Ishibashi:2004wx}, which is simply the statement that the lowest moment of gravitational radiation is the quadrupole, so the lowest energy modes have total angular momentum $\ell=2$. The mode of interest is thus $h_{\mu\nu,n=0,\ell=2,\mathfrak{m}=0}$ where $\mathfrak{m}$ labels azimuthal angular momenta. To solve the linearized Einstein equations for the scalar modes, one manipulates the equations such that they reduce to an equation for a single scalar field, a so called ``master variable''. The procedure is very complicated so we relegate the details to App.~\ref{sec:EOMsoln} and here we simply state the result:
\begin{align}\label{eqn:hsoln}
\begin{split}
    \mathfrak{h}_{0,2}^{S,\mathrm{trless}} &= \mathcal{N}_{0,2}^{S}e^{-i\Omega_{0,2}^{S}t}\Bigg[\frac{2d}{3}\frac{r^{4}}{r^{2}+1}\,{}_{2}{F}_{1}\left(1-\frac{d}{2},\frac{3}{2},\frac{5}{2};\frac{r^{2}}{r^{2}+1}\right)+\frac{r^{2}}{(r^{2}+1)^{\frac{d}{2}}}
    \\  &\qquad \qquad \qquad \qquad \qquad \qquad \qquad \qquad \qquad \qquad -2\frac{\Gamma(\frac{3}{2})\Gamma(\frac{d+2}{2})}{\Gamma(\frac{d+3}{2})}r\sqrt{r^{2}+1}\Bigg]
    \\ \mathfrak{h}_{0,2}^{S,\mathrm{tr}} &= \mathcal{N}_{0,2}^{S}e^{-i\Omega_{0,2}^{S}t}\frac{2d}{(d-1)}\Bigg[{-}\frac{(d+1)}{3}\frac{r^{4}}{r^{2}+1}\,{}_{2}{F}_{1}\left(1-\frac{d}{2},\frac{3}{2},\frac{5}{2};\frac{r^{2}}{r^{2}+1}\right)-\frac{r^{2}}{(r^{2}+1)^{\frac{d}{2}}}
    \\  &\qquad \qquad \qquad \qquad \qquad \qquad \qquad \qquad \qquad \qquad \qquad+\frac{\Gamma(\frac{3}{2})\Gamma(\frac{d}{2})}{\Gamma(\frac{d+1}{2})}r\sqrt{r^{2}+1}\Bigg]
    \\ \mathfrak{h}_{t,0,2} &= \mathcal{N}_{0,2}^{S}e^{-i\Omega_{0,2}^{S}t}\frac{i}{(r^{2}+1)^{\frac{d-2}{2}}}
    \\  \mathfrak{h}_{tt,0,2}  &= \mathcal{N}_{0,2}^{S}e^{-i\Omega_{0,2}^{S}t}2d\Bigg[\frac{(d+1)(d-1)}{3}\frac{r^{4}}{r^{2}+1}\,{}_{2}{F}_{1}\left(1-\frac{d}{2},\frac{3}{2},\frac{5}{2};\frac{r^{2}}{r^{2}+1}\right)+\frac{dr^{2}+1}{(r^{2}+1)^{\frac{d}{2}}}
    \\  &\qquad \qquad \qquad \qquad \qquad \qquad \qquad \qquad \qquad \qquad -\frac{4}{d}\frac{\Gamma(\frac{3}{2})\Gamma(\frac{d+2}{2})}{\Gamma(\frac{d-1}{2})}r\sqrt{r^{2}+1}\Bigg]
\end{split}
\end{align}
where $\mathcal{N}_{0,2}^{S}$ is a normalisation constant which will be computed later and $\Omega_{n,\ell}^{S}$ are the frequencies determined by normalisability to be
\begin{equation}\label{eqn:scalarmode_freq}
\Omega_{n,\ell}^{S} = d-2+\ell+2n.
\end{equation}
These are the energies of a given mode as measured by quasi-local Brown-York stress-tensor living on the boundary. Observe that our mode of interest has frequency $\Omega_{0,2}=d$ which is precisely the conformal dimension of the CFT stress-tensor, giving an early hint that this is the mode corresponding the stress-tensor, which will be proven later.

The solution in \eqref{eqn:hsoln} is labelled by $n$ and $\ell$ with the $\mathfrak{m}$ dependence in $h_{\mu\nu,0,2,\mathbf{0}}^{S}$ only coming from the spherical harmonic, which is
\begin{equation}\label{eqn:sphharm_l=2}
    \mathbb{S}_{\ell=2,\mathfrak{m}=\mathbf{0}}(\Omega) = \sqrt{\frac{(d+2)}{2(d-1)}}\left(d\cos^{2}\theta-1\right)
\end{equation}
where we normalise the spherical harmonics so that $\int d\Omega\,\sqrt{g_{S^{d-1}}}\, \mathbb{S}_{\ell,\mathfrak{m}}^{\ast}(\Omega)\mathbb{S}_{\ell',\mathfrak{m}'}(\Omega) = \delta_{\ell,\ell'}\delta_{\mathfrak{m},\mathfrak{m}'}V_{S^{d-1}}$.

\paragraph{Canonical quantization.}

The canonical quantization of the free graviton theory proceeds based on the decomposition of the graviton into tensor, vector, and scalar parts. Since they form independent representations, we expand the graviton in creation and annihilation operators for each type of mode
\begin{equation}\label{eqn:quantgravitonexp}
\hat{h}_{\mu\nu}(x) = \sum_{n,\mathbf{k}_{T}}h_{\mu\nu,n,\mathbf{k}_{T}}^{T}(x)a_{n,\mathbf{k}_{T}}^{T}+ \sum_{n,\mathbf{k}_{V}}h_{\mu\nu,n,\mathbf{k}_{V}}^{V}(x)a_{n,\mathbf{k}_{V}}^{V} + \sum_{n,\mathbf{k}_{S}}h_{\mu\nu,n,\mathbf{k}_{S}}^{S}(x)a_{n,\mathbf{k}_{S}}^{S} + \mathrm{h.c.}
\end{equation}
In particular, we will see that the different types of modes are orthogonal with respect to the generalized Klein-Gordon inner product. The commutator is constructed from the Dirac bracket: $\{\cdot,\cdot\}_{\mathrm{D.B.}} \rightarrow i[\cdot,\cdot]$. We now require that each mode of the tensor, vector, and scalar parts of the solution for the graviton are each unit normalised with respect to the generalized Klein-Gordon inner product so that the corresponding annihilation and creation operators have normalised commutation relations:
\begin{equation}\label{eqn:normalisedgravitonIP}
\langle h_{n,\mathbf{k}_{A}}^{A},h_{n',\mathbf{k}_{B}'}^{B}\rangle = \delta_{nn'}\delta^{AB}\delta_{\mathbf{k}_{A}\mathbf{k}_{B}'} \implies [a_{n,\mathbf{k}_{A}}^{A},a_{n',\mathbf{k}_{B}'}^{B\dagger}] = \delta_{nn'}\delta^{AB}\delta_{\mathbf{k}_{A},\mathbf{k}_{B}'}
\end{equation}
with all other commutators vanishing. The Hilbert space is a tensor product of the three Fock spaces $\mathbb{H} = \mathbb{H}^{S} \otimes \mathbb{H}^{V} \otimes \mathbb{H}^{T}$.

To obtain the generalized Klein-Gordon inner product, we use the symplectic form on phase space \eqref{eqn:symplecticform_graviton} which defines an inner-product on solutions of the equations of motion that is independent of choice of spacelike slice $\Sigma$ on-shell. The desired inner product can thus be defined by
\begin{equation}\label{eqn:MaxIP}
\langle h,\tilde{h}\rangle = i\Omega_{\mathfrak{R}}(h^{\ast},\tilde{h}).
\end{equation}
Recall that that symplectic form $\Omega_{\mathfrak{R}}$ is defined from the restriction of the presymplectic form $\Omega_{\mathfrak{P}}$ to the reduced phase space $\mathfrak{R}$ so we must impose all the constraints $C_{a}=0$ in the Klein-Gordon inner product. From the explicit expression for the symplectic form \eqref{eqn:symplecticform_graviton}, we can compute the generalized Klein-Gordon inner products on the second line of \eqref{eqn:normalisedgravitonIP}, then using divergenceless and tracelessness properties of $\mathbb{V}_{i}$ and $\mathbb{T}_{ij}$ along with integration by parts, one can confirm that these inner products indeed vanish.\footnote{Alternatively, one could prove that the generators of $SO(d)$ are self-adjoint with respect to the generalized Klein-Gordon inner product and then use that these modes form independent representations of $SO(d)$.} Requiring that our mode of interest $h_{\mu\nu,0,2,\mathbf{0}}^{S}$ be unit normalised gives the normalisation constant
\begin{equation}\label{eqn:graviton_normconstant}
    \mathcal{N}_{0,2}^{S} = \sqrt{\frac{\Gamma(d+2)}{4d(d-1)\Gamma(\frac{d}{2}+1)\Gamma(\frac{d}{2}+2)V_{S^{d-1}}}},
\end{equation}
with the explicit computation provided in App.~\ref{sec:norm}. 

The single-particle graviton state corresponding to this lowest-energy mode is given by
\begin{equation}\label{eqn:singleparticlestate}
    \ket{\mathfrak{g}} = a_{0,2,\mathbf{0}}^{S\dagger}\ket{0}.
\end{equation}
To prove that the state \eqref{eqn:singleparticlestate} is dual to the stress-tensor primary state $\ket{\tilde{\epsilon} \cdot \tilde{T}}$ that we considered in the CFT, we observe that this state and the state $\ket{\mathfrak{g}}$ are the unique states in the boundary and bulk Hilbert spaces, respectively, that have the eigenvalues $(d,2,\mathbf{0})$ for the Hamiltonian, total angular momentum, and azimuthal angular momenta operators, and are primary (annihilated by special conformal transformation) in the two dual theories.

\paragraph{Backreaction of the graviton.}

The final piece we need to compute the area contribution to the generalized entropy is the backreacted metric $\hat{g}_{\mu\nu}^{(2)}$. As explained in \S\ref{sec:obs}, this operator is a quadratic function \eqref{eqn:g2_formalsoln} of $\hat{h}_{\mu\nu}$, which we normal-order to make it well-defined. Thus, it takes the general form 
\begin{equation}\label{eqn:g2exp}
\hat{g}_{\mu\nu}^{(2)} = \sum_{q,q'} g^{(2)}_{\mu\nu,q,q'}a_{q}^{\dagger}a_{q'}.
\end{equation}
To determine a coefficient $g^{(2)}_{\mu\nu,q,q'}$ one computes the $qq'$ matrix element of the Einstein equations. We are interested in the $q=q'=(0,2,\mathbf{0})$ matrix element given by $\langle \hat{g}_{\mu\nu}^{(2)} \rangle_{\mathfrak{g}}$, which is obtained by solving
\begin{equation}\label{eqn:g2eqn_T}
E_{\mu\nu}^{(1)}[\langle \hat{g}^{(2)}\rangle_{\mathfrak{g}}] = \langle \hat{T}_{\mu\nu}^{\mathrm{grav}}\rangle_{\mathfrak{g}}.
\end{equation}
Recall from \S\ref{sec:HWformula} that we define the gravitational ``stress-tensor'' to be $\hat{T}_{\mu\nu}^{\mathrm{grav}}=-\normord{\hat{E}_{\mu\nu}^{(2)}[\hat{h}]}$. Here we will summarize how to obtain the solution and leave the details to App.~\ref{sec:gravbackreact}.

Using the explicit result for $T_{\mu\nu}^{\mathrm{grav}}$ in \eqref{eqn:Tgrav}, we find that its expectation value takes the following form
\begin{align}\label{eqn:Tgravexpval}
\begin{split}
    \bra{\mathfrak{g}} T_{tt}^{\mathrm{grav}} \ket{\mathfrak{g}} &= \mathcal{T}_{tt}(r)+\mathcal{S}_{tt}(r)\cos(2\theta)+\mathcal{V}_{tt}(r)\cos(4\theta)
    \\  \bra{\mathfrak{g}} T_{rr}^{\mathrm{grav}} \ket{\mathfrak{g}} &= \mathcal{T}_{rr}(r)+\mathcal{S}_{rr}(r)\cos(2\theta)+\mathcal{V}_{rr}(r)\cos(4\theta)
    \\  \bra{\mathfrak{g}} T_{\theta\theta}^{\mathrm{grav}} \ket{\mathfrak{g}} &= \mathcal{T}_{\theta\theta}(r)+\mathcal{S}_{\theta\theta}(r)\cos(2\theta)+\mathcal{V}_{\theta\theta}(r)\cos(4\theta)
    \\  \bra{\mathfrak{g}} T_{\phi_{i}\phi_{i}}^{\mathrm{grav}} \ket{\mathfrak{g}} &= \left(\mathcal{T}_{\phi_{i}\phi_{i}}(r)+\mathcal{S}_{\phi_{i}\phi_{i}}(r)\cos(2\theta)+\mathcal{V}_{\phi_{i}\phi_{i}}(r)\cos(4\theta)\right)(g_{S^{d-1}})_{\phi_{i}\phi_{i}}, \qquad 2 \leq i \leq d-1
    \\  \bra{\mathfrak{g}} T_{r\theta}^{\mathrm{grav}} \ket{\mathfrak{g}} &= \mathcal{S}_{r\theta}(r)\sin(2\theta)+\mathcal{V}_{r\theta}(r)\sin(4\theta)
\end{split}
\end{align}
for some tensors $\mathcal{T}_{\mu\nu}(r)$, $\mathcal{S}_{\mu\nu}(r)$, and $\mathcal{V}_{\mu\nu}(r)$ that depend on the dimension $d$, while all other expectation values vanish. Motivated by this, we make the following ansatz for the backreacted metric\footnote{The minus sign in the radial component of the metric may look funny, but it simply comes from expanding $g_{rr}=\frac{1}{(r^2+1)(1+\kappa^{2}\mathcal{F}_{2}(r,\theta))}$ to $O(\kappa^{2})$.}
\begin{equation}\label{eqn:g2ansatz}
    ds_{g^{(2)}}^{2} = -(r^{2}+1)\mathcal{F}_{1}(r,\theta)dt^{2}-\mathcal{F}_{2}^{\mathrm{g}}(r,\theta)\frac{dr^{2}}{(r^{2}+1)}+r^{2}\mathcal{F}_{3}(r,\theta)d\Omega_{d-1}^{2}.
\end{equation}
with
\begin{equation}\label{eqn:calFansatz}
    \mathcal{F}_{i}(r,\theta) = \mathscr{F}_{i,0}(r)+\mathscr{F}_{i,2}(r)\cos(2\theta)+\mathscr{F}_{i,4}(r)\cos(4\theta).
\end{equation}
Note that we assumed here that $g_{r\theta}^{(2)}=0$, which is not a priori an obvious choice, but will nevertheless turn out to give a non-trivial solution.

Plugging this ansatz for $g_{\mu\nu}^{(2)}$ into the linearized Einstein tensor \eqref{eqn:gravitonEOM} (with $h$ replaced by $g^{(2)}$) and setting it equal to the expectation values in \eqref{eqn:Tgravexpval}, we obtain a large set of coupled second-order linear ODEs for the $\mathscr{F}_{i,j}(r)$. With a great deal of effort, these can be solved explicitly for any given dimension $d$. We require that the solutions are such that the full metric $g_{\mu\nu}$ is asymptotically AdS and that there are no curvature singularities, which fixes all of the undetermined constants in the solution. It turns out that this can be done by allowing $g_{rr}^{(2)}$ to have a mild coordinate singularity and imposing that all other components have no singularities whatsoever. The explicit result for $\langle \hat{g}_{\mu\nu}^{(2)} \rangle_{\mathfrak{g}}$ in $d=4$ dimensions can be found in App \ref{sec:gravbackreact}.


\subsection{Area operator, perturbed extremal surface, and large diffeomorphisms}
\label{sec:areaop}

Now that we have the solution for the graviton $\hat{h}_{\mu\nu}$ and the backreacted metric $\hat{g}_{\mu\nu}^{(2)}$, we can compute the contribution from the area operator at $O(\kappa^{-1})$ and $O(\kappa^{0})$. In the process, we will show explicitly how to address the problems of large diffeomorphisms and IR divergences that were discussed in \S\ref{sec:HWgaugemore}.

We begin with the unperturbed classical extremal surface $\gamma^{(0)}$. Given the induced background metric $q_{\mu\nu}[G]$ on a codimension-$2$ surface $\gamma^{(0)}$ anchored on the boundary polar cap $B$ of size $\theta_{0}$, the area functional is given by
\begin{equation}\label{eqn:classicalarea}
    A[G,\gamma^{(0)}] = \int_{\gamma^{(0)}} \sqrt{q[G]}.
\end{equation}
Parametrizing the surface by $\theta^{(0)}(r)$ in global coordinates, we find the surface that extremizes the area:
\begin{equation}\label{eq:extremalsurf}
\theta^{(0)}(r) = \arccos\left(\cos(\theta_{0})\frac{\sqrt{r^{2}+1}}{r}\right).
\end{equation}
The deepest point that the surface reaches in the bulk is the radius $r_{\mathrm{min}}=\cot\theta_{0}$. The classical entanglement wedge is an AdS-Rindler wedge with bifurcation surface $\gamma^{(0)}$.

\paragraph{First-order area operator.}

We now consider the leading order part of the area operator $\hat{A}[\hat{h},\gamma^{(0)}]$, which gives an $O(1/\kappa)$ contribution to the generalized entropy. As discussed in \S\ref{sec:obs}, the expectation value of this operator will be zero for the single particle state $\ket{\mathfrak{g}}$ because it is a three-point function in a free theory, but it will give a non-zero answer for the superposition state
\begin{equation}\label{eqn:gravitonsuperposstate}
    \ket{\mathfrak{s}} = \frac{1}{\sqrt{2}}\left(\ket{0}+\ket{\mathfrak{g}}\right),
\end{equation}
which is dual to the CFT superposition state \eqref{eqn:CFTsuperposstate}.


The first order area operator is written as 
\begin{equation}\label{eqn:firstorderareaop}
    A[\hat{g},\hat{\gamma}]\Big|_{O(\kappa)} = \sum_q \left(A^{\rm lin}[h_q,\gamma^{(0)}] a_q + \text{h.c.} \right).
\end{equation}
The term $\hat{A}[G,\hat{\gamma}^{(1)}]$  does not receive an $O(\kappa)$ contribution as long as the surface is classically extremal and the endpoint stays ``unmoved'' due to the procedure discussed in \S\ref{sec:HWgaugemore}. We explicitly compute $A^{\rm lin}[h_q,\gamma^{(0)}]$ here and verify $A[G,\gamma^{(1)}]|_{O(\kappa)} = 0$ in the following subsubsections.
 
As stated in \S\ref{sec:HWgaugemore}, $A^{\rm lin}[h_q,\gamma^{(0)}]$ should be understood as a term $A^{\rm lin}_{\rm no\ cut-off}[h_q,\gamma^{(0)}]$ as if there were no IR divergence issue plus a term (possibly zero) coming from properly regulating and subtracting IR divergences. We emphasize again that the area is finite, but to extract the correct finite answer requires this careful regulation.
The $A^{\rm lin}_{\rm no\;cut-off}[h_{0,2,\mathbf{0}},\gamma^{(0)}]$ computation is standard\footnote{We take the integration bound to infinity here. The reader may wonder why the cut-off surface does not play a role here. The reason is that the integral is actually finite, and introducing a cut-off and then taking it to infinity is of no difference from directly integrating to asymptotic infinity. The cut-off is only relevant when we discuss subtracting off the IR divergences.}
\begin{equation}
\begin{aligned}
	& A^{\mathrm{lin}}_{\rm no\;cut-off} [h_{0,2,\mathbf{0}},\gamma^{(0)}] = \int_{\gamma^{(0)}}\sqrt{q[G+\kappa h_{0,2,\mathbf{0}}]} \Big|_{O(\kappa)}
\\   \qquad &= V_{S^{d-2}}\int_{r_{\mathrm{min}}}^{\infty}dr\,\sin^{d-2}\theta_{\rm ext}^{(0)}\left[\tilde{g}_{\theta\theta}(r,\theta_{\rm ext}^{(0)})\right]^{\frac{(d-2)}{2}}\left[\tilde{g}_{rr}(r,\theta_{\rm ext}^{(0)}) + \left((\theta_{\rm ext}^{(0)})^{\prime}\right)^2 \tilde{g}_{\theta\theta}(r,\theta_{\rm ext}^{(0)})\right]^{\frac{1}{2}}\bigg|_{O(\kappa)}
\end{aligned}
\end{equation}
where $\tilde{g}= G + \kappa h_{0,2,\mathbf{0}}$ and 
we dropped the possible $t$ dependence of the metric by setting $t=0$. The result for $d=4$ is
\begin{equation}\label{eq:linearareares}
	A^{\mathrm{lin}}[h_{0,2,\mathbf{0}},\gamma^{(0)}]|_{d=4} = V_{S^{2}}|\mathcal{N}_{0,2}^{S}| \frac{4\sin^4\theta_0}{5}\,.
\end{equation}
  
Let us address how to deal with the IR divergences. Following \S\ref{sec:HWgaugemore}, the first step is to determine the cut-off surface to the linear order, by requiring that the induced metric on the cut-off surface is fixed. 
More explicitly, we take the the cut-off surface to be parametrized by $\{t_{\rm bdy},\theta_{\rm bdy},\phi_{i,\text{bdy}}\}$, and require that the induced metric on the cut-off surface is 
\begin{equation}
	ds^2_{\rm cut-off} = -(r_c^2+1 + \kappa O(r_c^{-2})) dt_{\rm bdy}^2 + r_c^2 d \Omega^2_{d-1,\rm bdy}, 
\end{equation}
%
which determines the location of the cut-off surface to be\footnote{The reader may find it a bit strange that the location of the cut-off surface is not a real number. This should be understood as also the expectation value of an operator, that, e.g., $\hat{t}(t_{\rm bdy},\theta_{\rm bdy}) = t_{\rm bdy} + \kappa\sum_q (\delta t_q a_q + \text{h.c.}) + O(\kappa^2)$. }
\begin{equation}\label{eqn:1stordercutoff}
\begin{aligned}
	t(t_{\rm bdy},\theta_{\rm bdy}) = t_{\rm bdy} + \kappa \frac{2(-i e^{-i\Omega_{0,2,\mathbf{0}}^{S} t_{\rm bdy}}
    )\cos 2 \theta_{\rm bdy}}{r_c^4}\Big(1-\frac{10}{3} r_c^{-2} + O(r_c^{-4})\Big) \\
	r(t_{\rm bdy},\theta_{\rm bdy}) = r_c + \kappa \frac{(e^{-i\Omega_{0,2,\mathbf{0}}^{S} t_{\rm bdy}}
    )(1+2\cos 2 \theta_{\rm bdy})}{r_c^3}\Big(1-\frac{11}{6} r_c^{-2} + O(r_c^{-4})\Big) \\
	\theta(t_{\rm bdy},\theta_{\rm bdy}) = \theta_{\rm bdy} - \kappa\frac{2( e^{-i\Omega_{0,2,\mathbf{0}}^{S} t_{\rm bdy}}
    )\sin 2 \theta_{\rm bdy}}{r_c^4}\Big(1-\frac{5}{3} r_c^{-2} + O(r_c^{-4})\Big).  \\
\end{aligned}
\end{equation}

Taking the intersection of the cut-off surface and the extremal surface, with $t_{\rm bdy}$ chosen such that $t_{\rm bdy} = 0$ at this intersection, gives the new endpoint of the extremal surface $r_c^{\rm new}=r_c+ \kappa \frac{1+2\cos 2\theta_0}{r_c^3}+O(r_c^{-5})$, so that the shift of the endpoint gives a contribution to the area 
\begin{equation}
	A[G,\gamma^{(0)}_{r_c^{\rm new}}] - A[G,\gamma^{(0)}_{r_c}] = \int^{r_{c}^{\rm new}}_{r_c} \sqrt{q[G]} = V_{S^{d-2}} \int^{r_{c}^{\rm new}}_{r_c} r dr \sqrt{\frac{r^2-r_{\rm min}^2}{r^2+1}} = O(r_c^{-2})
\end{equation}
which vanishes after taking $r_c$ to infinity. Therefore, \eqref{eq:linearareares} is indeed the correct answer for the first order area correction. We will see in the next subsection that it matches the leading CFT entropy in \eqref{eqn:DeltaKCFT_superpos}.

We now consider the second-order contribution to the area operator, for which we will focus on the single-particle state $\ket{\mathfrak{g}}$ as no new qualitative features appear for the superposition states at that order.

\paragraph{Second-order area operator I: $\Delta A$ from the backreaction $g^{(2)}$.}
The simplest piece of the second-order area operator to compute is that coming from the backreacted metric $\hat{g}^{(2)}$ as it only involves the unperturbed classical extremal surface $\gamma^{(0)}$. There will be two contributions, one from ignoring the IR divergences and the other from properly treating the cut-offs. The same feature appeared in the backreaction of photons in \cite{Colin-Ellerin:2024npf} and comes from the slow fall-off the backreacted metric ($\mathfrak{g}_{\mu\nu}^{(2)} \sim 1/r^{d-4}$).

The area operator from the backreacted metric is given by 
\begin{equation}
	\normord{A[\hat{g},\hat{\gamma}]}\big|_{O(\kappa^2),\,\rm backreaction} = \sum_q A^{\rm lin}[g^{(2)}_q,\gamma^{(0)}] a^{\dagger}_q a_q.
\end{equation} 
We will compute the coefficient $A^{\rm lin}[g_{0,2,\mathbf{0}}^{(2)},\gamma^{(0)}]$ which results from taking the expectation value in the state $\ket{\mathfrak{g}}$. 

First, let us compute this area without the IR-cut-off:
\begin{align}\label{eqn:linearizedarea_backreactmetric}
\begin{split}
    & A_{\mathrm{no\;cut-off}}^{\mathrm{lin}}[g^{(2)}_{0,2,\mathbf{0}},\gamma^{(0)}] = \int_{\gamma^{(0)}}\sqrt{q[G+\kappa^2 g^{(2)}_{0,2,\mathbf{0}}]} \bigg|_{O(\kappa^2)}
\\  \qquad &= V_{S^{d-2}}\int_{r_{\mathrm{min}}}^{\infty}dr\,\sin^{d-2}\theta_{\rm ext}^{(0)}\left[\tilde{\tilde{g}}_{\theta\theta}(r,\theta_{\rm ext}^{(0)})\right]^{\frac{(d-2)}{2}}\left[\tilde{\tilde{g}}_{rr}(r,\theta_{\rm ext}^{(0)}) + \left((\theta_{\rm ext}^{(0)})^{\prime}\right)^2 \tilde{\tilde{g}}_{\theta\theta}(r,\theta_{\rm ext}^{(0)})\right]^{\frac{1}{2}}\bigg|_{O(\kappa^2)}
\end{split}
\end{align}
where $\tilde{\tilde{g}}=G+\kappa^2 g^{(2)}_{0,2,\mathbf{0}}$. This can be computed in any dimension $d$. For $d=4$, using the metric solution in App.~\ref{sec:gravbackreact}, we find
\begin{align}\label{eqn:linearizedarea_backreactmetric_explicit}
\begin{split}
    A_{\mathrm{no\;cut-off}}^{\mathrm{lin}}[g^{(2)}_{0,2,\mathbf{0}},\gamma^{(0)}] = & V_{S^{2}}|\mathcal{N}_{0,2}^{S}|^{2} \Big(  - \frac{44}{75} \theta_{0}^{2}+\frac{628}{225}\theta_{0}^{4}-\frac{13192}{7875}\theta_{0}^{6}+\frac{16148}{23625}\theta_{0}^{8}-\frac{586504}{1670625}\theta_{0}^{10}\\
     & -  \frac{221484392}{1064188125} \theta_0^{12}  + 
 \frac{404376656}{3192564375}\theta_0^{14} - \frac{3948064636}{74009446875}\theta_0^{16} + {O}(\theta_{0}^{18})\Big).
\end{split}
\end{align}
Here we have expanded to $O(\theta_{0}^{16})$ because that is the first order at which the relative entropy appears in $d=4$, so this is needed to test beyond the modular Hamiltonian formula to obtain the generalized entropy.

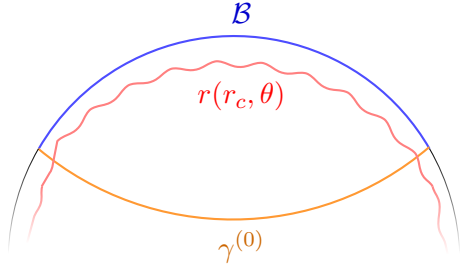
\begin{figure}[h]
\begin{center}
\begin{tikzpicture}[path fading=south]

\draw[thick,color=blue!70!white] (2.5,0) arc (30:150:3);
\draw[thick,color=orange!80!white] (-2.7,0) arc (230:310.5:4.02);
\draw[path fading] (2.5,0) arc (30:2:3);
\draw[path fading] (-2.7,0) arc (150:178:3);
\node[blue!80!black] at (0,1.8) {$\mathcal{B}$};
\node[orange!80!black] at (0,-1.3) {$\gamma^{(0)}$};

\draw[thick,color=red!50!white,decorate,decoration={snake,segment length=6mm,amplitude=.4mm}] (2.54,-0.5) arc (30:150:3);
\draw[thick,color=red!50!white,decorate,decoration={snake,segment length=7mm,amplitude=.4mm},path fading] (2.54,-0.5) -- (2.74,-1.4);
\draw[thick,color=red!50!white,decorate,decoration={snake,segment length=7mm,amplitude=.4mm},path fading] (-2.65,-0.5) -- (-2.85,-1.25);
\node at (0,0.7) {$\color{red}{r(r_{c},\theta)}$};

\end{tikzpicture}
\end{center}
\caption{Fixed-time Cauchy slice of AdS$_{d+1}$ (with $S^{d-2}$ directions suppressed) zoomed in on boundary subregion $\mathcal{B}$ (blue) and classical extremal surface $\gamma^{(0)}$ (orange) with wiggly cutoff surface $\mathcal{R}_{\mathrm{cut-off}}$ (red).}
\label{fig:wigglycutoff}
\end{figure}

We now need to analyse the contribution coming from properly dealing with the IR cut-off. The cut-off surface is defined such that the induced metric on the the $t=0$ slice of the cut-off surface is the same for the background metric $G$ and the perturbed metric $\tilde{\tilde{g}}$: $ds^{2}|_{\mathrm{cut-off},t=0} = r_{c}^{2}d\Omega_{d-1,\rm bdy}^{2}$. 
This ensures that we have a well-defined variational principle when comparing the area in one metric and its variation, namely the two surfaces have the same boundary anchor points. 
It turns out that the correction to the position of the surface at $O(\kappa^{2})$ has the behaviour $r_{c}^{-(d-2)}$ at large $r_{c}$, which, combining with the $r_{c}^{d-2}$ divergence in the area, gives an important finite contribution. We find the cut-off surface to be given by\footnote{This simple form of the cut-off surface location is because of the simple form of $g^{(2)}$: it is diagonal, time independent, and spherically symmetric.}
\begin{equation}\label{eqn:cutoff}
  \mathcal{R}_{\rm cut-off}: \; t=t_{\rm bdy}, \quad  \theta=\theta_{\rm bdy}, \quad r^2 = r_{c}^{2} - \kappa^{2}g_{\theta\theta}^{(2)}(r_{c},\theta),
\end{equation}
as illustrated in figure \ref{fig:wigglycutoff}.

Observe that this expression seems to ignore the $O(\kappa)$ contribution worked out in \eqref{eqn:1stordercutoff}.
One can view this from two perspectives: first, using the alternative form of the JLMS formula in  \S\ref{sec:altprescript}, the only cut-off involved quantity is the area $A^{\rm lin}[g^{(2)},\gamma^{(0)}]$, so that it is reasonable to isolate the part of the cut-off surface \eqref{eqn:cutoff} originating from $g^{(2)}$, which is a direct analog of the cut-off surface one introduces in the Maxwell case \cite{Colin-Ellerin:2024npf}; on the other hand, as will be discussed when summarizing the area calculation result, there is a consistent and complete cut-off surface including both the $O(\kappa)$ result in \eqref{eqn:1stordercutoff} and the  $O(\kappa^2)$ result in \eqref{eqn:cutoff}. 
It is just the $O(\kappa)$ cut-off surface does not end up giving finite contribution at this order. 

With this new cut-off surface, one simply finds the new endpoint of $\gamma^{(0)}$ by taking the intersection between $\gamma^{(0)}$ and $\mathcal{R}_{\rm cut-off}$. The result is 
\begin{equation}\label{eqn:2ndordercutoff}
   r(t_{\rm bdy},\theta_{\rm bdy}) =  r_c -\frac{1}{2 r_c }\kappa^2 g_{\theta\theta}^{(2)}(r_c,\theta_{\rm bdy}) + O(r_c^{-3})
   \,,
\end{equation}
so that the new end point for the extremal surface is $ r_c^{\rm new}= r_c -\frac{1}{2 r_c }\kappa^2 g_{\theta\theta}^{(2)}(r_c,\theta_0) + O(r_c^{-3})$.
Hence the area correction from this difference in cut-off surface is 
\begin{equation}\label{eqn:backreactcutoff}
\begin{aligned}
    	A[G,\gamma^{(0)}_{r_c^{\rm new}}] - A[G,\gamma^{(0)}_{r_c}] & = \int d\Omega_{d-2}\int^{r_{c}^{\rm new}}_{r_c} dr\, \sqrt{q[G]} \\
    & = V_{S^{d-2}} \int^{r_{c}^{\rm new}}_{r_c} dr\, \frac{r}{\sqrt{r^2+1}}(r^{2}-r_{\mathrm{min}}^2)^{\frac{(d-3)}{2}} 
    \\  &= - \frac{\kappa^2}{2}V_{S^{d-2}} \lim_{r_c\rightarrow \infty} r_{c}^{d-4}g_{\theta\theta}^{(2)}(r_c,\theta_0) + O(r_c^{-2}).
\end{aligned}
\end{equation}
Thus, the area with the IR cut-off is equal to the sum of \eqref{eqn:linearizedarea_backreactmetric} and \eqref{eqn:backreactcutoff}, which in $d=4$ is found to be
\begin{align}\label{eqn:linearizedarea_backreactmetricwcutoff_explicit}
\begin{split}
    A^{\mathrm{lin}}[g^{(2)},\gamma^{(0)}]|_{d=4}  = & V_{S^{2}}|\mathcal{N}_{0,2}^{S}|^{2}\Big(\frac{296}{225}\theta_{0}^{4}-\frac{3088}{4725}\theta_{0}^{6}+\frac{248}{875}\theta_{0}^{8}-\frac{557968}{2338875}\theta_{0}^{10} \\
    &+ \frac{585069872}{3192564375} \theta_0^{12} - \frac{26020768}{212837625}\theta_0^{14} + \frac{42918566632}{814103915625}
  \theta_0^{16}+{O}(\theta_{0}^{18})\Big).
\end{split}
\end{align}

Next, we turn to the quadratic area term for $h$ and $\gamma^{(1)}$.
This requires, of course, to first find $\gamma^{(1)}$. 
\paragraph{Second-order area operator II: solve for $\gamma^{(1)}$.}
Here we determine the perturbed extremal surface via directly extremizing the area. Given that the boundary polar cap and the graviton wavefunction $h_{0,2,\mathbf{0},\mu\nu}$ are symmetric under rotations of the $S^{d-2}$, the same will be true for the perturbed surface $\gamma^{(0)}+\kappa \gamma^{(1)}$. However, since the graviton $h_{\mu\nu}$ is time-dependent, the perturbed surface $\gamma^{(1)}$ can move in the timelike direction, unlike the classical extremal surface $\gamma^{(0)}$ which lives in the $t=0$ Cauchy slice. We thus take the components of $\kappa \gamma^{(1)}$ to be $\kappa \gamma^{(1)t}(r)$ and $\kappa \gamma^{(1)\theta}(r)$, and solve the corresponding Euler-Lagrange equations 
\begin{equation}
\begin{aligned}
    \frac{\partial}{\partial \gamma^{(1)t} }\sqrt{q[G+\kappa h,\gamma^{(0)}+\kappa \gamma^{(1)}]} - \frac{d}{dr} \frac{\partial}{\partial (\gamma^{(1)t})^{\prime}}\sqrt{q[G+\kappa h,\gamma^{(0)}+\kappa \gamma^{(1)}]} = 0\\
    \frac{\partial}{\partial \gamma^{(1)\theta} }\sqrt{q[G+\kappa h,\gamma^{(0)}+\kappa \gamma^{(1)}]} - \frac{d}{dr}\frac{\partial}{\partial (\gamma^{(1)\theta})^{\prime}}\sqrt{q[G+\kappa h,\gamma^{(0)}+\kappa \gamma^{(1)}]} = 0 \,.
\end{aligned}
\end{equation}
where $'$ denotes an $r$-derivative and the determinant of the induced metric on the surface is given by
\begin{equation}
    q[G+\kappa h;\gamma^{(0)}+\kappa \gamma^{(1)}] = (G+\kappa h)_{\mu\nu}{\gamma^{\mu}}'{\gamma^{\nu}}'\prod_{i=2}^{d-1} (G_{\phi_i\phi_i} +\kappa h_{\phi_i\phi_i}).
\end{equation}

\begin{figure}
\begin{center}
\begin{tikzpicture}
\node at (0,0) {\includegraphics{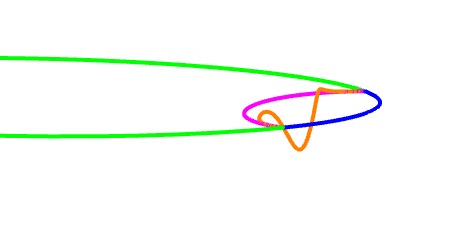}};
\node at (-0.2,0.2) {$\color{magenta}\gamma^{(0)}$};
\node at (3,0.2) {$\color{blue}B$};
\node at (1.2,-1.1) {$\color{orange}\gamma^{(0)}+\kappa\gamma_{0,2,\mathbf{0}}^{(1)}$};
\end{tikzpicture}
\end{center}
\caption{The perturbed extremal surface $\gamma^{(0)}+\kappa\gamma_{0,2,\mathbf{0}}^{(1)}$ (orange) for $\kappa=10$ (to make the perturbation visible) anchored on the boundary of the polar cap $B$ (blue) on the asymptotic boundary (green) of the $t=0$ slice of $AdS_{5}$ (with two azimuthal angles suppressed). The unperturbed extremal surface (magenta) lies in the $t=0$ bulk Cauchy slice, but the perturbed one bends down in the time direction. (Only a partial boundary Cauchy slice is displayed because we zoomed in on the extremal surfaces.)}
\label{fig:pertsurface}
\end{figure}

The resulting equations are two second-order ODEs, and the solution is uniquely determined, once the endpoint is fixed as a boundary condition. 
These can be solved in any dimension and here we simply state the result for the relevant $0,2,\mathbf{0}$ mode in $d=4$ dimensions:
\begin{equation}\label{eq:surfacepertexplicit}
\begin{aligned}
 &\gamma^{(1)\theta}_{0,2,\mathbf{0}}(r)|_{d=4} = \frac{1}{\sqrt{r^2-r_\mathrm{min}^2}}
    \frac{4 r_\mathrm{min}}{15  r^3 (1 + r_\mathrm{min}^2)} \\
    &\times\bigg(
    \frac{ r (2 r^2 (12 + 19 r^2 + 8 r^4)-(15 + 23 r^2 + 
          10 r^4) r_\mathrm{min}^2)}{(1 + r^2)^{3/2}} + 
    2 (-r^2(8r^2+7) +  r_\mathrm{min}^2 (5 r^2+4)) \bigg)\\
    &\gamma^{(1)t}_{0,2,\mathbf{0}}(r)|_{d=4} = \frac{4 i
    }{15(1+r_\mathrm{min}^2)} \\
    &\times\bigg( \frac{r^2(4r^2+2)+r_\mathrm{min}^2(15+33 r^2 +20r^4)}{r^3(1+r^2)^3} +     \frac{8(-4r_\mathrm{min}^2+r^2(1+r_\mathrm{min}^2))}{r^2(1 + r^2)^{3/2}} -8(1-3r_\mathrm{min}^2) \bigg).
\end{aligned}
\end{equation}
A plot of this surface can be found in figure \ref{fig:pertsurface}. We will use this explicit form to evaluate the area correction quadratic in $h$ and $\gamma^{(1)}$.

Alternatively, we can use $\theta$ instead of $r$ to parametrize the perturbation. 
This turns out to be computationally convenient later when we discuss the quadratic contribution to area from $h$ and $\gamma^{(1)}$, with the result ($C_\theta =\cos\theta$ and $S_\theta=\sin \theta$) 
\begin{equation}\label{eq:surfacepertexplicittheta}
\begin{aligned}
    \gamma^{(1)t}_{0,2,\mathbf{0} }|_{d=4} & = \frac{4i 
    }{15} \frac{(C_\theta-C_{\theta_0})^2}{C_\theta^4}  ( C_\theta^2 (3 C_\theta - C_{\theta_0})(5 C_{\theta}+ C_{\theta_0}) -2 (3 C_{\theta}^2 +2 C_{\theta} C_{\theta_0} + C_{\theta_0}^2) )\\
    \gamma^{(1)r}_{0,2,\mathbf{0} }|_{d=4} & = \frac{2 
    (C_{\theta_0}-C_{\theta})
    }{15(S^2_{\theta_0} - S^2_\theta)^{\frac{3}{2} }} \Big((C_\theta-C_{\theta_0})(8C^3_\theta +  C_\theta^2 C_{\theta_0} + C_{\theta} C_{\theta_0}^2 + C_{\theta_0}^3 ) - 2 (7 C_{\theta}^2 - 5 C_\theta C_{\theta_0} -4 C_{\theta_0}^2 ) \Big).
\end{aligned}
\end{equation}

%

%

\paragraph{Second-order area operator III: $\Delta A$  quadratic in $h$ and $\gamma^{(1)}$.}
Now we compute the quadratic contribution to the expectation value of the area operator from $\hat{h}$ and $\hat{\gamma}^{(1)}$. 
The relevant piece of the area operator is 
\begin{equation}\label{eqn:quadareaop}
    \normord{{A}[G+\kappa \hat{h},\gamma^{(0)}+\kappa \hat{\gamma}^{(1)}]}\big|_{O(\kappa^2)} =  \sum_q ( A[G+\kappa h_q, \gamma^{(0)} + \kappa \gamma^{(1)\ast}_q] + \text{c.c} \big)|_{O(\kappa^2)}a^{\dagger}_q a_q
\end{equation}
and the contributions to the area wavefunctional $A[G+\kappa h_q, \gamma^{(0)} + \kappa \gamma^{(1)*}_q]$ are\footnote{By $A^{\mathrm{quad}}$, we mean expansion of $A$ as a functional of $h$ to second-order in $h$, which is $\frac{1}{2}\frac{\delta^{2} A[G+\kappa h]}{\delta h^{2}}$.}
\begin{equation}
\begin{aligned}
    A[G+\kappa h_q, \gamma^{(0)} + \kappa \gamma^{(1)*}_q] = &  A^{\mathrm{quad}}[G,\gamma^{(1)}_q \gamma^{(1)*}_q]+ A^{\mathrm{quad}}[h_q h^{*}_q,\gamma^{(0)}]+A^{\mathrm{lin}}[h_q,\gamma^{(1)*}_q] \\
    &+ \text{(possible) cut-off contribution}.
\end{aligned}
\end{equation}
%

We repeat once again the procedure in \S\ref{sec:HWgaugemore} to define the IR-regulated area.
The new cut-off surface depends only on the metric $G+\kappa h_{0,2,\mathbf{0}}$, and the result in \eqref{eqn:1stordercutoff} can be extended to $O(\kappa^2)$.
One finds that the answer in \eqref{eqn:1stordercutoff} also holds to $O(\kappa^2)$.\footnote{To understand this statement, we need to remember how the cut-off surface gives an area contribution. The classical unperturbed area $A[G,\gamma^{(0)}]$ is divergent, and a shift in the cut-off surface, although it is $\kappa r_c^{-\#}$ suppressed, can make $A[G,\gamma^{(0)}]$ change by a finite amount. This is what we saw in the $O(\kappa)$ calculation. Now when we consider $O(\kappa^2)$, we have the same divergent quantity, but the shift of the cut-off surface is more suppressed in $r_c$ (it is roughly $\kappa^2  r_c^{-2\#}$). This means no finite contribution appears at $O(\kappa^2)$ due to the $O(\kappa)$ perturbation of the metric. There is, of course, the $O(\kappa^2)$ cut-off contribution due to the $O(\kappa^2)$ metric perturbation $g^{(2)}$, which is exactly \eqref{eqn:cutoff}.}

With both the cut-off surface and the (perturbed) extremal surface, we can determine the endpoint of the surface, and there should be no obstacles in obtaining the correct answer for the quadratic area difference. 
One has to be careful, though, about one last point regarding a switch of the parametrization of the surface: 
inspecting the surface perturbation result obtained in \eqref{eq:surfacepertexplicit}, we see that when approaching $r=r_{\mathrm{min}}$, which is the deepest point that the unperturbed extremal surface $\gamma^{(0)}$ can reach in the bulk, the perturbation diverges. This simply means that for the perturbed surface, the minimal radial coordinate that the surface can reach is greater than $r_{\mathrm{min}}$. 
One can solve for the perturbed minimal radius of the surface to find the shift of the lower bound of the radial integral, but it complicates the computation. 
In practice, we simply switch to the alternative $(t(\theta),r(\theta),\phi_{i})$ parametrization for $\gamma^{(1)}$ in practice. We find the cut-off in $\theta$ to be $\theta_{c}^{\rm new}=\theta_0 -\frac{\cot \theta_0}{2 r_c^2} - \frac{\cot \theta_0-\cot \theta_0^3}{r_c^4} - \kappa \frac{2 \sin 2\theta_0 }{5 r_c^4}  + O(r_c^{-6})$.

Now the quadratic area can be computed in any dimension and we find for $d=4$ 
\begin{equation}
\begin{aligned}
   &A[G+\kappa h_{0,2,\mathbf{0}},  \gamma^{(0)} + \kappa \gamma^{(1)*}_{0,2,\mathbf{0}}]|_{d=4} \\
   &= V_{S^2} |\mathcal{N}_{0,2}^S|^2 \left( -\frac{128}{525} \theta_0^8 + \frac{1024}{3465}\theta_0^{10}- \frac{233984}{1126125} \theta_0^{12} +\frac{604928}{4729725} \theta_0^{14}  - \frac{970016128}{18091198125}\theta_0^{16} + O(\theta_0^{18})\right),
\end{aligned}
\end{equation}
where there is no contribution from the cut-off surface. 

\paragraph{Final answer for area correction.}
We summarize the final result for the first and second order area correction in $d=4$ dimensions, for the superposition state $| \mathfrak{s}  \rangle$ and the single particle state $|\mathfrak{g} \rangle $ respectively
\begin{equation}\label{eqn:arearesfinal}
\begin{aligned}
    \langle \mathfrak{s} | A[\hat{g},\hat{\gamma}] | \mathfrak{s} \rangle |_{O(\kappa),d=4} &=  V_{S^2} |\mathcal{N}_{0,2}^S| \sqrt{2} \frac{4 \sin^4 \theta_0}{15}\,,\\
    \langle \mathfrak{g} | A[\hat{g},\hat{\gamma}]| \mathfrak{g} \rangle |_{O(\kappa^2),d=4} &= V_{S^2} |\mathcal{N}_{0,2}^S|^2  \Big( \frac{296}{225} \theta_0^4- \frac{3088}{4725} \theta_0^6 + \frac{104}{2625}\theta_0^8  + \frac{12112}{212625}\theta_0^{10} \\
    & - \frac{547376}{22325625}\theta_0^{12} + \frac{92384}{16372125}\theta_0^{14} - \frac{43068184}{47888465625}\theta_0^{16} + O(\theta^{18}) \Big).
\end{aligned}
\end{equation}
Recall that the superposition state $\ket{\mathfrak{s}}$ certainly has a non-zero area expectation value at $O(\kappa^2)$, but it is not included in our analysis as it is no more interesting than the single-particle state $\ket{\mathfrak{g}}$ at that order.

Finally, let us remark that although we have separated the area calculation into multiple pieces to make it understandable, one can simply define the cut-off surface up to $O(\kappa^2)$, which is simply a summation of \eqref{eqn:1stordercutoff} and \eqref{eqn:2ndordercutoff}, and the extremal surface $\gamma$ up to $O(\kappa)$ for the full perturbed metric $g=G+\kappa \gamma^{(1)}+\kappa^2 g^{(2)}$, and compute the area $A[g,\gamma_{\rm cut}] - A[G,\gamma_{\rm cut}^{(0)}]$ using the procedure described in \S\ref{sec:HWgaugemore}, and it will give \eqref{eqn:arearesfinal} in a very clean way.

\paragraph{Large diffeomorphisms and the extremal gauge.} 

Now we want to understand the large diffeomorphisms mentioned in \S\ref{sec:HWgaugemore}, which we will present from a general way of determining the perturbation of extremal surfaces. 
We use the results in \cite{Mosk:2017vsz} to describe the perturbed extremal surface anchored on a polar cap boundary subregion in vacuum AdS. 
Moreover, we restrict to perturbations normal to the unperturbed surface (any tangent component of $\gamma^{(1)}$ can be removed by reparametrization of the surface).\footnote{This is to get a closed formula since the tangent direction perturbation is arbitrary.} 
Therefore, we can expand the perturbation in terms of the two normal vectors:
\begin{equation}\label{eqn:gamma_exp}
    \gamma_{\mu}^{(1)} = \sum_{m=1}^{2}\varpi_{m}n_{(m)\mu}.
\end{equation}
Then, in our simple case of interest, the equations for the different $\varpi_{m}$ decouple and we find
\begin{equation}\label{eqn:perteqn}
    \left(\nabla_{H^{d-1}}^{2}-(d-1)\right)\varpi_{m} + (-1)^{m}\delta_{\kappa h}K_{(m)}|_{\kappa=0} = 0, \qquad m=1,2,
\end{equation}
where 
\begin{equation}\label{eqn:firstorderK_term2}
    \delta_{\kappa h}K_{(m)}|_{\kappa=0} = 
    -q^{(0)\mu\nu}{\Gamma^{(1)\eta}_{\mu\nu}}n_{(m)\eta}^{(0)},
\end{equation}
and $\nabla_{H^{d-1}}^{2}$ is the Laplacian for the extremal surface which is a copy of $H^{d-1}$.
Note that to obtain this rather simple form for the result, we have used the conditions that the vacuum AdS spacetime is maximally symmetric and $K_{\mu\nu}^{(0)}=0$ for the classical extremal surface $\gamma^{(0)}$ have been used.

The extremality equation \eqref{eqn:perteqn} is simply the massive Klein-Gordon equation for a scalar $\varpi_{m}$ on $H^{d-1}$ with $m^{2}=d-1$ with a source term $\delta_{\kappa h}{^{(m)}K}|_{\kappa=0}$. This source term \eqref{eqn:firstorderK_term2} is not simple as it depends on $h_{\mu\nu}$ and its derivatives. Nevertheless, we can always invert this to obtain a solution
\begin{equation}\label{eqn:pertsoln}
    \varpi_{m}(\alpha) = (-1)^{m+1}\int_{H^{d-1}}d^{d-1}\beta \, G(\alpha,\beta) \left(\delta_{\kappa h}K_{(m)}(\beta)|_{\kappa=0}\right), \qquad m=1,2.
\end{equation}
where $G(\alpha,\beta)$ is the Green's function for the massive Klein-Gordon operator defined by
\begin{equation}\label{eqn:Greensfn}
    \left(\nabla_{H^{d-1}}^{2}-(d-1)\right)G(\alpha,\beta) = \frac{1}{\sqrt{G_{H^{d-1}}}}\delta^{d-1}(\alpha-\beta).
\end{equation}
This formula for the perturbed surface as a function of the graviton allows us to understand general properties of this surface in terms of general properties of the graviton. For instance, the graviton in AdS has fall-off at asymptotic infinity dictated by the extrapolate dictionary. One can check from \eqref{eqn:gamma_exp}, \eqref{eqn:pertsoln}, \eqref{eqn:firstorderK_term2}, using only large-$r$ behavior of the background metric and the normal vectors to the unperturbed surface, that the perturbed surface will have the same fall-off as the graviton:
\begin{equation}\label{eqn:surfacepert_falloff}
    h_{\mu\nu} \sim \frac{1}{r^{d-2}} \quad \mathrm{and} \quad h_{\mu r} \sim \frac{1}{r^{d-1}} \implies \gamma_{t}^{(1)} \sim \frac{1}{r^{d-2}}, \quad \gamma_{r}^{(1)} \sim \frac{1}{r^{d+1}}, \quad \gamma_{\theta}^{(1)} \sim \frac{1}{r^{d}}.
\end{equation}
This means that the diffeomorphism to go to Hollands-Wald gauge will be large, in particular, it is given by
\begin{equation}
    v^{\mu} = \underbrace{\underbrace{-\gamma^{(1)\mu}}_{\mathrm{ext.\;gauge}} + s^{\mu}}_{\mathrm{HW\;gauge}}
\end{equation}
where\footnote{We have fixed a numerical factor in \cite{Mosk:2017vsz} that was incorrect.}  $s^{\mu} = -\frac{1}{2\pi}\xi_{\nu}h^{\mathrm{ext},\mu\nu}+\frac{1}{4\pi}\sigma^{\mu\nu}h_{\nu\alpha}^{\mathrm{ext}}\xi^{\alpha}$ vanishes on $\gamma^{(0)}$, with $h^{\mathrm{ext}}$ the graviton in extremal gauge, but it is necessary to impose the second condition of Hollands-Wald \eqref{eqn:HWcondition2}. We will henceforth ignore $s^{\mu}$ and focus on the extremal gauge part of the diffeomorphism $v^{\mu}$.

One can explicitly check \eqref{eqn:surfacepert_falloff} using our global radial gauge calculation, where the extremal surface deviation is given by \eqref{eq:surfacepertexplicit} and \eqref{eq:surfacepertexplicittheta}, and one only need subtract off the tangent part. 
From \eqref{eq:surfacepertexplicit}, we get that $\tilde{\gamma}^{(1)\mu}=\gamma^{(1)\mu} - t^{\mu} (t_{\nu} \gamma^{(1)\nu})$ has 
\begin{equation}
    \tilde{\gamma}^{(1)t}\sim \frac{1}{r^{d}}, \quad    \tilde{\gamma}^{(1)r}\sim \frac{1}{r^{d-1}}, \quad \tilde{\gamma}^{(1)\theta} \sim \frac{1}{r^{d+2}}\,.
\end{equation}

Therefore, to go to the extremal gauge (or Hollands-Wald gauge), we need to pick a vector field ${v}$ to counter this $\gamma^{(1)}$ (at least the normal components). 
We can see that $v$ is a large diffeomorphism, because the corresponding metric perturbation, to the linear order in $\kappa$, $\nabla_{(\mu}v_{\nu)}$ gives a non-trivial profile under the extrapolate dictionary, with the result $\nabla_{(t}v_{t)}\sim \frac{2}{15}{r^{-2}}(8-17\cos 2\theta)$ and $\nabla_{(t}v_{\theta)}\sim \frac{12}{5}{r^{-2}}\sin2\theta$. The new graviton wavefunction is $h^{\rm radial}+\pounds_{v}G$, and one can use this to directly reproduce the area results in \eqref{eqn:arearesfinal}, with the benefit that there is no need to worry about $\gamma^{(1)}$.

\subsection{Canonical energy and JLMS formula}
\label{sec:JLMStest}

Now that we have the expectation values of the area operator, we can test the JLMS formula for gravitons \eqref{eqn:JLMSgraviton} proposed in \S\ref{sec:JLMSgraviton}, as well as the alternate prescription involving $\int T^{\mathrm{grav}}$ in \eqref{eqn:JLMSgraviton2}. 

First, we analyse the JLMS formula at $O(\kappa^{-1})$ for the superposition state $\ket{\mathfrak{s}}$. The expectation value of the bulk vacuum modular Hamiltonian (or equivalently, the canonical energy) is $O(\kappa^{0})$ in this state so only the piece of the area operator $\bra{\mathfrak{s}} {A}[\hat{h},\gamma^{(0)}]\ket{\mathfrak{s}}$ contributes, with the result for $d=4$ given in \eqref{eqn:arearesfinal}. On the CFT side, the vacuum-subtracted expectation value of the CFT vacuum modular Hamiltonian in the CFT superposition state $\ket{S}$ was computed in \eqref{eqn:DeltaKCFT_superpos}. Using the following standard AdS/CFT relation between $C_{T}$ and $G_{N}$ \cite{Kovtun:2008kw}:\footnote{Note that in using their writing of the relation, we have $c=C_{T}/2$).}
\begin{equation}\label{eqn:CT_GNreln}
    C_{T} = \frac{V_{S^{d-1}}}{4\pi G_{N}}\frac{\Gamma(d+2)}{(d-1)\Gamma(\frac{d}{2})^{2}},
\end{equation}
and the normalisation constant \eqref{eqn:graviton_normconstant}, we find that the JLMS formula holds at this order for the superposition state
\begin{equation}\label{eqn:JLMSleadingorder}
(\Delta K^{S}_{B})_{\Delta_{\mathrm{gap}} \to \infty} = \frac{\bra{\mathfrak{s}}A^{\mathrm{lin}}[\hat{h},\gamma^{(0)}]\ket{\mathfrak{s}}}{\kappa}.
\end{equation}

Next, we examine the JLMS formula for gravitons \eqref{eqn:JLMSgraviton} at $O(\kappa^{0})$ for the single-particle state $\ket{\mathfrak{g}}$. This requires that we compute the expectation value of the bulk (gauge-invariant) vacuum modular Hamiltonian $K_{0}$. As proven in \S\ref{sec:JLMSgraviton}, $K_{0}$ is equal to the canonical energy in Hollands-Wald gauge, which can be written in a gauge-invariant way in terms of the canonical energy in any gauge by including a crucial boundary term localized on the entangling surface $\gamma^{(0)}$ given in \eqref{eqn:Ecan_gaugetrans}. 
We thus define the gauge-invariant bulk vacuum modular Hamiltonian operator in terms of global Hilbert space operators as 
\begin{equation}\label{eqn:Ecanop}
   \hat{K}_{b} = \, \normord{\hat{E}_{\rm can}[\hat{h}^{\mathrm{HW}}]} \; = \sum_q \Big(E_{\rm can}[h_q] + \int_{\gamma^{(0)}} \Upsilon_q\Big) a^{\dagger}_q a_q
\end{equation}
where we have relied on the argument in \S\ref{sec:restoregaugeinv} that $E_{\rm can} + \int \Upsilon$ is gauge-invariant. 

By taking the expectation value of $\hat{K}_{b}$ in the bulk state $\ket{\mathfrak{g}}$, dual to the boundary state $\ket{\tilde{\epsilon}\cdot {T}}$ in \eqref{eqn:specialpol}, one can pick out the $q=(0,2,\mathbf{0})$ mode and find the bulk canonical energy, which for $d=4$ is given by\footnote{The calculation can be performed for any fixed $d$ but here we only present $d=4$.}
\begin{equation}\label{eqn:Ecan1}
\begin{aligned}
       &E_{\rm can}[h_{0,2,\mathbf{0}}]|_{d=4}  = \int_b \omega(h_{0,2,\mathbf{0}},\pounds_\xi h_{0,2,\mathbf{0}})  \\
    & = V_{S^2}|\mathcal{N}_{0,2}^S|^2 \Big( - \frac{128}{63} \theta_0^8 + \frac{53888}{10395} \theta_0^{10} - \frac{1440896}{405405}\theta_0^{12}  + \frac{9238784}{8513505}\theta_0^{14} - \frac{540224}{2631447}\theta_0^{16} + O(\theta^{18}) \Big).
\end{aligned}
\end{equation}
and the entangling surface contribution is\footnote{One can do an independent calculation of the canonical energy in the Hollands-Wald gauge with no boundary term and find
\begin{equation}
        E_{\rm can}[h^{\rm HW}_{0,2,\mathbf{0}}]|_{d=4} = \int_b \omega(h^{\rm HW}_{0,2,\mathbf{0}},\pounds_\xi h^{\rm HW}_{0,2,\mathbf{0}}) =  \eqref{eqn:Ecan1}+ \eqref{eqn:Ecan2},
\end{equation}
as expected.} 
\begin{align}\label{eqn:Ecan2}
\begin{split}
    &\int_{\gamma^{(0)}} \Upsilon_{0,2,\mathbf{0}}|_{d=4} 
    \\  &= V_{S^2}|\mathcal{N}_{0,2}^S|^2 \Big( \frac{3968}{1575} \theta_0^8- \frac{11392}{2079} \theta_0^{10} + \frac{36870112}{10135125}\theta_0^{12}  - \frac{47029504}{42567525}\theta_0^{14} + \frac{289899712}{1391630625}\theta_0^{16} + O(\theta^{18}) \Big).
\end{split}
\end{align}

We can now compare with the CFT result \eqref{eqn:K0dif} for $d=4$ in the $\theta_0$ expansion 
\begin{equation}
\begin{aligned}
 \Delta K_{B}^{\tilde{\epsilon}\cdot \tilde{T} }|_{d=4} = & \frac{74}{135} \theta_0^4 - \frac{772}{2835} \theta_0^6 +
 \frac{62}{525}\theta_0^8 - \frac{53092}{1403325} \theta_0^{10} + \\
    &    
 \frac{17747468}{1915538625} \theta_0^{12} - 
 \frac{221992}{127702575} \theta_0^{14} + 
 \frac{124401658}{488462349375} \theta_0^{16} + O(\theta_0^{18}).
\end{aligned}
\end{equation}
Combining our results \eqref{eqn:Ecan1} plus \eqref{eqn:Ecan2} for $\Delta K_{b}^{0}$ with the expectation value of the area operator at $O(\kappa^{2})$ in \eqref{eqn:arearesfinal}, we find an exact match
\begin{equation}\label{eqn:JLMSsecondorder}
    \Delta K_{B}^{\tilde{\epsilon}\cdot \tilde{T} } = 4\pi\frac{ \langle  \mathfrak{g} | A[\hat{g},\hat{\gamma}]| \mathfrak{g} \rangle}{\kappa^2} + 
 \Delta K_{b}^{\mathfrak{g}} 
\end{equation}
using the normalisation factor $|\mathcal{N}_{0,2}^S|$ in \eqref{eqn:graviton_normconstant}.

For completeness, we also present the alternative version of the JLMS formula in \S\ref{sec:altprescript}, in which we need $A^{\rm lin}[g^{(2)},\gamma^{(0)}]$ already computed in \eqref{eqn:linearizedarea_backreactmetric_explicit} and $\int T^{\rm grav}$ given explicitly in $d=4$ as follows 
\begin{equation}
\begin{aligned}
    \langle \mathfrak{g} | \int_{b} d^dx & \sqrt{-G_b}:T^{\rm grav}_{\mu\nu}:\tau^{\mu}\xi^{\nu} | \mathfrak{g} \rangle \Big|_{O(\kappa^2),d=4}  \\
    = & V_{S^2} |\mathcal{N}_{0,2}^S|^2  \Big( - \frac{1024}{3465} \theta_0^{10} - \frac{8704}{27027} \theta_0^{12} + \frac{8704}{36855}\theta_0^{14}  - \frac{15083008}{144729585}\theta_0^{16}
     + O(\theta^{18}) \Big),
\end{aligned}
\end{equation}
and again we find an exact match 
\begin{equation}
		\Delta K_{\rm B}^{\tilde{\epsilon}\cdot \tilde{T} } = 4\pi\frac{ \langle \mathfrak{g} | {A}^{\rm lin}[\hat{g}^{(2)},\gamma^{(0)}] | \mathfrak{g} \rangle }{ \kappa^2} +  \langle \mathfrak{g} | \int_{b} d^dx \sqrt{-G_b}:T^{\rm grav}_{\mu\nu}:\tau^{\mu}\xi^{\nu} | \mathfrak{g} \rangle.
\end{equation}



\subsection{AdS-Rindler quantization}
\label{sec:AdS-Rindler_graviton}

The background entanglement wedge $\mathcal{D}(b)$ is an AdS-Rindler wedge whose bifurcation surface is the unperturbed extremal surface $\gamma^{(0)}$. To obtain the vN entropy for the graviton excited state $\ket{\mathfrak{g}}$ in the subregion $b$, we need to quantize the graviton in this AdS-Rindler wedge and relate the global modes analysed in \S\ref{sec:gravitonglobal} to the AdS-Rindler modes.

The AdS-Rindler metric is given by
\begin{equation}\label{eqn:AdSRindlermetric}
ds^{2} = -(\rho^{2}-1)d\tau^{2} + \frac{d\rho^{2}}{\rho^{2}-1} + \rho^{2}\underbrace{\left(du^{2}+\sinh^{2}u\,d\Omega_{d-2}^{2}\right)}_{(dH^{d-1})^{2}},
\end{equation}
where $\tau \in (-\infty,\infty)$, $\rho \in (1,\infty)$ and $\gamma^{(0)}$ lies at $\{\tau=0,\rho=1\}$. The domain of dependence $D(B)$ of $B$ in the asymptotic boundary lies at $\rho = \infty$. The transformation from AdS-Rindler to global AdS coordinates can be found in App.~\ref{sec:AdS-Rindler}, and one can see that the size $\theta_{0}$ of the polar cap $B$ on the boundary controls the size of this AdS-Rindler wedge.

We will quantize the graviton in the AdS-Rindler wedge in the same way as global AdS: we find the constraints and gauge-fix classically to obtain the reduced phase space, then construct the Dirac bracket which is promoted to a commutator. We will be brief as we already gave a detailed analysis in the global case.

\paragraph{Phase space.}

The unconstrained phase space consists of the space of initial conditions $\mathfrak{P}^{R} = \{\mathpzc{h}_{\mu\nu},\pi^{\mu\nu}\}$ on a fixed-$\tau$ slice. There are $d+1$ primary constraints on phase space coming from the fact that $\mathpzc{h}_{\mu\tau}$ acts as a Lagrange multiplier in the Lagrangian
\begin{equation}\label{eqn:primaryconstraints_Rindler}
    \mathpzc{C}^{\mu} \equiv \pi^{\mu \tau} \approx 0.
\end{equation}
The primary Hamiltonian obtained from Legendre transformation of the Lagrangian and adding the primary constraints times Lagrange multipliers gives
\begin{align}\label{eqn:primH_Rindler}
\begin{split}
&H_{p}^{R} = \int_{b} d^{d}x\,\bigg[{-}\frac{G_{\tau\tau}}{\sqrt{-G}}\left(\pi^{\mathpzc{\mathfrak{i}}\mathpzc{\mathfrak{j}}}\pi_{\mathfrak{i}\mathfrak{j}}-\frac{1}{(d-1)}(\pi_{\mathfrak{k}}^{\mathfrak{k}})^{2}\right)+\sqrt{-G}\,\mathpzc{h}^{\tau\tau}E_{\tau\tau}^{(1)}[h]-2\hat{\nabla}_{\mathfrak{i}}\pi^{\mathfrak{i}\mathfrak{j}}\mathpzc{h}_{\mathfrak{j}\tau}+u_{\mu}^{R(p)}\pi^{\mu t}
\\  &+\sqrt{G_{\Sigma}}\left(\frac{1}{4}\mathpzc{h}_{\mathfrak{i}}^{\mathfrak{i}}\hat{\nabla}^{2}\mathpzc{h}_{\mathfrak{j}}^{\mathfrak{j}}-\frac{1}{4}\mathpzc{h}^{\mathfrak{i}\mathfrak{j}}\hat{\nabla}^{2}\mathpzc{h}_{\mathfrak{i}\mathfrak{j}}-\frac{1}{2}\mathpzc{h}_{\mathfrak{i}}^{\mathfrak{i}}\hat{\nabla}^{\mathfrak{j}}\hat{\nabla}^{\mathfrak{k}}\mathpzc{h}_{\mathfrak{j}\mathfrak{k}}+\frac{1}{2}\mathpzc{h}^{\mathfrak{i}\mathfrak{j}}\hat{\nabla}^{\mathfrak{i}}\hat{\nabla}_{\mathfrak{i}}\mathpzc{h}_{\mathfrak{k}\mathfrak{i}}+(d-1)\mathpzc{h}_{\mathfrak{i}\mathfrak{j}}\mathpzc{h}^{\mathfrak{i}\mathfrak{j}}-\frac{(d-1)}{2}\mathpzc{h}_{\mathfrak{i}}^{\mathfrak{i}}\mathpzc{h}_{\mathfrak{j}}^{\mathfrak{j}}\right)\bigg]
\end{split}
\end{align}
where $\mathpzc{u},\mathpzc{v},\mathpzc{w},\ldots$ are spatial indices, $u_{\mu}^{R(p)}$ are Lagrange multipliers, and $\hat{\nabla}$ is covariant derivative for $b$. Requiring that the primary constraints hold for all times leads to $d+1$ secondary constraints from $\{\mathpzc{C}^{\mu},H_{p}^{R}\}=0$, viz.,
\begin{align}\label{eqn:diffconstr_Rindler}
\begin{split}
    \mathrm{Hamiltonian\;constraint:\;}& \tilde{\mathpzc{C}}_{\tau} \equiv \sqrt{-G}E_{\tau}^{(1)\tau} = 0
    \\ \mathrm{Momentum\;constraints:\;}& \tilde{\mathpzc{C}}_{\rho} \equiv \sqrt{-G}E_{\rho}^{(1)\tau} = 0, \; \tilde{\mathpzc{C}}_{\mathfrak{i}} \equiv E_{\mathfrak{i}}^{(1)\tau} = 0,
\end{split}
\end{align}
where $\mathfrak{i}$ are coordinates on the hyperbolic space $H^{d-1}$. We will choose Rindler holographic gauge to obtain $d+1$ more constraints
\begin{equation}\label{eqn:gauge-fixing_Rindler}
    \mathpzc{G}_{t} \equiv \pi_{\rho\rho}-\frac{1}{(d-1)}G_{\rho\rho}\pi_{\mathfrak{j}}^{\mathfrak{j}} = 0, \qquad \mathpzc{G}_{\mathfrak{i}} \equiv \mathpzc{h}_{\rho \mathfrak{i}} = 0,
\end{equation}
and an argument analogous to \eqref{eqn:htr=0} gives $\mathpzc{G}_{t} = 0 \implies \mathpzc{h}_{\rho\tau}=0$, and we get the final $d+1$ constraints by requiring this hold for all times $\{H_{\mathrm{p}}^{R},\mathpzc{G}_{\mu}\}=0$, implying
\begin{align}\label{eqn:gauge-fixingalltimes_Rindler}
\begin{split}
    \tilde{\mathpzc{G}}_{\tau} &\equiv {-}\frac{\rho}{(\rho^{2}-1)^{2}}\hat{\nabla}_{\rho}\mathpzc{h}_{\tau\tau}+\frac{\rho}{2}\hat{\nabla}_{\rho}\mathpzc{h}_{\rho\rho}-(d-1)\mathpzc{h}_{\rho\rho}
    \\  &+\frac{1}{(d-1)}\left(\frac{(d-2)}{2}(\hat{\nabla}_{\alpha}\hat{\nabla}^{\alpha}\mathpzc{h}_{\rho\rho}+\hat{\nabla}_{\rho}\hat{\nabla}_{\rho}\mathpzc{h}_{\alpha}^{\alpha}+G_{\rho\rho}\mathpzc{h}_{\alpha}^{\alpha}-2\hat{\nabla}_{\alpha}\hat{\nabla}_{\rho}\mathpzc{h}^{\alpha}_{\rho})+\frac{1}{2}G_{\rho\rho}\hat{\nabla}_{\alpha}\hat{\nabla}_{\beta}\mathpzc{h}^{\alpha\beta}\right) = 0
    \\  \tilde{\mathpzc{G}}_{\mathfrak{i}} &\equiv 2\sqrt{-G}\,\hat{\nabla}_{\mathfrak{(i}}\mathpzc{h}_{\rho)\tau}-G_{\tau\tau}\left(\pi_{\mathfrak{i}\rho}-\frac{1}{(d-1)}\pi_{\mathfrak{k}}^{\mathfrak{k}}G_{\mathfrak{i}\rho}\right) = 0,
\end{split}
\end{align}
where $\alpha$ labels coordinates on $H^{d-1}$. Note that this is not the same gauge as the one used for the graviton in global AdS, but it will not matter because we will compute the Bogoliubov coefficients relating the two sets of modes in a manifestly gauge-invariant way. The reduced phase space is thus
\begin{equation}\label{eqn:reducedphasespace_Rindler}
    \mathfrak{R}^{R} = \{(\mathpzc{h}_{\mu\nu},\pi^{\mu\nu})\,|\,\mathpzc{C}^{\mu}=\tilde{\mathpzc{C}}_{\mu}=\mathscr{G}_{\mu}=\tilde{\mathpzc{G}}^{\mu}=0\}.
\end{equation}

We then construct the Dirac bracket from the Poisson bracket matrix of all the constraints $\mathpzc{C}_{ab}=\{\mathpzc{C}_{a},\mathpzc{C}_{b}\}$ just as we did in \eqref{eqn:Diracbracket}, but we do not present the details here. The presymplectic form given in \S\ref{sec:HWformula} on the unconstrained phase space $\mathfrak{P}^{R}$ is given by
\begin{align}\label{eqn:symplecticform_graviton_Rindler}
\begin{split}
\Omega_{\mathfrak{P}^{R}}(\mathpzc{h}_{1},\mathpzc{h}_{2}) &= \int_{b}d^{d}x\,\sqrt{-G}\bigg(\frac{1}{2} \mathpzc{h}_{1} \nabla_{\nu} \mathpzc{h}_{2}^{\tau\nu}-\frac{1}{2}  \mathpzc{h}_{1} \nabla^{\tau} \mathpzc{h}_{2}+\frac{1}{2} \mathpzc{h}_{1}^{\mu\nu} \nabla^{\tau} \mathpzc{h}_{2,\mu\nu}
\\  &\qquad \qquad \qquad \qquad \qquad \qquad - \mathpzc{h}_{1,\nu\gamma} \nabla^{\gamma} \mathpzc{h}_{2}^{\nu\tau}+\frac{1}{2} \mathpzc{h}_{1}^{\tau\nu} \boldsymbol{\wedge} \nabla_{\nu} \mathpzc{h}_{2} - (1 \leftrightarrow 2)\bigg).
\end{split}
\end{align}
with the symplectic form on the reduced phase space $\mathfrak{R}^{R}$ obtained from restriction $\Omega_{\mathfrak{R}^{R}}=\Omega_{\mathfrak{P}^{R}}|_{\mathpzc{C}^{\mu}=\tilde{\mathpzc{C}}_{\mu}=\mathscr{G}_{\mu}=\tilde{\mathpzc{G}}^{\mu}=0}$.

\paragraph{Equations of motion.}

To solve the linearized Einstein equations, we will follow the same method as we did for global case. We decompose the graviton $\mathpzc{h}_{\mu\nu}$ in AdS-Rindler based on representations of the $SO(d-1,1)$ isometry group of the $H_{d-1}$ part of the metric, namely into tensor, vector, and scalar parts. Whether such a decomposition is the most general possible form of a symmetric $2$-tensor is not known by mathematicians due to the non-compactness of $H_{d-1}$, but it is expected to be true and to be unique so we will assume this is the case.\footnote{We thank Akihiro Ishibashi for discussions on this.}

Let $\mathfrak{v},\mathfrak{w}\ldots$ denote indiceds for the two-dimensional $(\tau,\rho)$ part of the AdS-Rindler spacetime and $\alpha,\beta,\gamma$ denote indices for $H^{d-1}$. We can write the graviton as
\begin{equation}\label{eqn:gravitonexp_Rindler}
    \mathpzc{h}_{\mu\nu}dx^{\mu}dx^{\nu} = \mathpzc{h}_{\mathfrak{v}\mathfrak{w}}dx^{\mathfrak{v}}dx^{\mathfrak{w}}+2\mathpzc{h}_{\mathfrak{v}\alpha}dx^{\mathfrak{v}}dx^{\alpha}+\mathpzc{h}_{\alpha\beta}dx^{\alpha\beta}.
\end{equation}
The $\mathpzc{h}_{\mathfrak{v}\mathfrak{w}}$ part transforms as a scalar under the action of $SO(d-1,1)$ so\footnote{By an abuse of notation, we denote the eigenvalues of the scalar Laplacian on $H^{d-1}$ by $\lambda \equiv \lambda_{S}$.}
\begin{equation}\label{eqn:hvwexp}
    \mathpzc{h}_{\mathfrak{v}\mathfrak{w}}(\tau,\rho,\Xi) = \sum_{\lambda}\tilde{\mathfrak{h}}_{\mathfrak{v}\mathfrak{w},\lambda}(\tau,\rho)H_{\lambda}(\Xi),
\end{equation}
where $H_{\lambda}(\Xi)$ are the eigenfunctions of the scalar Laplacian discussed in App.~\ref{sec:hyperbolicLaplacian}, the $\mathpzc{h}_{\mathfrak{v}\alpha}$ part transforms as a vector so we decompose it as
\begin{equation}\label{eqn:hviexp}
    \mathpzc{h}_{\mathfrak{v}\alpha}(\tau,\rho,\Xi) = \sum_{\lambda_{V}}\tilde{\mathfrak{h}}_{\mathfrak{v},\lambda_{V}}^{V_{H}}(\tau,\rho)V_{\alpha,\lambda_{V}}(\Xi)+\sum_{\lambda}\tilde{\mathfrak{h}}_{\mathfrak{v},\lambda}^{S_{H}}(\tau,\rho)D_{\alpha}H_{\lambda}(\Xi),
\end{equation}
where $V_{\alpha,\lambda_{V}}(\Xi)$ are eigenfunctions of the vector Laplacian on $H^{d-1}$, and $\mathpzc{h}_{\alpha\beta}$ transforms as a tensor so we decompose this as
\begin{align}\label{eqn:hijexp_Rindler}
\begin{split}
    &\mathpzc{h}_{\alpha\beta}(\tau,\rho,\Xi) = \sum_{\lambda_{T}}\tilde{\mathfrak{h}}_{\lambda_{T}}^{T_{H}}(\tau,\rho)T_{\alpha\beta,\lambda_{T}}^{H}(\Xi) +2\sum_{\lambda_{V}}\tilde{\mathfrak{h}}_{\lambda_{V}}^{V_{H}}(\tau,\rho)D_{(\alpha}V_{\beta),\lambda_{V}}^{H}(\Xi)
    \\  &+\sum_{\lambda}\left(\tilde{\mathfrak{h}}_{\lambda}^{S_{H},\mathrm{tr}}(\tau,\rho)(G_{H^{d-1}})_{\alpha\beta}H_{\lambda}(\Xi)+\tilde{\mathfrak{h}}_{\lambda}^{S_{H},\mathrm{trless}}(\tau,\rho)\left(D_{\alpha}D_{\beta}-\frac{1}{d-1}(G_{H^{d-1}})_{\alpha\beta}D_{\mathfrak{k}}D^{\mathfrak{k}}\right)H_{\lambda}(\Xi)\right),
\end{split}
\end{align}
where $T_{\alpha\beta,\lambda_{T}}^{H}(\alpha)$ are eigenfunctions of the tensor Laplacian on $H^{d-1}$.

Therefore, the scalar part of the graviton is given by
\begin{align}\label{eqn:gravscalar_Rindler}
\begin{split}
    \mathpzc{h}_{\mu\nu}^{S_{H}}dx^{\mu}dx^{\nu} &= \sum_{\lambda}\bigg[\tilde{\mathfrak{h}}_{\mathfrak{v}\mathfrak{w},\lambda}H_{\lambda}dx^{\mathfrak{v}}dx^{\mathfrak{w}}+2\tilde{\mathfrak{h}}_{\mathfrak{v},\lambda}^{S_{H}}D_{\alpha}H_{\lambda}dx^{\mathfrak{v}}dx^{\alpha}
    \\  &\qquad +\left(\tilde{\mathfrak{h}}_{\lambda}^{S_{H},\mathrm{tr}}(g_{H^{d-1}})_{\alpha\beta}+\tilde{\mathfrak{h}}_{\lambda}^{S_{H},\mathrm{trless}}\left(D_{\alpha}D_{\beta}-\frac{1}{(d-1)}(g_{H^{d-1}})_{\alpha\beta}D_{\mathfrak{k}}D^{\mathfrak{k}}\right)\right)H_{\lambda}dx^{\alpha}dx^{\beta}\bigg],
\end{split}
\end{align}
the vector part is
\begin{equation}\label{eqn:gravvector_Rindler}
    \mathpzc{h}_{\mu\nu}^{V_{H}}dx^{\mu}dx^{\nu} = \sum_{\lambda_{V}}\left(\tilde{\mathfrak{h}}_{\mathfrak{v},\lambda_{V}}^{V_{H}}V_{\alpha,\lambda_{V}}^{H}dx^{\mathfrak{v}}dx^{\alpha}+2\tilde{\mathfrak{h}}_{\lambda_{V}}^{V_{H}}D_{(\alpha}V_{\beta),\lambda_{V}}^{H}dx^{\alpha}dx^{\beta}\right),
\end{equation}
and the tensor part is
\begin{equation}\label{eqn:gravtensor_Rindler}
    \mathpzc{h}_{\mu\nu}^{T_{H}}dx^{\mu}dx^{\nu} = \sum_{\lambda_{T}}\tilde{\mathfrak{h}}_{\lambda_{T}}^{T_{H}}T_{\alpha\beta,\lambda_{T}}^{H}dx^{\alpha}dx^{\beta}.
\end{equation}
The sums over eigenvalues is actually an integral because they are continuous so $``\sum_{\lambda_{A}}" = \int \frac{d\lambda_{A}}{2\pi}$.

The eigenfunctions of the vector and tensor Laplacians on $H^{d-1}$ are not known explicitly, but the eigenvalues for the vector eigenfunctions are known. Thankfully, we will only need the scalar part $\mathpzc{h}_{\mu\nu}^{S_{H}}$ as these are the only modes with non-zero inner product with the global mode of interest $h_{0,2,\mathbf{0}}^{S}$, which we prove in App.~\ref{sec:scalaroverlaps}.

The Einstein equations can be now be solved in the same way as the global case: use the ``master variable'' formalism to reduce the equations to a differential equation for a single scalar field $\phi_{S_{H}}$ and then use the solution to this equation to determine the scalar part of the graviton $\mathpzc{h}_{\mu\nu}^{S_{H}}$. While the solution for $\phi_{S_{H}}$ is simple, the $\rho$-dependence of the resulting solutions for $\tilde{\mathfrak{h}}_{\mathfrak{v}\mathfrak{w},\lambda}, \tilde{\mathfrak{h}}_{\mathfrak{v},\lambda}^{S_{H}}, \tilde{\mathfrak{h}}_{\lambda}^{S_{H},\mathrm{tr}},$ and $\tilde{\mathfrak{h}}_{\lambda}^{S_{H},\mathrm{trless}}$ is very complicated so we do not write them explicitly here. The wavefunction solutions take the form
\begin{equation}
\tilde{\mathfrak{h}}_{\mathfrak{v}\mathfrak{w},\lambda}(\tau,\rho) = \int \frac{d\omega}{2\pi}\tilde{\mathfrak{h}}_{\mathfrak{v}\mathfrak{w},\omega,\lambda}(\tau,\rho), \qquad \tilde{\mathfrak{h}}_{\mathfrak{v}\mathfrak{w},\omega,\lambda}(\tau,\rho) = e^{-i\omega\tau}\mathfrak{f}_{\mathfrak{v}\mathfrak{w},\lambda}(\rho)
\end{equation}
with the radial wavefunctions $\mathfrak{f}_{\mathfrak{v}\mathfrak{w},\lambda}(\rho)$ given by a finite sum of infinite sums of incomplete Beta functions, and similarly for $\tilde{\mathfrak{h}}_{\mathfrak{v},\lambda}^{S_{H}}, \tilde{\mathfrak{h}}_{\lambda}^{S_{H},\mathrm{tr}},$ and $\tilde{\mathfrak{h}}_{\lambda}^{S_{H},\mathrm{trless}}$. The details can be found in App.~\ref{sec:AdSRindlerEOM}. 

Nevertheless, in spite of the fact that the wavefunctions are very complicated, the normalisation of these modes are determined by the behavior of the graviton wavefunctions near the Rindler horizon ($\rho=1$) and the Bogoliubov coefficients can be found from the behavior near asymptotic infinity ($\rho=\infty$), and in both regions the wavefunctions simplify dramatically. This is what makes our computations analytically tractable.

\paragraph{Canonical quantization.}

The canonical quantization of the reduced phase space can be performed in the same way as was done in global AdS: the decomposition of the graviton into tensor, vector, and scalar parts means that we expand the graviton in creation and annihilation operators for each type of mode
\begin{align}\label{eqn:quantgravitonexp_Rindler}
\begin{split}
    \hat{\mathpzc{h}}_{\mu\nu}(x) &= \hat{\mathpzc{h}}_{\mu\nu}^{S_{H}}(x)+\hat{\mathpzc{h}}_{\mu\nu}^{V_{H}}(x)+\hat{\mathpzc{h}}_{\mu\nu}^{T_{H}}(x)
    \\  \hat{\mathpzc{h}}_{\mu\nu}^{A_{H}}(x) &= \sum_{I \in \{L,R\}}\int_{\omega>0}\frac{d\omega}{2\pi}\sum_{\lambda_{A}}\left(\mathpzc{h}_{\mu\nu,\omega,\lambda_{A},I}^{A_{H}}(x)b_{\omega,\lambda_{A},I}^{A_{H}} + \mathrm{h.c.}\right), \qquad A_{H}=S_{H},V_{H},T_{H}
\end{split}
\end{align}
with the Hilbert space given by a tensor product of the three Fock spaces $\mathbb{H}^{R} = \mathbb{H}^{S_{H},R} \otimes \mathbb{H}^{V_{H},R} \otimes \mathbb{H}^{T_{H},R}$. By $L$ and $R$ here, we mean `left' and `right' Rindler wedges, corresponding to $b^{c}$ and $b$, respectively.

Indeed, one can check that the different types of modes are orthogonal with respect to the generalized Klein-Gordon inner product obtained from the symplectic form via $\langle\mathpzc{h},\tilde{\mathpzc{h}}\rangle = i\Omega_{\mathfrak{R}^{R}}(\mathpzc{h}^{\ast},\tilde{\mathpzc{h}})$. The commutator comes from promoting the Dirac bracket $\{\cdot,\cdot\}_{\mathrm{D.B.}} \rightarrow i[\cdot,\cdot]$. We require that each mode of the tensor, vector, and scalar parts of the solution for the graviton are each delta-function normalisable with respect to the generalized Klein-Gordon inner product so that the corresponding annihilation and creation operators have normalised commutation relations:
\begin{equation}\label{eqn:deltafnnormalisability}
\langle h_{\omega,\lambda_{A},I}^{A_{H}},h_{\omega,\lambda_{B}',J}^{B_{H}}\rangle = \delta_{I,J}\delta^{A_{H},B_{H}}(2\pi)^{2}\delta(\omega-\omega')\delta^{(d-1)}(\lambda_{A}-\lambda_{B}')
\end{equation}
leading to
\begin{equation}\label{eqn:gravladderops_CCR_Rindler}
[b_{\omega,\lambda_{A},I}^{A_{H}},b_{\omega',\lambda_{B}',I}^{B_{H}\dagger}] = \delta_{I,J}\delta^{A_{H},B_{H}}(2\pi)^{2}\delta(\omega-\omega')\delta^{(d-1)}(\lambda_{A}-\lambda_{B}'), 
\end{equation}
with all other commutators vanishing. The delta function for quantum numbers on $H^{d-1}$ is defined by $\delta^{(d-1)}(\lambda_{A}-\lambda_{B}')\equiv \delta(\lambda_{A}-\lambda_{B}')\delta_{\mathfrak{m}_{A},\mathfrak{m}_{B}'}$ where $\mathfrak{m}_{A}$ labels the eigenfunctions of the $A$ Laplacian on $S^{d-2}$ with $A=S,V,T$.

Observe that the frequencies $\omega$ of the AdS-Rindler modes are continuous so the theory does not have a finite density of states. This is closely related to the fact that the global Hilbert space does not factorize into a tensor product of the Hilbert space for the left and right Rindler wedges in any quantum field theory due to infinite UV entanglement coming from modes near the entangling surface. These issues can be resolved by inserting a brick-wall cut-off at $\rho=\epsilon$ for some small $\epsilon > 0$. There is an additional subtlety coming from the diffeomorphism constraints \eqref{eqn:diffconstr} on the global Hilbert space that relate modes in $L$ and $R$. This obstruction to factorization is often remedied by the inclusion of edge modes on the entangling surface. In practice, however, we will compute vacuum-subtracted vN entropies which are not sensitive to any of these issues and give finite answers that are independent of $\epsilon$ and any edge modes,\footnote{It can be explicitly shown that for Maxwell theory in a Rindler wedge of Minkowski spacetime for Gaussian wavepackets \cite{Colin-Ellerin} and in AdS-Rindler spacetimes for single-particles states \cite{Colin-Ellerin:2024npf} that these edge modes never contribute to vacuum-subtracted vN entropies. We will see that in the graviton case we find agreement between the vacuum-subtracted generalized entropy and CFT vacuum-subtracted vN entropy without their inclusion.} that is, we expect our results to be valid in the continuum. Therefore, we will continue to work with continuous frequencies and ignore all of these issues in what follows.

With these caveats in mind, the (normalised) global vacuum in AdS is equal to the thermofield double state in AdS-Rindler
\begin{equation}\label{eqn:MinkvactoRindlerTFD}
\ket{0}_{S} \otimes \ket{0}_{V} \otimes \ket{0}_{T} = \otimes_{A_{H} \in \{S_{H},V_{H},T_{H}\}}\bigotimes_{\omega,\lambda_{A}^{\ast}}\sqrt{1-e^{-2\pi\omega}}\sum_{n}e^{-\pi E_{n}}\ket{n,\omega,\lambda_{A}^{\ast}}_{L,A_{H}}\ket{n,\omega,\lambda_{A}}_{R,A_{H}},
\end{equation}
where $E_{n}=\omega n$. Henceforth we will drop the product and simply write $\ket{0} = \sum_{n}e^{-\pi E_{n}}\ket{n^{\ast}}_{L}\ket{n}_{R}$ where $\ket{n^{\ast}}$ is the CRT conjugate of $\ket{n}$.

For our computations of the vN entropy, we will need the two-point functions of the $R$ Rindler annihilation and creation operators in the global AdS vacuum, which are given by
\begin{align}\label{eqn:AdSRindler2ptfns}
\begin{split}
\bra{0}b_{\omega,\lambda_{A},R}^{A_{H}}b_{\omega',\lambda_{A}',R}^{A_{H}}\ket{0} &= \bra{0}b_{\omega,\lambda_{A},R}^{A_{H}\dagger}b_{\omega',\lambda_{A}',R}^{A_{H}\dagger}\ket{0} = 0
\\	\bra{0}b_{\omega,\lambda_{A},R}^{A_{H}}b_{\omega',\lambda_{A}',R}^{A_{H}\dagger}\ket{0} &= \frac{(2\pi)^{2}}{1-e^{-2\pi\omega}}\delta(\omega-\omega')\delta^{d-1}(\lambda_{A}-\lambda_{A}')
\\	\bra{0}b_{\omega,\lambda_{A},R}^{A_{H}\dagger}b_{\omega',\lambda_{A}',R}^{A_{H}}\ket{0} &= \frac{(2\pi)^{2}}{e^{2\pi\omega}-1}\delta(\omega-\omega')\delta^{d-1}(\lambda_{A}-\lambda_{A}').
\end{split}
\end{align}

\paragraph{Bogoliubov coefficients.}

\begin{figure}
\begin{center}
\includegraphics[scale=0.6]{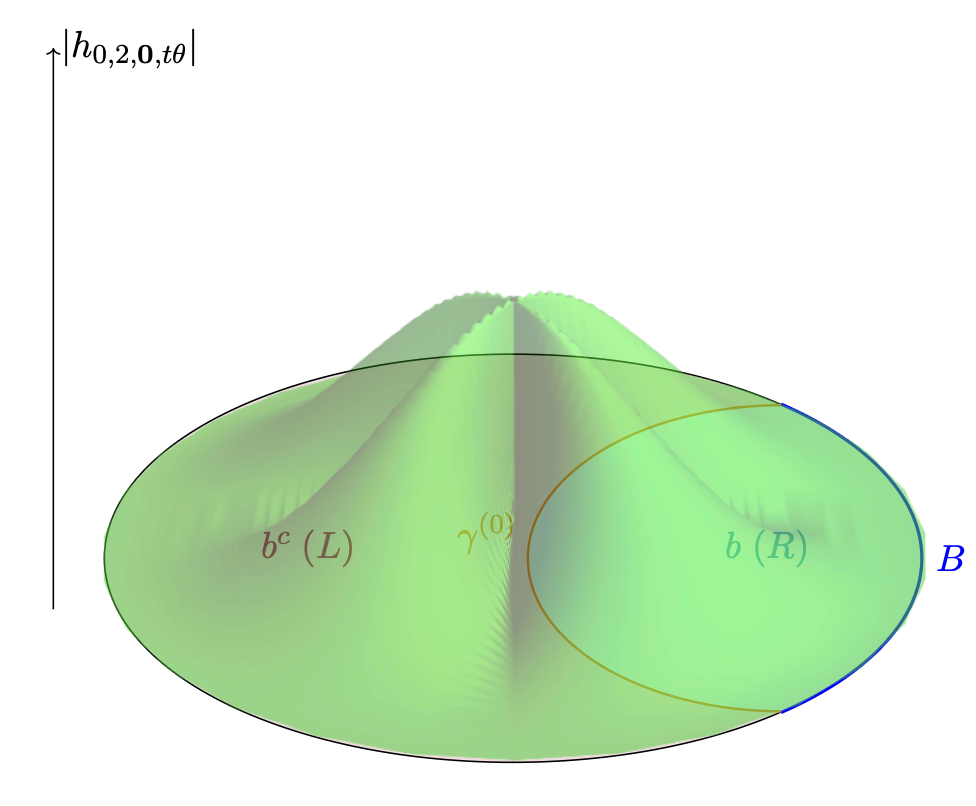}
\end{center}
\caption{The $t\theta$ component of the wavefunction for the single-particle state $\ket{\mathfrak{g}}$ (the wavefunction is smooth, the kinks come from taking the magnitude). The vN entropy for the $R$ Rindler wedge is obtained by tracing out the $L$ wedge for the excited state density matrix.}
\label{fig:gravitonwavefunction}
\end{figure}

Given the global graviton excited state $\ket{\mathfrak{g}}$, we want to trace out the $L$ Rindler wedge to construct the reduced density matrix for the $R$ wedge, see figure \ref{fig:gravitonwavefunction} for an illustration. To do so, we need the relationship between the global modes and AdS-Rindler modes. The global annihilation and creation operators can be related to the Rindler ones by
\begin{equation}\label{eqn:globaltoRindlerops}
a_{n,\ell,\mathfrak{m}}^{A} = \sum_{B_{H}}\sum_{I}\int_{0}^{\infty}\frac{d\omega}{2\pi}\sum_{\lambda}\left(\alpha_{n,\ell,\mathfrak{m};\omega,\lambda_{B},I}^{A,B_{H}}b_{\omega,\lambda_{B},I}^{B_{H}}+\beta_{n,\ell,\mathfrak{m};\omega,\lambda_{B},I}^{A,B_{H}\ast}b_{\omega,\lambda_{B},I}^{B_{H}\dagger}\right)
\end{equation}
where $\alpha_{n,\ell,\mathfrak{m};\omega,\lambda,I}^{A,B_{H}}$, $\beta_{n,\ell,\mathfrak{m};\omega,\lambda,I}^{A,B_{H}}$ are Bogoliubov coefficients. Using the commutators \eqref{eqn:normalisedgravitonIP} and \eqref{eqn:gravladderops_CCR_Rindler}, we find a set of conditions on these coefficients
\begin{align}\label{eqn:Bogcoeffcondition}
\begin{split}
\delta_{n,n'}\delta_{\ell,\ell'}\delta_{\mathfrak{m},\mathfrak{m}'} &= [a_{n,\ell,\mathfrak{m}}^{A},a_{n',\ell',\mathfrak{m}'}^{A\dagger}] 
\\  &= \sum_{B_{H}}\sum_{I}\int_{0}^{\infty}\frac{d\omega}{2\pi}\sum_{\lambda_{B}}\left(\alpha_{n,\ell,\mathfrak{m};\omega,\lambda_{B},I}^{A,B_{H}}\alpha_{n',\ell',\mathfrak{m'};\omega,\lambda_{B},I}^{A,B_{H}\ast}-\beta_{n,\ell,\mathfrak{m};\omega,\lambda_{B},I}^{A,B_{H}\ast}\beta_{n',\ell',\mathfrak{m'};\omega,\lambda_{B},I}^{A,B_{H}}\right).
\end{split}
\end{align}
Henceforth, we will focus only on the mode we care about ($n=0$, $\ell=2$, $\mathfrak{m}=\mathbf{0}$), and as mentioned above, it is proven in App.~\ref{sec:scalaroverlaps} that this mode has no overlap with the vector and tensor modes on the hyperboloid, so
\begin{equation}\label{eqn:vanishingBogcoeffs} \alpha_{0,2,\mathbf{0};\omega,\lambda_{V},I}^{S,V_{H}}=\alpha_{0,2,\mathbf{0};\omega,\lambda_{T},I}^{S,T_{H}}=\beta_{0,2,\mathbf{0};\omega,\lambda_{V},I}^{S,V_{H}}=\beta_{0,2,\mathbf{0};\omega,\lambda_{T},I}^{S,T_{H}}=0.
\end{equation}
We thus denote $\alpha_{\omega,\lambda,I} \equiv \alpha_{0,2,\mathbf{0};\omega,\lambda,I}^{S,S_{H}}$ and $\beta_{\omega,\lambda,I} \equiv \beta_{0,2,\mathbf{0};\omega,\lambda,I}^{S,S_{H}}$ since these are the only non-vanishing Bogoliubov coefficients. 

An important constraint is obtained from $a_{0,2,\mathbf{0}}^{S}\ket{0}=0$, which implies
\begin{equation}\label{eqn:vacann}
\alpha_{\omega,\lambda,L}^{S,S_{H}} = -e^{\pi\omega}\beta_{\omega,\lambda^{\ast},R}, \qquad \beta_{\omega,\lambda,L}^{\ast} = -e^{-\pi\omega}\alpha_{\omega,\lambda^{\ast},R}.
\end{equation}
As a result we can rewrite all $L$ Bogoliubov coefficients in terms of only the $R$ Bogoliubov coefficients, so the constraint \eqref{eqn:Bogcoeffcondition} becomes
\begin{equation}\label{eqn:BogcoeffRcondition}
\int_{0}^{\infty}\frac{d\omega}{2\pi}\sum_{\lambda}\left(|\alpha_{\omega,\lambda,R}|^{2}(1-e^{-2\pi\omega})-|\beta_{\omega,\lambda,R}|^{2}(1-e^{2\pi\omega})\right) = 1.
\end{equation}
We can also use \eqref{eqn:MinkvactoRindlerTFD} to write left creation and annihilation operators acting on $\ket{0}$ in terms of only right operators
\begin{equation}\label{eqn:AdSRindlerops_LtoR}
b_{\omega,\lambda,L}\ket{0} = e^{-\pi\omega}b_{\omega,\lambda^{\ast},R}^{\dagger}\ket{0}, \qquad b_{\omega,\lambda,L}^{\dagger}\ket{0} = e^{\pi\omega}b_{\omega,\lambda^{\ast},R}\ket{0}.
\end{equation}
The single-particle state can thus be written in terms of purely right modes
\begin{equation}\label{eqn:singlepartstate_AdSRindler}
\ket{\mathfrak{g}} = a_{0,2,\mathbf{0}}^{\dagger}\ket{0} = \int\frac{d\omega}{2\pi}\sum_{\lambda}\left((1-e^{-2\pi\omega})\alpha_{\omega,\lambda,R}^{\ast}b_{\omega,\lambda,R} + (1-e^{2\pi\omega})\beta_{\omega,\lambda,R}^{\ast}b_{\omega,\lambda,R}^{\dagger}\right)\ket{0}
\end{equation}
Having obtained an expression for the state only in terms of $R$ objects, we will drop the $R$ labels: $\alpha_{\omega,\lambda} \equiv \alpha_{\omega,\lambda,R}$ and $b_{\omega,\lambda} \equiv b_{\omega,\lambda,R}$. The graviton state density matrix is thus
\begin{align}\label{eqn:psidensitymatrix}
\begin{split}
\rho^{\mathfrak{g}} = | \mathfrak{g}\rangle\langle\mathfrak{g} | = &\int\frac{d\omega}{2\pi}\sum_{\lambda}\left((1-e^{-2\pi\omega})\alpha_{\omega,\lambda}^{\ast}b_{\omega,\lambda}^{\dagger} + (1-e^{2\pi\omega})\beta_{\omega,\lambda}b_{\omega,\lambda}\right)| 0\rangle\langle0 |
\\	&\int\frac{d\omega'}{2\pi}\sum_{\lambda'}\left((1-e^{-2\pi\omega'})\alpha_{\omega',\lambda'}b_{\omega',\lambda'} + (1-e^{2\pi\omega'})\beta_{\omega',\lambda'}^{\ast}b_{\omega',\lambda'}^{\dagger}\right).
\end{split}
\end{align}

The reduced density matrix for the global vacuum state is the thermal state for the $R$ wedge
\begin{equation}
    \rho_{b}^{0} = e^{-2\pi H_{R}}
\end{equation}
where $H_{R}$ is the Rindler Hamiltonian obtained in \eqref{eqn:primH_Rindler} on the reduced phase space, that is, setting the constraints strongly to zero, which can be written in terms of modes as
\begin{equation}\label{eqn:RindlerH_modes}
    H_{R} = \sum_{A_{H}}\int_{0}^{\infty}\frac{d\omega}{2\pi}\sum_{\lambda_{A}}\omega b_{\omega,\lambda_{A}}^{A_{H}\dagger}b_{\omega,\lambda_{A}}^{A_{H}}. 
\end{equation}
It is now trivial to obtain the reduced density matrix
\begin{align}\label{eqn:AdSRindler_scalarreducedrho}
\begin{split}
\rho_{b}^{\mathfrak{g}} = \Tr_{b^{c}}\rho^{\mathfrak{g}} = &\int\frac{d\omega}{2\pi}\sum_{\lambda}\left((1-e^{-2\pi\omega})\alpha_{\omega,\lambda}^{\ast}b_{\omega,\lambda}^{\dagger} + (1-e^{2\pi\omega})\beta_{\omega,\lambda}b_{\omega,\lambda}\right)e^{-K_b^{0}}
\\	&\int\frac{d\omega'}{2\pi}\sum_{\lambda'}\left((1-e^{-2\pi\omega'})\alpha_{\omega',\lambda'}b_{\omega',\lambda'} + (1-e^{2\pi\omega'})\beta_{\omega',\lambda'}^{\ast}b_{\omega',\lambda'}^{\dagger}\right),
\end{split}
\end{align}
where $K_{b}^{0}=2\pi H_{R}$.

The Bogoliubov coefficients are computed explicitly in App.~in a gauge-invariant way so it is fine that we computed the global graviton $h_{\mu\nu}^{S}$ and Rindler graviton $\mathpzc{h}_{\mu\nu}$ wavefunctions in different gauges. At leading order in $\theta_{0}$, the result is
\begin{align}\label{eqn:Bogcoeff_betasimpl_grav_final2}
\begin{split}
\beta_{\omega,\tilde{\lambda}}^{S,S_{H}} &\sim -\frac{\mathcal{N}_{0,1}^{S}\mathcal{N}_{\lambda,0}^{H}}{\mathcal{N}_{\omega,\lambda}^{S,R\ast}}\sqrt{\frac{(d+2)}{2(d-1)}}\frac{2^{d-2}d^{2}\pi^{\frac{d}{2}-1}}{\Gamma(d+2)\Gamma(\frac{d-2}{2})}\left|\Gamma\left(\frac{\zeta+i(\omega+\tilde{\lambda})}{2}\right)\right|^{2}\left|\Gamma\left(\frac{\zeta+i(\omega-\tilde{\lambda})}{2}\right)\right|^{2}\theta_{0}^{d}
\\  \alpha_{\omega,\tilde{\lambda}}^{S,S_{H}} &\sim -\beta_{\omega,\tilde{\lambda}}^{S,S_{H}},
\end{split}
\end{align}
where $\zeta \equiv \frac{(d-2)}{2}$ and $\lambda = \zeta^{2}+\tilde{\lambda}^{2}$, $\mathcal{N}_{\lambda,0}^{H}$ is the normalisation constant for $H_{\lambda}$ in \eqref{eq:hyperbolicevnorm1}, and $\mathcal{N}_{\omega,\lambda}^{S,R}$ is the normalisation for the Rindler wavefunctions in \eqref{eqn:Rindlernormconstant}. The notation $\sim$ means that we are missing non-perturbative corrections in $\theta_{0}$ that behave as $e^{-\theta_{0}\omega}$, which are important for the condition \eqref{eqn:Bogcoeffcondition} to hold since that $\omega$ integral converges slowly at large $\omega$. However, they are unimportant for the $\omega$ integrals needed for the vN entropy which converge very fast so we do not include these corrections here. There are also subleading corrections that are perturbative in $\theta_{0}$ starting at $O(\theta_{0}^{d+2})$ that can be included, but we content ourselves with matching the leading order contribution in $\theta_{0}$ to the generalized entropy with the dual CFT entropy. 

With the reduced density matrix in hand, we are now ready to compute the vN entropy for the state $\mathfrak{g}$ in the subregion $b$.

\subsection{Entanglement entropy of gravitons}
\label{sec:gravitonvN}

We will compute vN Neumann entropy for the reduced density matrix $\rho_{b}^{\mathfrak{g}}$ via the replica trick
\begin{equation}\label{eqn:vNentropy}
    S(\rho_{b}^{\mathfrak{g}}) = -\Tr\left(\rho_{b}^{\mathfrak{g}}\log(\rho_{b}^{\mathfrak{g}})\right) = -\partial_{n}\Tr\left((\rho_{b}^{\mathfrak{g}})^{n}\right)\big|_{n=1},
\end{equation}
where in the second equality one computes the moments of the density matrix and analytically continues in $n$. These moments can be computed perturbatively in the difference between the excited state and the vacuum: (in this subsection, we denote the vacuum density matrix $\omega_b$ by $\rho_b^0$ to avoid confusion with the frequency $\omega$)
\begin{equation}\label{eqn:rhosmallparam}
\delta \rho = \rho_{b}^{\mathfrak{g}} - \rho_{b}^{0},
\end{equation}
which is small in the small $\theta_{0}$ limit.

\paragraph{First-order contribution.}

At first-order in $\delta \rho$, we find
\begin{equation}\label{eqn:traceexp_1storder}
\Tr\left((\rho_{b}^{\mathfrak{g}})^{n}\right) = \Tr\left((\rho_b^0+\delta \rho)^{n}\right) = \Tr\left((\rho_b^0)^{n}\right)+n\Tr\left((\rho_b^0)^{n-1}\delta\rho\right)+O\left((\delta\rho)^{2}\right).
\end{equation}
Therefore, the vacuum-subtracted vN entropy is
\begin{equation}\label{eqn:bulkvacsubtractEE_1storder}
\Delta S(\rho_{b}^{\mathfrak{g}})\big|_{O(\delta\rho)}  = -\partial_{n}\Tr\left((\rho_{b}^{0})^{n-1}\delta\rho\right)\big|_{n=1} = -\Tr\left(\delta \rho \log \omega_{b}\right) = \Delta K_{b}^{\mathfrak{g}},
\end{equation}
where we used the fact that $\Tr(\rho_{b}^{\mathfrak{g}})=\Tr(\rho_{b}^0)=1$, so once again we have the first law of entanglement entropy.

Using \eqref{eqn:RindlerH_modes}, \eqref{eqn:AdSRindler_scalarreducedrho}, and \eqref{eqn:Bogcoeff_betasimpl_grav_final2}, we obtain
\begin{equation}\label{eqn:DeltaK0bulk}
    \Delta K_{b}^{\mathfrak{g}} = \int d\omega\,\omega\sum_{\lambda}\left(|\alpha_{\omega,\lambda}|^{2}+|\beta_{\omega,\lambda}|^{2}\right) = \frac{V_{S^{d-2}}}{V_{S^{d-1}}}\frac{\pi d(d+2)\Gamma(\frac{d-1}{2})\Gamma(d+1)}{8\Gamma(d+\frac{3}{2})\Gamma(\frac{d}{2}+2)}\theta_{0}^{2d}\left(1+O(\theta_{0}^{2})\right)
\end{equation}
where the integrals over $\omega$ and $\lambda$ can be performed using eq. 6.413.1 in \cite{GradRyz} and eq. 5.13.5 in \cite{NIST:DLMF}, respectively. This agrees with the canonical energy in Hollands-Wald gauge computed in \S\ref{sec:JLMStest}, and hence provides a nice check of our argument in \S\ref{sec:JLMSgraviton} for why this particular canonical energy is the modular Hamiltonian for the background graviton state which in this case is the vacuum state.

\paragraph{Second-order contribution.}

Let us now go beyond the first law of entanglement and compute the second-order contribution to the vN entropy. At second-order in the small $\delta\rho$ expansion, the trace gives 
\begin{align}\label{eqn:traceexp_2ndorder}
\begin{split}
\Tr\left((\rho_{b}^{\mathfrak{g}})^{n}\right)\Big|_{O(\delta\rho^2)} &= \frac{n}{2}\sum_{a=0}^{n-2}\Tr\left(\delta\rho(\rho_{b}^{0})^{a}\delta\rho(\rho_{b}^{0})^{n-a-2}\right)
\\	&= \frac{n}{2}\Tr\left(\rho_{b}^{\mathfrak{g}}\tilde{\rho}(n)(\rho_{b}^{0})^{n-2}\right)+\frac{n(n-1)}{2}\Tr\left((\rho_{b}^{0})^{n}\right)-n(n-1)\Tr\left(\rho_{b}^{\mathfrak{g}}(\rho_{b}^{0})^{n-1}\right)
\end{split}
\end{align}
where we have defined
\begin{equation}\label{eqn:tilderhodef}
\tilde{\rho}(n) \equiv \sum_{a=0}^{n-2}(\rho_{b}^{0})^{a}\rho_{b}^{\mathfrak{g}}(\rho_{b}^{0})^{-a}.
\end{equation}
We evaluate $\tilde{\rho}(n)$ using the Baker-Campbell-Hausdorff formula, leading to 
\begin{align}\label{eqn:tilderho}
\begin{split}
\tilde{\rho}(n) &= \int \frac{d\omega_{1}}{2\pi}\,\frac{d\omega_{2}}{2\pi}\,\sum_{\lambda_{1},\lambda_{2}}\bigg[\frac{\mathfrak{E}\left((n-1)(\omega_{2}-\omega_{1})\right)\mathfrak{E}(-\omega_{1})\mathfrak{E}(-\omega_{2})}{\mathfrak{E}(\omega_{2}-\omega_{1})}\alpha_{1}^{\ast}\alpha_{2}b_{1}^{\dagger}\rho_{b}^{0}b_{2}
\\	&+\frac{\mathfrak{E}\left((n-1)(-\omega_{1}-\omega_{2})\right)\mathfrak{E}(-\omega_{1})\mathfrak{E}(\omega_{2})}{\mathfrak{E}(-\omega_{1}-\omega_{2})}\alpha_{1}^{\ast}\beta_{2}^{\ast}b_{1}^{\dagger}\rho_{b}^{0}b_{2}^{\dagger}
\\	&+\frac{\mathfrak{E}\left((n-1)(\omega_{1}+\omega_{2})\right)\mathfrak{E}(\omega_{1})\mathfrak{E}(-\omega_{2})}{\mathfrak{E}(\omega_{1}+\omega_{2})}\beta_{1}\alpha_{2}b_{1}\rho_{b}^{0}b_{2}
\\	&+\frac{\mathfrak{E}\left((n-1)(\omega_{1}-\omega_{2})\right)\mathfrak{E}(\omega_{1})\mathfrak{E}(\omega_{2})}{\mathfrak{E}(\omega_{1}-\omega_{2})}\beta_{1}\beta_{2}^{\ast}b_{1}\rho_{b}^{0}b_{2}^{\dagger}\bigg]
\end{split}
\end{align}
where we have abbreviated $\alpha_{1} \equiv \alpha_{\omega_{1},\lambda_{1}}$ and the same for $\beta$, and we have defined
\begin{equation}\label{eqn:Edef}
\mathfrak{E}(x) = 1-e^{2\pi x}.
\end{equation}
Observe that $\tilde{\rho}(n)$ is now manifestly analytic in $n$ and has a simple $n$ derivative. Thus, we can compute the second-order contribution to the vacuum-subtracted vN entropy:
\begin{align}\label{eqn:bulkvN_2ndorder}
\begin{split}
\Delta S(\rho_{b}^{\mathfrak{g}})\big|_{O(\delta\rho^2)} &= -\partial_{n}\Tr\left((\rho_{b}^{\mathfrak{g}})^{n}\right)\big|_{O(\delta\rho^2),\,n=1}
\\  &= \pi \int \frac{d\omega_{1}}{2\pi}\,\frac{d\omega_{2}}{2\pi}\,\sum_{\lambda_{1},\lambda_{2}}\left((\omega_{1}+\omega_{2})\frac{\mathfrak{E}(\omega_{1})\mathfrak{E}(\omega_{2})}{\mathfrak{E}(\omega_{1}+\omega_{2})}+(\omega_{1}-\omega_{2})\frac{\mathfrak{E}(-\omega_{1})\mathfrak{E}(\omega_{2})}{\mathfrak{E}(\omega_{2}-\omega_{1})}\right)
\\	&\qquad \times \left(|\alpha_{1}|^{2}|\beta_{2}|^{2}+|\beta_{1}|^{2}|\alpha_{2}|^{2}+2\alpha_{1}^{\ast}\beta_{1}\alpha_{2}\beta_{2}^{\ast}\right)
\\  &= -\frac{\sqrt{\pi}\Gamma(2d+1)}{2\Gamma(2d+\frac{3}{2})}\theta_{0}^{4d}\left(1+O(\theta_{0}^{2})\right).
\end{split}
\end{align}
Higher-order contributions in $\delta \rho$ give larger powers of $\theta_{0}$ which are subleading. The result \eqref{eqn:bulkvN_2ndorder} gives the leading contribution to the relative entropy
\begin{equation}
    S_{\mathrm{rel}}(\rho_{b}^{\mathfrak{g}}|\rho_{b}^{0}) = \Delta K_{b}^{0} - \Delta S(\rho_{b}^{\mathfrak{g}}) = -\Delta S(\rho_{b}^{\mathfrak{g}})\big|_{O(\delta\rho^2)} + O((\delta \rho)^{3}).
\end{equation}
Therefore, we find exact agreement between the graviton relative entropy given by (minus) \eqref{eqn:bulkvN_2ndorder} and the CFT relative entropy in \eqref{eqn:SrelTfinal}
\begin{equation}\label{Srelmatch}
    S_{\mathrm{rel}}(\rho_{b}^{\mathfrak{g}}|\rho_{b}^{0}) = S_{\mathrm{rel}}(\rho^{\tilde{\epsilon} \cdot \tilde{T}}|\rho^{0})_{\mathrm{GFF}}.
\end{equation}
Combined with the JLMS formula verified in \S\ref{sec:JLMStest}, we find the desired match between bulk and boundary
\begin{equation}
    \left(\Delta S^{\tilde{\epsilon} \cdot \tilde{T}}\right)_{1 \ll \Delta_{\mathrm{gap}} \ll C_{T}} = \Delta S_{\mathrm{gen}}(\rho_{b}^{\mathfrak{g}}).
\end{equation}

\section{Discussion}
\label{sec:discussion}

The primary goal of this work was to understand how to include gravitational fluctuations into the generalized entropy of spacetime subregions in the context of the quantum extremal surface formula. We were guided by the principle of gauge-invariance with respect to perturbative diffeomophisms and sought a formula that took the form of areas plus graviton von Neumann entropy and agreed with the CFT entropy. This was shown to require that the area be treated as a quantum object. We first proposed that one use the classical extremal surface $\gamma^{(0)}$, define a quantum extremal gauge and then extremize the generalized entropy over all gauge-equivalent states. At $O(1)$, this was argued to be equivalent to solving for a new extremal surface $\gamma^{(1)}$ and treating this perturbed surface as an operator $\hat{\gamma}^{(1)}$ on the graviton Hilbert space. 

Concretely, we then explored our prescription at $O(1)$. Using the covariant phase space analysis of stationary black hole spacetimes for second-order perturbations in the metric, which we made suitably gauge-invariant and adapted to asymptotically AdS spacetimes, we argued for our prescription for first-order perturbations of the state. This required promoting the classical phase space results to quantum formulae, whilst preserving gauge-invariance. Next, we provided a non-trivial example of stress-tensor excited states (and superpositions) in $d$-dimensional CFTs for $d>2$, satisfying the necessary conditions for a holographic CFT, where we explicitly computed the von Neumann entropy for polar cap subregions, going beyond first-order perturbations in the state. The generalized entropy for the dual graviton states using our prescription was then computed and found to agree precisely with the CFT result, providing a crucial check of our prescription. We also observed that any state for which the graviton has non-zero expecation value will give $S_{\mathrm{gen}} \sim G_{N}^{-\frac{1}{2}}$, which is parametrically larger than any perturbative state in an ordinary QFT, and checked consistency of this fact with the CFT entropy.

\paragraph{Path integral derivation.} The most immediate next step is to derive our prescription from the gravitational path integral using the techniques first developed by Lewkowycz and Maldacena \cite{Lewkowycz:2013nqa,Faulkner:2013ana,Dong:2017xht}. This is more non-trivial for gravitons. For instance, the usual replica trick for the vN entropy involves computing a partition function for quantum fields living on a replicated manifold with a conical singularity. This is an off-shell background for the graviton so perturbation theory on such manifolds does not make sense.\footnote{Beyond the issue of whether quantum perturbation theory around an off-shell classical field is well-defined, there is the problem that the free graviton action is not even gauge-invariant with respect to perturbative diffeomorphisms when the background is off-shell.} Furthermore, it is not clear how to relate diffeomorphisms on the covering space to those on the quotient space. We expect that all of this can be dealt with by an appropriate partial gauge-fixing of the path integral, but we leave this to future work.

\paragraph{Higher orders in $G_N$.} It is not clear to what extent the prescription \eqref{eq:genentfluctmaintext} makes sense beyond $O(1)$ corrections. Beyond this order $\log \rho_b$ will be highly gauge-dependent and it is not clear whether the condition \eqref{eq:quantumextremalmaintext} can even be satisfied. It seems likely that understanding the path integral derivation better will clarify matters significantly here.

\paragraph{Edge modes and renormalisation of $G_{N}$.} If we regularise entanglement entropy in gauge theories by placing them on a lattice, an important role is played by ``edge modes'' at the boundary of the region $b$ that are typically included in the center of the algebras associated to both $b$ and its complement. In the continuum theory, these modes do not exist, in the sense that any densely defined operator needs to be smeared in either space or time and so the center of the two algebras is trivial unless there are topological charges in the theory. This is not expected to be the case in quantum gravity where all charges allowed by the gauge group are supposed to be present \cite{Polchinski:2003bq, Banks:2010zn, Harlow:2018tng}. However, it has been argued that the entanglement entropy of edge modes is crucial for generalised entropy to be invariant under RG flow (when taking into account the renormalization of $G_N$) and hence to be UV finite \cite{Donnelly:2015hxa}. Since relative entropy is well defined in the continuum theory, it obviously cannot include any contribution from edge modes. However, we were also able to match the CFT answer for vacuum-subtracted entropies without including them. Intuitively, this is presumably because any excited state looks like the vacuum at sufficiently small scales and so any contribution from edge modes cancels, but it would be good to understand this better.\footnote{To support this intuition, we mention that it can be explicitly shown that for Maxwell theory in a Rindler wedge of Minkowski spacetime for Gaussian wavepackets \cite{Colin-Ellerin} and in AdS-Rindler spacetimes for single-particles states \cite{Colin-Ellerin:2024npf} that these edge modes never contribute to vacuum-subtracted vN entropies.}

\paragraph{Higher-derivative corrections.} When higher-derivative corrections to Einstein gravity are present, the area in the quantum extremal surface formula gets replaced by the Dong-Camps entropy \cite{Dong:2013qoa, Camps:2013zua}. For the generalized entropy of graviton fluctuations in such theories, we expect that the area operator will get replaced by a Dong-Camps entropy operator and then extremization over gauge-equivalent states for a quantum extremal gauge will again lead to a gauge-invaraint prescription (or equivalently, at $O(1)$, extremization over surfaces and defining a codimension-$2$ surface operator).

\paragraph{A tensor network picture for graviton entanglement.} One of the most important pieces of progress in understanding the QES prescription was the discovery of simple tensor network toy models \cite{Swingle:2009bg, Pastawski:2015qua, Hayden:2016cfa} that obey an analogous prescription. Those models have since been upgraded to include fluctuations in the geometry by replacing the edges of the tensor network with submaximally entangled quantum states \cite{Akers:2018fow, Dong:2018seb, Bao:2018pvs, Cheng:2022ori}. However, a tensor network is only a toy model of a spatial slice of a holographic spacetime, so those approaches can only model spatial (and not temporal) fluctuations. In particular, this makes all operators describing the fluctuations commute. It would no doubt be very helpful to find some analogous but more sophisticated toy model that can provide intuition for the graviton generalized entropies calculated in this paper.

\paragraph{Large fluctuations in the geometry and min-/max-EWs.} In this paper, we only considered perturbative metric fluctuations about the fixed classical background metric $G_{\mu\nu}$. However, there is no obvious reason why we could not extend the definition of quantum extremal gauge to allow nonperturbative diffeomorphisms and hence $O(1)$ metric fluctuations, and thereby construct a definition of an entanglement wedge in wildly fluctuating geometries. Relatedly, it was observed in \cite{Akers:2020pmf,Akers:2023fqr} that, even when the classical background geometry is fixed, the entanglement wedge may still have large $O(1)$ fluctuations, in the sense that operators within a large intermediate region $b'$ can only be partially, but not fully reconstructed from the boundary region $B$. This effect can be quantified in terms of a ``min-EW'' capturing the largest region that can be at least partially reconstructed from $B$, and a smaller ``max-EW'' where everything can be perfectly reconstructed. When the min- and max-EWs differ, the naive QES prescription appears to break down, even potentially at leading order in $G_N$. Optimistically it may instead be possible to analyse such states using a quantum extremal gauge constructed by acting with nonperturbatively large, state-dependent diffeomorphisms, so that the entanglement wedge becomes fixed but the classical geometry has large fluctuations.

\acknowledgments

The authors would like to thank Don Marolf, Arvin Shahbazi-Moghaddam and Pratik Rath for useful discussions. SC-E, GL, and GP are supported by the Department of Energy, Office of Science, Office of High Energy Physics under QuantISED Award DE-SC0019380. GP was also supported by an Early Career Award DE-FOA-0002563, by AFOSR award FA9550-22-1-0098 and by a Sloan Fellowship.

\appendix

\section{Perturbative curvature tensors}
\label{sec:pertcurv}

This appendix provides the expansion in $\kappa$ of curvature tensors used throughout the main text. We consider the metric expanded around its background value as in \eqref{eqn:backgdexp_metric}. We ignore the $g^{(2)}$ part of the metric as only linearized tensors for $g^{(2)}$ are needed and these can be obtained from linearized tensors below, replacing $h$ with $g^{(2)}$.

Let us start by expanding the Riemann tensor
\begin{align}\label{eqn:Riemanntensorexp}
\begin{split}
    {R_{\mu\nu\rho}}^{\sigma} &= {R_{\mu\nu\rho}}^{\sigma(G)}+\kappa{R_{\mu\nu\rho}}^{\sigma(1)}+\kappa^{2}{R_{\mu\nu\rho}}^{\sigma(2)}+O(\kappa^{3})
\\    {R_{\mu\nu\rho}}^{\sigma(1)} &= \nabla_{\nu}{\Gamma^{\sigma(1)}}_{\mu\rho}-\nabla_{\mu}{\Gamma^{\sigma(1)}}_{\nu\rho}
\\    {R_{\mu\nu\rho}}^{\sigma(2)} &= \nabla_{\nu}{\Gamma^{\sigma(2)}}_{\mu\rho}-\nabla_{\mu}{\Gamma^{\sigma(2)}}_{\nu\rho}+{\Gamma^{\sigma(1)}}_{\alpha\nu}{\Gamma^{\alpha(1)}}_{\mu\rho}-{\Gamma^{\sigma(1)}}_{\alpha\mu}{\Gamma^{\alpha(1)}}_{\nu\rho},
\end{split}
\end{align}
where we can act with covariant derivatives on the connection because even though ${\Gamma^{\sigma}}_{\mu\rho}$ is not a tensor, $\delta{\Gamma^{\sigma}}_{\mu\rho}$ is actually a tensor. The covariant derivative is defined with respect to the background metric so $\nabla_{\mu} = \nabla^{(G)}_{\mu}$ and we always use the background metric $G_{\mu\nu}$ to raise and lower indices. We can compute the difference of Christoffel symbols and we find
\begin{align}\label{eqn:Christoffeldiff}
\begin{split}
   {\Gamma^{\sigma(1)}}_{\mu\rho} &= \frac{1}{2}G^{\sigma\alpha}\left(\nabla_{\mu}h_{\rho\alpha}+\nabla_{\rho}h_{\mu\alpha}-\nabla_{\alpha}h_{\mu\rho}\right) 
   \\ {\Gamma^{\sigma(2)}}_{\mu\rho} &=  -\frac{1}{2}h^{\sigma\alpha}\left(\nabla_{\mu}h_{\rho\alpha}+\nabla_{\rho}h_{\mu\alpha}-\nabla_{\alpha}h_{\mu\rho}\right).
\end{split}
\end{align}
We thus obtain the expansion of the Ricci tensor
\begin{align}\label{eqn:Riccitensorexp}
\begin{split}
    R_{\mu\rho} &= R_{\mu\rho}^{(G)}+\kappa R_{\mu\rho}^{(1)}+\kappa^{2}R_{\mu\rho}^{(2)}+O(\kappa^{3})
\\    R_{\mu\rho}^{(1)} &= \frac{1}{2}\left(\nabla^{\alpha}\nabla_{\mu}h_{\rho\alpha}+\nabla^{\alpha}\nabla_{\rho}h_{\mu\alpha}-\nabla_{\mu}\nabla_{\rho}h - \nabla^{2}h_{\mu\rho}\right)
\\    R_{\mu\rho}^{(2)} &= -\frac{1}{2}\bigg(\nabla_{\nu}h^{\nu\alpha}\left(\nabla_{\mu}h_{\rho\alpha}+\nabla_{\rho}h_{\mu\alpha}-\nabla_{\alpha}h_{\mu\rho}\right)-\nabla_{\mu}h^{\nu\alpha}\nabla_{\rho}h_{\nu\alpha}
\\  &\qquad +h^{\nu\alpha}\left(\nabla_{\nu}\nabla_{\mu}h_{\rho\alpha}+\nabla_{\nu}\nabla_{\rho}h_{\mu\alpha}-\nabla_{\mu}\nabla_{\rho}h_{\nu\alpha} - \nabla_{\nu}\nabla_{\alpha}h_{\mu\rho}\right)
\\  &\qquad +\frac{1}{2}G^{\nu\beta}G^{\alpha\lambda}\left(\nabla_{\alpha}h_{\mu\beta}+\nabla_{\mu}h_{\alpha\beta}-\nabla_{\beta}h_{\alpha\mu}\right)\left(\nabla_{\nu}h_{\rho\lambda}+\nabla_{\rho}h_{\nu\lambda}-\nabla_{\lambda}h_{\nu\rho}\right)
\\  &\qquad -\frac{1}{2}\nabla^{\sigma}h\left(\nabla_{\mu}h_{\rho\sigma}+\nabla_{\rho}h_{\mu\sigma}-\nabla_{\sigma}h_{\mu\rho}\right)\bigg),
\end{split}
\end{align}
and hence the Ricci scalar has the expansion
\begin{align}\label{eqn:Ricciscalaarexp}
\begin{split}
    R &= R^{(G)}+\kappa R^{(1)}+\kappa^{2}R^{(2)}+O(\kappa^{3})
\\  R^{(1)} &= \nabla^{\alpha}\nabla^{\mu}h_{\mu\alpha}-\nabla^{2}h-h^{\mu\rho}R_{\mu\rho}^{(G)}
\\  R^{(2)} &= h^{\mu\alpha}h_{\alpha\beta}G^{\beta\rho}R_{\mu\rho}^{(G)}-\nabla_{\nu}h^{\nu\alpha}\left(\nabla^{\rho}h_{\rho\alpha}-\nabla_{\alpha}h\right)-\frac{1}{2}\nabla^{\alpha}h^{\rho\nu}\nabla_{\nu}h_{\rho\alpha}-\frac{1}{4}\nabla^{\alpha}h\nabla_{\alpha}h
\\  &\qquad +\frac{3}{4}\nabla^{\rho}h^{\nu\alpha}\nabla_{\rho}h_{\nu\alpha}-h^{\nu\alpha}\left(\nabla^{\rho}\nabla_{\nu}h_{\rho\alpha}+\nabla_{\nu}\nabla^{\rho}h_{\rho\alpha}-\nabla^{2}h_{\nu\alpha} - \nabla_{\nu}\nabla_{\alpha}h\right).
\end{split}
\end{align}
We can also expand the volume element as
\begin{equation}\label{eqn:voleltexp}
    \sqrt{-g} = \sqrt{-G}\left(1+\frac{\kappa}{2}h-\frac{\kappa^{2}}{4}h_{\mu\nu}h^{\mu\nu}+\frac{\kappa^{2}}{8}h^{2}\right)+O(\kappa^{3}).
\end{equation}
%

\section{Hollands-Wald gauge v.s. extremal gauge}
\label{sec:areacorrection+canonenergy}

In this appendix, we explain and derive some of the details needed in the discussion of the canonical energy and area corrections in \S\ref{sec:HW}.

\subsection{$E_{\rm can}$ has no boundary term}
\label{sec:detailcanonicalenergy}
Here we would like to argue that any extremal gauge $h^{\mathrm{ext}}$, satisfying $\delta K^{(1)}[G+\kappa h^{\mathrm{ext}}]|_{\gamma^{(0)}}=0$, will give the same $E_{\rm can}$ without any boundary term. 
The motivations for this are twofold: (1) it is not a priori clear that all of the Hollands-Wald gauges gives the same $E_{\rm can}$, i.e., if one performs a large gauge transformation (with respect to the horizon) for the subregion $b$ that preserves the Hollands-Wald gauge conditions (2) according to the argument in the introduction, the extremal gauge should be sufficient in defining the generalized entropy, thus the second gauge condition \eqref{eqn:HWcondition2} must be redundant. 


In fact, the argument can be made rather general: since the $\Upsilon$ term is the difference between $E_{\mathrm{can}}$ in different gauges, we examine that under what conditions does $\Upsilon=0$.
Suppose $h$ is a metric perturbation  in arbitrary gauge and the only condition that we impose is that $h$ has a smooth profile near the horizon in the two-sided coordinate, such as the global AdS coordinate for AdS-Rindler.
We pick any one realization of the Hollands-Wald gauge to be $h^{\rm HW}$. 
Defining $\Delta h = h-h^{\rm HW} = \pounds_v G$ as the diffeomorphism for the difference between a general gauge and the Hollands-Wald gauge, we write explicitly
\begin{equation}\label{eqn:Upsilonexplicit}
\Upsilon = \Omega(h^{\rm HW},\pounds_{\xi}\Delta h) + \Omega(\Delta h,\pounds_{\xi} h^{\rm HW}) + \Omega(\Delta h,\pounds_{\xi} \Delta h),
\end{equation}
We will show 
\begin{equation}\label{eqn:Upsilonexplicit_again}
    \Upsilon=\underbrace{\Omega(h^{\mathrm{HW}},\pounds_{\xi}\Delta h) +  \Omega(\Delta h ,\pounds_{\xi}h^{\mathrm{HW}})}_{\rm always\;zero} \;\; + \underbrace{\Omega(\Delta h,\pounds_{\xi}\Delta h)}_{\rm only\; zero \;in \; extremal \;gauge}\,.
\end{equation}


We first prove that $\Omega(h^{\mathrm{HW}},\pounds_{\xi}\Delta h)=0$, as long as $v$ corresponds to a smooth vector field in a globally well-defined coordinate system, such as the global AdS coordinate for AdS-Rindler. The second term in \eqref{eqn:Upsilonexplicit_again} can be dealt with in the exact same way by moving the Lie derivative to the first argument.
Note that $\Delta h=\pounds_v G$ so that $\pounds_\xi \Delta h=\pounds_{[\xi,v]} G$, and we can explicitly evaluate $\Omega(h^{\mathrm{HW}},\pounds_{[\xi,v]} G)$ using the phase space identity of diffeomorphism charges $\Omega(\delta g,\pounds_u G) =$ boundary charges with $u=[\xi,v]$ (see \eqref{eqn:symplecticform_Noetherchargereln}):
\begin{equation}
    \Omega(h,\pounds_{[\xi,v]}G) =\int_{\gamma^{(0)}}  \frac{d}{d\lambda} \mathbf{Q}_{[\xi,v]}[G+\lambda h]-\iota_{[\xi,v]} \boldsymbol{\theta}[G;h].
\end{equation}
We compute both quantities directly, using the definitions of $\boldsymbol{\theta}$ and $\mathbf{Q}$  in \eqref{eqn:symppotcurrent} and \eqref{eqn:Noethercharge}. For $\mathbf{Q}$, we find
\begin{equation}
    \mathbf{Q}_{[\xi,v]}[G+\lambda h] = - \frac{1}{\kappa^2}\tilde{\nabla}^{\mu}[\xi,v]^{\nu} \boldsymbol{\epsilon}_{\mu\nu}
\end{equation}
where $\boldsymbol{\epsilon}_{\mu\nu}=\sigma_{\mu\nu}\sqrt{q}\, dx^{1} \wedge \ldots \wedge dx^{d-1}$ is the binormal form for the surface $\gamma^{(0)}$ in the metric $G+\lambda h$ and $\tilde{\nabla}_{\mu} \equiv \nabla_{\mu}^{(G+\lambda h)}$. To evaluate the RHS, we observe that 
\begin{equation}
\begin{aligned}
    \tilde{\nabla}^{\mu}[\xi,v]^{\nu} & = \tilde{\nabla}^{\mu} (\xi^{\alpha} \tilde{\nabla}_\alpha v^{\nu}) - \tilde{\nabla}^{\mu} ( v^{\alpha} \tilde{\nabla}_\alpha \xi^{\nu}
    ) \\
    & = (\tilde{\nabla}^{\mu} \xi^{\alpha})(\tilde{\nabla}_\alpha v^{\nu})- (\tilde{\nabla}^{\mu} v^{\alpha})(\tilde{\nabla}_\alpha \xi^{\nu}) - v^\alpha \tilde{\nabla}^{\mu} \tilde{\nabla}_{\alpha}\xi^{\nu}
\end{aligned}
\end{equation}
and the following fact can be used to simplify the calculation\footnote{Recall that in the definition of $\xi$ there is a surface gravity factor $\mathfrak{s}$. This is where the $\mathfrak{s}$ factor comes from.}
\begin{equation}
    \nabla^{\mu} \nabla_{\alpha}\xi^{\nu}|_{\gamma^{(0)}} = [(\nabla^{\mu} \xi^{\alpha})(\nabla_\alpha v^{\nu})- (\nabla^{\mu} v^{\alpha})(\nabla_{\alpha} \xi^{\nu})] \sigma^{(0)}_{\mu\nu}|_{\gamma^{(0)}} = 0, \quad \partial_{\mu}\xi^{\alpha}= \mathfrak{s}\sigma^{(0),\alpha}_{\mu}\,.
\end{equation}

After some calculation, we see that the contributions from $\mathbf{Q}$ and $\iota \boldsymbol{\theta}$ give\footnote{To make the calculation simple, one can collect the contribution from $v^{\alpha}$ and $v^{i}$ separately. 
The $v^{\alpha}$ contribution cancels between the last two terms. 
One can also keep track of all the $v_i$ contributions and see that they cancel. 
Note that in $\delta \epsilon^{\mu}_{\nu}= \delta(k^{\mu}l_{\nu}-k_{\nu}l^{\mu})$, only the upper index can have components along the tangent direction $X_i^{\mu}$ and the lower index does not, because $\delta (k_\mu X_i^{\mu})=0$ and $\delta X_i$=0 give $\delta (k_\mu) X^{\mu}_i=0$. (The same is true for $l$).}
\begin{align}
\begin{split}
     \Omega(h,\pounds_{[\xi,v]}G) &= \frac{\mathfrak{s}}{\kappa^2}\int_{\gamma^{(0)}}\sqrt{q}\bigg(\frac{d}{d\lambda } ( {-} \sigma^{(0),\alpha}_{\mu} \nabla_{\alpha} v_{\nu} +  \nabla_{\mu} v_{\beta} g^{\alpha\beta} \sigma_{\alpha}^{(0),\kappa} g_{\kappa \nu} ) \sigma^{\mu \nu} 
     \\ &- v^{\alpha} \tilde{\nabla}_{\alpha}(\sigma_{\mu}^{(0),\nu}) \sigma_{\nu}^{(0),\mu} - v^{\alpha} \sigma^{(0),\mu}_{\alpha} \sigma^{(0)}_{\mu\nu} G^{\nu \beta} G^{\sigma \rho} (\nabla_{\sigma}h_{\beta\rho} - \nabla_{\beta}h_{\sigma\rho})\bigg),
\end{split}
\end{align}
which in the end gives
\begin{equation}\label{eqn:omegaresult}
   \Omega(h,\pounds_{[\xi,v]}G) = \frac{\mathfrak{s}}{\kappa^2}\int_{\gamma^{(0)}} \sqrt{q} \Big(2(\nabla_k v_{k} h_{ll}+\nabla_l v_{l} h_{kk})
    + P^{\alpha\beta} v_{\alpha} \nabla_{\beta} h_{ij} q^{ij} 
    \Big)\,.
\end{equation}
where $P$ is the projection onto the normal directions to $\gamma^{(0)}$ and $q$ is the projection to the tangent directions. 
We also use the unit null vectors, satisfying $k^{\mu}l_{\mu}=1$ and $k^{\mu} k_{\mu}=l^{\mu}l_{\mu}=0$, to contract with $v,h$ and $\nabla$. 
Now we see that if $h$ is specified to be $h^{\rm HW}$, then $ \nabla_{k} h^{\rm HW}_{ij} q^{ij} = \delta K_{(k)} = \nabla_{l} h^{\rm HW}_{ij} q^{ij} = \delta K_{(l)}=0$.
At the same time, one can also check that $h^{\rm HW}_{kk}= h^{\rm HW}_{ll}=0$ for the Gaussian null coordinate condition \eqref{eqn:HWcondition2}.
This proves that the first two terms in \eqref{eqn:Upsilonexplicit_again} are zero. 

When specializing to $h=\Delta h = \pounds_v G$, \eqref{eqn:omegaresult} gives
\begin{equation}\label{eqn:omegashiftresult}
   \Omega(\pounds_v G,\pounds_{[\xi,v]}G) =  \frac{\mathfrak{s}}{\kappa^2}\int_{\gamma^{(0)}} \sqrt{q} \Big(4(\nabla_k v_{k} \nabla_l v_{l})
    + 2 P^{\alpha\beta} v_{\alpha} \nabla_{\beta} \nabla_i v_j q^{ij}
    \Big)\,.
\end{equation}
As an application of such a calculation, we prove the statement given at the end of \S\ref{sec:restoregaugeinv}, that $E_{\rm can}[h^{\mathrm{ext}}]=E_{\rm can}[h^{\rm HW}]$, even when $h^{\mathrm{ext}}$ does not satisfy the second Hollands-Wald gauge condition \eqref{eqn:HWcondition2}.
We perform a gauge transformation $v^{\mu}$ from $h^{\mathrm{ext}}$ to $h^{\rm HW}$. Since $h^{\mathrm{ext}}$ satisfies the first Hollands-Wald condition \eqref{eqn:HWgauge1}, $v$ cannot move the extremal surface,  i.e. $v^\alpha  P_{\alpha\beta}|_{\gamma^{(0)}} = 0$. 
And we will prove below that $\nabla_k v_k = \nabla_l v_l =0 $, and hence all the terms in \eqref{eqn:omegashiftresult} vanish.

Here we prove, rather inelegantly, that $\nabla_k v_k = \nabla_l v_l =0 $ if $v$ is a gauge transformation going between extremal gauge $h^{\mathrm{ext}}$ and Hollands-Wald gauge $h^{\rm HW}$.
The argument requires the following assumptions, which are true in a general stationary black hole background: (1) there exists a near-horizon Rindler-like coordinate, and (2) the metric perturbations ($h^{\mathrm{ext}}$ and $h^{\rm HW}$) are smooth in a neighborhood of the extremal surface. 
We start by making an ansatz for the vector field $v$ in the Rindler-like coordinate. For the background metric, we have 
\begin{equation}
    ds^2 = - \left(\delta\rho\right)^2 d\tau^2 + d\left(\delta\rho\right)^2 + q_{ij}(x_i)dx^i dx^j + O((\delta\rho)^2) ,
\end{equation}
where $q_{ij}$ is the induced metric on the bifurcation surface.
We write $v^{\mu}$ as 
$v^{\mu}= v^{(-1),\mu} \delta \rho^{-1} +  v^{(0),\mu} + v^{(1)\mu} \delta \rho + O(\delta\rho^2)$, and the requirement is that, when doing the coordinate transform back to the global coordinate which is smoothly defined in a neighborhood the horizon, $\nabla_{(\mu} v_{\nu)}$ be continuous and differentiable. 
After some calculation, one finds that the following must be true
\begin{equation}\label{eqn:smoothvinRindler}
\begin{aligned}
   & v^{(-1) \delta\rho}= v^{(-1) i}=0, \\
   & \partial_\tau^2 v^{(0) \delta\rho} = v^{(0) \delta\rho}, \quad \partial_\tau v^{(0) \delta\rho} =   v^{(-1) \tau}, \quad \partial_{\tau}v^{(0)i} = 0, \\
   & \partial_{\tau}^2 v^{(1)\delta\rho} = 4 v^{(1)\delta\rho}, \quad \partial_\tau v^{(1) \delta\rho} =  v^{(0) \tau}, \quad \partial_{\tau}^2 v^{(1)i} = v^{(1)i}.
\end{aligned}
\end{equation}
The first line just rules out some $\delta\rho$ order in some of the components, while the second line tells us that $v^{\tau}=\frac{e^{\tau} f_+(x_i)+e^{-\tau} f_-(x_i)}{\delta\rho} + \ldots$ is not exactly well-defined on the horizon $\delta\rho=0$, indicating that it is actually a large gauge transformation. 
Indeed, one can check that, e.g., $ \delta^{(1)}K_{(\rho)} \propto (\nabla_{i}^2-(d-1)) v^{(0),\delta\rho}$ where $\nabla_{i}^2$ is the Laplacian for the bifurcation surface, so that $v^{(-1)\tau}$ (and $v^{(0)\rho}$ accordingly) is a large gauge transformation changing the extrinsic curvature. 
In our case of extremal gauge, we do not want to have $v^{(-1)\tau}$ and $v^{(0)\rho}$.
Lastly, we need to actually compute $\nabla_k v_k$ and  $\nabla_l v_l$ from the last line of \eqref{eqn:smoothvinRindler}, and the compuatation shows that $\nabla_k v_k$ and $\nabla_l v_l$ are zero. 
Higher orders of $\delta\rho$ will certainly give a vanishing result for $\nabla_k v_k$ and $\nabla_l v_l$.


\subsection{Area is Noether charge}
\label{sec:areaequalsNoethercharge}
Here we would like to argue that for any extremal gauge $h^{\mathrm{ext}}$ the (variation of) Noether charge on the bifurcation surface satisfies $\kappa^2 \delta^2 \int \mathbf{Q} =2\mathfrak{s}\delta^2 A$ for the unperturbed surface $\gamma^{(0)}$.
Here $\delta^2 A[\gamma^{(0)}]$ is $A^{\mathrm{lin}}[g^{(2)},\gamma^{(0)}] + A^{\mathrm{quad}}[h^{\mathrm{ext}},\gamma^{(0)}]$. 
Since the Noether charge takes the form $\int \mathbf{Q} = \kappa^{-2}\int_{\gamma^{(0)}}\nabla_{\mu}\xi^{\nu}\boldsymbol{\epsilon}^{\mu}_{~\nu}\sqrt{q}$ and  $\nabla_{\mu}\xi^{\nu}|_{\gamma^{(0)}}= \partial_{\mu}\xi^{\nu}$, we see that first
\begin{equation}
      \kappa^{2}\delta \int \mathbf{Q}- 2\mathfrak{s} \delta A = \mathfrak{s}\int_{\gamma^{(0)}} \boldsymbol{\epsilon}^{(0),\nu}_{\mu}\delta(\boldsymbol{\epsilon}^{\mu}_{~\nu}) \sqrt{q} = \mathfrak{s}\int_{\gamma^{(0)}} (k^{(0),\mu} \delta k_{\mu} + k^{(0)}_{\mu} \delta k^{\mu} + (k\leftrightarrow l) ) \sqrt{q} 
\end{equation}
which vanishes because $\delta(n^{\mu}n_{\nu})=\delta(t^{\mu}t_{\nu})=0$, and gives the byproduct  \eqref{eqn:secondorderQandA} if one replaces $\delta$ with $\delta_{\kappa^2 g^{(2)}}$;
and then
\begin{equation}
    \kappa^{2}\delta^2 \int_{\gamma^{(0)}} \mathbf{Q}- 2\mathfrak{s} \delta^2 A[\gamma^{(0)}] 
    = \mathfrak{s}\int_{\gamma^{(0)}}\left(\delta k_{\mu}\delta l^{\mu} + \delta k^{\mu}\delta l_{\mu} + 2 k^{\mu}\delta k_{\mu} l_{\nu}\delta l^{\nu} - 2 k^{\mu}\delta l_{\mu} k_{\nu}\delta l^{\nu} \right)\sqrt{q}.
\end{equation}
Now we perform a more detailed decomposition of such an expression, by projecting $\delta n$ and $\delta t$ on the unperturbed vectors $t^{(0),\mu},n^{(0),\mu},X_i^{\mu}=X_i^{(0),\mu}$ (here $X_i$ are the tangent vectors to $\gamma^{(0)}$ which do not depend on the metric). 
Using the relations
$\delta (n_{\mu} n_{\nu}g^{\mu\nu})=0$, $\delta (n_{\mu} t_{\nu}g^{\mu\nu})=0$, $\delta (n_{\mu} X_i^\mu)=0$, one finds that 
\begin{equation}\label{eqn:Q-A_2ndorder}
     \kappa^{2}\delta^2 \int \mathbf{Q} - 2\mathfrak{s} \delta^2 A[\gamma^{(0)}] = \mathfrak{s}\int_{\gamma^{(0)}} (h_{kk}^{\mathrm{ext}} h_{ll}^{\mathrm{ext}}) \sqrt{q}\,.
\end{equation}
which vanishes, again resulting from $h_{kk}^{\mathrm{ext}}=h_{ll}^{\mathrm{ext}}=0$, so that $\kappa^2 \delta^2 \int \mathbf{Q} -\delta^2 A$ is equal to zero, as desired. 

As one interesting check of all the results here, we can see that for any gauge, from our results \S\ref{sec:restoregaugeinv} we have
\begin{equation}
\begin{aligned}
     \delta^2 \int_B \mathbf{Q}   & = \frac{2 \mathfrak{s}}{\kappa^2} ( \delta^2 A[\gamma^{(0)}] + A^{\mathrm{lin}}[h,\gamma^{(1)}] + A^{\mathrm{quad}}[G,\gamma^{(1)}]) + E_{\rm can}[h] + \Upsilon \\
     & = \underbrace{\delta^2 \int_{\gamma^{(0)}} \mathbf{Q} + E_{\rm can}[h] }_{ \delta^2 \int_B \mathbf{Q} }  + \frac{\mathfrak{s}}{\kappa^{2}}\int_{\gamma^{(0)}} P^{\alpha\beta} v_{\alpha} \nabla_{\beta} h_{ij} q^{ij} \sqrt{q} + \frac{2\mathfrak{s}}{\kappa^2}(A^{\mathrm{lin}}[h,\gamma^{(1)}] + A^{\mathrm{quad}}[G,\gamma^{(1)}])\,.
\end{aligned}
\end{equation}
To make this consistent, $ \int_{\gamma^{(0)}} P^{\alpha\beta} v_{\alpha} \nabla_{\beta} h_{ij} q^{ij} \sqrt{q} + 2(A^{\mathrm{lin}}[h,\gamma^{(1)}] + A^{\mathrm{quad}}[G,\gamma^{(1)}])$ should be zero. 
It is indeed the case, and one can prove this by noticing that $\int_{\gamma^{(0)}} P^{\alpha\beta} v_{\alpha} \nabla_{\beta} h_{ij} q^{ij} \sqrt{q}$ is exactly $-  A^{\mathrm{lin}}[h,\gamma^{(1)}] = - 2 \frac{\delta^2 A[g,\gamma]}{\delta g \delta \gamma}$, because $v^{\alpha}$ cancels $\gamma^{(1)\alpha}$.
We also observe that, 
in AdS-Rindler, $-\frac{\delta^2 A[g,\gamma]}{\delta g \delta \gamma}=\frac{\delta^2 A[g,\gamma]}{\delta \gamma^2}\equiv A^{\rm quad}[G,\gamma^{(1)}]$. It would be interesting to know if this fact is true whenever the unperturbed surface $\gamma^{(0)}$ is extremal.
Putting everything together, we show that $ \int_{\gamma^{(0)}} P^{\alpha\beta} v_{\alpha} \nabla_{\beta} h_{ij} q^{ij} \sqrt{q} + 2(A^{\mathrm{lin}}[h,\gamma^{(1)}] + A^{\mathrm{quad}}[G,\gamma^{(1)}])$ is $ \int_{\gamma^{(0)}} P^{\alpha\beta} v_{\alpha} \nabla_{\beta} h_{ij} q^{ij} \sqrt{q} + 2\frac{\delta^2 A[g,\gamma]}{\delta g \delta \gamma}=0$. 

\subsection{Boundary conditions}
\label{sec:bdyconditions}

Let us now address the proof in \S\ref{sec:JLMSgraviton} regarding the boundary condition and gauge choice when introducing the symplectic form in \eqref{eqn:introduceOmega}.

When introducing the basis of solutions to the Einstein equations and the symplectic form, one should in general be careful because (1) once we must pick some boundary conditions on the horizon, for which there will be corresponding boundary terms in the action in order to have a well-defined variational principle; (2) with a boundary term in the action, certain boundary terms must also be added to the symplectic form, see 
e.g. \cite{Harlow:2019yfa}.
In our case, one choice of boundary condition would be to have a brick wall at $\delta\rho=\epsilon$ on which we impose Dirichlet boundary conditions $h_{\delta \rho \mu}|_{\delta\rho=\epsilon}=0$,
and then introduce a GHY boundary term. 
This leads to a boundary term in the symplectic form given by $\mathbf{c}=\iota_c \boldsymbol{\epsilon}_{\gamma^{(0)}}$ with $c^{\nu} = h_{\mu\alpha}n^{\alpha}q^{\mu\nu}|_{\gamma^{(0)}}$ \cite{Harlow:2019yfa}, which will vanish for our choice of boundary conditions. 
The Rindler radial gauge $\mathpzc{h}_{\delta\rho}^\mu=0$ that we used in \S\ref{sec:AdS-Rindler_graviton} implicitly imposed those boundary conditions because we imposed $\mathpzc{h}_{\delta\rho}^{\mu}=0$ everywhere in the wedge. 

One might be unsatisfied about the brick wall way of ``defining" the boundary near the horizon and the boundary conditions, and may try to define some more general boundary terms for the symplectic form.
But by inspecting the Einstein equations near the horizon, we show that solutions in the Rindler radial gauge behave as 
\begin{equation}
   \mathpzc{h}_{\tau \tau} \sim \delta\rho^4 \delta\rho^{\pm {i\omega}} , \quad  \mathpzc{h}_{\tau i} \sim \delta\rho^2 \delta\rho^{\pm {i\omega}} , \quad \mathpzc{h}^i_{i} \sim \delta\rho^2 \delta\rho^{\pm {i\omega}}, \quad \mathpzc{h}_{(ij)} \sim \delta\rho^{\pm {i\omega}} \,.
\end{equation}
and one can argue that even after summing over modes, this behavior remains true.\footnote{Technically, what is expected to be true is (here $h^{\rm R, sum}= \sum_{\lambda}\int d \omega \beta_{\omega,\lambda} \mathpzc{h}^R_{\omega,\lambda}$, which is supposed to be the Bogoliubov transform of the global mode)
\begin{equation*}
    \lim_{\delta\rho\rightarrow 0} \delta \rho^\varepsilon h^{\rm R,sum}_{(ij)} = 0 , \quad   \lim_{\delta\rho\rightarrow 0} \delta \rho^{2-\varepsilon} \big(h^{\rm R,sum} \big)^i_i = 0, \quad \lim_{\delta\rho\rightarrow 0} \delta\rho^{2-\varepsilon} h^{\rm R,sum}_{\tau i} = 0 , \quad \lim_{\delta\rho\rightarrow 0} \delta\rho^{4-\varepsilon} h^{\rm R,sum}_{\tau \tau} = 0 
\end{equation*} 
with $\varepsilon$ any infinitesimally small positive number. 
This requires, of course, knowledge of the analytic structure of $\beta_{\omega,\lambda}$. In the specific case we are considering, $\beta_{\omega,\lambda}$ is analytic and power-law in $\omega$ near $\omega\rightarrow 0$ and exponentially decaying as $\omega \rightarrow \infty$. This means that $h^{R,\rm sum}$, as the Fourier transform of $\beta_{\omega,\lambda}$, satisfies the above equation.
}
This implies that even if there is a different boundary term, it is likely that in the radial gauge the boundary term is still zero because of this near-horizon fall-off. 

Generalizing the argument, we expect that for a general stationary black hole background, one can still go to the near-horizon Rindler-like coordinate and define a gauge like the Rindler radial gauge.
To be even more general, the proper requirement on the boundary condition for $h_{\mu\nu}$ is to have $h_{\mu\rho}$ have some uniform decay near the horizon behaving as $|h_{\mu\rho}|<\rho^{\alpha}$ for some $\alpha>0$. 
From the near-horizon Einstein equations, the other metric components (except for those only in the tangent directions to the extremal surface) should also have a similar fall-off.
Then the argument above still holds true that the symplectic form does not need to have a boundary term in this gauge, i.e., it vanishes.

\section{CFT stress-tensor three-point function}
\label{sec:CFTstresstensor}

The three-point function of three stress-tensors on $\mathbb{R}^{d}$ is fixed by conformal symmetry and the symmetries of the stress-tensor to be
\begin{equation}\label{eqn:TTT3ptfn}
\bra{0}T_{\mu\nu}(x_{1})T_{\rho\sigma}(x_{2})T_{\alpha\beta}(x_{3})\ket{0}_{\mathbb{R}^{d}} = \frac{1}{x_{12}^{d}x_{23}^{d}x_{13}^{d}}{I_{\mu\nu}}^{\mu'\nu'}(x_{13}){I_{\rho\sigma}}^{\rho'\sigma'}(x_{23})t_{\mu'\nu'\rho'\sigma'\alpha\beta}(X),
\end{equation}
where $I_{\mu\nu,\rho\sigma}(x)$ is given in \eqref{eqn:TTtwoptfn} and
\begin{align}\label{eqn:t2def}
\begin{split}
X_{\mu} &= \frac{x_{13,\mu}}{x_{13}^{2}}-\frac{x_{23,\mu}}{x_{23}^{2}}, \qquad \hat{X}_{\mu} = \frac{X_{\mu}}{|X|}
\\ t_{\mu\nu\rho\sigma\alpha\beta}(X) &= \check{a}\Big(h^{5}_{\mu\nu\rho\sigma\alpha\beta}-2h^{4}_{\mu\nu\rho\sigma\alpha\beta}(\hat{X})-2h^{4}_{\rho\sigma\mu\nu\alpha\beta}(\hat{X})
\\	&\qquad \qquad \qquad +8h^{2}_{\rho\sigma\alpha\beta}(\hat{X})h^{1}_{\mu\nu}(\hat{X})+8h^{2}_{\mu\nu\alpha\beta}(\hat{X})h^{1}_{\rho\sigma}(\hat{X})\Big)
\\	&+\check{b}\Big(h^{4}_{\alpha\beta\mu\nu\rho\sigma}(\hat{X})-h^{4}_{\mu\nu\rho\sigma\alpha\beta}(\hat{X})-h^{4}_{\rho\sigma\mu\nu\alpha\beta}(\hat{X})
\\	&\qquad \qquad \qquad +4h^{2}_{\rho\sigma\alpha\beta}(\hat{X})h^{1}_{\mu\nu}(\hat{X})+4h^{2}_{\mu\nu\alpha\beta}(\hat{X})h^{1}_{\rho\sigma}(\hat{X})\Big)
\\	&+\check{c}\Big(h^{3}_{\mu\nu\rho\sigma}h^{1}_{\alpha\beta}(\hat{X})+h^{3}_{\rho\sigma\alpha\beta}h^{1}_{\mu\nu}(\hat{X})+h^{3}_{\mu\nu\alpha\beta}h^{1}_{\rho\sigma}(\hat{X})
\\	&\qquad \qquad \qquad -2h^{2}_{\rho\sigma\alpha\beta}(\hat{X})h^{1}_{\mu\nu}(\hat{X})-2h^{2}_{\mu\nu\alpha\beta}(\hat{X})h^{1}_{\rho\sigma}(\hat{X})\Big)
\\	&+\check{e}\left(h^{2}_{\mu\nu\rho\sigma}(\hat{X})h^{1}_{\alpha\beta}(\hat{X})-h^{2}_{\rho\sigma\alpha\beta}(\hat{X})h^{1}_{\mu\nu}(\hat{X})-h^{2}_{\mu\nu\alpha\beta}(\hat{X})h^{1}_{\rho\sigma}(\hat{X})\right)
\\	&+\check{f}h^{1}_{\mu\nu}(\hat{X})h^{1}_{\rho\sigma}(\hat{X})h^{1}_{\alpha\beta}(\hat{X})
\\  h^{1}_{\rho\sigma}(\hat{X}) &= \hat{X}_{\rho}\hat{X}_{\sigma}-\frac{1}{d}\delta_{\rho\sigma}
\\	h^{2}_{\rho\sigma\alpha\beta}(\hat{X}) &= \hat{X}_{\rho}\hat{X}_{\alpha}\delta_{\sigma\beta}+\hat{X}_{\sigma}\hat{X}_{\alpha}\delta_{\rho\beta}+\hat{X}_{\rho}\hat{X}_{\beta}\delta_{\sigma\alpha}+\hat{X}_{\sigma}\hat{X}_{\beta}\delta_{\rho\alpha} - \frac{4}{d}\hat{X}_{\rho}\hat{X}_{\sigma}\delta_{\alpha\beta} - \frac{4}{d}\hat{X}_{\alpha}\hat{X}_{\beta}\delta_{\rho\sigma} 
\\  &+ \frac{4}{d^{2}}\delta_{\rho\sigma}\delta_{\alpha\beta}
\\	h^{3}_{\rho\sigma\alpha\beta} &= \delta_{\rho\alpha}\delta_{\sigma\beta}+\delta_{\rho\beta}\delta_{\sigma\alpha}-\frac{2}{d}\delta_{\rho\sigma}\delta_{\alpha\beta}.
\\	h^{4}_{\mu\nu\rho\sigma\alpha\beta}(\hat{X}) &= h^{3}_{\mu\nu\sigma\alpha}\hat{X}_{\rho}\hat{X}_{\beta}+h^{3}_{\mu\nu\rho\alpha}\hat{X}_{\sigma}\hat{X}_{\beta}+h^{3}_{\mu\nu\rho\beta}\hat{X}_{\sigma}\hat{X}_{\alpha}+h^{3}_{\mu\nu\sigma\beta}\hat{X}_{\rho}\hat{X}_{\alpha} - \frac{2}{d}h^{2}_{\mu\nu\alpha\beta}(\hat{X})\delta_{\rho\sigma} 
\\  &- \frac{2}{d}h^{2}_{\mu\nu\rho\sigma}(\hat{X})\delta_{\alpha\beta} - \frac{8}{d^{2}}\delta_{\rho\sigma}\delta_{\alpha\beta}h^{1}_{\mu\nu}(\hat{X})
\\	h^{5}_{\mu\nu\rho\sigma\alpha\beta} &= \delta_{\mu\sigma}\delta_{\nu\alpha}\delta_{\rho\beta}+\delta_{\nu\sigma}\delta_{\mu\alpha}\delta_{\rho\beta}+\delta_{\mu\rho}\delta_{\nu\alpha}\delta_{\sigma\beta}+\delta_{\mu\sigma}\delta_{\nu\beta}\delta_{\rho\alpha}+\delta_{\nu\rho}\delta_{\mu\alpha}\delta_{\sigma\beta}+\delta_{\nu\sigma}\delta_{\mu\beta}\delta_{\rho\alpha}
\\	&+\delta_{\mu\rho}\delta_{\nu\beta}\delta_{\sigma\alpha}+\delta_{\nu\rho}\delta_{\mu\beta}\delta_{\sigma\alpha}-\frac{4}{d}\delta_{\mu\nu}h^{3}_{\rho\sigma\alpha\beta}-\frac{4}{d}\delta_{\rho\sigma}h^{3}_{\mu\nu\alpha\beta}-\frac{4}{d}\delta_{\alpha\beta}h^{3}_{\mu\nu\rho\sigma}-\frac{8}{d^{2}}\delta_{\mu\nu}\delta_{\rho\sigma}\delta_{\alpha\beta}.
\end{split}
\end{align}
The conservation of the stress-tensor requires
\begin{equation}\label{eqn:TTTconservationconstraint}
\check{e} = (d+2)\left(d\check{a}+\check{b}-\check{c}\right), \qquad \check{f} = (d-2)(d+4)\left(4\check{a}+2\check{b}-\check{c}\right),
\end{equation}
and the Ward identity gives
\begin{equation}\label{eqn:WardIdcondTTT}
(d-2)(d+3)\check{a}-2\check{b}-(d+1)\check{c} = \frac{d(d+2)}{4V_{S^{d-1}}^{3}}C_{T},
\end{equation}
leaving two theory-dependent constants $\check{a}$, $\check{b}$. The anisotropy constants $t_{2},t_{4}$ are related to $\check{a},\check{b},\check{c}$ by \cite{Buchel:2009sk}
\begin{align}\label{eqn:t2t4toabc}
\begin{split}
t_{2} &= 2\frac{(d+1)}{d}\frac{\left((d-1)(d^{2}+8d+4)\check{a}+3d^{2}\check{b}-d(2d+1)\check{c}\right)}{\left((d+3)(d-2)\check{a}-2\check{b}-(d+1)\check{c}\right)}
\\ t_{4} &= \frac{(d+1)(d+2)}{d}\frac{\left({-}3(d-1)(2d+1)\check{a}-2d^{2}\check{b}+d(d+1)\check{c}\right)}{\left((d+3)(d-2)\check{a}-2\check{b}-(d+1)\check{c}\right)}.
\end{split}
\end{align}
Using \eqref{eqn:WardIdcondTTT}, we arrive at
\begin{equation}\label{eqn:Einsteinabc}
t_{2}=t_{4}=0 \implies \check{a} = -\frac{d^{3}}{4(d+1)(d-1)^{2}V_{S^{d-1}}^{3}}C_{T}, \; \check{b} = \frac{(d^{3}-3d^{2}+1)}{d^{2}}\check{a}, \; \check{c} = \frac{(2d^{2}-2d-1)}{d}\check{a}.
\end{equation}

To compute the expectation value of the stress-tensor in the stress-tensor state, we use the standard radial quantization map $r=e^{t_{E}}$ from the cylinder to $\mathbb{R}^{d}$ so the bra stress-tensor state on $\mathbb{R}^{d}$ is given by
\begin{equation}\label{eqn:Tdualstate}
\bra{\epsilon^{\ast} \cdot T}_{\mathbb{R}^{d}} = \lim_{|x| \to \infty}|x|^{2d}\bra{0}\epsilon^{\ast\mu\nu}{I_{\mu}}^{\rho}(x){I_{\nu}}^{\sigma}(x)T_{\rho\sigma}(x),
\end{equation}
and the ket state is $\ket{\epsilon \cdot T}_{\mathbb{R}^{d}} = \epsilon \cdot T(0)\ket{0}_{\mathbb{R}^{d}}$. Then we can use the above results for the three-point function to compute $\bra{\epsilon^{\ast} \cdot T}_{\mathbb{R}^{d}}T_{\mu\nu}(x)\ket{\epsilon \cdot T}_{\mathbb{R}^{d}}$. Mapping to the cylinder, we find
\begin{align}\label{eqn:K0Tdif}
\begin{split}
\Delta K_{0}^{\epsilon \cdot \tilde{T}} &\equiv \frac{\bra{\epsilon^{\ast} \cdot \tilde{T}}K_{\mathrm{cyl}}^{0}\ket{\epsilon \cdot \tilde{T}}}{\langle\epsilon^{\ast} \cdot T|\epsilon \cdot T\rangle} - \bra{0}K_{\mathrm{cyl}}^{0}\ket{0}
\\	&= 2\pi\frac{1}{|\epsilon|^{2}}\frac{V_{S^{d-1}}^{2}}{C_{T}}\int_{S^{d-2}} d\Omega_{d-2}\,\int_{0}^{\theta_{0}} d\theta\, \sin^{d-2}\theta\frac{\left(\cos\theta-\cos\theta_{0}\right)}{\sin\theta_{0}}
\\	&\Bigg[4\left(\frac{2}{d}|\epsilon|^{2}-(d+3)(d-2)|\epsilon \cdot \hat{x}_{2}|^{2}+(d^{2}+d-8)|\epsilon \times \hat{x}_{2}|^{2}\right)\check{a}
\\	&\qquad -2\left((d+3)(d-2)|\epsilon \cdot \hat{x}_{2}|^{2}-2(d-3)|\epsilon \times \hat{x}_{2}|^{2}\right)\check{b}
\\	&\qquad -\left(2\frac{(d-1)}{d}|\epsilon|^{2}-(d-2)(d+3)|\epsilon \cdot \hat{x}_{2}|^{2}+4(d-1)|\epsilon \times \hat{x}_{2}|^{2}\right)\check{c}\Bigg],
\end{split}
\end{align}
where $\hat{x}_{2}$ is a unit vector on $S^{d-1}$ and we have defined $|\epsilon|^{2} \equiv \epsilon_{\mu\nu}^{\ast}\epsilon^{\mu\nu}$, $\epsilon \cdot x \equiv \epsilon_{\mu\nu}x^{\mu}x^{\nu}$, $(\epsilon \times x)^{2} \equiv \epsilon^{\mu\nu}\epsilon_{\alpha\nu}x_{\mu}x^{\alpha}$. Specializing to the polarization tensor \eqref{eqn:specialpol}, using \eqref{eqn:Einsteinabc}, and performing the integrals, one finds \eqref{eqn:K0dif}.

Let us finally discuss the $d=3$ case. In $d \geq 4$, there are three parity-even conformal structures with coefficients $\check{a},\check{b},\check{c}$ that are related by the Ward identity \eqref{eqn:WardIdcondTTT} leaving two theory-dependent constants. However, in $d=3$ dimensions, these structures become linearly dependent, related by
\begin{equation}\label{eqn:lindep}
-17\check{a}-6\check{b}+3\check{c} = 0.
\end{equation}
So there is now only one theory-dependent constant for the parity-even part of the three-point function, but there is also one parity-odd conformal structure that is not present in higher dimensions leading to there still being two theory-dependent constants. The coefficient of the parity-even structure can be related to $t_{4}$ but $t_{2}$ is no longer present, which is reflected in the AdS action \eqref{eqn:AdSgravityaction} by there being no $\alpha_{2}W^{2}$ in AdS$_{4}$. The coefficient $n_{\mathrm{odd}}$ of the parity-odd part was shown in \cite{Afkhami-Jeddi:2018own} to behave as $\log \Delta_{\mathrm{gap}}/\Delta_{\mathrm{gap}}^{4}$ so it vanishes as $\Delta_{\mathrm{gap}} \to \infty$. This parity-odd structure appears in the AdS action as a parity odd term $\tilde{\alpha}_{4}\tilde{W}_{\mu\nu\alpha\beta}W^{\mu\nu\rho\sigma}{W_{\rho\sigma}}^{\alpha\beta}$ where $\tilde{W}_{\mu\nu\alpha\beta} = \frac{1}{2}\epsilon_{\mu\nu\rho\sigma}{W^{\rho\sigma}}_{\alpha\beta}$. As in higher dimensions, $t_{4}$ is related to $\alpha_{4}$  and goes as $\log \Delta_{\mathrm{gap}}/\Delta_{\mathrm{gap}}^{4}$ so taking $\Delta_{\mathrm{gap}} \to \infty$ gives $t_{4}=0$ and corresponds to Einstein gravity in AdS$_{4}$. 

These conditions completely fix the three-point function and so we find
\begin{equation}\label{eqn:DeltaK0_d=3}
\left(\Delta K_{0}^{\epsilon \cdot \tilde{T}}\right)_{d=3,t_{4}=n_{\mathrm{odd}}=0} = \frac{3\pi}{1280} \frac{V_{S^{1}}}{V_{S^{2}}}\left(-886\sin\theta_{0}-41\sin(3\theta_{0})-3\sin(5\theta_{0})+2048\tan\left(\frac{\theta_{0}}{2}\right)\right).
\end{equation}
%

\section{Dirac bracket}
\label{sec:Diracbracket}

Here we present the explicit result for the constraint matrix $C_{ab}(x,y)$ needed to construct the Dirac bracket for the reduced phase space in global AdS in \S\ref{sec:gravitonglobal}.

Recall that the constraint matrix is given by
\begin{equation}\label{eqn:constraintmatrix_again}
    C_{ab}(x,y) = \{C_{a}(x),C_{b}(y)\}_{\mathrm{P.B.}} =
                \left(\begin{matrix}
                \mathbf{0}_{2(d+1)} & M_{1}(x,y) \\ 
                M_{1}^{T}(x,y) & M_{2}(x,y) \\
                \end{matrix}\right).
\end{equation}
We find
\begin{equation}\label{eqn:M1def}
M_{1}(x,y) = \left(\begin{matrix}
                \{\mathcal{C}_{\mu}(x),\mathcal{G}_{\nu}(y)\}_{\mathrm{P.B.}} &  \{\mathcal{C}_{\mu}(x),\tilde{\mathcal{G}}_{\nu}(y)\}_{\mathrm{P.B.}} \\
                \{\tilde{\mathcal{C}}_{\mu}(x),\mathcal{G}_{\nu}(y)\}_{\mathrm{P.B.}} &  \{\tilde{\mathcal{C}}_{\mu}(x),\tilde{\mathcal{G}}_{\nu}(y)\}_{\mathrm{P.B.}} \\
                \end{matrix}\right)
                = \left(\begin{matrix}
                \mathbf{0}_{d+1} & \mathscr{B}(x,y) \\
                \mathscr{D}(x,y) & \mathscr{E}(x,y) \\
                \end{matrix}\right)
\end{equation}
where
\begin{align}\label{eqn:BDE}
\begin{split}
\mathscr{B}_{t\mathpzc{i}}(x,y) &= \mathscr{B}_{\mathpzc{i}t}(x,y) = 0, \quad \mathscr{B}_{tt}(x,y) = \frac{r}{(r^{2}+1)}\left(\partial_{r,1}-2\frac{r}{(r^{2}+1)}\right)\delta^{d}(x-y) 
\\  \mathscr{B}_{r\mathpzc{i}}(x,y) &= \sqrt{-G}G_{rr}(1+\delta_{\mathpzc{i}r})\partial_{x,\mathpzc{i}}\delta^{d}(x-y), \quad \mathscr{B}_{\phi_{i}r}(x,y) = 0
\\  \mathscr{B}_{\phi_{i}\phi_{j}}(x,y) &= \sqrt{-G}G_{\phi_{i}\phi_{j}}\left(\partial_{x,r}+\frac{2}{r}\right)\delta^{d}(x-y)
\\  \mathscr{D}_{tt}(x,y) &= -\frac{1}{2}\sqrt{-G}\Bigg(\left(\partial_{r}^{2}+\Gamma_{\mathpzc{i}r}^{\mathpzc{i}}\left(\partial_{r}-\frac{r}{(r^{2}+1)}+\Gamma_{\mathpzc{j}r}^{\mathpzc{j}}\right)\right)\delta^{d}(x-y)
\\  &+\partial_{r}\left(\left(\Gamma_{rr}^{r}+\Gamma_{\mathpzc{i}r}^{\mathpzc{i}}\right)\delta^{d}(x-y)\right)-\frac{1}{\sqrt{G_{\Sigma}}}\partial_{\mathpzc{i}}\left(\sqrt{G_{\Sigma}}G^{\mathpzc{i}\mathpzc{j}}\partial_{\mathpzc{j}}\right)\left(G_{rr}\delta^{d}(x-y)\right)
\\  &-\frac{1}{(d-1)}G_{rr}G^{\mathpzc{i}\mathpzc{j}}\bigg(\left(\partial_{\mathpzc{i}}\partial_{\mathpzc{j}}-\Gamma_{\mathpzc{k}\mathpzc{i}}^{\mathpzc{k}}\partial_{\mathpzc{j}}+\Gamma_{\mathpzc{k}\mathpzc{l}}^{\mathpzc{k}}\Gamma_{\mathpzc{i}\mathpzc{j}}^{\mathpzc{l}}+\Gamma_{\mathpzc{k}\mathpzc{i}}^{\mathpzc{k}}\Gamma_{\mathpzc{l}\mathpzc{j}}^{\mathpzc{l}}\right)\delta^{d}(x-y)
\\  &+\partial_{\mathpzc{k}}\left(\Gamma_{\mathpzc{i}\mathpzc{j}}^{\mathpzc{k}}\delta^{d}(x-y)\right)+\partial_{\mathpzc{i}}\left(\Gamma_{\mathpzc{j}\mathpzc{k}}^{\mathpzc{k}}\delta^{d}(x-y)\right)-\frac{1}{\sqrt{G_{\Sigma}}}\partial_{\mathpzc{k}}\left(\sqrt{G_{\Sigma}}G^{\mathpzc{k}\mathpzc{l}}\partial_{\mathpzc{l}}\right)\left(G_{\mathpzc{i}\mathpzc{j}}\delta^{d}(x-y)\right)\bigg)\Bigg)
\\  \mathscr{D}_{\mathpzc{i}t}(x,y) &= \mathscr{D}_{t\mathpzc{i}}(x,y) = 0, \qquad \mathscr{D}_{\mathpzc{i}\mathpzc{j}}(x,y) = \delta_{\mathpzc{i}\mathpzc{j}}G_{\mathpzc{i}\mathpzc{i}}\partial_{x,r}\delta^{d}(x-y)
\\  \mathscr{E}_{tt}(x,y) &= \mathscr{E}_{\mathpzc{i}\mathpzc{j}}(x,y) = 0, \quad \mathscr{E}_{tr}(x,y) = -G_{tt}\mathscr{D}_{tt}(x,y)
\\  \mathscr{E}_{t\phi_{i}}(x,y) &= \frac{1}{2}\sqrt{-G}G_{tt}\left(\partial_{\phi_{i}}\partial_{r}+\frac{1}{r}\partial_{\phi_{i}}+\Gamma_{\mathpzc{j}\phi_{i}}^{\mathpzc{j}}\partial_{r}+\left(\Gamma_{\mathpzc{j}r}^{\mathpzc{j}}+\frac{1}{r}\right)\Gamma_{\mathpzc{j}\phi_{i}}^{\mathpzc{j}}\right)\delta^{d}(x-y), \quad \mathscr{E}_{\mathpzc{i}t}(x,y) \neq 0,
\end{split}       
\end{align}
where we have not written $\mathscr{E}_{\mathpzc{i}t}(x,y)$ explicitly as it is very unwieldy but it can be obtained so $\mathscr{M}_{tt}(x,y)$ given below with now third-order derivatives owing the derivative in $\tilde{\mathcal{C}}_{\mathpzc{i}}$, and we find 
\begin{align}\label{eqn:M2def}
\begin{split}
M_{2}(x,y) &= \left(\begin{matrix}
                \mathscr{P}(x,y) & \mathscr{M}(x,y) \\
                -\mathscr{M}(x,y) & \mathscr{N}(x,y) \\
                \end{matrix}\right)
\\  \mathscr{P}_{\mu\nu}(x,y) &= G_{rr}^{2}\left(\delta_{\mu t}\delta_{\nu r}-\delta_{\mu r}\delta_{\nu t}\right)
\\  \mathscr{M}_{tt}(x,y) &= \frac{(d-2)}{(d-1)}G_{rr}^{2}\bigg(\frac{(r^{2}+1)}{2}\partial_{x,r}^{2}\delta^{d}(x-y)+\left(\frac{((d-1)r^{2}+d-2)}{2r^{2}}-\frac{(3r^{2}+1)}{2}\right)\partial_{x,r}\delta^{d}(x-y)
\\  &+\left(\frac{1}{(r^{2}+1)}-(d-2)\frac{(r^{2}+1)}{r^{2}}-\frac{1}{2}\right)\delta^{d}(x-y)+\frac{(d-2)}{2(d-1)}G^{\phi_{i}\phi_{i}}\partial_{x,\phi_{i}}^{2}\delta^{d}(x-y)\bigg)
\\  &+\frac{1}{2(d-1)}G_{rr}G^{\phi_{i}\phi_{i}}\Bigg(\bigg(\partial_{x,\phi_{i}}^{2}-\Gamma_{\phi_{i}\phi_{i}}^{\phi_{j}}\partial_{x,\phi_{j}}-\Gamma_{\phi_{j}\phi_{i}}^{\phi_{j}}\partial_{x,\phi_{i}}
\\  &+\Gamma_{\phi_{i}\phi_{i}}^{\phi_{m}}\left(\Gamma_{\phi_{j}\phi_{m}}^{\phi_{j}}+\Gamma_{\phi_{j}\phi_{j}}^{\phi_{m}}G^{\phi_{j}\phi_{j}}G_{\phi_{m}\phi_{m}}\right)+\Gamma_{\phi_{i}\phi_{j}}^{\phi_{m}}\left(\Gamma_{\phi_{i}\phi_{m}}^{\phi_{j}}+\Gamma_{\phi_{i}\phi_{j}}^{\phi_{m}}G^{\phi_{j}\phi_{j}}G_{\phi_{m}\phi_{m}}\right)\bigg)\delta^{d}(x-y)
\\  &-\partial_{\phi_{i}}\left(\Gamma_{\phi_{j}\phi_{i}}^{\phi_{j}}\delta^{d}(x-y)\right)-\partial_{\phi_{j}}\left(\Gamma_{\phi_{i}\phi_{i}}^{\phi_{j}}\delta^{d}(x-y)\right)\Bigg)
\\  &-G_{rr}\left(\frac{(r^{2}+1)}{2r}\partial_{x,r}\delta^{d}(x-y)+d\frac{(r^{2}+1)}{r^{2}}\delta^{d}(x-y)\right)
\\  \mathscr{M}_{t\mathpzc{i}}(x,y) &= \mathscr{M}_{\mathpzc{i}t}(x,y) = 0, \qquad \mathscr{M}_{\mathpzc{i}\mathpzc{j}}(x,y) = \left({-}\delta_{\mathpzc{i}\mathpzc{j}}G_{\mathpzc{i}\mathpzc{j}}+\frac{1}{(d-1)}\delta_{\mathpzc{i}r}G_{rr}\right)\delta^{d}(x-y)
\\  \mathscr{N}_{tt}(x,y) &= \mathscr{N}_{\mathpzc{i}\mathpzc{j}}(x,y) = 0, \qquad \mathscr{N}_{rt}(x,y) = -\mathscr{N}_{tr}(y,x) = -G_{tt}\mathscr{M}_{tt}(x,y) 
\\  \mathscr{N}_{t\phi_{i}}(x,y) &= -\mathscr{N}_{\phi_{i}t}(y,x) = \frac{(d-2)}{2(d-1)}\left(\frac{2}{r}\partial_{x,\phi_{i}}-2G_{\phi_{i}\phi_{i}}G^{\phi_{k}\phi_{k}}\Gamma_{\phi_{k}\phi_{k}}^{\phi_{i}}\left(\partial_{x,r}+\frac{r}{(r^{2}+1)}\right)\right)\delta^{d}(x-y)
\\  &-\frac{1}{2(d-1)}G_{rr}\frac{(r^{2}+1)}{r}\left(2d\partial_{x,\phi_{i}}-3\Gamma_{\phi_{j}\phi_{i}}^{\phi_{j}}-(2d+1)G_{\phi_{i}\phi_{i}}G^{\phi_{j}\phi_{k}}\Gamma_{\phi_{j}\phi_{k}}^{\phi_{i}}\right)\delta^{d}(x-y).
\end{split}       
\end{align}
The key point, which is crucial for the construction of the Dirac bracket and means that we have chosen a good gauge-fixing, is that the constraint matrix is invertible with inverse given by
\begin{equation}\label{eqn:Poissonbracketinvmatrix_again}
C_{ab}^{-1} = \left(\begin{matrix} 
                        -(M_{1}^{T})^{-1}M_{2}M_{1}^{-1} & (M_{1}^{T})^{-1} \\
                        M_{1}^{-1} & \mathbf{0}_{2(d+1)} 
                    \end{matrix}\right),
\end{equation}
where
\begin{equation}\label{eqn:M1inv}
M_{1}^{-1} = \left(\begin{matrix} 
                        -\mathscr{D}^{-1}\mathscr{E}\mathscr{B}^{-1} & \mathscr{D}^{-1} \\
                        \mathscr{B}^{-1} & \mathbf{0}_{(d+1)} 
                    \end{matrix}\right)
\end{equation}
with
\begin{align}\label{eqn:BDinv}
\begin{split}
\mathscr{D}_{\mathpzc{i}t}^{-1}(x,y) &= \mathscr{D}_{t\mathpzc{i}}^{-1}(x,y) = 0, \quad, \mathscr{D}_{tt}^{-1}(x,y) = \mathscr{D}_{tt}(x,y)^{-1}, \quad \mathscr{D}_{\mathpzc{i}\mathpzc{j}}^{-1}(x,y) =  \delta_{\mathpzc{i}\mathpzc{j}}D_{\mathpzc{i}\mathpzc{i}}(x,y)^{-1}
\\  \mathscr{B}_{\mathpzc{i}t}^{-1}(x,y) &= \mathscr{B}_{t\mathpzc{i}}^{-1}(x,y) = \mathscr{B}_{\phi_{i}r}^{-1}(x,y) = 0, \quad \mathscr{B}_{tt}^{-1}(x,y) = \mathscr{B}_{tt}(x,y)^{-1}, \quad \mathscr{B}_{rr}^{-1}(x,y) = \mathscr{B}_{rr}(x,y)^{-1}
\\  \mathscr{B}_{r\phi_{i}}^{-1}(x,y) &= -\mathscr{B}_{rr}(x,y)^{-1}\mathscr{B}_{r\phi_{i}}(x,y)\mathscr{B}_{\phi_{i}\phi_{i}}(x,y)^{-1}, \quad \mathscr{B}_{\phi_{i}\phi_{j}}^{-1}(x,y) = \delta_{ij}\mathscr{B}_{\phi_{i}\phi_{i}}(x,y)^{-1}.
\end{split}
\end{align}
The inverse of a matrix element, say $\mathscr{B}_{rr}(x,y)^{-1}$, is obtained from the integral definition of the inverse of such functions \eqref{eqn:Poissonbracketmatrixinvrel}, which means we write the element as a differential operator acting on the Dirac delta function $D_{x}\delta^{d}(x-y)$, and then find the Green's function $G_{D_{x}}(x,y)$ of the differential operator $D_{x}$.

\section{Details on solving equations of motion}
\label{sec:EOMsoln}

This appendix provides the details of solving the linearized Einstein equations for the graviton in global AdS and AdS-Rindler spacetimes needed in \S\ref{sec:gravitonglobal} and \S\ref{sec:AdS-Rindler_graviton}, respectively.

\subsection{Global AdS}
\label{sec:globalAdSEOM}

To solve the linearized Einstein equations, we use the ``master variable'' formalism developed in \cite{Ishibashi:2004wx,Kodama:2000fa}, which reduces all of these coupled differential equations to a single differential equation for a single master variable $\phi_{S}$. They state everything in terms of gauge-invariant variables and use all of the linearized Einstein equations. We will adapt their formalism to the case at hand where our equations of motion are $E_{\mu\nu}^{(1)}[h]=0$ for $\mu \neq r$ and $\nu \neq r$ to obtain the master equation, solve this equation for $\phi_{S}$ and then use $\phi_{S}$ to solve for $h_{\mu\nu}$ in the gauge $h_{\mu r}=0$.

\paragraph{Gauge-invariant variables.}

To construct gauge-invariant variables, we examine how the different parts of the scalar part of the graviton \eqref{eqn:gravscalar} transform under linearized diffeomorphisms $h_{\mu\nu} \to h_{\mu\nu} + 2\nabla_{(\mu}\chi_{\nu)}$. We can decompose the gauge parameter into scalar and vector parts
\begin{equation}\label{eqn:linearizedgaugeparam_decomp}
    \chi_{\mu}^{V}dx^{\mu} = \sum_{\mathbf{k}_{V}}\chi_{\mathbf{k}_{V}}^{V}\mathbb{V}_{i,\mathbf{k}_{V}}dx^{i}, \qquad \chi_{\mu}^{S}dx^{\mu} = \sum_{\mathbf{k}_{S}}\chi_{a,\mathbf{k}_{S}}\mathbb{S}_{\mathbf{k}_{S}}\,dx^{a}+\chi_{\mathbf{k}_{S}}^{S}D_{i}\mathbb{S}_{\mathbf{k}_{S}}dx^{i}.
\end{equation}
The scalar part of the graviton then transforms as
\begin{align}\label{eqn:gravscalar_gaugetrans}
\begin{split}
    \mathfrak{h}_{ab,\mathbf{k}_{S}} &\to \mathfrak{h}_{ab,\mathbf{k}_{S}}+2\widehat{\nabla}_{(a}\chi_{b),\mathbf{k}_{S}}
    \\ \mathfrak{h}_{a,\mathbf{k}_{S}}^{S} &\to \mathfrak{h}_{a,\mathbf{k}_{S}}^{S}+\chi_{a,\mathbf{k}_{S}}+r^{2}\widehat{\nabla}_{a}\left(\frac{\chi_{\mathbf{k}_{S}}^{S}}{r^{2}}\right)
    \\  \mathfrak{h}_{\mathbf{k}_{S}}^{S,\mathrm{tr}} &\to \mathfrak{h}_{\mathbf{k}_{S}}^{S,\mathrm{tr}} - \frac{2k_{S}^{2}}{(d-1)}\chi_{\mathbf{k}_{S}}^{S}+2r\left(\widehat{\nabla}^{a}r\right)\chi_{a,\mathbf{k}_{S}}
    \\  \mathfrak{h}_{\mathbf{k}_{S}}^{S,\mathrm{trless}} &\to \mathfrak{h}_{\mathbf{k}_{S}}^{S,\mathrm{trless}} + 2\chi_{\mathbf{k}_{S}}^{S},
\end{split}
\end{align}
where $\widehat{\nabla}_{a}$ is the covariant derivative on the $(t,r)$ part of global AdS. Therefore, the following two variables are gauge-invariant
\begin{align}\label{eqn:gaugeinvvars}
\begin{split}
    Z_{\mathbf{k}_{S}} &= (d-1)r^{d-5}\left(\mathfrak{h}_{\mathbf{k}_{S}}^{S,\mathrm{tr}}+\frac{k_{S}^{2}}{(d-1)}\mathfrak{h}_{\mathbf{k}_{S}}^{S,\mathrm{trless}}+2r\left(\widehat{\nabla}^{a}r\right)X_{a,\mathbf{k}_{S}}\right)
    \\ Z_{ab,\mathbf{k}_{S}} &= r^{d-3}\left(\mathfrak{h}_{ab,\mathbf{k}_{S}}+2\widehat{\nabla}_{(a}X_{b),\mathbf{k}_{S}}\right)+\frac{(d-2)}{(d-1)}Z_{\mathbf{k}_{S}}G_{ab},
\end{split}
\end{align}
where
\begin{equation}\label{eqn:Xdef}
    X_{a,\mathbf{k}_{S}} = -\mathfrak{h}_{a,\mathbf{k}_{S}}^{S}+\frac{1}{2}r^{2}\widehat{\nabla}_{a}\left(\frac{\mathfrak{h}_{\mathbf{k}_{S}}^{S,\mathrm{trless}}}{r^{2}}\right).
\end{equation}

\paragraph{Reduction to single scalar field equation.} Observe that even though we do not have the $E_{\mathpzc{i} r}^{(1)}[h]=0$ equations, the Bianchi identity $\nabla^{\mu}E_{\mu\nu}^{(1)}[h]=0$ and imposing all the other Einstein equations still gives
\begin{align}\label{eqn:Bianchi}
\begin{split}
E^{(1)r\phi_{i}}[h] &= \sum_{\mathbf{k}_{S}}\frac{\mathfrak{f}_{\mathbf{k}_{S}}(t)}{r^{d+1}}g^{\phi_{i}\phi_{j}}D_{j}\mathbb{S}_{\mathbf{k}_{S}}(\Omega)
\\ E^{(1)rr}[h] &= -\sum_{\mathbf{k}_{S}}\mathfrak{c}_{\mathbf{k}_{S}}(t)\frac{\sqrt{r^{2}+1}}{r^{d-1}}\mathbb{S}_{\mathbf{k}_{S}}(\Omega)
\end{split}
\end{align}
for some functions $\mathfrak{f}_{\mathbf{k}_{S}}(t)$ and $\mathfrak{c}_{\mathbf{k}_{S}}(t)$.

Let us rewrite the linearized Einstein tensor \eqref{eqn:gravitonEOM} in terms of the gauge-invariant variables. The $ai$ equations become
\begin{equation}\label{eqn:Eai_gaugeinv}
    2E_{ai}^{(1)}[h] = \sum_{\mathbf{k}_{S}}\left[\frac{1}{r^{d-3}}\left(\widehat{\nabla}^{b}Z_{ab,\mathbf{k}_{S}}-\widehat{\nabla}_{a}{Z_{b,\mathbf{k}_{S}}}^{b}\right) + \frac{(d-2)}{r^{d-2}}\widehat{\nabla}_{a}r\left({Z_{b,\mathbf{k}_{S}}}^{b}-Z_{\mathbf{k}_{S}}\right)\right]D_{i}\mathbb{S}_{\mathbf{k}_{S}}
\end{equation}
and the traceless part of the $ij$ equations becomes
\begin{align}\label{eqn:Eijtrless_gaugeinv}
\begin{split}
   2\bigg(E_{ij}^{(1)}[h]-&\frac{1}{(d-1)}(G_{S^{d-1}})_{ij}E_{i}^{(1)i}[h]\bigg) 
   \\   &= -\sum_{\mathbf{k}_{S}}\frac{1}{r^{d-3}}\left(Z_{a,\mathbf{k}_{S}}^{a}-Z_{\mathbf{k}_{S}}\right)\left(D_{i}D_{j}-\frac{1}{(d-1)}(g_{S^{d-1}})_{ij}D_{m}D^{m}\right)\mathbb{S}_{\mathbf{k}_{S}} = 0.
\end{split}
\end{align}
Therefore, we can use $ti$ and $ij$ Einstein equations applied to \eqref{eqn:Eai_gaugeinv} and \eqref{eqn:Eijtrless_gaugeinv} along with the $ri$ equation in \eqref{eqn:Bianchi} to obtain the following equations for the gauge-invariant variables
\begin{align}
    \widehat{\nabla}^{b}Z_{tb,\mathbf{k}_{S}}-\widehat{\nabla}_{t}{Z_{b,\mathbf{k}_{S}}}^{b} &= 0 \label{eqn:gaugeinveqn1}
    \\ \widehat{\nabla}^{b}Z_{rb,\mathbf{k}_{S}}-\widehat{\nabla}_{r}{Z_{b,\mathbf{k}_{S}}}^{b} &= 2\frac{\mathfrak{f}_{\mathbf{k}_{S}}(t)}{(r^{2}+1)r^{d+1}} \label{eqn:gaugeinveqn2}
    \\  Z_{a,\mathbf{k}_{S}}^{a}-Z_{\mathbf{k}_{S}} &= 0 \label{eqn:gaugeinveqn3}.
\end{align}
Finally, we need the $ab$ linearized Einstein tensor for which we define the combination
\begin{equation}\label{eqn:Znew}
    \mathcal{Z}_{ab} = \frac{1}{r^{d-3}}\left(Z_{ab}-\frac{(d-2)}{(d-1)}ZG_{ab}\right),
\end{equation}
and then we find
\begin{align}\label{eqn:Eab_gaugeinv}
\begin{split}
    2E_{ab}^{(1)}[h] &= \sum_{\mathbf{k}_{S}}\Bigg({-}\widehat{\nabla}^{2}\mathcal{Z}_{ab,\mathbf{k}_{S}}+2\widehat{\nabla}_{(a}\widehat{\nabla}^{c}\mathcal{Z}_{cb),\mathbf{k}_{S}}+(d-1)\frac{\widehat{\nabla}^{c}r}{r}\left(2\widehat{\nabla}_{(a} \mathcal{Z}_{cb),\mathbf{k}_{S}}-\widehat{\nabla}_{c} \mathcal{Z}_{ab,\mathbf{k}_{S}}\right)
    \\  &-\widehat{\nabla}_{a}\widehat{\nabla}_{b}{\mathcal{Z}_{c,\mathbf{k}_{S}}}^{c}+\left(2(d-2)+\frac{k_{S}^{2}}{r^{2}}\right)\mathcal{Z}_{ab,\mathbf{k}_{S}}-\widehat{\nabla}_{a}\widehat{\nabla}_{b}\left(\frac{Z_{\mathbf{k}_{S}}}{r^{d-3}}\right)-2\frac{\widehat{\nabla}_{(a}r}{r}\widehat{\nabla}_{b)} \left(\frac{Z_{\mathbf{k}_{S}}}{r^{d-3}}\right)
    \\  &-G_{ab}\bigg[\widehat{\nabla}_{c}\widehat{\nabla}_{d}\mathcal{Z}_{\mathbf{k}_{S}}^{cd}+2(d-1)\frac{\widehat{\nabla}^{c}r}{r}\widehat{\nabla}^{d} \mathcal{Z}_{cd,\mathbf{k}_{S}}-\widehat{\nabla}^{2}{\mathcal{Z}_{c,\mathbf{k}_{S}}}^{c}-(d-1)\frac{\widehat{\nabla}^{c}r}{r}\widehat{\nabla}_{c} {\mathcal{Z}_{d,\mathbf{k}_{S}}}^{d}
    \\  &+ \left(\frac{k_{S}^{2}}{r^{2}}-1\right){\mathcal{Z}_{c,\mathbf{k}_{S}}}^{c}+(d-1)\left(2\frac{\widehat{\nabla}^{c}\widehat{\nabla}^{d}r}{r}+(d-2)\frac{\widehat{\nabla}^{c}r\widehat{\nabla}^{d}r}{r^{2}}\right)\mathcal{Z}_{cd,\mathbf{k}_{S}}-\widehat{\nabla}^{2}\left(\frac{Z_{\mathbf{k}_{S}}}{r^{d-3}}\right)
    \\  &-d\frac{\widehat{\nabla}^{c}r}{r}\widehat{\nabla}_{c}\left(\frac{Z_{\mathbf{k}_{S}}}{r^{d-3}}\right)-\frac{(d-2)(d-1-k_{S}^{2})}{(d-1)}\frac{Z_{\mathbf{k}_{S}}}{r^{d-1}}\bigg]\Bigg)\mathbb{S}_{\mathbf{k}_{S}}.
\end{split}
\end{align}

Now, we will use the proof in \cite{Ishibashi:2004wx} to show that $Z_{ab}$ can be computed in terms of a single scalar field $\phi_{S}$, but we need to adjust the proof because we do not impose $E_{\mu r}^{(1)}[h]=0$. We will use their notation. By \eqref{eqn:gaugeinveqn3}, we have that $Z$ can also be computed in terms of $\phi_{S}$. We define a scalar field
\begin{equation}\label{eqn:phiSexp}
    \phi_{S}(t,r,\Omega) = \sum_{\mathbf{k}_{S}}\phi_{S,\mathbf{k}_{S}}(t,r)\mathbb{S}_{\mathbf{k}_{S}}(\Omega)
\end{equation}
by
\begin{equation}\label{eqn:phiSdef}
    \left(\widehat{\nabla}^{2}-2\right)\phi_{S,\mathbf{k}_{S}} = {Z_{a,\mathbf{k}_{S}}}^{a}.
\end{equation}
This is possible because we have a hyperbolic linear PDE in two-dimensions and the metric $-G_{ab}$ is globally hyperbolic (even though $G_{ab}$ is not), so if initial data is specified on a timelike slice ($r=r_{0}$), then the Cauchy problem is well-posed and we can find $\phi_{S}$ everywhere. 

Choose the initial data $(\phi_{S,\mathbf{k}_{S}}(t,r_{0}),\partial_{r}\phi_{S,\mathbf{k}_{S}}(t,r_{0}))$ such that
\begin{equation}\label{eqn:initdata}
  t^{a}\left(\widehat{\nabla}_{a}\widehat{\nabla}_{b}-G_{ab}\right)\phi_{S,\mathbf{k}_{S}}\Big|_{r=r_{0}} = t^{a}{Z_{ab,\mathbf{k}_{S}}}\big|_{r=r_{0}}. 
\end{equation}
where $t^{a}$ is the timelike Killing vector of global AdS. This gives two linear ODEs in $t$ which can be solved for $(\phi_{S}(t,r_{0}),\partial_{r}\phi_{S}(t,r_{0}))$.

Define
\begin{equation}
    \mathcal{S}_{ab,\mathbf{k}_{S}} = Z_{ab,\mathbf{k}_{S}}-\left(\widehat{\nabla}_{a}\widehat{\nabla}_{b}-G_{ab}\right)\phi_{S,\mathbf{k}_{S}},
\end{equation}
which is symmetric, traceless, and the $t$ component is divergenceless ($\widehat{\nabla}^{a}\mathcal{S}_{at,\mathbf{k}_{S}}=0$) by \eqref{eqn:gaugeinveqn1} and \eqref{eqn:phiSdef}. The goal is to show that $\mathcal{S}_{ab,\mathbf{k}_{S}}=0$. To demonstrate this, we only need the two-dimensional $(t,r)$ spacetime so we ignore the $S^{d-1}$ in what follows (and drop the $\mathbf{k}_{S}$ label).

Consider the vector 
\begin{equation}\label{eqn:vdef}
    v_{a} = \mathcal{S}_{ab}t^{b},
\end{equation}
which is divergenceless by the Killing equation, and hence 
\begin{equation}\label{eqn:sdef}
    v^{a} = \epsilon^{ab}\widehat{\nabla}_{b}s
\end{equation}
for some scalar $s$. Our choice of initial data \eqref{eqn:initdata} means that $v^{a}|_{r=r_{0}} = 0$ and thus $\widehat{\nabla}_{a}s|_{r=r_{0}}=0$. Furthermore, we can shift $s$ by a constant without changing $v^{a}$ so we can choose this constant such that $s|_{r=r_{0}}=0$. We want to show that $s=0$ everywhere. To do this, observe that
\begin{align}\label{eqn:videntity}
\begin{split}
    \widehat{\nabla}_{a}(\epsilon^{ab}v_{b}) = \widehat{\nabla}_{a}(\epsilon^{ab}{\mathcal{S}_{b}}^{c}t_{c}) &= \widehat{\nabla}_{a}(\epsilon^{bc}{\mathcal{S}_{b}}^{a}t_{c}) 
    \\  &= \epsilon^{bc}{\mathcal{S}_{b}}^{a}\widehat{\nabla}_{a}t_{c}+\epsilon^{bc}t_{c}\widehat{\nabla}_{a}{\mathcal{S}_{b}}^{a} 
    \\  &= -\frac{1}{2}\epsilon^{bc}{\mathcal{S}_{b}}^{a}\epsilon_{ac}\epsilon^{de}\widehat{\nabla}_{d}t_{e}+\epsilon^{bc}t_{c}\widehat{\nabla}_{a}{\mathcal{S}_{b}}^{a} 
    \\  &= \epsilon^{bc}t_{c}\widehat{\nabla}_{a}{\mathcal{S}_{b}}^{a}
\end{split}
\end{align}
where in the second equality we used that any traceless $2$-tensor on our spacetime satisfies $\epsilon^{ab}{\mathcal{S}_{b}}^{c}=\epsilon^{bc}{\mathcal{S}_{b}}^{a}$, in the fourth equality we used that any Killing vector on our spacetime satisfies $\widehat{\nabla}_{a}t_{c} = -\frac{1}{2}\epsilon_{ac}\epsilon^{de}\widehat{\nabla}_{d}t_{e}$, and in the final equality we used $\epsilon^{bc}\epsilon_{ac} = \delta_{a}^{b}$ and tracelessness of $\mathcal{S}_{ab}$. Using \eqref{eqn:Eai_gaugeinv}, this implies
\begin{equation}\label{eqn:seqn}
\widehat{\nabla}^{2}s = -\epsilon^{bc}t_{c}\widehat{\nabla}_{a}{\mathcal{S}_{b}}^{a} = 2\frac{\mathfrak{f}_{\mathbf{k}_{S}}(t)}{r^{4}}.
\end{equation}
We want the graviton to have oscillatory behavior in time $e^{-i\Omega t}$, which means so will $Z_{ab}$ and hence so will $s=e^{-i\Omega t}\varrho(r)$. Then by separation of variables
\begin{equation}\label{eqn:sepvars}
(r^{2}+1)\partial_{r}^{2}\varrho(r)+2r\partial_{r}\varrho(r)+\frac{(\Omega)^{2}}{(r^{2}+1)}\varrho(r) = \frac{\mathfrak{f}_{\mathbf{k}_{S}}}{r^{4}}, \qquad 2e^{i\Omega t}\mathfrak{f}_{\mathbf{k}_{S}}(t) = \mathfrak{f}_{\mathbf{k}_{S}}
\end{equation}
for some constant $\mathfrak{f}_{\mathbf{k}_{S}}$. This inhomogeneous second-order linear differential equation has solution
\begin{align}\label{eqn:varrhosoln}
\begin{split}
    &\varrho(r) = C_{1}\cos\left(\Omega\tan^{-1}(r)\right)+C_{2}\sin\left(\Omega\tan^{-1}(r)\right)
    \\  &-\frac{\mathfrak{f}_{\mathbf{k}_{S}}}{\Omega}\int^{r}d\mathfrak{r}\,\frac{(\mathfrak{r}^{2}+1)}{\mathfrak{r}^{4}}\left[\cos\left(\Omega\tan^{-1}(\mathfrak{r})\right)\sin\left(\Omega\tan^{-1}(r)\right)-\sin\left(\Omega\tan^{-1}(\mathfrak{r})\right)\cos\left(\Omega\tan^{-1}(r)\right)\right].
\end{split}
\end{align}
One can perform this integral exactly in terms of incomplete Beta functions which goes as $\frac{1}{r^{2}}$ as $r \to 0$ and this singular behavior cannot be canceled by the homogeneous solution which goes to $C_{1}$ as $r \to 0$. We conclude that $\mathfrak{f}_{\mathbf{k}_{S}}=0$. With the initial data for $s$ being zero, there is no solution for $\varrho(r)$ except $\varrho(r)=0$, and hence $s=0$. Therefore, $v=0$. Now we use
\begin{equation}\label{eqn:2dmetricformulae}
    G_{ab} = \frac{1}{(r^{2}+1)}\left(-t_{a}t_{b}+\epsilon_{ac}\epsilon_{bd}t^{c}t^{d}\right), \qquad \epsilon_{ab}\epsilon_{cd}t^{d} = t_{a}G_{bc}-t_{b}G_{ac}
\end{equation}
to obtain
\begin{equation}
    \mathcal{S}_{ab} = {\mathcal{S}_{a}}^{c}G_{cb} = -\frac{1}{(r^{2}+1)}\left(v_{a}t_{b}+v_{b}t_{a}-G_{ab}v_{c}t^{c}\right) = 0
\end{equation}
which is the desired result. Therefore,
\begin{equation}\label{eqn:Zab_phiS}
    Z_{ab} = \left(\widehat{\nabla}_{a}\widehat{\nabla}_{b}-G_{ab}\right)\phi_{S},
\end{equation}
which has the gauge freedom that we can shift $\phi_{S} \to \phi_{S} + \phi_{0}$ for
\begin{equation}\label{eqn:phiSgaugefreedom}
    \left(\widehat{\nabla}_{a}\widehat{\nabla}_{b}-G_{ab}\right)\phi_{0}=0
\end{equation}
without changing $Z_{ab}$. 

Finally, we plug \eqref{eqn:Zab_phiS} into the $ab$ components of the Einstein tensor \eqref{eqn:Eab_gaugeinv} to obtain 
\begin{equation}\label{eqn:EphiSeqn}
\left(\widehat{\nabla}_{a}\widehat{\nabla}_{b}-G_{ab}\right)E(\phi_{S}) = -2r^{d-1}\left(E_{ab}^{(1)}[h]-{E_{r}}^{r(1)}[h]G_{ab}\right)
\end{equation}
where
\begin{equation}\label{eqn:EphiSdef}
    E(\phi_{S}) = r^{2}\left(\widehat{\nabla}^{2}-(d-1)\frac{\widehat{\nabla}^{a}r}{r}\widehat{\nabla}_{a}-\frac{(k_{S}^{2}-(d-1))}{r^{2}}+(d-3)\right)\phi_{S}.
\end{equation}
We can solve this set of equations. The $a=t$, $b=r$ equation gives
\begin{equation}\label{eqn:Eeqn_tr}
    \partial_{t}\left(\partial_{r}-\frac{r}{(r^{2}+1)}\right)E(t,r) = 0 \implies \left((r^{2}+1)\partial_{r}-r\right)E(t,r) = \tilde{f}(r)
\end{equation}
for some function $\tilde{f}(r)$. We can take an $r$ derivative to obtain
\begin{equation}\label{eqn:tildefrderiv}
    \left((r^{2}+1)\partial_{r}^{2}+r\partial_{r}-1\right)E(t,r) = \partial_{r}\tilde{f}(r).
\end{equation}
The $a=b=r$ equation gives
\begin{equation}\label{eqn:Eeqnrr}
    \left(\partial_{r}^{2}+\frac{r}{(r^{2}+1)}\partial_{r}-\frac{1}{(r^{2}+1)}\right)E(t,r) = 0 \implies \tilde{f}(r) = d_{1}
\end{equation}
for some constant $d_{1}$. Now we can solve the second equation in \eqref{eqn:Eeqn_tr} to obtain
\begin{equation}\label{eqn:Esolnagain}
    E(t,r) = d_{1}r+\tilde{C}(t)\sqrt{r^{2}+1}
\end{equation}
for some function $\tilde{C}(t)$. Finally, we use the $a=b=t$ equation to find
\begin{align}\label{eqn:tildeC(t)}
\begin{split}
    &\left[\partial_{t}^{2}+(r^{2}+1)(-r\partial_{r}+1)\right]\tilde{C}(t)\sqrt{r^{2}+1} = -2\mathfrak{c}(t)\sqrt{r^{2}+1} 
    \\  &\implies \left(\partial_{t}^{2}+1\right)\tilde{C}(t) = -2\mathfrak{c}(t)\sqrt{r^{2}+1} \\ &\implies \tilde{C}(t) = d_{2}\cos t + d_{3}\sin t, \;\; \mathfrak{c}(t)=0
\end{split}
\end{align}
for some constants $d_{2}$, $d_{3}$, where we used the fact that the lefthand side of the third equation is independent of $r$, while the righthand side has $r$ dependence, which can only be made $r$-independent by $\mathfrak{c}(t)=0$. Therefore,
\begin{equation}\label{eqn:Esolagain}
    E_{\mathrm{sol}}(t,r) = d_{1}r + \sqrt{r^{2}+1}\left(d_{2}\cos t + d_{3}\sin t\right).
\end{equation}

Now all of the analysis from \cite{Ishibashi:2004wx} carries over, which goes as follows. Notice that we have the same equation for $\phi_{0}$ in \eqref{eqn:phiSgaugefreedom} so
\begin{equation}\label{eqn:phi0soln}
    \phi_{0}(t,r) = \tilde{c}_{1}r + \sqrt{r^{2}+1}\left(\tilde{c}_{2}\cos t + \tilde{c}_{3}\sin t\right).
\end{equation}

Let us see how $E(\phi_{S})$ transforms under the gauge transformation $\phi_{S} \to \phi_{S} + \phi_{0}$. We find
\begin{align}
\begin{split}
    E(\phi_{S}+\phi_{0}) &= E_{\mathrm{sol}} + \left((d-1)r^{2}-(k_{S}^{2}-(d-1))-(d-1)r(r^{2}+1)\partial_{r}\right)\phi_{0} 
    \\  &= \tilde{\tilde{c}}_{1}r+\sqrt{r^{2}+1}\left(\tilde{\tilde{c}}_{2}\cos t + \tilde{\tilde{c}}_{3}\sin t\right)
\end{split}
\end{align}
where
\begin{equation}\label{eqn:tildetildec}
    \tilde{\tilde{c}}_{1} = c_{1}-k_{S}^{2}\tilde{c}_{1}, \qquad \tilde{\tilde{c}}_{2} = c_{2}-(k_{S}^{2}-(d-1))\tilde{c}_{2}, \qquad \tilde{\tilde{c}}_{3} = c_{3}-(k_{S}^{2}-(d-1))\tilde{c}_{3}.
\end{equation}
Therefore, we can choose $\tilde{c}_{i}$ such that all $\tilde{\tilde{c}}_{i}=0$ and hence make $E(\phi_{S}+\phi_{0})=0$. Defining $\Phi = r^{\frac{(d-1)}{2}}(\phi_{S}+\phi_{0})$, then \eqref{eqn:EphiSdef} gives a simple differential equation akin to the Klein-Gordon equation for a scalar field in AdS:
\begin{equation}\label{eqn:Phieqn}
    \widehat{\nabla}^{2}\Phi-\left[\frac{(d-3)(d-5)}{4}+\left(\frac{(d-1)(d-3)}{4}+k_{S}^{2}\right)\frac{1}{r^{2}}\right]\Phi = 0.
\end{equation}
This is how the linearized Einstein equations reduce to that of a scalar field which is the aforementioned ``master variable''.

\paragraph{Solutions.}
\label{sec:gravitonsolns}

Let us now solve \eqref{eqn:Phieqn} for $\Phi(t,r)$. We find
\begin{align}\label{eqn:phisoln}
\begin{split}
    \phi_{S;n,\ell}(t,r)&+\phi_{0;n,\ell}(t,r) = \mathcal{A}_{S}e^{-i\Omega_{n,\ell}^{S}t}\left(\frac{r}{\sqrt{r^{2}+1}}\right)^{d-2+\ell}r
    \\  &\times {}_{2}{F}_{1}\left(\frac{d-2+\ell+\Omega_{n,\ell}^{S}}{2},\frac{d-2+\ell-\Omega_{n,\ell}^{S}}{2},\ell+\frac{d}{2};\frac{r^{2}}{r^{2}+1}\right),
\end{split}
\end{align}
where the second solution of the differential equation must be thrown away in order for $\mathfrak{h}_{n,\mathbf{k}_{S}}^{S,\mathrm{trless}}(t,r)$ to be non-singular at $r=0$. The extrapolate dictionary \eqref{eqn:extrapolatedictionary} imposes the quantization of the frequencies $\Omega_{n,\ell}^{S}$ as in \eqref{eqn:scalarmode_freq}.

Armed with this solution, we now solve for the graviton wavefunctions. Observe that we have four independent equations: $Z_{ab} = \left(\widehat{\nabla}_{a}\widehat{\nabla}_{b}-G_{ab}\right)\phi_{S}$ and $Z = Z_{a}^{a} = \left(\widehat{\nabla}^{2}-2\right)\phi_{S}$ and we have four functions for the gauge-fixed graviton: $\mathfrak{h}_{tt}$, $\mathfrak{h}_{t}$, $\mathfrak{h}^{S,\mathrm{trless}}$, and $\mathfrak{h}^{S,\mathrm{tr}}$ so we expect to be able to solve uniquely for the graviton, subject to the appropriate boundary conditions.

First, we solve for $\mathfrak{h}^{S,\mathrm{trless}}$ using $Z_{rr}$. We can explicitly compute $Z_{rr}$ and $Z$ from \eqref{eqn:phisoln} and then we get the following differential equation for $\mathfrak{h}^{S,\mathrm{trless}}$:
\begin{equation}\label{eqn:htrlesseqn}
\frac{1}{\sqrt{r^{2}+1}}\partial_{r}\left[r^{2}\sqrt{r^{2}+1}\widehat{\nabla}_{r}\left(\frac{\mathfrak{h}_{n,\mathbf{k}_{S}}^{S,\mathrm{trless}}}{r^{2}}\right)\right] = \frac{1}{r^{d-3}}\left(Z_{rr,n,\mathbf{k}_{S}}-\frac{(d-2)}{(d-1)}Z_{n,\mathbf{k}_{S}}G_{rr}\right),
\end{equation}
leading to the solution in \eqref{eqn:hsoln}, where the integration constants are fixed by \eqref{eqn:extrapolatedictionary}. Then we can use $Z$ to compute $\mathfrak{h}_{0,2}^{S,\mathrm{tr}}(t,r)$
\begin{equation}\label{eqn:htreqn}
    \mathfrak{h}_{0,2}^{S,\mathrm{tr}} = \frac{r^{5-d}}{(d-1)}Z_{0,2}-\frac{2d}{(d-1)}\mathfrak{h}_{0,2}^{S,\mathrm{trless}}-r^{3}(r^{2}+1)\partial_{r}\left(\frac{\mathfrak{h}_{0,2}^{S,\mathrm{trless}}}{r^{2}}\right)
\end{equation}
leading to the solution in \eqref{eqn:hsoln}. Next, we use $Z_{tr}$ to compute $\mathfrak{h}_{t,0,2}(t,r)$
\begin{equation}\label{eqn:Ztr_ht}
    Z_{tr,0,2} = r^{d-3}\left[-(r^{2}+1)\partial_{r}\left(\frac{\mathfrak{h}_{t,0,2}}{(r^{2}+1)}\right)+r\sqrt{r^{2}+1}\,\partial_{r}\left(\frac{1}{r\sqrt{r^{2}+1}}\,\partial_{t}\mathfrak{h}_{0,2}^{S,\mathrm{trless}}\right)\right],
\end{equation}
giving the solution in \eqref{eqn:hsoln} with the integration constant again fixed by \eqref{eqn:extrapolatedictionary}. Finally, we can use $Z_{tt}$ to compute $\mathfrak{h}_{tt,0,2}(t,r)$, and we find
\begin{equation}\label{eqn:htteqn}
    \mathfrak{h}_{tt,0,2}(t,r) = \frac{1}{r^{d-3}}\left(Z_{tt,0,2}-\frac{(d-2)}{(d-1)}Z_{0,2}G_{tt}\right)+2\partial_{t}\mathfrak{h}_{t,0,2}-\partial_{t}^{2}\mathfrak{h}_{0,2}^{S,\mathrm{trless}}+r^{3}(r^{2}+1)\partial_{r}\left(\frac{\mathfrak{h}_{0,2}^{S,\mathrm{trless}}}{r^{2}}\right)
\end{equation}
which gives the solution in \eqref{eqn:hsoln}.

\paragraph{Normalisation.}
\label{sec:norm}

We now compute $\mathcal{N}_{0,2}^{S}$ by imposing unit norm.
We find
\begin{align}\label{eqn:graviton_normalisationconstant}
\begin{split}
    \langle h_{0,2,\mathfrak{m}}^{S},h_{0,2,\mathfrak{m}}^{S}\rangle &= i\Omega\left(h_{0,2,\mathfrak{m}}^{S\ast},h_{0,2,\mathfrak{m}}^{S}\right)
    \\  &= \Omega_{0,2}^{S}V_{S^{d-1}}(d-2)\int_{0}^{\infty} dr\,G^{tt}r^{d-5}\bigg[{-}(d-1)|\mathfrak{h}_{0,2}^{\mathrm{tr}}|^{2}+\frac{2d(d+1)}{(d-1)}|\mathfrak{h}_{0,2}^{\mathrm{trless}}|^{2}
    \\  &-\frac{i}{\Omega_{0,2}^{S}}4d\left(\mathfrak{h}_{0,2}^{\mathrm{tr}\ast}+\frac{(d+1)}{(d-1)}\mathfrak{h}_{0,2}^{\mathrm{trless}\ast}\right)\mathfrak{h}_{t,0,2}\bigg].
\end{split}
\end{align}
This looks very difficult, but we can simplify this integral by using the following nice fact:
\begin{equation}\label{eqn:nicefact}
    \mathfrak{h}_{0,2}^{\mathrm{tr}} + \frac{(d+1)}{(d-1)}\mathfrak{h}_{0,2}^{\mathrm{trless}} = -\mathcal{N}_{0,2}^{S}e^{-i\Omega_{0,2}^{S}t}\frac{r^{2}}{(r^{2}+1)^{\frac{d}{2}}},
\end{equation}
which gives
\begin{align}\label{eqn:graviton_normalisationconstant_simpl}
\begin{split}
    \langle h_{0,2,\mathfrak{m}}^{S},h_{0,2,\mathfrak{m}}^{S}\rangle &= -\Omega_{0,2}^{S}V_{S^{d-1}}(d-2)|\mathcal{N}_{0,2}^{S}|^{2}\int_{0}^{\infty} dr\,G^{tt}r^{d-5}\bigg[(d+1)d^{2}r^{2}(r^{2}+1)B_{\frac{1}{r^{2}+1}}\left(\frac{d}{2},\frac{3}{2}\right)^{2}
    \\  &\qquad {-}2\frac{r^{2}((d+2)r^{2}+2)}{(r^{2}+1)^{d}}\bigg].
\end{split}
\end{align}
The integral on the first line can be computed via integration by parts as follows
\begin{align}\label{eqn:graviton_normalisationconstant_hardint}
\begin{split}
   \int_{0}^{\infty} dr\,r^{d-3}B_{\frac{1}{r^{2}+1}}\left(\frac{d}{2},\frac{3}{2}\right)^{2} &= -\frac{2}{(d+1)(d-2)}\int_{0}^{1}dw\,w^{\frac{d}{2}-1}(1-w)^{\frac{d-1}{2}}{}_{2}{F}_{1}\left(1-\frac{d}{2},\frac{3}{2},\frac{5}{2};y\right)
   \\   &= \frac{4\Gamma(\frac{d}{2})\Gamma(\frac{d}{2}+2)}{(d+1)(d-2)\Gamma(d+2)},
\end{split}
\end{align}
where $w \equiv \frac{1}{r^{2}+1}$. Therefore, we obtain the final result for the inner product
\begin{equation}\label{eqn:graviton_normalisationconstant_final}
    \langle h_{0,2,\mathfrak{m}}^{S},h_{0,2,\mathfrak{m}}^{S}\rangle = 4(d-1)d\frac{\Gamma(\frac{d}{2}+1)\Gamma(\frac{d}{2}+2)}{\Gamma(d+2)}V_{S^{d-1}}|\mathcal{N}_{0,2}^{S}|^{2}.
\end{equation}
Requiring unit norm fixes our normalisation constant to be the one given in \eqref{eqn:graviton_normconstant}.

\subsection{AdS-Rindler}
\label{sec:AdSRindlerEOM}

The AdS-Rindler equations of motion can be solved by similar methods so we will be brief, focusing as always on the scalar mode $\mathpzc{h}_{\mu\nu}^{S_{H}}$ whose decomposition is given in \eqref{eqn:gravscalar_Rindler}. The gauge-invaraint variables are
\begin{align}\label{eqn:gaugeinvvars_Rindler}
\begin{split}
    \tilde{Z}_{\lambda} &= (d-1)\rho^{d-5}\left(\tilde{\mathfrak{h}}_{\lambda}^{S_{H},\mathrm{tr}}+\frac{\lambda}{(d-1)}\tilde{\mathfrak{h}}_{\lambda}^{S_{H},\mathrm{trless}}+2\rho\left(\widetilde{\nabla}^{\mathfrak{v}}\rho\right)\tilde{X}_{\mathfrak{v},\lambda}\right)
    \\ \tilde{Z}_{\mathfrak{v}\mathfrak{w},\lambda} &= \rho^{d-3}\left(\tilde{\mathfrak{h}}_{\mathfrak{v}\mathfrak{w},\lambda}+2\widetilde{\nabla}_{(\mathfrak{v}}\tilde{X}_{\mathfrak{w}),\lambda}\right)+\frac{(d-2)}{(d-1)}\tilde{Z}_{\lambda}G_{\mathfrak{v}\mathfrak{w}},
\end{split}
\end{align}
where $\widetilde{\nabla}$ is the covariant derivative on the $(\tau,\rho)$ part of the geometry and the Einstein equations give the following relations
\begin{align}\label{eqn:Zrelns_Rindler}
\begin{split}
    \widetilde{\nabla}^{\mathfrak{w}}\tilde{Z}_{\mathfrak{v}\mathfrak{w},\lambda}-\widetilde{\nabla}_{\mathfrak{v}}{\tilde{Z}_{\mathfrak{w},\lambda}}^{\mathfrak{w}} &= 0
    \\  \tilde{Z}_{\mathfrak{v},\lambda}^{\mathfrak{v}}-\tilde{Z}_{\lambda} &= 0.
\end{split}
\end{align}
All of the Einstein equations reduce to the equation for a single ``master'' scalar field $\phi_{S_{H}}^{R}$ related to the gauge-invariant variables by
\begin{equation}\label{eqn:Zab_phiSR}
    \tilde{Z}_{\mathfrak{v}\mathfrak{w},\lambda} = \left(\widetilde{\nabla}_{\mathfrak{v}}\widetilde{\nabla}_{\mathfrak{w}}-G_{\mathfrak{v}\mathfrak{w}}\right)\phi_{S_{H},\lambda}^{R},
\end{equation}
up to the gauge freedom that we can shift $\phi_{S_{H},\lambda}^{R} \to \Phi_{\lambda}^{R} = \phi_{S_{H},\lambda}^{R}+\phi_{0,\lambda}^{R}$ for
\begin{equation}\label{eqn:phiSRgaugefreedom}
    \left(\widetilde{\nabla}_{\mathfrak{v}}\widetilde{\nabla}_{\mathfrak{w}}-G_{\mathfrak{v}\mathfrak{w}}\right)\phi_{0,\lambda}^{R}=0
\end{equation}
without changing $\tilde{Z}_{\mathfrak{v}\mathfrak{w}}$. The Einstein equations imply that the master field satisfies the differential equation
\begin{equation}\label{eqn:PhiReqn}
    \widetilde{\nabla}^{2}\Phi_{\lambda}^{R}-(d-1)\frac{(\rho^{2}-1)}{\rho}\partial_{\rho}\Phi_{\lambda}^{R}-\left(\frac{\lambda+(d-1)}{\rho^2}-(d-3)\right)\Phi_{\lambda}^{R} = 0.
\end{equation}
Performing the separation of variables $\Phi_{\lambda}^{R}(\tau,\rho)= \int\frac{d\omega}{2\pi}\,\Phi_{\omega,\lambda}^{R}(\tau,\rho)$ with $\Phi_{\omega,\lambda}^{R}(\tau,\rho)=e^{-i\omega\tau}\tilde{\Phi}_{\omega,\lambda}^{R}(\rho)$ and imposing the AdS-Rindler extrapolate dictionary
\begin{equation}\label{eqn:extrapolatedictionary_Rindler}
\lim_{\rho \to \infty} \rho^{d-2}\mathpzc{h}_{\mu\nu}(\tau,\rho,\alpha) \propto \tilde{T}_{\mu\nu}(\tau,\alpha),
\end{equation}
we find the solution to be
\begin{equation}\label{eqn:PhiRfinal}
\tilde{\Phi}_{\omega,\lambda}^{R}(\rho) = c^{(1)}\left(1-\frac{1}{\rho^{2}}\right)^{-i\frac{\omega}{2}}\rho\,{}_{2}{F}_{1}\left(\frac{1}{2}\left(\frac{d-2}{2}-i\omega_{-}\right),\frac{1}{2}\left(\frac{d-2}{2}-i\omega_{+}\right),\frac{d-2}{2};\frac{1}{\rho^{2}}\right)
\end{equation}
with
\begin{equation}\label{eqn:omegapmdef}
i\omega_{\pm} \equiv i\omega\pm\sqrt{\frac{(d-2)^{2}}{4}-\lambda}.
\end{equation}
Observe that the solution satisfies ingoing boundary conditions at the horizon $\tilde{\Phi}_{\omega,\lambda}^{R}(\rho \to 1) \propto (\rho-1)^{-\frac{\omega}{2}}$.

Armed with the solution for the ``master'' scalar field, we can now solve for the graviton in Rindler radial gauge $h_{\mu\rho}=0$. We can explicitly compute $\tilde{Z}_{\rho\rho}$ and $\tilde{Z}$ from \eqref{eqn:PhiRfinal}, \eqref{eqn:Zab_phiSR}, and \eqref{eqn:Zrelns_Rindler}, and then we get the following differential equation for $\tilde{\mathfrak{h}}^{S_{H},\mathrm{trless}}$:
\begin{equation}\label{eqn:htrlesseqn_Rindler}
\frac{1}{\sqrt{\rho^{2}-1}}\partial_{\rho}\left[\rho^{2}\sqrt{\rho^{2}-1}\widetilde{\nabla}_{\rho}\left(\frac{\tilde{\mathfrak{h}}_{\omega,\lambda}^{S_{H},\mathrm{trless}}}{\rho^{2}}\right)\right] = \frac{1}{\rho^{d-3}}\left(\tilde{Z}_{\rho\rho,\omega,\lambda}-\frac{(d-2)}{(d-1)}\tilde{Z}_{\omega,\lambda}G_{\rho\rho}\right).
\end{equation}
This can be solved by switching to the $\eta=\rho^{-2}$ variable, which leads to an integral over $\eta$ of a hypergeometric function of $\eta$, which can be series expanded for $0 \leq \eta 1$ (corresponding to $1<\rho$), and then integrated term-by-term. The result is a finite sum of infinite sums of incomplete Beta functions, with the integration constants fixed by imposing \eqref{eqn:extrapolatedictionary_Rindler}, although the result is rather ugly so we do not write it explicitly here. We can now use the solution for $\tilde{\mathfrak{h}}_{\omega,\lambda}^{S_{H},\mathrm{trless}}$ to obtain the other parts of the scalar mode from their governing equations:
\begin{align}\label{eqn:otherRindlerwavefneqns}
\begin{split}
\tilde{\mathfrak{h}}_{\omega,\lambda}^{S_{H},\mathrm{tr}}(\tau,\rho) &= \frac{\rho^{5-d}}{(d-1)}\tilde{Z}_{\omega,\lambda}-\frac{2d}{(d-1)}\tilde{\mathfrak{h}}_{\omega,\lambda}^{S_{H},\mathrm{trless}}-\rho^{3}(\rho^{2}-1)\partial_{\rho}\left(\frac{\tilde{\mathfrak{h}}_{\omega,\lambda}^{S_{H},\mathrm{trless}}}{\rho^{2}}\right)
\\  \partial_{\rho}\left(\frac{\tilde{\mathfrak{h}}_{\tau,\omega,\lambda}}{(\rho^{2}-1)}\right) &= \frac{\rho}{\sqrt{\rho^{2}-1}}\,\partial_{\rho}\left(\frac{1}{\rho\sqrt{\rho^{2}-1}}\,\partial_{\tau}\tilde{\mathfrak{h}}_{\omega,\lambda}^{S_{H},\mathrm{trless}}\right)-\frac{1}{\rho^{d-3}(\rho^{2}-1)}\tilde{Z}_{\tau\rho,\omega,\lambda}
\\  \tilde{\mathfrak{h}}_{\tau\tau,\omega,\lambda}(\tau,\rho) &= \frac{1}{\rho^{d-3}}\left(\tilde{Z}_{\tau\tau,\omega,\lambda}-\frac{(d-2)}{(d-1)}\tilde{Z}_{\omega,\lambda}G_{\tau\tau}\right)+2\partial_{\tau}\tilde{\mathfrak{h}}_{\tau,\omega,\lambda}^{S_{H}}-\partial_{\tau}^{2}\tilde{\mathfrak{h}}_{\omega,\lambda}^{S_{H},\mathrm{trless}}
\\  &\qquad +\rho^{3}(\rho^{2}-1)\partial_{\rho}\left(\frac{\tilde{\mathfrak{h}}_{\omega,\lambda}^{S_{H},\mathrm{trless}}}{\rho^{2}}\right)
\end{split}
\end{align}
where the solutions again take the form of finite sums of infinite sums of incomplete Beta functions, with the integration constants again determined by \eqref{eqn:extrapolatedictionary_Rindler}.

Despite the fact that the solutions are very complicated, they are not complicated in two important limits: (1) the first few terms in an expansion near the asymptotic boundary, given by small $\eta$ (large $\rho$), are simple and these are what we use in \S\ref{sec:Bogcoeffs_calc} to compute the Bogoliubov coefficients; (2) the expansion near the horizon ($\rho=1$) is simple and this is what we use to determine the normalization constant for the wavefunctions.

Let us now explain in more detail how to compute this normalization constant. We require delta-function normalizability of the wavefunctions with respect to the generalized Klein-Gordon inner product as given in \eqref{eqn:deltafnnormalisability}. This delta-function can only come from the oscillatory behavior in $\rho$ of the wavefunction near $\rho=1$ with phase proportional to $\omega$. Thus, it suffices to focus only on the expansion of the wavefunctions near the horizon when computing the inner product. We find their behavior to be
\begin{align}\label{eqn:Rindlerwavefn_horlimit}
\begin{split}
    \tilde{\mathfrak{h}}_{\omega,\lambda}^{S_{H},\mathrm{trless}}(\rho \to 1) &= \mathcal{N}_{\omega,\lambda}^{S,R}e^{-i\omega\tau}\left(\tilde{\mathcal{X}}\left(1-\frac{1}{\rho^{2}}\right)^{-i\frac{\omega}{2}}+\mathrm{c.c.}\right), 
    \\  \tilde{\mathfrak{h}}_{\omega,\lambda}^{S_{H},\mathrm{tr}}(\rho \to 1) &\sim \tilde{\mathfrak{h}}_{\tau,\omega,\lambda}^{S_{H}}(\rho \to 1) \sim (\rho-1)^{1 \pm i\frac{\omega}{2}},\\
    \tilde{\mathfrak{h}}_{\tau\tau,\omega,\lambda}(\rho \to 1) & \sim (\rho-1)^{2 \pm i\frac{\omega}{2}}
\end{split}
\end{align}
where $\mathcal{N}_{\omega,\lambda}^{S,R}$ is the sought after normalization constant, and we have defined
\begin{equation}\label{eqn:tildeXdef}
	\tilde{\mathcal{X}} \equiv \frac{(d-1)\Gamma\left(\frac{d-2}{2}\right)\Gamma(i\omega)}{\Gamma\left(\frac{1}{2}\left(\frac{d-2}{2}+i\omega_{-}\right)\right)\Gamma\left(\frac{1}{2}\left(\frac{d-2}{2}+i\omega_{+}\right)\right)}.
\end{equation}
This means that only $\tilde{\mathfrak{h}}_{\omega,\lambda}^{S_{H},\mathrm{trless}}$ is needed to compute $\mathcal{N}_{\omega,\lambda}^{S,R}$ and, using \eqref{eqn:symplecticform_graviton_Rindler}, we obtain
\begin{align}\label{eqn:graviton_normalizationconstant_Rindler}
\begin{split}
    \langle \tilde{h}_{\omega,\lambda}^{S_{H}},\tilde{h}_{\omega',\lambda'}^{S_{H}}\rangle &= 2\pi \delta^{(d-1)}(\lambda-\lambda') (d-2)|\mathcal{N}_{\omega,\lambda}^{S,R}|^{2} \frac{(\omega+\omega')}{2}
    \\  \times \lambda&\frac{(\lambda+d-1)}{(d-1)}\int_{1}^{1+\epsilon}d\rho\,\left(\tilde{\mathcal{X}}|_{\omega\to\omega'}(\rho^{2}-1)^{-i\frac{\omega'}{2}-1}+\mathrm{c.c.}\right)\left(\tilde{\mathcal{X}}(\rho^{2}-1)^{-i\frac{\omega}{2}}+\mathrm{c.c.}\right)
    \\  &= (2\pi)^{2}\delta(\omega-\omega')\delta^{(d-1)}(\lambda-\lambda')|\mathcal{N}_{\omega,\lambda}^{S,R}|^{2}|\tilde{\mathcal{X}}|^{2}\frac{(d-2)}{(d-1)}\omega\lambda(\lambda+d-1),
\end{split}
\end{align}
where we dropped all non-singular terms as $\rho \to 1$ in the first equality.

Therefore, we obtain the normalization constant
\begin{align}\label{eqn:Rindlernormconstant}
\begin{split}
    \mathcal{N}_{\omega,\lambda}^{S,R} &= \bigg|\frac{\Gamma\left(\frac{1}{2}\left(\frac{d-2}{2}+i\omega_{-}\right)\right)\Gamma\left(\frac{1}{2}\left(\frac{d-2}{2}+i\omega_{+}\right)\right)}{\Gamma\left(\frac{d-2}{2}\right)\Gamma(i\omega)}\bigg|\frac{1}{\sqrt{(d-1)(d-2)\omega\lambda(\lambda+d-1)}}.
\end{split}
\end{align}
by requiring that \eqref{eqn:graviton_normalizationconstant_Rindler} gives a standard delta-function normalisation.

\section{Backreaction of gravitons}
\label{sec:gravbackreact}

This appendix provides the details of solving \eqref{eqn:g2eqn_T} to obtain $\langle \hat{g}_{\mu\nu}^{(2)}\rangle_{\mathfrak{g}}$ from the backreaction of $\hat{h}_{\mu\nu}$. The quadratic Einstein tensor for $h$ can be written explicitly using the formulas in App.~\ref{sec:pertcurv}:
\begin{align}\label{eqn:Tgrav}
\begin{split}
    T_{\mu\nu}^{\mathrm{grav}} &= -E_{\mu\nu}^{\mathrm{quad}}[h] = \frac{1}{2}\bigg(\nabla_{\beta}h^{\beta\alpha}\left(\nabla_{\mu}h_{\nu\alpha}+\nabla_{\nu}h_{\mu\alpha}-\nabla_{\alpha}h_{\mu\nu}\right)-\nabla_{\mu}h^{\beta\alpha}\nabla_{\nu}h_{\beta\alpha}
\\  & +h^{\beta\alpha}\left(\nabla_{\beta}\nabla_{\mu}h_{\nu\alpha}+\nabla_{\beta}\nabla_{\nu}h_{\mu\alpha}-\nabla_{\mu}\nabla_{\nu}h_{\beta\alpha} - \nabla_{\beta}\nabla_{\alpha}h_{\mu\nu}\right)
\\  & +\frac{1}{2}G^{\sigma\beta}G^{\alpha\lambda}\left(\nabla_{\alpha}h_{\mu\beta}+\nabla_{\mu}h_{\alpha\beta}-\nabla_{\beta}h_{\alpha\mu}\right)\left(\nabla_{\sigma}h_{\nu\lambda}+\nabla_{\nu}h_{\sigma\lambda}-\nabla_{\lambda}h_{\sigma\nu}\right)
\\  & -\frac{1}{2}\nabla^{\sigma}h\left(\nabla_{\mu}h_{\nu\sigma}+\nabla_{\nu}h_{\mu\sigma}-\nabla_{\sigma}h_{\mu\nu}\right)\bigg) + \frac{1}{2}h_{\mu\nu}\left(\nabla^{\alpha}\nabla^{\beta}h_{\beta\alpha}-\nabla^{2}h+dh\right)
\\  &+ \frac{1}{2}G_{\mu\nu}\bigg({-}\nabla_{\beta}h^{\beta\alpha}\left(\nabla^{\rho}h_{\rho\alpha}-\nabla_{\alpha}h\right)-\frac{1}{2}\nabla^{\alpha}h^{\rho\beta}\nabla_{\beta}h_{\rho\alpha}-\frac{1}{4}\nabla^{\alpha}h\nabla_{\alpha}h+\frac{3}{4}\nabla^{\rho}h^{\beta\alpha}\nabla_{\rho}h_{\beta\alpha}
\\  & -h^{\beta\alpha}\left(\nabla^{\rho}\nabla_{\beta}h_{\rho\alpha}+\nabla_{\beta}\nabla^{\rho}h_{\rho\alpha}-\nabla^{2}h_{\beta\alpha} - \nabla_{\beta}\nabla_{\alpha}h\right)-dh^{\alpha\beta}h_{\alpha\beta}\bigg). 
\end{split}
\end{align}
The simplest way to compute the linearized Einstein tensor for $g^{(2)}$ given the ansatz \eqref{eqn:g2ansatz} is to use the perturbed curvature tensors in App.~D of \cite{Colin-Ellerin:2024npf}. Given the further ansatz \eqref{eqn:calFansatz}, the equation \eqref{eqn:g2eqn_T} for $\langle\hat{g}^{(2)}\rangle_{\mathfrak{g}}$ can be solved.

Consider the following combination of Einstein's equations (with no sum on $i$):
\begin{equation}\label{eqn:Etheta+Ephi}
    {E_{\theta}}^{\theta}[\langle\hat{g}^{(2)}\rangle_{\mathfrak{g}}]-{E_{\phi_{i}}}^{\phi_{i}}[\langle\hat{g}^{(2)}\rangle_{\mathfrak{g}}] = \left(\langle \hat{T}_{\theta}^{\mathrm{grav},\theta} \rangle_{\mathfrak{g}} - \langle \hat{T}_{\phi_{i}}^{\mathrm{grav},\phi_{i}} \rangle_{\mathfrak{g}}\right),
\end{equation}
which gives
\begin{equation}\label{eqn:Etheta+Ephi2}
    \sin\theta\partial_{\theta}\left(\csc\theta\partial_{\theta}U(r,\theta)\right) = 2r^{2}\left(\langle \hat{T}_{\theta}^{\mathrm{grav},\theta} \rangle_{\mathfrak{g}} - \langle \hat{T}_{\phi_{i}}^{\mathrm{grav},\phi_{i}} \rangle_{\mathfrak{g}}\right),
\end{equation}
where
\begin{equation}\label{eqn:Udef}
    U(r,\theta) \equiv \mathcal{F}_{2}(r,\theta)-\mathcal{F}_{1}(r,\theta)-(d-3)\mathcal{F}_{3}(r,\theta).
\end{equation}
Using \eqref{eqn:Tgravexpval}, we can integrate up the righthand side \eqref{eqn:Etheta+Ephi2} to obtain
\begin{align}\label{eqn:Usoln}
\begin{split}
    U(r,\theta) &= 2r^{2}\Bigg\{{-}\frac{1}{2}\left(\mathcal{T}_{\theta}^{\theta}-\mathcal{T}_{\phi_{i}}^{\phi_{i}}\right)-\left(\frac{1}{2}\cos(2\theta)+1\right)\left(\mathcal{S}_{\theta}^{\theta}-\mathcal{S}_{\phi_{i}}^{\phi_{i}}\right)
    \\  &\qquad -\frac{1}{3}\left(\frac{1}{4}\cos(4\theta)+\cos(2\theta)+\frac{5}{2}\right)\left(\mathcal{V}_{\theta}^{\theta}-\mathcal{V}_{\phi_{i}}^{\phi_{i}}\right)+u_{1}(r)\cos\theta+u_{2}(r)\Bigg\},
\end{split}
\end{align}
where $u_{1}(r)$, $u_{2}(r)$ are integration `constants', and we used the fact that 
\begin{equation}\label{eqn:Tfact}
    \left(\langle \hat{T}_{\theta}^{\mathrm{grav},\theta} \rangle_{\mathfrak{g}} - \langle \hat{T}_{\phi_{i}}^{\mathrm{grav},\phi_{i}} \rangle_{\mathfrak{g}}\right)\Big|_{\theta=0} = 0.
\end{equation}
To prove this fact, observe that $\partial_{\theta}\langle \hat{T}_{\theta\theta}\rangle_{\mathfrak{g}}|_{\theta=0}=\langle \hat{T}_{r\theta}\rangle_{\mathfrak{g}}|_{\theta=0}=0$ so the conservation of the stress-tensor (Bianchi identity) implies
\begin{align}\label{eqn:theta=0proof}
\begin{split}
    0 &= \nabla_{\mu}\langle \hat{T}^{\mathrm{grav},\mu}_{\theta}\rangle_{\mathfrak{g}} = \partial_{r}\langle \hat{T}_{\theta}^{\mathrm{grav},r}\rangle_{\mathfrak{g}}+\partial_{\theta}\langle \hat{T}_{\theta}^{\mathrm{grav},\theta}\rangle_{\mathfrak{g}}+(d-2)\cot\theta\left(\langle \hat{T}_{\theta}^{\mathrm{grav},\theta}\rangle_{\mathfrak{g}} - \langle \hat{T}_{\phi_{i}}^{\mathrm{grav},\phi_{i}}\rangle_{\mathfrak{g}}\right)
    \\ &\implies \left(\langle \hat{T}_{\theta}^{\mathrm{grav},\theta}\rangle_{\mathfrak{g}} - \langle \hat{T}_{\phi_{i}}^{\mathrm{grav},\phi_{i}}\rangle_{\mathfrak{g}}\right)\Big|_{\theta=0} = -\frac{1}{(d-2)}\tan\theta\left(\partial_{r}\langle \hat{T}_{\theta}^{\mathrm{grav},r}\rangle_{\mathfrak{g}}+\partial_{\theta}\langle \hat{T}_{\theta}^{\mathrm{grav},\theta}\rangle_{\mathfrak{g}}\right)|_{\theta=0}=0.
\end{split}
\end{align}
Thus, \eqref{eqn:Usoln} gives
\begin{align}\label{eqn:Usolnconsequence}
\begin{split}
    \mathscr{F}_{2,2}-\mathscr{F}_{1,2}-(d-3)\mathscr{F}_{3,2} &= -2 r^{2}\left(\frac{1}{2}\left(\mathcal{S}_{\theta}^{\theta}-\mathcal{S}_{\phi_{i}}^{\phi_{i}}\right)+\frac{1}{3}\left(\mathcal{V}_{\theta}^{\theta}-\mathcal{V}_{\phi_{i}}^{\phi_{i}}\right)\right)
    \\  \mathscr{F}_{2,4}-\mathscr{F}_{1,4}-(d-3)\mathscr{F}_{3,4} &= -\frac{1}{6} r^{2}\left(\mathcal{V}_{\theta}^{\theta}-\mathcal{V}_{\phi_{i}}^{\phi_{i}}\right).
\end{split}
\end{align}
Next, the $r\theta$ Einstein equation gives
\begin{align}\label{eqn:EE_rtheta}
\begin{split}
    E_{r\theta}^{(1)}[\langle \hat{g}^{(2)}\rangle_{\mathfrak{g}}] &= \langle \hat{T}_{r\theta}^{\mathrm{grav}} \rangle_{\mathfrak{g}} 
    \\  &\implies  \mathcal{F}_{1}-r(r^{2}+1)\mathcal{F}_{1}'+\left(r^{2}+2-d(r^{2}+1)\right)\mathcal{F}_{2}-(d-2)r(r^{2}+1)\mathcal{F}_{3}' 
    \\  &\qquad \qquad \qquad = -2 r(r^{2}+1)\left(\frac{1}{2}\mathcal{S}_{r\theta}(r)\cos(2\theta)+\frac{1}{4}\mathcal{V}_{r\theta}(r)\cos(4\theta)+\mathfrak{f}(r)\right),
\end{split}
\end{align}
where $\mathfrak{f}(r)$ is an integration `constant' coming from integrating both sides with respect to $\theta$, and thus
\begin{align}\label{eqn:EE_rtheta2}
\begin{split}
    \mathscr{F}_{1,2}-r(r^{2}+1)\mathscr{F}_{1,2}'+\left(r^{2}+2-d(r^{2}+1)\right)\mathscr{F}_{2,2}-(d-2)r(r^{2}+1)\mathscr{F}_{3,2}' &= - r(r^{2}+1)\mathcal{S}_{r\theta}
    \\ \mathscr{F}_{1,4}-r(r^{2}+1)\mathscr{F}_{1,4}'+\left(r^{2}+2-d(r^{2}+1)\right)\mathscr{F}_{2,4}-(d-2)r(r^{2}+1)\mathscr{F}_{3,4}' &= -\frac{1}{2} r(r^{2}+1)\mathcal{V}_{r\theta}.
\end{split}
\end{align}
The $tt$ Einstein equation gives
\begin{align}\label{eqn:EE_tt}
\begin{split}
    &E_{tt}[\langle \hat{g}^{(2)} \rangle_{\mathfrak{g}}] = \langle \hat{T}_{tt}^{\mathrm{grav}} \rangle_{\mathfrak{g}}
\\  &\implies -(d-1)(d-2+dr^{2})\mathcal{F}_{2}^{\mathrm{g}}-(d-2)(d-1)\mathcal{F}_{3}^{\mathrm{g}}-(d-1)r(r^{2}+1)\left(\mathcal{F}_{2}^{\mathrm{g}'}+r\mathcal{F}_{3}^{\mathrm{g}''}\right)
\\  &-(d-1)r\left(d+(d+1)r^{2}\right)\mathcal{F}_{3}^{\mathrm{g}'}+\left((d-2)\cot\theta\partial_{\theta}+\partial_{\theta}^{2}\right)\left(\mathcal{F}_{2}^{\mathrm{g}}-(d-2)\mathcal{F}_{3}^{\mathrm{g}}\right) = 2\frac{r^{2}}{(r^{2}+1)}\langle \hat{T}_{tt}^{\mathrm{grav}} \rangle_{\mathfrak{g}}
\end{split}
\end{align}
so we obtain
\begin{align}\label{eqn:EE_tt2}
\begin{split}
    &2\frac{r^{2}}{(r^{2}+1)}\mathcal{T}_{tt} = -(d-1)(d-2+dr^{2})\mathscr{F}_{2,0}-(d-2)(d-1)\mathscr{F}_{3,0}-(d-1)r(r^{2}+1)\left(\mathscr{F}_{2,0}'+r\mathscr{F}_{3,0}''\right)
\\  &-(d-1)r\left(d+(d+1)r^{2}\right)\mathscr{F}_{3,0}'-2(d-2)\left(\mathscr{F}_{2,2}+2\mathscr{F}_{2,4}-(d-2)\left(\mathscr{F}_{3,2}+2\mathscr{F}_{3,4}\right)\right)
\\  &2\frac{r^{2}}{(r^{2}+1)}\mathcal{S}_{tt}= -(d^{2}-d+2+(d-1)dr^{2})\mathscr{F}_{2,2}+(d-2)(d+1)\mathscr{F}_{3,2}
\\  &-(d-1)r(r^{2}+1)\left(\mathscr{F}_{2,2}'+r\mathscr{F}_{3,2}''\right)-(d-1)r\left(d+(d+1)r^{2}\right)\mathscr{F}_{3,2}'-8(d-2)\left(\mathscr{F}_{2,4}-(d-2)\mathscr{F}_{3,4}\right)
\\  &2\frac{r^{2}}{(r^{2}+1)}\mathcal{V}_{tt} = -(d^{2}+d+10+(d-1)dr^{2})\mathscr{F}_{2,4}+3(d-2)(d+3)\mathscr{F}_{3,4}
\\  &-(d-1)r(r^{2}+1)\left(\mathscr{F}_{2,4}'+r\mathscr{F}_{3,4}''\right)-(d-1)r\left(d+(d+1)r^{2}\right)\mathscr{F}_{3,4}'.
\end{split}
\end{align}
The $rr$ Einstein equation gives
\begin{align}\label{eqn:EE_rr}
\begin{split}
&E_{rr}[\langle \hat{g}^{(2)} \rangle_{\mathfrak{g}}] = \langle \hat{T}_{rr}^{\mathrm{grav}} \rangle_{\mathfrak{g}}
\\  &\implies (d-1)r(r^{2}+1)\mathcal{F}_{1}^{\mathrm{g}'}+(d-1)(d-2+dr^{2})\mathcal{F}_{2}^{\mathrm{g}}+(d-1)(d-2)\left(r\mathcal{F}_{3}^{\mathrm{g}}\right)'+(d-1)^{2}r^{3}\mathcal{F}_{3}^{\mathrm{g}'}
\\	&+\left((d-2)\cot\theta\partial_{\theta}+\partial_{\theta}^{2}\right)\left(\mathcal{F}_{1}^{\mathrm{g}}+(d-2)\mathcal{F}_{3}^{\mathrm{g}}\right) = 2r^{2}(r^{2}+1)\langle\hat{T}_{rr}^{\mathrm{grav}}\rangle_{\mathfrak{g}}
\end{split}
\end{align}
so we obtain
\begin{align}\label{eqn:EE_rr2}
\begin{split}
2r^{2}(r^{2}+1)\mathcal{T}_{rr} &= (d-1)r(r^{2}+1)\mathscr{F}_{1,0}'+(d-1)(d-2+dr^{2})\mathscr{F}_{2,0}+(d-1)(d-2)\left(r\mathscr{F}_{3,0}\right)'
\\  &+(d-1)^{2}r^{3}\mathscr{F}_{3,0}'-2(d-2)\left(\mathscr{F}_{1,2}+2\mathscr{F}_{1,4}+(d-2)\left(\mathscr{F}_{3,2}+2\mathscr{F}_{3,4}\right)\right)
\\  2r^{2}(r^{2}+1)\mathcal{S}_{rr} &= (d-1)r(r^{2}+1)\mathscr{F}_{1,2}'+(d-1)(d-2+dr^{2})\mathscr{F}_{2,2}+(d-1)(d-2)\left(r\mathscr{F}_{3,2}\right)'
\\  &+(d-1)^{2}r^{3}\mathscr{F}_{3,2}'-2\left(d\mathscr{F}_{1,2}+4(d-2)\mathscr{F}_{1,4}+(d-2)\left(d\mathscr{F}_{3,2}+4(d-2)\mathscr{F}_{3,4}\right)\right)
\\  2r^{2}(r^{2}+1)\mathcal{V}_{rr} &= (d-1)r(r^{2}+1)\mathscr{F}_{1,4}'+(d-1)(d-2+dr^{2})\mathscr{F}_{2,4}+(d-1)(d-2)\left(r\mathscr{F}_{3,4}\right)'
\\  &+(d-1)^{2}r^{3}\mathscr{F}_{3,4}'-4(d+2)\left(\mathscr{F}_{1,4}+(d-2)\mathscr{F}_{3,4}\right).
\end{split}
\end{align}
Finally, the $\theta\theta$ Einstein equation gives
\begin{align}\label{eqn:EE_thetatheta}
\begin{split}
&E_{\theta\theta}[\langle \hat{g}^{(2)} \rangle_{\mathfrak{g}}] = \langle \hat{T}_{\theta\theta}^{\mathrm{grav}} \rangle_{\mathfrak{g}}
\\    &\implies \left((d-2)(d-3)+d(d-1)r^{2}\right)\mathcal{F}_{2}^{\mathrm{g}}+(d-2)(d-3)\mathcal{F}_{3}^{\mathrm{g}}+r\left(d-2+(d+1)r^{2}\right)\mathcal{F}_{1}^{\mathrm{g}'}
\\	&+r\left(d-2+(d-1)r^{2}\right)\mathcal{F}_{2}^{\mathrm{g}'}+r(d-2)\left(d-1+(d+1)r^{2}\right)\mathcal{F}_{3}^{\mathrm{g}'}+r^{2}(r^{2}+1)\left(\mathcal{F}_{1}^{\mathrm{g}}+(d-2)\mathcal{F}_{3}^{\mathrm{g}}\right)''
\\	&+(d-2)\cot\theta\partial_{\theta}\left(\mathcal{F}_{1}^{\mathrm{g}}-\mathcal{F}_{2}^{\mathrm{g}}+(d-3)\mathcal{F}_{3}^{\mathrm{g}}\right) = 2\langle \hat{T}_{\theta\theta}^{\mathrm{grav}} \rangle_{\mathfrak{g}}
\end{split}
\end{align}
so we obtain
\begin{align}\label{eqn:EE_thetatheta2}
\begin{split}
2\mathcal{T}_{\theta\theta} &= \left((d-2)(d-3)+d(d-1)r^{2}\right)\mathscr{F}_{2,0}+(d-2)(d-3)\mathscr{F}_{3,0}+r\left(d-2+(d+1)r^{2}\right)\mathscr{F}_{1,0}'
\\	&+r\left(d-2+(d-1)r^{2}\right)\mathscr{F}_{2,0}'+r(d-2)\left(d-1+(d+1)r^{2}\right)\mathscr{F}_{3,0}'
\\  &+r^{2}(r^{2}+1)\left(\mathscr{F}_{1,0}+(d-2)\mathscr{F}_{3,0}\right)''-2(d-2)\bigg(\mathscr{F}_{1,2}+2\mathscr{F}_{1,4}-\mathscr{F}_{2,2}-2\mathscr{F}_{2,4}
\\  &+(d-3)\left(\mathscr{F}_{3,2}+2\mathscr{F}_{3,4}\right)\bigg) 
\\  2\mathcal{S}_{\theta\theta} &= \left((d-2)(d-3)+d(d-1)r^{2}\right)\mathscr{F}_{2,2}+(d-2)(d-3)\mathscr{F}_{3,2}+r\left(d-2+(d+1)r^{2}\right)\mathscr{F}_{1,2}'
\\	&+r\left(d-2+(d-1)r^{2}\right)\mathscr{F}_{2,2}'+r(d-2)\left(d-1+(d+1)r^{2}\right)\mathscr{F}_{3,2}'
\\  &+r^{2}(r^{2}+1)\left(\mathscr{F}_{1,2}+(d-2)\mathscr{F}_{3,2}\right)''-2(d-2)\bigg(\mathscr{F}_{1,2}+4\mathscr{F}_{1,4}-\mathscr{F}_{2,2}-4\mathscr{F}_{2,4}
\\  &+(d-3)\left(\mathscr{F}_{3,2}+4\mathscr{F}_{3,4}\right)\bigg) 
\\  2\mathcal{V}_{\theta\theta} &= \left((d-2)(d-3)+d(d-1)r^{2}\right)\mathscr{F}_{2,4}+(d-2)(d-3)\mathscr{F}_{3,4}+r\left(d-2+(d+1)r^{2}\right)\mathscr{F}_{1,4}'
\\	&+r\left(d-2+(d-1)r^{2}\right)\mathscr{F}_{2,4}'+r(d-2)\left(d-1+(d+1)r^{2}\right)\mathscr{F}_{3,4}'
\\  &+r^{2}(r^{2}+1)\left(\mathscr{F}_{1,4}+(d-2)\mathscr{F}_{3,4}\right)''-4(d-2)\left(\mathscr{F}_{1,4}-\mathscr{F}_{2,4}+(d-3)\mathscr{F}_{3,4}\right).
\end{split}
\end{align}
We will not need the Einstein equations for $\phi_{i}\phi_{i}$ ($2 \leq i \leq d-1$).

To solve all of these coupled second-order linear ordinary differential equations, notice that the equations obtained from matching coefficients of $\cos(4\theta)$ (i.e., those involving $\mathcal{V}_{\mu\nu}$) only contain $\mathscr{F}_{i,4}$. Furthermore, the equations from matching coefficients of $\cos(2\theta)$ (i.e., those involving $\mathcal{S}_{\mu\nu}$) only contain $\mathscr{F}_{i,4}$ and $\mathscr{F}_{i,2}$. Therefore, the strategy is clear: first we solve the equations containing $\mathcal{V}_{\mu\nu}$ for the $\mathscr{F}_{i,4}$, then we plug these results into the equations containing $\mathcal{S}_{\mu\nu}$ to solve for the $\mathscr{F}_{i,2}$, and finally we plug these results into the equations containing $\mathcal{T}_{\mu\nu}$ to solve for the $\mathscr{F}_{i,0}$.

Let us collect all the $\mathcal{V}_{\mu\nu}$ equations here for convenience:
\begin{align}\label{eqn:Veqns}
\begin{split}
    -\frac{1}{2}r(r^{2}+1)\mathcal{V}_{r\theta} &= \mathscr{F}_{1,4}-r(r^{2}+1)\mathscr{F}_{1,4}'+\left(r^{2}+2-d(r^{2}+1)\right)\mathscr{F}_{2,4}-(d-2)r(r^{2}+1)\mathscr{F}_{3,4}'
\\  2\frac{r^{2}}{(r^{2}+1)}\mathcal{V}_{tt} &= -(d^{2}+d+10+(d-1)dr^{2})\mathscr{F}_{2,4}+3(d-2)(d+3)\mathscr{F}_{3,4}
\\  &-(d-1)r(r^{2}+1)\left(\mathscr{F}_{2,4}'+r\mathscr{F}_{3,4}''\right)-(d-1)r\left(d+(d+1)r^{2}\right)\mathscr{F}_{3,4}'
\\  2r^{2}(r^{2}+1)\mathcal{V}_{rr} &= (d-1)r(r^{2}+1)\mathscr{F}_{1,4}'+(d-1)(d-2+dr^{2})\mathscr{F}_{2,4}+(d-1)(d-2)\left(r\mathscr{F}_{3,4}\right)'
\\  &+(d-1)^{2}r^{3}\mathscr{F}_{3,4}'-4(d+2)\left(\mathscr{F}_{1,4}+(d-2)\mathscr{F}_{3,4}\right)
\\  2\mathcal{V}_{\theta\theta} &= \left((d-2)(d-3)+d(d-1)r^{2}\right)\mathscr{F}_{2,4}+(d-2)(d-3)\mathscr{F}_{3,4}
\\  &+r\left(d-2+(d+1)r^{2}\right)\mathscr{F}_{1,4}'
\\	&+r\left(d-2+(d-1)r^{2}\right)\mathscr{F}_{2,4}'+r(d-2)\left(d-1+(d+1)r^{2}\right)\mathscr{F}_{3,4}'
\\  &+r^{2}(r^{2}+1)\left(\mathscr{F}_{1,4}+(d-2)\mathscr{F}_{3,4}\right)''-4(d-2)\left(\mathscr{F}_{1,4}-\mathscr{F}_{2,4}+(d-3)\mathscr{F}_{3,4}\right).
\end{split}
\end{align}
We first add the second and third equations in \eqref{eqn:Veqns} to obtain
\begin{align}\label{eqn:Fi4_eq1}
\begin{split}
    2r^{2}\left(\frac{\mathcal{V}_{tt}}{(r^{2}+1)}+(r^{2}+1)\mathcal{V}_{rr}\right) &= -4(d+2)\left(\mathscr{F}_{1,4}+\mathscr{F}_{2,4}\right)
    \\  &\qquad +(d-1)r(r^{2}+1)\left(\mathscr{F}_{1,4}'-\mathscr{F}_{2,4}'-2\mathscr{F}_{3,4}'-r\mathscr{F}_{3,4}''\right).
\end{split}
\end{align}
Now we solve the second equation of \eqref{eqn:Usolnconsequence} for $\mathscr{F}_{1,4}$ and plug it into \eqref{eqn:Fi4_eq1}, which allows us to solve for $\mathscr{F}_{2,4}$ in terms of $\mathscr{F}_{3,4}$ and its derivatives, viz.,
\begin{align}\label{eqn:F24eqn}
\begin{split}
    \mathscr{F}_{2,4} &= -\frac{1}{(d+2)}\Bigg[\frac{r^{2}}{4}\Bigg(\frac{\mathcal{V}_{tt}}{(r^{2}+1)}+(r^{2}+1)\mathcal{V}_{rr}-\frac{(d-1)}{12}r(r^{2}+1)\left(\mathcal{V}_{\theta}^{\theta}-\mathcal{V}_{\phi_{i}}^{\phi_{i}}\right)'
    \\  &\qquad -\frac{1}{6}\left((d-1)r^{2}-(d+5)\right)\left(\mathcal{V}_{\theta}^{\theta}-\mathcal{V}_{\phi_{i}}^{\phi_{i}}\right)\bigg)+\frac{(d-1)}{8}r(r^{2}+1)\left((d-1)\mathscr{F}_{3,4}'+r\mathscr{F}_{3,4}''\right)\Bigg]
    \\  &+\frac{(d-3)}{2}\mathscr{F}_{3,4}.
\end{split}
\end{align}
Next, we add $(d-1)$ times the first equation in \eqref{eqn:Veqns} to the third equation to obtain
\begin{equation}\label{eqn:Fi4_eq2}
     \frac{1}{2}r(r^{2}+1)\left({-}(d-1)\mathcal{V}_{r\theta}+4 r\mathcal{V}_{rr}\right) = -3(d+3)(\mathscr{F}_{1,4}+(d-2)\mathscr{F}_{3,4})+(d-1)r^{2}(\mathscr{F}_{2,4}+r\mathscr{F}_{3,4}').
\end{equation}
Now we plug into this the equation for $\mathscr{F}_{1,4}$ obtained from \eqref{eqn:Usolnconsequence} and $\mathscr{F}_{2,4}$ from \eqref{eqn:F24eqn} to find an equation purely in terms of $\mathscr{F}_{3,4}$ which we can solve:
\begin{align}\label{eqn:F34diffeq}
\begin{split}
    \mathbb{S}_{4} &= 4(d+2)\left((d-3)r^{2}-3(d+3)\right)\mathscr{F}_{3,4}
    \\  &\qquad +r\left({-}(d-1)^{2}r^{4}+2(d^{2}+8d+3)r^{2}+3(d-1)(d+3)\right)\mathscr{F}_{3,4}'
    \\  &\qquad -r^{2}(r^{2}+1)\left({-}3(d+3)+(d-1)r^{2}\right)\mathscr{F}_{3,4}''
\end{split}  
\end{align}
where the source term $\mathbb{S}_{4}$ is given by
\begin{align}\label{eqn:source4}
\begin{split}
    \mathbb{S}_{4} &= \frac{4 r}{(d-1)}\bigg((d+2)(r^{2}+1)\left({-}(d-1)\mathcal{V}_{r\theta}+4 r\mathcal{V}_{rr}\right)
    \\  &+\frac{r}{12}\left({-}(d-1)^{2}r^{4}+2(d-1)(2d+7)r^{2}+9(d+3)(d+1)\right)\left(\mathcal{V}_{\theta}^{\theta}-\mathcal{V}_{\phi_{i}}^{\phi_{i}}\right)
    \\  &+\frac{r(r^{2}+1)}{2}\left({-}3(d+3)+(d-1)r^{2}\right)\left(\mathcal{V}_{rr}+\frac{\mathcal{V}_{tt}}{(r^{2}+1)^{2}}-\frac{(d-1)r}{12}\left(\mathcal{V}_{\theta}^{\theta}-\mathcal{V}_{\phi_{i}}^{\phi_{i}}\right)'\right)\bigg).
\end{split}
\end{align}
We now have an inhomogeneous second-order differential equation for $\mathscr{F}_{3,4}$ that we need to solve. We cannot solve \eqref{eqn:F34diffeq} as an arbitrary function of $d$, but for any fixed $d$, it can be solved. For $d=4$, we find
\begin{align}\label{eqn:F34soln}
\begin{split}
    &\mathscr{F}_{3,4}(r)|_{d=4} = \frac{C_{1}}{r^{6}}\left(r^{4}-3r^{2}-6\right)+\frac{C_{2}}{2r^{6}}\left(2(r^{4}-3r^{2}-6)\log(r^{2}+1)-(r^{6}-3r^{2}(r^{2}+1)+18)\right)
    \\  &-\frac{8}{675}|\mathcal{N}_{0,2}^{S}|^{2}\bigg[2304\frac{(r^{4}-3r^{2}-6)\log(r^{2}+1)}{r^{6}}-\frac{4}{r(r^{2}+1)^{\frac{5}{2}}}(135+745r^{2}+196r^{4}+288r^{6})
    \\  &+\frac{1}{r^{6}(r^{2}+1)^{4}}(-26382-104895r^{2}-151106r^{4}-83915r^{6}+2220r^{8}+15900r^{10}+2304r^{12})\bigg].
\end{split}
\end{align}

We need to fix the coefficients $C_{1}$ and $C_{2}$. As mentioned in the main text, we do this by imposing that the full metric $g_{\mu\nu}$ be asymptotically AdS and that there are no curvature singularities, which can be obtained by letting $g_{rr}^{(2)}$ have a mild coordinates singularity and requiring no singularities at all for the other components. 

The requirement that the backreacted metric be asymptotically AdS fixes $C_{2}$ and requiring finiteness at the origin ($r=0$) fixes $C_{1}$ so we arrive at
\begin{align}\label{eqn:F34_final}
\begin{split}
    \mathscr{F}_{3,4}(r)|_{d=4} &= -\frac{8}{675}|\mathcal{N}_{0,2}^{S}|^{2}\bigg[{-}\frac{4}{r(r^{2}+1)^{\frac{5}{2}}}(135+745r^{2}+196r^{4}+288r^{6})
    \\  &+\frac{1}{r^{2}(r^{2}+1)^{4}}(257+1379r^{2}+4296r^4+4591r^{6}+2515r^{8}+1152r^{10})\bigg].
\end{split}
\end{align}
We can plug this into \eqref{eqn:F24eqn} to obtain $\mathscr{F}_{2,4}$:
\begin{align}\label{eqn:F24_final}
\begin{split}
    \mathscr{F}_{2,4}(r)|_{d=4} &= \frac{8}{675}|\mathcal{N}_{0,2}^{S}|^{2}\bigg[{-}\frac{16r}{(r^{2}+1)^{\frac{7}{2}}}(10-109r^{2}+286r^{4})
    \\  &+\frac{1}{r^{2}(r^{2}+1)^{5}}(353+1765r^{2}+4880r^{4}+4447r^{6}+5063r^{8}+4570r^{10})\bigg],
\end{split}
\end{align}
which we see leads to a mild coordinate singularity $\langle \hat{g}_{rr}^{(2)}\rangle_{\mathfrak{g}} \sim 1/r^{2}$, and then use the second equation in \eqref{eqn:Usolnconsequence} to obtain $\mathscr{F}_{1,4}$:
\begin{align}\label{eqn:F14_final}
\begin{split}
    \mathscr{F}_{1,4}(r)|_{d=4} &= \frac{8}{675}|\mathcal{N}_{0,2}^{S}|^{2}\bigg[{-}\frac{4r}{(r^{2}+1)^{\frac{7}{2}}}(8865+30895r^{2}+35324r^{4}+13024r^{6})
    \\  &+\frac{1}{(r^{2}+1)^{5}}(10073+93565r^{2}+271214r^{4}+355009r^{6}+219437r^{8}+52096r^{10})\bigg].
\end{split}
\end{align}
One can check that these solutions satisfy the fourth equation in \eqref{eqn:Veqns}.

Next, we use the $\mathcal{S}_{\mu\nu}$ equations to solve for the $\mathscr{F}_{2,i}$, which we collect here for convenience:
\begin{align}\label{eqn:Seqns}
\begin{split}
-r(r^{2}+1)\mathcal{S}_{r\theta} &= \mathscr{F}_{1,2}-r(r^{2}+1)\mathscr{F}_{1,2}'+\left(r^{2}+2-d(r^{2}+1)\right)\mathscr{F}_{2,2}-(d-2)r(r^{2}+1)\mathscr{F}_{3,2}'
\\  2\frac{r^{2}}{(r^{2}+1)}\mathcal{S}_{tt} &= -(d^{2}-d+2+(d-1)dr^{2})\mathscr{F}_{2,2}+(d-2)(d+1)\mathscr{F}_{3,2}
\\  -(d-1)r(r^{2}+1)&\left(\mathscr{F}_{2,2}'+r\mathscr{F}_{3,2}''\right)-(d-1)r\left(d+(d+1)r^{2}\right)\mathscr{F}_{3,2}'-8(d-2)\left(\mathscr{F}_{2,4}-(d-2)\mathscr{F}_{3,4}\right)
\\  2r^{2}(r^{2}+1)\mathcal{S}_{rr} &= (d-1)r(r^{2}+1)\mathscr{F}_{1,2}'+(d-1)(d-2+dr^{2})\mathscr{F}_{2,2}+(d-1)(d-2)\left(r\mathscr{F}_{3,2}\right)'
\\  +(d-1)^{2}r^{3}\mathscr{F}_{3,2}'&-2\left(d\mathscr{F}_{1,2}+4(d-2)\mathscr{F}_{1,4}+(d-2)\left(d\mathscr{F}_{3,2}+4(d-2)\mathscr{F}_{3,4}\right)\right)
\\  2\mathcal{S}_{\theta\theta} &= \left((d-2)(d-3)+d(d-1)r^{2}\right)\mathscr{F}_{2,2}+(d-2)(d-3)\mathscr{F}_{3,2}
\\  &+r\left(d-2+(d+1)r^{2}\right)\mathscr{F}_{1,2}'+r\left(d-2+(d-1)r^{2}\right)\mathscr{F}_{2,2}'
\\  &+r(d-2)\left(d-1+(d+1)r^{2}\right)\mathscr{F}_{3,2}'+r^{2}(r^{2}+1)\left(\mathscr{F}_{1,2}+(d-2)\mathscr{F}_{3,2}\right)''
\\  &-2(d-2)\left(\mathscr{F}_{1,2}+4\mathscr{F}_{1,4}-\mathscr{F}_{2,2}-4\mathscr{F}_{2,4}+(d-3)\left(\mathscr{F}_{3,2}+4\mathscr{F}_{3,4}\right)\right).
\end{split}
\end{align}
We first add the second and third equations in \eqref{eqn:Seqns} to obtain
\begin{align}\label{eqn:Fi2_eq1}
\begin{split}
    2r^{2}\left(\frac{\mathcal{S}_{tt}}{(r^{2}+1)}+(r^{2}+1)\mathcal{S}_{rr}\right)&+8(d-2)(\mathscr{F}_{1,4}+\mathscr{F}_{2,4}) = -2d\left(\mathscr{F}_{1,2}+\mathscr{F}_{2,2}\right)
    \\  &+(d-1)r(r^{2}+1)\left(\mathscr{F}_{1,2}'-\mathscr{F}_{2,2}'-2\mathscr{F}_{3,2}'-r\mathscr{F}_{3,2}''\right).
\end{split}
\end{align}
Now we solve the second equation of \eqref{eqn:Usolnconsequence} for $\mathscr{F}_{1,2}$ and plug it into \eqref{eqn:Fi2_eq1}, which allows us to solve for $\mathscr{F}_{2,2}$ in terms of $\mathscr{F}_{3,2}$ and its derivatives, viz.,
\begin{align}\label{eqn:F22eqn}
\begin{split}
    &\mathscr{F}_{2,2} = -\frac{1}{d}\Bigg[\frac{r^{2}}{2}\Bigg(\frac{\mathcal{S}_{tt}}{(r^{2}+1)}+(r^{2}+1)\mathcal{S}_{rr}-(d-1)r(r^{2}+1)\left(\frac{1}{3}\left(\mathcal{V}_{\theta}^{\theta}-\mathcal{V}_{\phi_{i}}^{\phi_{i}}\right)'+\frac{1}{2}\left(\mathcal{S}_{\theta}^{\theta}-\mathcal{S}_{\phi_{i}}^{\phi_{i}}\right)'\right)
    \\  &\qquad -\left((d-1)r^{2}-1\right)\left(\frac{2}{3}\left(\mathcal{V}_{\theta}^{\theta}-\mathcal{V}_{\phi_{i}}^{\phi_{i}}\right)+\left(\mathcal{S}_{\theta}^{\theta}-\mathcal{S}_{\phi_{i}}^{\phi_{i}}\right)\right)\bigg)+2(d-2)(\mathscr{F}_{1,4}+\mathscr{F}_{2,4})
    \\  &\qquad +\frac{(d-1)}{4}r(r^{2}+1)\left((d-1)\mathscr{F}_{3,2}'+r\mathscr{F}_{3,2}''\right)\Bigg]+\frac{(d-3)}{2}\mathscr{F}_{3,2}.
\end{split}
\end{align}
Next, we add $(d-1)$ times the first equation in \eqref{eqn:Seqns} to the third equation to obtain
\begin{align}\label{eqn:Fi2_eq2}
\begin{split}
    r(r^{2}+1)\left({-}(d-1)\mathcal{S}_{r\theta}+2 r\mathcal{S}_{rr}\right)+8(d-2)&(\mathscr{F}_{1,4}+(d-2)\mathscr{F}_{3,4}) = -(d+1)\mathscr{F}_{1,2}+(d-1)r^{2}\mathscr{F}_{2,2}
    \\  &-(d+1)(d-2)\mathscr{F}_{3,2}+(d-1)r^{3}\mathscr{F}_{3,2}'.
\end{split}
\end{align}
Now we plug into this the equation for $\mathscr{F}_{1,2}$ obtained from \eqref{eqn:Usolnconsequence} and $\mathscr{F}_{2,2}$ from \eqref{eqn:F22eqn} to find an equation purely in terms of $\mathscr{F}_{3,2}$ which we can solve:
\begin{align}\label{eqn:F32diffeq}
\begin{split}
    \mathbb{S}_{2} &= r^{2}\left({-}(d-1)r^{4}+2r^{2}+d+1\right)\mathscr{F}_{3,2}''-r\left({-}(d+1)(d-1)-2(3d-1)r^{2}+(d-1)^{2}r^{4}\right)\mathscr{F}_{3,2}'
    \\  &+2d\left(-(d+1)+(d-3)r^{2}\right)\mathscr{F}_{3,2}
\end{split}
\end{align}
where
\begin{align}\label{eqn:source2}
\begin{split}
    \mathbb{S}_{2} &= \frac{8(d-2)}{(d-1)}\left(4d(d-2)\mathscr{F}_{3,4}+\left({-}(d+1)+(d-1)r^{2}\right)\mathscr{F}_{2,4}+\left((d-1)r^{2}+3d-1\right)\mathscr{F}_{1,4}\right)
    \\  &+\frac{r}{(d-1)}\bigg[4d(r^{2}+1)\left({-}(d-1)\mathcal{S}_{r\theta}+2 r\mathcal{S}_{rr}\right)
    \\  &-2r\left((d-1)^{2}r^{4}+(d+2)(d-1)r^{2}-(2d-1)(d+1)\right)\left(\frac{2}{3}\left(\mathcal{V}_{\theta}^{\theta}-\mathcal{V}_{\phi_{i}}^{\phi_{i}}\right)+\left(\mathcal{S}_{\theta}^{\theta}-\mathcal{S}_{\phi_{i}}^{\phi_{i}}\right)\right)
    \\  &+r(r^{2}+1)\left({-}(d+1)+(d-1)r^{2}\right)
    \\  &\qquad \times \left(2\mathcal{S}_{rr}+2\frac{\mathcal{S}_{tt}}{(r^{2}+1)^{2}}-(d-1)r\left(\frac{2}{3}\left(\mathcal{V}_{\theta}^{\theta}-\mathcal{V}_{\phi_{i}}^{\phi_{i}}\right)'+\left(\mathcal{S}_{\theta}^{\theta}-\mathcal{S}_{\phi_{i}}^{\phi_{i}}\right)'\right)\right)\bigg].
\end{split}
\end{align}
We thus find the solution
\begin{align}\label{eqn:F32soln}
\begin{split}
    &\mathscr{F}_{3,2}(r)|_{d=4} = \frac{C_{3}}{r^{4}}(r^{2}-1)+\frac{C_{4}}{2r^{4}}\left(r^{2}(3r^{2}-2)-2(r^{2}-1)\log(r^{2}+1)\right)
    \\  &-\frac{16}{675}|\mathcal{N}_{0,2}^{S}|^{2}\bigg[3584\frac{(r^{2}-1)\log(r^{2}+1)}{r^{4}}-\frac{4}{r(r^{2}+1)^{\frac{5}{2}}}(405+1235r^{2}+2444r^{4}+1344r^{6})
    \\  &+\frac{1}{r^{4}(r^{2}+1)^{4}}(-3616-6846r^{2}+5638r^{4}+20567r^{6}+14689r^{8}+3584r^{10})\bigg]
\end{split}
\end{align}
Requiring the backreacted metric be asymptotically AdS fixes $C_{4}$ and requiring finiteness at the origin ($r=0$) fixes $C_{3}$ so we arrive at
\begin{align}\label{eqn:F32_final}
\begin{split}
    \mathscr{F}_{3,2}(r)|_{d=4} &= -\frac{16}{675}|\mathcal{N}_{0,2}^{S}|^{2}\bigg[{-}\frac{4}{r(r^{2}+1)^{\frac{5}{2}}}(405+1235r^{2}+2444r^{4}+1344r^{6})
    \\  &+\frac{1}{r^{2}(r^{2}+1)^{4}}(418+3190r^{2}+13335r^4+21761r^{6}+17888r^{8}+5376r^{10})\bigg].
\end{split}
\end{align}
Plugging this into \eqref{eqn:F22eqn}, we obtain
\begin{align}\label{eqn:F22_final}
\begin{split}
    \mathscr{F}_{2,2}(r)|_{d=4} &= \frac{16}{675}|\mathcal{N}_{0,2}^{S}|^{2}\bigg[{-}\frac{80r}{(r^{2}+1)^{\frac{7}{2}}}(-14+149r^{2}+82r^{4})
    \\  &+\frac{1}{r^{2}(r^{2}+1)^{5}}(130+650r^{2}+3175r^4+12551r^{6}+21220r^{8}+6464r^{10})\bigg],
\end{split}
\end{align}
which again gives a mild coordinate singularity $\langle \hat{g}_{rr}^{(2)}\rangle_{\mathfrak{g}} \sim 1/r^{2}$, and then plugging \eqref{eqn:F32_final} and \eqref{eqn:F22_final} into the first equation of \eqref{eqn:Usolnconsequence}, we find
\begin{align}\label{eqn:F12_final}
\begin{split}
    \mathscr{F}_{1,2}(r)|_{d=4} &= \frac{16}{675}|\mathcal{N}_{0,2}^{S}|^{2}\bigg[{-}\frac{20r}{(r^{2}+1)^{\frac{7}{2}}}(927+3961r^{2}+4964r^{4}+1984r^{6})
    \\  &+\frac{1}{r^{2}(r^{2}+1)^{5}}(4090+51200r^{2}+170167r^4+242669r^{6}+158752r^{8}+39680r^{10})\bigg].
\end{split}
\end{align}
One can check that these solutions satisfy the fourth equation in \eqref{eqn:Seqns}.

Finally, we need to solve the $\mathcal{T}_{\mu\nu}$ equations for the $\mathscr{F}_{i,0}$, which we collect here for convenience:
\begin{align}\label{eqn:Teqns}
\begin{split}
2\frac{r^{2}}{(r^{2}+1)}\mathcal{T}_{tt} &= -(d-1)\left((d-2+dr^{2})\mathscr{F}_{2,0}+(d-2)\mathscr{F}_{3,0}\right)-(d-1)r(r^{2}+1)\left(\mathscr{F}_{2,0}'+r\mathscr{F}_{3,0}''\right)
\\  &-(d-1)r\left(d+(d+1)r^{2}\right)\mathscr{F}_{3,0}'-2(d-2)\left(\mathscr{F}_{2,2}+2\mathscr{F}_{2,4}-(d-2)\left(\mathscr{F}_{3,2}+2\mathscr{F}_{3,4}\right)\right)
\\ 2r^{2}(r^{2}+1)\mathcal{T}_{rr} &= (d-1)r(r^{2}+1)\mathscr{F}_{1,0}'+(d-1)(d-2+dr^{2})\mathscr{F}_{2,0}+(d-1)(d-2)\left(r\mathscr{F}_{3,0}\right)'
\\  &+(d-1)^{2}r^{3}\mathscr{F}_{3,0}'-2(d-2)\left(\mathscr{F}_{1,2}+2\mathscr{F}_{1,4}+(d-2)\left(\mathscr{F}_{3,2}+2\mathscr{F}_{3,4}\right)\right)
\\ 2\mathcal{T}_{\theta\theta} &= \left((d-2)(d-3)+d(d-1)r^{2}\right)\mathscr{F}_{2,0}+(d-2)(d-3)\mathscr{F}_{3,0}
\\  &+r\left(d-2+(d+1)r^{2}\right)\mathscr{F}_{1,0}'+r\left(d-2+(d-1)r^{2}\right)\mathscr{F}_{2,0}'
\\  &+r(d-2)\left(d-1+(d+1)r^{2}\right)\mathscr{F}_{3,0}'+r^{2}(r^{2}+1)\left(\mathscr{F}_{1,0}+(d-2)\mathscr{F}_{3,0}\right)''
\\  &-2(d-2)\left(\mathscr{F}_{1,2}+2\mathscr{F}_{1,4}-\mathscr{F}_{2,2}-2\mathscr{F}_{2,4}+(d-3)\left(\mathscr{F}_{3,2}+2\mathscr{F}_{3,4}\right)\right).
\end{split}
\end{align}
At this point, we have the residual gauge freedom to eliminate $\mathscr{F}_{3,0}$ using the diffeomorphism $x^{\mu} \to x^{\mu} + \xi^{\mu}$ where $\xi^{r} = -G_{N}r\mathscr{F}_{3,0}(r)$, $\xi^{\mu \neq r} = 0$ and redefining $\mathscr{F}_{1,0}$, $\mathscr{F}_{2,0}$ by their shifted functions after the diffeomorphism (alternatively, we can just make this part of our ansatz). Then the first equation in \eqref{eqn:Teqns} gives
\begin{align}\label{eqn:F20soln}
\begin{split}
    \mathscr{F}_{2,0}(r)|_{d=4} &= \frac{C_{5}}{r^{2}(r^{2}+1)}-\frac{16}{225}|\mathcal{N}_{0,2}^{S}|^{2}\bigg[{-}\frac{64r}{(r^{2}+1)^{\frac{5}{2}}}({-}5+26r^{2})
    \\  &+\frac{1}{r^{2}(r^{2}+1)^{5}}(2601+10786r^{2}+17309r^4+13630r^{6}+6656r^{8}+1664r^{10})\bigg].
\end{split}
\end{align}
Then the second equation in \eqref{eqn:Teqns} gives
\begin{align}\label{eqn:F10soln}
\begin{split}
    \mathscr{F}_{1,0}(r)|_{d=4} &= \frac{C_{5}}{r^{2}(r^{2}+1)}+C_{6}+\frac{16}{225}|\mathcal{N}_{0,2}^{S}|^{2}\bigg[{-}\frac{16r}{(r^{2}+1)^{\frac{5}{2}}}(135+680r^{2}+576r^{4})
    \\  &+\frac{1}{(r^{2}+1)^{5}}({-}47+6245r^{2}+28157r^4+46515r^{6}+33920r^{8}+9216r^{10})\bigg].
\end{split}
\end{align}
We fix $C_{6}$ such that $g_{tt}$ is asymptotically AdS and fix $C_{5}$ by requiring smoothness of $g_{tt}$ at the origin ($r=0$). This gives
\begin{align}\label{eqn:F20soln_final}
\begin{split}
\mathscr{F}_{2,0}(r)|_{d=4} &= \frac{16}{225}|\mathcal{N}_{0,2}^{S}|^{2}\bigg[\frac{64r}{(r^{2}+1)^{\frac{5}{2}}}({-}5+26r^{2})
\\  &-\frac{1}{r^{2}(r^{2}+1)^{5}}({-}73+90r^{2}+1265r^4+2934r^{6}+3982r^{8}+1664r^{10})\bigg]
\end{split}
\end{align}
which again gives a mild coordinate singularity $\langle \hat{g}_{rr}^{(2)}\rangle_{\mathfrak{g}} \sim 1/r^{2}$, and
\begin{align}\label{eqn:F10soln_final}
\begin{split}
    \mathscr{F}_{1,0}(r)|_{d=4} &= \frac{16}{225}|\mathcal{N}_{0,2}^{S}|^{2}\bigg[{-}\frac{16r}{(r^{2}+1)^{\frac{5}{2}}}(135+680r^{2}+576r^{4})
    \\  &+\frac{1}{r^{2}(r^{2}+1)^{5}}({-}2674-19959r^{2}-55879r^4-74699r^{6}-48319r^{8}-12160r^{10})\bigg].
\end{split}
\end{align}
One can check that these solutions satisfy the third equation in \eqref{eqn:Teqns}. This completes our solution for the backreaction. Any other dimension $d \neq 4$ can also be obtained.

\section{AdS-Rindler}
\label{sec:AdS-Rindler}

In this appendix, we collect some useful facts about AdS-Rindler that we need for computations throughout this work. 

\subsection{Coordinate transformation}
\label{sec:globaltoAdS-Rindler}

The change of coordinates from AdS-Rindler to global AdS is given by
\begin{align}\label{eqn:Rindlertoglobal}
\begin{split}
t &= \arctan\left(\frac{\sqrt{\rho^{2}-1}\sinh\tau}{\rho\cosh u\cosh\eta + \sqrt{\rho^{2}-1}\cosh\tau\sinh\eta}\right)
\\ r &= \sqrt{\rho^{2}\sinh^{2}u + \left(\sqrt{\rho^{2}-1}\cosh\tau\cosh\eta+\rho\cosh u \sinh\eta\right)^{2}}
\\ \theta &= \arccos\left(\frac{ \sqrt{\rho^{2}-1}\cosh\tau\cosh\eta+\rho\cosh u \sinh\eta,}{\sqrt{\rho^{2}\sinh^{2}u + \left(\sqrt{\rho^{2}-1}\cosh\tau\cosh\eta+\rho\cosh u \sinh\eta\right)^{2}}}\right),
\end{split}
\end{align}
and all angular coordinates on the common $S^{d-2}$ are the same. The angular size of the polar cap $\theta_{0}$ is related to the Rindler parameter $\eta$ by
\begin{equation}\label{eqn:etatotheta}
\cosh\eta = \frac{1}{\sin\theta_{0}}.
\end{equation}
%
~~~~~~~~~~~~~~~~~~~~~~~~~~~~~~~~~~~~~~~~~~~
\subsection{Eigenfunctions of hyperboloid Laplacian}
\label{sec:hyperbolicLaplacian}

The eigenfunctions of the scalar Laplacian on the hyperboloid $H^{d-1}$ fiber of AdS-Rindler were worked out in \cite{Colin-Ellerin:2024npf}. They satisfy
\begin{equation}\label{eqn:hyperbolicLapeqn}
\nabla_{H^{d-1}}^{2}H_{\lambda,\ell,\mathfrak{m}}(u,\Omega) = -\lambda H_{\lambda,\ell,\mathfrak{m}}(u,\Omega)
\end{equation}
where
\begin{equation}\label{eqn:hyperbolicballeigfn}
H_{\lambda,\ell,\mathfrak{m}}(u,\Omega) = h_{\lambda,\ell}(u)Y_{\ell,\mathfrak{m}}^{(d-2)}(\Omega),
\end{equation}
with $Y_{\ell,\mathfrak{m}}^{(d-2)}(\Omega)$ a $(d-2)$-dimensional spherical harmonic. The solutions of \eqref{eqn:hyperbolicLapeqn} are
\begin{equation}\label{eqn:hyperbolicballeigneqn_soln}
h_{\lambda,\ell}(u) = \mathcal{N}_{\lambda,\ell}^{H}\tanh^{\ell}\left(\frac{u}{2}\right)\sech^{2\zeta-2i\tilde{\lambda}}\left(\frac{u}{2}\right){}_{2}{F}_{1}\left(\ell+\zeta-i\tilde{\lambda},\frac{1}{2}-i\tilde{\lambda},\ell+\zeta+\frac{1}{2};\tanh^{2}\left(\frac{u}{2}\right)\right)
\end{equation}
with eigenvalues
\begin{equation}
\lambda = \tilde{\lambda}^{2}+\zeta^{2}, \qquad \tilde{\lambda} \in (0,\infty), \qquad \zeta \equiv \frac{d-2}{2}.\label{eqn:lambdatotildelambda}
\end{equation}
Delta-function normalisability fixes the normalisation constant
\begin{equation}\label{eq:hyperbolicevnorm1}
\mathcal{N}_{\lambda,\ell}^{H} = \frac{2^{-\zeta}}{\sqrt{V_{S^{d-2}}}}\frac{\Gamma(\ell+\zeta-i\tilde{\lambda})\Gamma\left(\frac{1}{2}-i\tilde{\lambda}\right)}{\Gamma(\ell+\zeta+\frac{1}{2})\Gamma(-2i\tilde{\lambda})}.
\end{equation}
%

\section{Calculation of Bogoliubov coefficients}
\label{sec:Bogcoeffs}

In this appendix, we compute the Bogoliubov coefficients for the graviton in AdS-Rindler.

\subsection{Only scalar overlaps}
\label{sec:scalaroverlaps}

We begin by proving that for the global scalar mode $h_{\mu\nu,0,2,\mathbf{0}}^{S}$, only the scalar mode for the AdS-Rindler graviton can have non-zero overlap with this mode, i.e., $\alpha_{0,2,\mathbf{0};\omega,\lambda}^{S,V_{H}}=\alpha_{0,2,\mathbf{0};\omega,\lambda}^{S,T_{H}}=0$ and $\beta_{0,2,\mathbf{0};\omega,\lambda}^{S,V_{H}}=\beta_{0,2,\mathbf{0};\omega,\lambda}^{S,T_{H}}=0$. First, consider the inner product with the Rindler vector mode:
\begin{align}\label{eqn:scalarvec_IP}
\begin{split}
    \langle h_{0,2,\mathbf{0}}^{S},h_{\omega,\lambda_{V}}^{V_{H}}\rangle &= i\int_{\Sigma_{t}} \sqrt{-G}\bigg({-}h_{\omega,\lambda_{V}}^{V_{H},t\nu}\nabla_{\nu}h_{0,2,\mathbf{0}}^{S\ast}+\frac{1}{2}h_{0,2,\mathbf{0}}^{S\mu\nu\ast}\nabla^{t}h_{\omega,\lambda_{V},\mu\nu}^{V_{H}}-\frac{1}{2}h_{\omega,\lambda,\mu\nu}^{V_{H}}\nabla^{t}h_{0,2,\mathbf{0}}^{S\mu\nu\ast}
    \\  &-h_{0,2,\mathbf{0},\mu\nu}^{S\ast}\nabla^{\mu}h_{\omega,\lambda_{V}}^{V_{H}\nu t}+h_{\omega,\lambda_{V},\mu\nu}^{V_{H}}\nabla^{\mu}h_{0,2,\mathbf{0}}^{S\nu t\ast}\bigg)
\end{split}
\end{align}
where we used the fact that $h_{\omega,\lambda,\mu\nu}^{V_{H}}$ is traceless. Using divergenceless of the vector harmonics, one finds that all terms only depend $V_{\lambda_{V},u}$ or $\partial_{u}V_{\lambda_{V},u}$ and the integrand has no $\phi_{\mathfrak{i}}$ dependence ($2 \leq i \leq d-1$) except from the measure. The argument now follows in a similar way to the Maxwell case given in \cite{Colin-Ellerin:2024npf}, which we repeat here. 

We can decompose the $V_{\boldsymbol{\lambda}_{V},\alpha}$ in terms of how it transforms under the isometries of $S^{d-2}$:
\begin{equation}\label{eqn:vectoreigfn_decomp}
V_{\boldsymbol{\lambda}_{V},\alpha} = V_{\boldsymbol{\lambda}_{V},\alpha}^{S}+V_{\boldsymbol{\lambda}_{V},\alpha}^{V}
\end{equation}
where\footnote{When $d=3$ there is no vector part and when $d=4$ we can distinguish the scalar and vector parts by their parity.}
\begin{align}\label{eqn:vectoreigenfn_parts}
\begin{split}
V_{\boldsymbol{\lambda}_{V},\alpha}^{S}dx^{\alpha} &= \sum_{\mathbf{k}_{S}}V_{\boldsymbol{\lambda}_{V},\mathbf{k}_{S},u}^{S}(u)\mathbb{S}_{\mathbf{k}_{S}}^{(d-2)}dx^{u}+V_{\boldsymbol{\lambda}_{V},\mathbf{k}_{S}}^{S}(u)D_{\phi_{\mathfrak{i}}}\mathbb{S}_{\mathbf{k}_{S}}^{(d-2)}dx^{\phi_{\mathfrak{i}}}
\\  V_{\boldsymbol{\lambda}_{V},\alpha}^{V}dx^{\alpha} &= \sum_{\mathbf{k}_{V}}V_{\boldsymbol{\lambda}_{V},\mathbf{k}_{V}}^{V}(u)\mathbb{V}_{\mathbf{k}_{V},\alpha}^{(d-2)}dx^{\alpha}.
\end{split}
\end{align}
Now observe that $h_{0,2,\mathbf{0},\mu\nu}^{S}$ is constant on $S^{d-2}$ so the integral in \eqref{eqn:scalarvec_IP} is only non-zero for $V_{\boldsymbol{\lambda}_{V},\mathbf{0},u}^{S}(u)$ by the orthogonality of spherical harmonics on $S^{d-2}$. Finally, we consider the divergenceless property of $h_{\mu\nu}^{V_{H}}$, which can be applied separately to $V_{\boldsymbol{\lambda}_{V},\alpha}^{S}$ and to $V_{\boldsymbol{\lambda}_{V},\alpha}^{V}$ since the divergenceless condition is invariant under rotation of $S^{d-2}$ (and invariant under parity which is needed for $d=4$). Therefore, the divergenceless condition gives
\begin{align}\label{eqn:divlesscondition_vec}
\begin{split}
0 = \nabla_{\alpha}V_{\boldsymbol{\lambda}_{V},\mathbf{0}}^{S\alpha} = \frac{1}{\sqrt{g_{H^{d-1}}}}\partial_{\alpha}\left(\sqrt{g_{H^{d-1}}}V_{\boldsymbol{\lambda}_{V},\mathbf{0}}^{S\alpha}\right) &= \frac{1}{\sinh^{d-2}u}\partial_{u}\left(\sinh^{d-2}uV_{\boldsymbol{\lambda}_{V},\mathbf{0}}^{Su}\right) 
\\  \implies V_{\boldsymbol{\lambda}_{V},\mathbf{0}}^{Su} &= A\csch^{d-2}u,
\end{split}
\end{align}
which is not regular at the origin so we conclude that $A=0$, and hence the inner product \eqref{eqn:scalarvec_IP} vanishes.

Next, consider the inner product with the tensor mode on $H^{d-1}$ (recall that the tensor mode is only non-trivial for $d\geq 4$):
\begin{align}\label{eqn:scalartens_IP}
\begin{split}
    \langle h_{0,2,\mathbf{0}}^{S},h_{\omega,\lambda_{T}}^{T_{H}}\rangle &= i\int_{\Sigma_{t}} \sqrt{-G}\bigg({-}h_{\omega,\lambda_{T}}^{T_{H},t\nu}\nabla_{\nu}h_{0,2,\mathbf{0}}^{S\ast}+\frac{1}{2}h_{0,2,\mathbf{0}}^{S\nu\alpha\ast}\nabla^{t}h_{\omega,\lambda_{T},\nu\alpha}^{T_{H}}-\frac{1}{2}h_{\omega,\lambda_{T},\nu\alpha}^{T_{H}}\nabla^{t}h_{0,2,\mathbf{0}}^{S\nu\alpha\ast}
    \\  &-h_{0,2,\mathbf{0},\nu\alpha}^{S\ast}\nabla^{\alpha}h_{\omega,\lambda_{T}}^{T_{H}\nu t}+h_{\omega,\lambda_{T},\nu\alpha}^{T_{H}}\nabla^{\alpha}h_{0,2,\mathbf{0}}^{S\nu t\ast}\bigg)
\end{split}
\end{align}
where we used the fact that $h_{\omega,\lambda,\mu\nu}^{T_{H}}$ is traceless. We also find from tracelessness that all terms only depend $T_{\lambda_{T},uu}$ and the integrand has no $\phi_{i}$ dependence ($2 \leq i \leq d-1$) except from the measure. Now we decompose $T_{\lambda_{T},\alpha\beta}^{H}$ in terms of representations on $S^{d-2}$
\begin{equation}\label{eqn:tensoreigfn_decomp}
T_{\boldsymbol{\lambda}_{T},\alpha\beta} = T_{\boldsymbol{\lambda}_{T},\alpha\beta}^{S}+T_{\boldsymbol{\lambda}_{T},\alpha\beta}^{V}+T_{\boldsymbol{\lambda}_{T},\alpha\beta}^{T}
\end{equation}
where the different parts are given as in \eqref{eqn:gravscalar}, \eqref{eqn:gravvector}, and \eqref{eqn:gravtensor}, but on $S^{d-2}$. We only care about the scalar part because this is the only one with non-zero $uu$ component. Since $h_{0,2,\mathbf{0},\mu\nu}^{S}$ is constant on $S^{d-2}$, the integral in \eqref{eqn:scalartens_IP} is only non-zero for $T_{\boldsymbol{\lambda}_{T},\mathbf{0},uu}^{S}(u)$ by the orthogonality of spherical harmonics on $S^{d-2}$. Observe that the tracelessness and the divergenceless properties of $T_{\boldsymbol{\lambda}_{T},\alpha\beta}$ can be applied separately to $T_{\boldsymbol{\lambda}_{T},\alpha\beta}^{S}$, $T_{\boldsymbol{\lambda}_{T},\alpha\beta}^{S}$, and $T_{\boldsymbol{\lambda}_{T},\alpha\beta}^{T}$ since the these conditions are invariant under rotation of $S^{d-2}$. Thus, we find
\begin{equation}\label{eqn:divless+tracecondition_tens}
T_{\boldsymbol{\lambda}_{T},\mathbf{0},\alpha}^{S\alpha}=0, \qquad \nabla_{\alpha}T_{\boldsymbol{\lambda}_{T},\mathbf{0}}^{S\alpha\beta}=0.
\end{equation}
The first condition gives 
\begin{equation}\label{eqn:tracecondition_tens}
T_{\boldsymbol{\lambda}_{T},\mathbf{0},u}^{u}+\frac{(d-2)}{\sinh^{2}u}T_{\boldsymbol{\lambda}_{T},\mathbf{0}}^{S,\mathrm{tr}}=0,
\end{equation}
and the second condition for $\mathfrak{j}=u$ gives
\begin{equation}\label{eqn:divlesscondition_tens}
0 = \nabla_{\alpha}T_{\boldsymbol{\lambda}_{T},\mathbf{0}}^{S\alpha u} = \partial_{u}T_{\boldsymbol{\lambda}_{T},\mathbf{0}}^{uu}+(d-2)\coth u \left(T_{\boldsymbol{\lambda}_{T},\mathbf{0}}^{uu}-\frac{1}{\sinh^{2}u}T_{\boldsymbol{\lambda}_{T},\mathbf{0}}^{S,\mathrm{tr}}\right).
\end{equation}
Plugging in \eqref{eqn:tracecondition_tens}, we conclude
\begin{equation}
    0 = \frac{1}{\sinh^{d-3}u}\partial_{u}\left(\sinh^{d-3}uT_{\boldsymbol{\lambda}_{T},\mathbf{0}}^{uu}\right) \implies T_{\boldsymbol{\lambda}_{T},\mathbf{0}}^{uu} = B\csch^{d-3}u,
\end{equation}
which is singular at $u=0$ so we must have $B=0$. Therefore, only the scalar part of the Rindler graviton can appear in the expansion of $h_{0,2,\mathbf{0}}^{S}$ in Rindler modes.

\subsection{Gauge-invariant computation}
\label{sec:Bogcoeffs_calc}

We now proceed to compute the Bogoliubov coefficients $\alpha_{0,2,\mathbf{0};\omega,\lambda}^{S,S_{H}}$ and $\beta_{0,2,\mathbf{0};\omega,\lambda}^{S,S_{H}}$. The standard method of computing the Bogoliubov coefficients from the inner product between global and Rindler modes is too diffciult due to the Rindler wavefunctions being very complicated so instead we use a trick originally developed for scalars in AdS$_{3}$-Rindler \cite{Belin:2018juv} and then extended to scalars and photons in higher dimensions \cite{Colin-Ellerin:2024npf}.

The idea is to compute a two-point function of the global creation operator and some local operator defined in terms of the graviton, such as  $\bra{0}h_{\mu\nu}(x)a_{0,2,\mathbf{0}}^{S\dagger}\ket{0}$, and then send the position of the local operator to the asymptotic boundary where it simplifies. Since we solved for the graviton $h_{\mu\nu}$ in global AdS and AdS-Rindler using two different gauges, it is difficult to compare the two-point function $\bra{0}h_{\mu\nu}(x)a_{0,2,\mathbf{0}}^{S\dagger}\ket{0}$ in the two quantizations. Instead, we can replace $h_{\mu\nu}$ with a gauge-invariant quantity so that we do not need to make any gauge transformations in comparing the two-point function between the two quantizations. 

The simplest gauge-invariant quantity is the Weyl tensor:
\begin{align}\label{eqn:Weyltensor}
\begin{split}
    W_{\mu\nu\alpha\beta} &= R_{\mu\nu\alpha\beta} - \frac{2}{(d-1)}\left(g_{\mu[\alpha}R_{\beta]\nu} - g_{\nu[\alpha}R_{\beta]\mu}\right) + \frac{2}{d(d-1)}Rg_{\mu[\alpha}g_{\beta]\nu} 
    \\  &= R_{\mu\nu\alpha\beta} + 2g_{\mu[\alpha}g_{\beta]\nu}
\end{split}
\end{align}
where on the second line we used the (non-perturbative) Einstein equation. Gauge-invariance follows from the fact that the background Weyl tensor vanishes in AdS: $W_{\mu\nu\alpha\beta}^{(0)} = 0$. To see this, contract the Weyl tensor with any other tensor $A_{\mu\nu\alpha\beta}$ that has the same symmetries/antisymmetries and whose background value does not vanish: $A_{\mu\nu\alpha\beta}^{(0)} \neq 0$. For example, the Riemann tensor would be such a tensor. Now, since $W_{\mu\nu\alpha\beta}A^{\mu\nu\alpha\beta}$ is diffeomorphism invariant by construction, it must be invariant under linearized diffeomorphisms. Expanding to first-order in $\kappa$, we conclude that $W_{\mu\nu\alpha\beta}^{(0)}A^{(1)\mu\nu\alpha\beta}+W_{\mu\nu\alpha\beta}^{(1)}A^{(0)\mu\nu\alpha\beta} = W_{\mu\nu\alpha\beta}^{(1)}A^{(0)\mu\nu\alpha\beta}$ must be diffeomorphism-invariant. But, only $W_{\mu\nu\alpha\beta}^{(1)}$ transforms under linearized diffeomorphisms so it must be gauge-invariant.

From \eqref{eqn:Riemanntensorexp} and \eqref{eqn:Christoffeldiff}, the linearized Weyl tensor can be written explicitly as
\begin{equation}\label{eqn:Weyllin}
    W_{\mu\nu\alpha\beta}^{(1)} = 2\left(\nabla_{\ast[\mu}\nabla_{[\beta}h_{\alpha]\nu]\ast}+G_{\ast[\mu[\alpha}h_{\beta]\nu]\ast}\right)
\end{equation}
where $\ast[\cdot]\ast$ denotes the second symmetrization. One can check that indeed \eqref{eqn:Weyllin} is invariant under linearized diffeomorphisms.

Therefore, we will consider the collection of two-point functions
\begin{equation}\label{eqn:Bog2ptfn_grav}
    \mathfrak{F}_{\mu\nu\alpha\beta}^{\mathrm{grav}}(t,\Omega) = \lim_{r \to \infty}r^{d}\bra{0}W_{\mu\nu\alpha\beta}^{(1)S}(t,r,\Omega)a_{0,2,\mathbf{0}}^{S\dagger}\ket{0}.
\end{equation}
We will only need the two-point functions $\mathfrak{F}_{trtr}^{\mathrm{grav}}$, $\mathfrak{F}_{r\theta r\theta}^{\mathrm{grav}}$, and $\mathfrak{F}_{tr\theta r}^{\mathrm{grav}}$. The desired components of the Weyl tensor from the global quantization of the graviton are given by
\begin{align}\label{eqn:Bog2ptfn_grav_global}
\begin{split}
    \mathfrak{F}^{\mathrm{grav},trtr}(t,\Omega) &= -d(d-2)\mathcal{N}_{0,2}^{S}e^{-i\Omega_{0,2}^{S}t}Y_{2,\mathbf{0}}(\Omega)
    \\  \mathfrak{F}^{\mathrm{grav},r\theta r\theta}(t,\Omega) &= \frac{d(d-2)}{2}\mathcal{N}_{0,2}^{S}e^{-i\Omega_{0,2}^{S}t}\left(2Y_{2,\mathbf{0}}(\Omega)+\partial_{\theta}^{2}Y_{2,\mathbf{0}}(\Omega)\right)
    \\  \mathfrak{F}^{\mathrm{grav},t r\theta r}(t,\Omega) &= -i\frac{d(d-2)}{2}\mathcal{N}_{0,2}^{S}e^{-i\Omega_{0,2}^{S}t}\partial_{\theta}Y_{2,\mathbf{0}}(\Omega).    
\end{split}
\end{align}
To obtain these two-point functions in Rindler coordinates, we take $r \to \infty$ by taking $\rho \to \infty$ so the global coordinates appearing in \eqref{eqn:Bog2ptfn_grav_global} as a function of Rindler coordinates are the $\rho \to \infty$ limit of \eqref{eqn:Rindlertoglobal}, viz.,
\begin{align}\label{eqn:tvarphi_bdylimit}
\begin{split}
\lim_{\rho \to \infty}t(\tau,\rho,\Xi) &= \arctan\left(\frac{\sinh\tau}{\cosh\tau\sinh\eta+\cosh u\cosh\eta}\right) \equiv t(\tau,\Xi)
\\ \lim_{\rho \to \infty}\theta(\tau,\rho,\Xi) &= \arccos\left(\frac{\cosh\tau\cosh\eta+\cosh u \sinh\eta,}{\sqrt{\sinh^{2}u + \left(\cosh\tau\cosh\eta+\cosh u \sinh\eta\right)^{2}}}\right) \equiv \theta(\tau,\Xi).
\end{split}
\end{align}

Next, we compute the desired two-point functions \eqref{eqn:Bog2ptfn_grav} using the AdS-Rindler quantization of the graviton. The AdS-Rindler wavefunctions for the graviton are very complicated, but our method has the major advantage that we only need their boundary values. The result is still quite unwieldy so let us make some definitions:
\begin{align}\label{eqn:fdefs}
\begin{split}
    f_{1}(\eta,\tau,u) &\equiv \cosh\eta\sinh u\sinh\tau
    \\ f_{2}(\eta,\tau,u) &\equiv \cosh\eta\cosh u\cosh \tau+\sinh\eta
    \\ f_{3}(\eta,\tau,u) &\equiv \cosh u\sinh\eta+\cosh\tau\cosh\eta
    \\ f_{4}(\eta,\tau,u) &\equiv \cosh\tau\sinh\eta+\cosh u\cosh\eta
    \\ f_{5}(\eta,\tau,u) &\equiv \cosh\eta\cosh u\cosh \tau+\sinh\eta    
\end{split}
\end{align}
and
\begin{align}\label{eqn:Wwavefn}
\begin{split}
    &\mathscr{R}_{\omega,\lambda}^{(trtr)}(\tau,u) \equiv \mathcal{N}_{\omega,\lambda}^{S,R}\Bigg[\frac{\lambda}{2}\bigg((\omega^{2}-(d-2)-\lambda)f_{1}(\eta,\tau,u)^{2}-(d-2)(d-1+\lambda)f_{2}(\eta,\tau,u)^{2}\bigg)H_{\lambda}
    \\  &+i\omega(d-2)(d-1+\lambda)f_{1}(\eta,\tau,u)f_{2}(\eta,\tau,u)D_{u}H_{\lambda}-\frac{1}{2}\left(\lambda-(d-1)\omega^{2}\right)f_{1}(\eta,\tau,u)^{2}D_{u}^{2}H_{\lambda}\Bigg]
    \\  &\mathscr{R}_{\omega,\lambda}^{(r\theta r\theta)}(\tau,u) \equiv \mathcal{N}_{\omega,\lambda}^{S,R}\Bigg[\frac{\lambda}{2}\bigg((\omega^{2}-(d-2)-\lambda)f_{2}(\eta,\tau,u)^{2}-(d-2)(d-1+\lambda)f_{1}(\eta,\tau,u)^{2}\bigg)H_{\lambda}
    \\  &+i\omega(d-2)(d-1+\lambda)f_{1}(\eta,\tau,u)f_{2}(\eta,\tau,u) D_{u}H_{\lambda}-\frac{1}{2}\left(\lambda-(d-1)\omega^{2}\right)f_{2}(\eta,\tau,u)^{2} D_{u}^{2}H_{\lambda}\Bigg]
    \\  &\mathscr{R}_{\omega,\lambda}^{(t r\theta r)}(\tau,u) \equiv \mathcal{N}_{\omega,\lambda}^{S,R}\Bigg[{-}\frac{\lambda}{2}\left(\omega^{2}-d(d-2+\lambda)+\lambda\right)f_{1}(\eta,\tau,u)f_{2}(\eta,\tau,u)H_{\lambda}
    \\  &-i\frac{\omega}{2}(d-2)(d-1+\lambda)\left(f_{1}(\eta,\tau,u)^{2}+f_{2}(\eta,\tau,u)^{2}\right) D_{u}H_{\lambda}
    \\  &+\frac{1}{2}\left(\lambda-(d-1)\omega^{2}\right)f_{1}(\eta,\tau,u)f_{2}(\eta,\tau,u) D_{u}^{2}H_{\lambda}\Bigg]
\end{split}
\end{align}
Then, using 
\begin{equation}\label{eqn:globaltoRindler_bdylimit}
\lim_{\rho \to \infty}r(\tau,\rho,\Xi)^{d} = \lim_{\rho \to \infty}\rho^{d}\left(\sinh^2 u + f_{3}(\eta,\tau,u)^2\right)^{\frac{d}{2}}
\end{equation}
and the boundary limit of the AdS-Rindler graviton wavefunctions in App.~\ref{sec:AdSRindlerEOM}, the desired two-point functions in Rindler coordinates are given by
\begin{align}\label{eqn:Bog2ptfn_Rindler_grav}
\begin{split}
    \mathfrak{F}^{\mathrm{grav},trtr}(t,\Omega) &= \frac{\left(\sinh^2 u + f_{3}(\eta,\tau,u)^2\right)^{\frac{d}{2}+1}}{(\sinh^{2}\tau+f_{4}(\eta,\tau,u)^{2})^{2}}
    \\	&\qquad \times \int\frac{d\omega}{2\pi}\sum_{\lambda}\left(e^{-i\omega\tau}\mathscr{R}_{\omega,\lambda}^{(trtr)}(\tau,u)\alpha_{\omega,\lambda}^{S,S_{H}\ast}-e^{i\omega\tau}\mathscr{R}_{\omega,\lambda}^{(trtr)\ast}(\tau,u)\beta_{\omega,\lambda}^{S,S_{H}}\right)
    \\  \mathfrak{F}^{\mathrm{grav},r\theta r\theta}(t,\Omega) &= \left(\sinh^2 u + f_{3}(\eta,\tau,u)^2\right)^{\frac{d}{2}-1}
    \\	&\qquad \times \int\frac{d\omega}{2\pi}\sum_{\lambda}\left(e^{-i\omega\tau}\mathscr{R}_{\omega,\lambda}^{(r\theta r\theta)}(\tau,u)\alpha_{\omega,\lambda}^{S,S_{H}\ast}-e^{i\omega\tau}\mathscr{R}_{\omega,\lambda}^{(r\theta r\theta)\ast}(\tau,u)\beta_{\omega,\lambda}^{S,S_{H}}\right)
    \\  \mathfrak{F}^{\mathrm{grav},t r\theta r}(t,\Omega) &= \frac{\left(\sinh^2 u + f_{3}(\eta,\tau,u)^2\right)^{\frac{d}{2}}}{(\sinh^{2}\tau+f_{4}(\eta,\tau,u)^{2})}
    \\	&\qquad \times \int\frac{d\omega}{2\pi}\sum_{\lambda}\left(e^{-i\omega\tau}\mathscr{R}_{\omega,\lambda}^{(t r\theta r)}(\tau,u)\alpha_{\omega,\lambda}^{S,S_{H}\ast}-e^{i\omega\tau}\mathscr{R}_{\omega,\lambda}^{(t r\theta r)\ast}(\tau,u)\beta_{\omega,\lambda}^{S,S_{H}}\right).
\end{split}
\end{align}
Equating these with the results in global coordinates \eqref{eqn:Bog2ptfn_grav_global} gives
\begin{align}\label{eqn:Bog2ptfn_matching}
\begin{split}
    -d(d-2)\mathcal{N}_{0,2}^{S}e^{-i\Omega_{0,2}^{S}t}&Y_{2,\mathbf{0}}(\Omega) = \frac{\left(\sinh^2 u + f_{3}(\eta,\tau,u)^2\right)^{\frac{d}{2}+1}}{(\sinh^{2}\tau+f_{4}(\eta,\tau,u)^{2})^{2}}
    \\	&\times \int\frac{d\omega}{2\pi}\sum_{\lambda}\left(e^{-i\omega\tau}\mathscr{R}_{\omega,\lambda}^{(trtr)}(\tau,u)\alpha_{\omega,\lambda}^{S,S_{H}\ast}-e^{i\omega\tau}\mathscr{R}_{\omega,\lambda}^{(trtr)\ast}(\tau,u)\beta_{\omega,\lambda}^{S,S_{H}}\right)
    \\ \frac{d(d-2)}{2}\mathcal{N}_{0,2}^{S}e^{-i\Omega_{0,2}^{S}t}&\left(2Y_{2,\mathbf{0}}(\Omega)+\partial_{\theta}^{2}Y_{2,\mathbf{0}}(\Omega)\right) = \left(\sinh^2 u + f_{3}(\eta,\tau,u)^2\right)^{\frac{d}{2}-1}
    \\	&\times \int\frac{d\omega}{2\pi}\sum_{\lambda}\left(e^{-i\omega\tau}\mathscr{R}_{\omega,\lambda}^{(r\theta r\theta)}(\tau,u)\alpha_{\omega,\lambda}^{S,S_{H}\ast}-e^{i\omega\tau}\mathscr{R}_{\omega,\lambda}^{(r\theta r\theta)\ast}(\tau,u)\beta_{\omega,\lambda}^{S,S_{H}}\right)
    \\  -i\frac{d(d-2)}{2}\mathcal{N}_{0,2}^{S}e^{-i\Omega_{0,2}^{S}t}&\partial_{\theta}Y_{2,\mathbf{0}}(\Omega) = \frac{\left(\sinh^2 u + f_{3}(\eta,\tau,u)^2\right)^{\frac{d}{2}}}{(\sinh^{2}\tau+f_{4}(\eta,\tau,u)^{2})}
    \\	&\times \int\frac{d\omega}{2\pi}\sum_{\lambda}\left(e^{-i\omega\tau}\mathscr{R}_{\omega,\lambda}^{(t r\theta r)}(\tau,u)\alpha_{\omega,\lambda}^{S,S_{H}\ast}-e^{i\omega\tau}\mathscr{R}_{\omega,\lambda}^{(t r\theta r)\ast}(\tau,u)\beta_{\omega,\lambda}^{S,S_{H}}\right).
\end{split}
\end{align}
Hence we define
\begin{align}\label{eqn:Bdef_grav}
\begin{split}
    \mathcal{B}^{\mathrm{grav},(trtr)}(\tau,\Xi) &= -d(d-2)\mathcal{N}_{0,2}^{S}e^{-i\Omega_{0,2}^{S}t}Y_{2,\mathbf{0}}(\Omega)\frac{(\sinh^{2}\tau+f_{4}(\eta,\tau,u)^{2})^{2}}{\left(\sinh^2 u + f_{3}(\eta,\tau,u)^2\right)^{\frac{d}{2}+1}}
    \\ \mathcal{B}^{\mathrm{grav},(r\theta r\theta)}(\tau,\Xi) &= \frac{d(d-2)}{2}\mathcal{N}_{0,2}^{S}e^{-i\Omega_{0,2}^{S}t}\frac{\left(2Y_{2,\mathbf{0}}(\Omega)+\partial_{\theta}^{2}Y_{2,\mathbf{0}}(\Omega)\right)}{\left(\sinh^2 u + f_{3}(\eta,\tau,u)^2\right)^{\frac{d}{2}-1}}
    \\ \mathcal{B}^{\mathrm{grav},(t r\theta r)}(\tau,\Xi) &= -i\frac{d(d-2)}{2}\mathcal{N}_{0,2}^{S}e^{-i\Omega_{0,2}^{S}t}\partial_{\theta}Y_{2,\mathbf{0}}(\Omega)\frac{(\sinh^{2}\tau+f_{4}(\eta,\tau,u)^{2})}{\left(\sinh^2 u + f_{3}(\eta,\tau,u)^2\right)^{\frac{d}{2}}}
\end{split}
\end{align}
so that \eqref{eqn:Bog2ptfn_matching} becomes
\begin{align}\label{eqn:B}
\begin{split}
    \mathcal{B}^{\mathrm{grav},(trtr)}(\tau,\Xi) &= \int\frac{d\omega}{2\pi}\sum_{\lambda}\left(e^{-i\omega\tau}\mathscr{R}_{\omega,\lambda}^{(trtr)}(\tau,u)\alpha_{\omega,\lambda}^{S,S_{H}\ast}-e^{i\omega\tau}\mathscr{R}_{\omega,\lambda}^{(trtr)\ast}(\tau,u)\beta_{\omega,\lambda}^{S,S_{H}}\right)
    \\ \mathcal{B}^{\mathrm{grav},(r\theta r\theta)}(\tau,\Xi) &= \int\frac{d\omega}{2\pi}\sum_{\lambda}\left(e^{-i\omega\tau}\mathscr{R}_{\omega,\lambda}^{(r\theta r\theta)}(\tau,u)\alpha_{\omega,\lambda}^{S,S_{H}\ast}-e^{i\omega\tau}\mathscr{R}_{\omega,\lambda}^{(r\theta r\theta)\ast}(\tau,u)\beta_{\omega,\lambda}^{S,S_{H}}\right)
    \\ \mathcal{B}^{\mathrm{grav},(t r\theta r)}(\tau,\Xi) &= \int\frac{d\omega}{2\pi}\sum_{\lambda}\left(e^{-i\omega\tau}\mathscr{R}_{\omega,\lambda}^{(t r\theta r)}(\tau,u)\alpha_{\omega,\lambda}^{S,S_{H}\ast}-e^{i\omega\tau}\mathscr{R}_{\omega,\lambda}^{(t r\theta r)\ast}(\tau,u)\beta_{\omega,\lambda}^{S,S_{H}}\right).
\end{split}
\end{align}

Now the goal is to obtain a certain linear combination of the $\mathcal{B}^{\mathrm{grav},(\mu\nu\alpha\beta)}$ such that the resulting linear combination of $\mathscr{R}_{\omega,\lambda}^{(\mu\nu\alpha\beta)}$ only contains $H_{\lambda}$ and not its derivatives. We find
\begin{align}\label{eqn:Bcombo}
\begin{split}
    \mathcal{B}^{\mathrm{grav}}(\tau,\Xi) &\equiv \frac{2}{(f_{2}(\eta,\tau,u)^{2}-f_{1}(\eta,\tau,u)^{2})^{2}}\Big[f_{2}(\eta,\tau,u)^{2}\mathcal{B}^{\mathrm{grav},(trtr)}(\tau,\Xi) + f_{1}(\eta,\tau,u)^{2}\mathcal{B}^{\mathrm{grav},(r\theta r\theta)}(\tau,\Xi) 
    \\  &\qquad + 2f_{1}(\eta,\tau,u)f_{2}(\eta,\tau,u)\mathcal{B}^{\mathrm{grav},(t r\theta r)}(\tau,\Xi)\Big]
    \\  &= -\int\frac{d\omega}{2\pi}\sum_{\lambda}(d-2)\lambda(d-1+\lambda)\left(\mathcal{N}_{\omega,\lambda}^{S,R}e^{-i\omega\tau}H_{\lambda}\alpha_{\omega,\lambda}^{S,S_{H}\ast}-\mathcal{N}_{\omega,\lambda}^{S,R\ast}e^{i\omega\tau}H_{\lambda}\beta_{\omega,\lambda}^{S,S_{H}}\right).
\end{split}
\end{align}
Let us make one more definition
\begin{equation}\label{eqn:cdef}
    c_{\omega,\lambda} = -(d-2)\lambda(\lambda+d-1).
\end{equation}
We can now use orthogonality of the eigenfunctions of the Laplacian on $H^{d-1}$ to obtain the Bogoliubov coefficients
\begin{align}\label{eqn:Bogcoeffs_grav}
\begin{split}
    \beta_{\omega,\lambda}^{S,S_{H}} &= -\frac{1}{\mathcal{N}_{\omega,\lambda}^{S,R\ast}}\frac{1}{c_{\omega,\lambda}}\int_{-\infty}^{\infty} d\tau\,e^{-i\omega\tau}\int_{H^{d-1}}d^{d-1}x\,\sqrt{g_{H^{d-1}}}H_{\lambda}(\Xi)\mathcal{B}^{\mathrm{grav}}(\tau,\Xi)
    \\ \alpha_{\omega,\lambda}^{S,S_{H}} &= \frac{1}{\mathcal{N}_{\omega,\lambda}^{S,R\ast}}\frac{1}{c_{\omega,\lambda}}\int_{-\infty}^{\infty} d\tau\,e^{-i\omega\tau}\int_{H^{d-1}}d^{d-1}x\,\sqrt{g_{H^{d-1}}}H_{\lambda}(\Xi)\mathcal{B}^{\mathrm{grav}\ast}(\tau,\Xi).
\end{split}
\end{align}
Plugging in \eqref{eqn:Bdef_grav} and performing the integral over $S^{d-2}$, we obtain
\begin{align}\label{eqn:Bogcoeff_betasimpl_grav}
\begin{split}
&\beta_{\omega,\lambda}^{S,S_{H}} = -\frac{\mathcal{N}_{0,1}^{S}\mathcal{N}_{\lambda,0}^{H}V_{S^{d-2}}}{c_{\omega,\lambda}\mathcal{N}_{\omega,\lambda}^{S,R\ast}}\sqrt{\frac{2(d+2)}{(d-1)}}d(d-2)\int_{-\infty}^{\infty} d\tau\,e^{-i\omega\tau}\int_{0}^{\infty}du\,\sinh^{d-2}u\sech^{2\zeta-2i\tilde{\lambda}}\left(\frac{u}{2}\right)
\\  &{}_{2}{F}_{1}\left(\zeta-i\tilde{\lambda},\frac{1}{2}-i\tilde{\lambda},\zeta+\frac{1}{2};\tanh^{2}\left(\frac{u}{2}\right)\right) \frac{e^{-i\Omega_{0,2}^{S}t(\tau,\alpha)}}{\left(\sinh^2 u + f_{3}(\eta,\tau,u)^2\right)^{\frac{d}{2}}(f_{2}(\eta,\tau,u)^{2}-f_{1}(\eta,\tau,u)^{2})^{2}} 
\\	&\times \Bigg[\frac{\left(\sinh^{2}u-(d-1)\,f_{3}(\eta,\tau,u)^{2}\right)f_{5}(\eta,\tau,u)^{2}\left(\sinh^{2}\tau+f_{4}(\eta,\tau,u)^{2}\right)^{2}}{\left(\sinh^{2}u + f_{3}(\eta,\tau,u)^{2}\right)^{2}}
\\	&+f_{1}(\eta,\tau,u)^{2}\left((d-1)\sinh^{2}u - f_{3}(\eta,\tau,u)^{2}\right)
\\  &+2di\sinh u f_{1}(\eta,\tau,u)f_{5}(\eta,\tau,u)\left(\sinh^{2}\tau+f_{4}(\eta,\tau,u)\right) \frac{f_{3}(\eta,\tau,u)}{\left(\sinh^{2}u + f_{3}(\eta,\tau,u)^{2}\right)}\Bigg]
\\	&\sim \frac{\mathcal{N}_{0,1}^{S}\mathcal{N}_{\lambda,0}^{H}V_{S^{d-2}}}{c_{\omega,\lambda}\mathcal{N}_{\omega,\lambda}^{S,R\ast}}d(d-2)2^{d+1}\sqrt{\frac{(d+2)}{2(d-1)}}e^{-d\eta}\int_{-\infty}^{\infty} d\tau\,e^{-i\omega\tau}\int_{0}^{\infty}du\,\sinh^{d-2}u\sech^{2\zeta-2i\tilde{\lambda}}\left(\frac{u}{2}\right)
\\  &\qquad \times {}_{2}{F}_{1}\left(\zeta-i\tilde{\lambda},\frac{1}{2}-i\tilde{\lambda},\zeta+\frac{1}{2};\tanh^{2}\left(\frac{u}{2}\right)\right)\frac{\left(d\left(\cosh u\cosh\tau+1\right)^{2}-(\cosh u+\cosh\tau)^{2}\right)}{\left(\cosh u+\cosh\tau\right)^{d+2}},
\end{split}
\end{align}
where we have expanded at large $\eta$ to obtain $\sim$ whose precise meaning is explained below \eqref{eqn:Bogcoeff_betasimpl_grav_final2}, and we have dropped higher powers of $e^{-\eta}$. These integrals can be computed using very similar techniques to those in App.~E of \cite{Colin-Ellerin:2024npf} and we find
\begin{equation}\label{eqn:Bogcoeff_betasimpl_grav_final}
\beta_{\omega,\tilde{\lambda}}^{S,S_{H}} \sim -\frac{\mathcal{N}_{0,1}^{S}\mathcal{N}_{\lambda,0}^{H}}{\mathcal{N}_{\omega,\lambda}^{S,R\ast}}\sqrt{\frac{(d+2)}{2(d-1)}}e^{-d\eta}\frac{2^{2d-2}d^{2}\pi^{\frac{d}{2}-1}}{\Gamma(d+2)\Gamma(\frac{d-2}{2})}\left|\Gamma\left(\frac{\zeta+i(\omega+\tilde{\lambda})}{2}\right)\right|^{2}\left|\Gamma\left(\frac{\zeta+i(\omega-\tilde{\lambda})}{2}\right)\right|^{2}.
\end{equation}
Similarly, we find $\alpha_{\omega,\tilde{\lambda}}^{S,S_{H}} \sim -\beta_{\omega,\tilde{\lambda}}^{S,S_{H}}$.

\bibliography{GeneralizedEntropyGravitons-refs} 
\bibliographystyle{JHEP}


\end{document}